\documentclass[acmtog,nonacm]{acmart}
\usepackage{graphicx}
\usepackage{overpic}
\usepackage{caption}
\usepackage{subcaption}
\usepackage{float}
\usepackage{xcolor}

\acmJournal{TOG}

\begin{document}

\title[Twisted Edges]{Twisted Edges: A Unified Framework for Designing Linked Knot (LK) Structures Using Labeled Non-Manifold Surface Meshes}

\author{Tolga Talha Yıldız}
\affiliation{%
  \institution{Texas A\&M University}
  \city{College Station}
  \state{Texas}
  \country{USA}
}
\authornote{Corresponding Author}
\email{tolgayildiz@tamu.edu}

\author{Uğur Önal}
\affiliation{%
  \institution{Texas A\&M University}
  \city{College Station}
  \state{Texas}
  \country{USA}
}

\author{Vinayak R. Krishnamurthy}
\affiliation{%
  \institution{Texas A\&M University}
  \city{College Station}
  \state{Texas}
  \country{USA}
}

\author{Ergun Akleman}
\affiliation{%
  \institution{Texas A\&M University}
  \city{College Station}
  \state{Texas}
  \country{USA}
}

\makeatletter
\let\@authorsaddresses\@empty
\makeatother
\renewcommand{\shortauthors}{Yildiz, et al.}

\begin{abstract}
We present Twisted Edges, a unified framework for designing Linked Knot (LK) structures using labeled non-manifold surface meshes. While the concept of edge twists, originating in topological graph theory, is foundational to these designs, prior approaches have been strictly limited to binary states. We identify this restriction as a critical barrier; binary twisting fails to capture the full spectrum of topological possibilities, rendering a vast class of structural and dynamic behaviors inaccessible. 

To overcome this limitation, we generalize the twist formulation to support arbitrary integer twist labels. This expansion reveals that while zero twists may introduce disconnections, applying even twists to 2-manifold meshes robustly preserves connectivity, transforming surfaces into fully connected, chainmail-like structures where faces form consistently linked cycles. Furthermore, we extend this framework to non-manifold meshes, where specific integer assignments prevent cycle merging. This capability, unattainable with binary methods, enables the design of partial connectivity and functional hinges, supporting dynamic folding and articulation. Theoretically, we show that these integer-twisted meshes correspond to knotted surfaces in four dimensions, with LK structures arising as their immersions into $\mathbb{R}^3$. By breaking the binary constraint, this work establishes a coherent paradigm for the systematic exploration of previously unstudied woven and articulated structures.
\end{abstract}


\begin{teaserfigure}
    \centering
    \begin{subfigure}[t]{0.16\textwidth}
        \includegraphics[width=\textwidth]{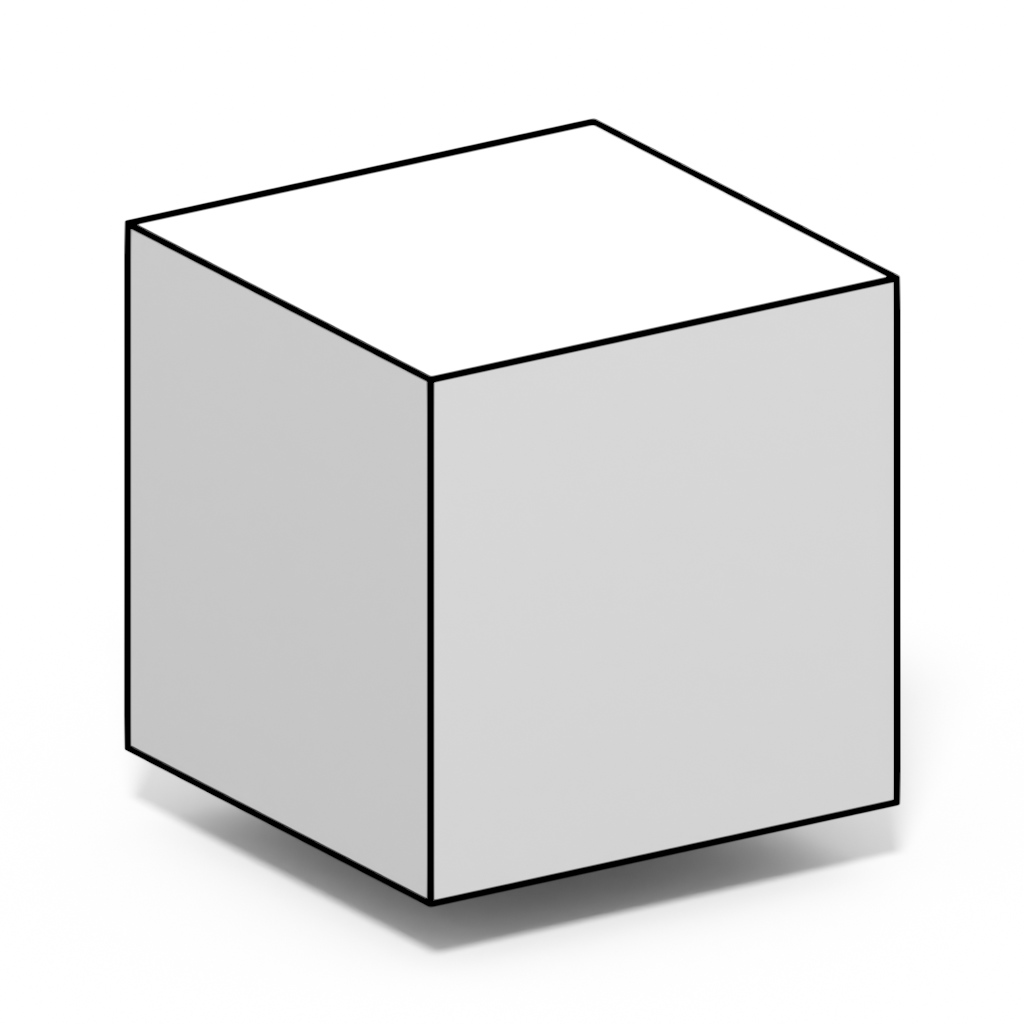}
        \includegraphics[width=\textwidth]{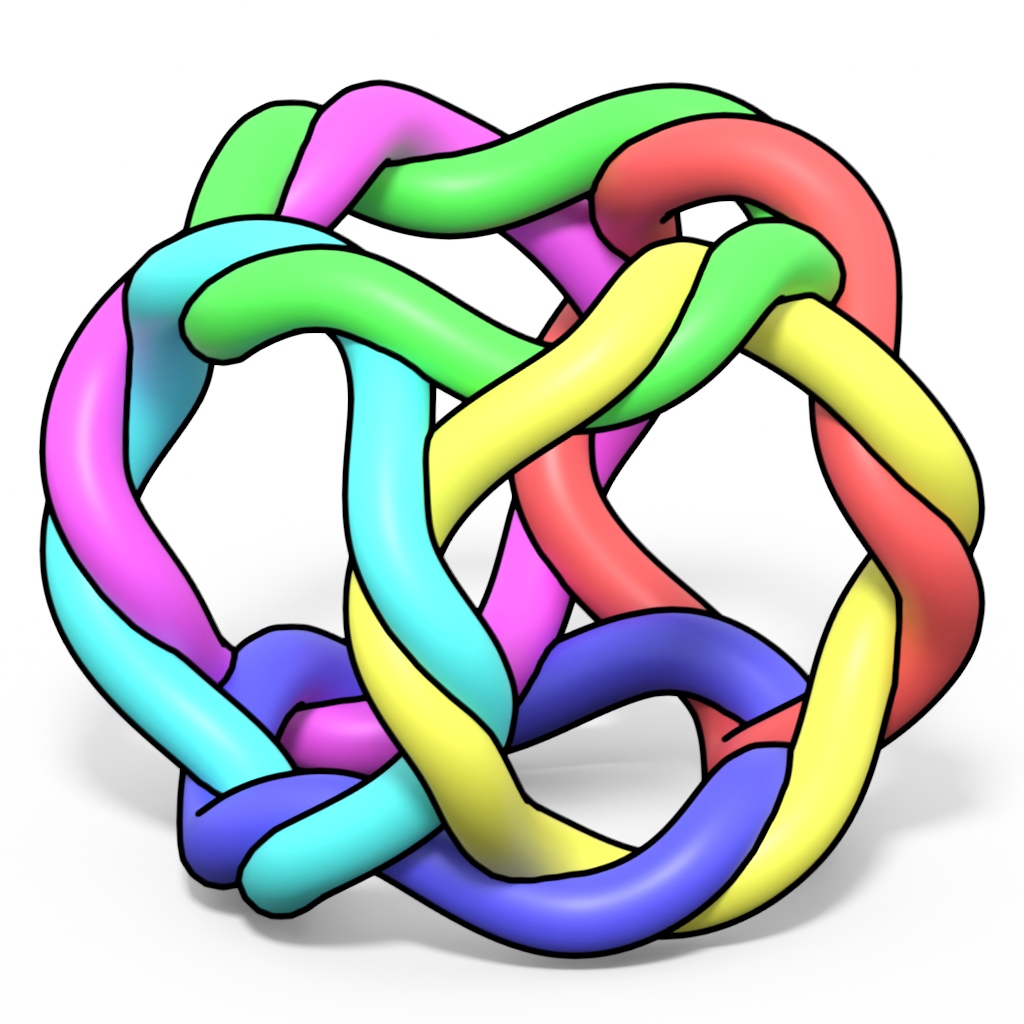}
        \caption{A link.}
        \label{teaser0}
    \end{subfigure}
    \hfill
        \begin{subfigure}[t]{0.16\textwidth}
        \includegraphics[width=\textwidth]{images/teaser/cube/cube_base.jpg}
        \includegraphics[width=\textwidth]{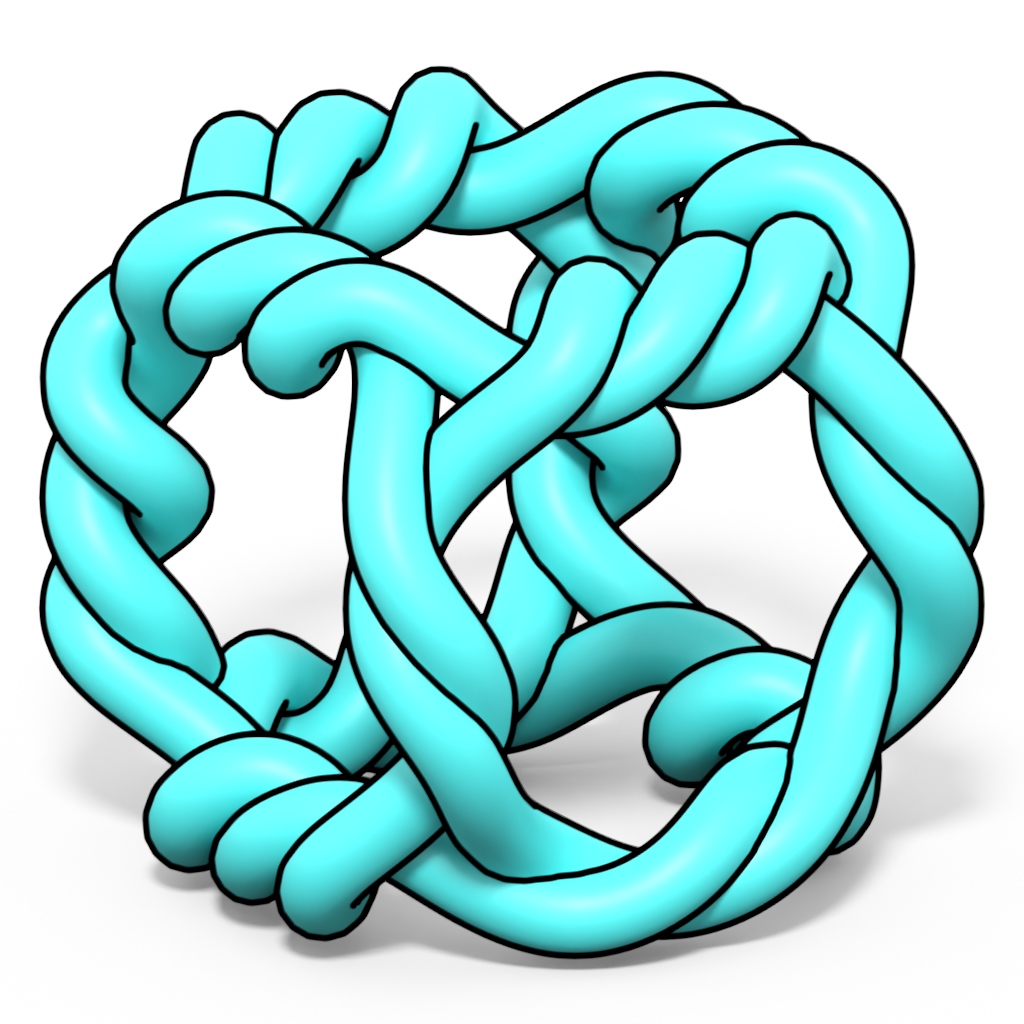}
        \caption{A knot.}
        \label{teaser1}
    \end{subfigure}
    \hfill
    \begin{subfigure}[t]{0.16\textwidth}
        \includegraphics[width=\textwidth]{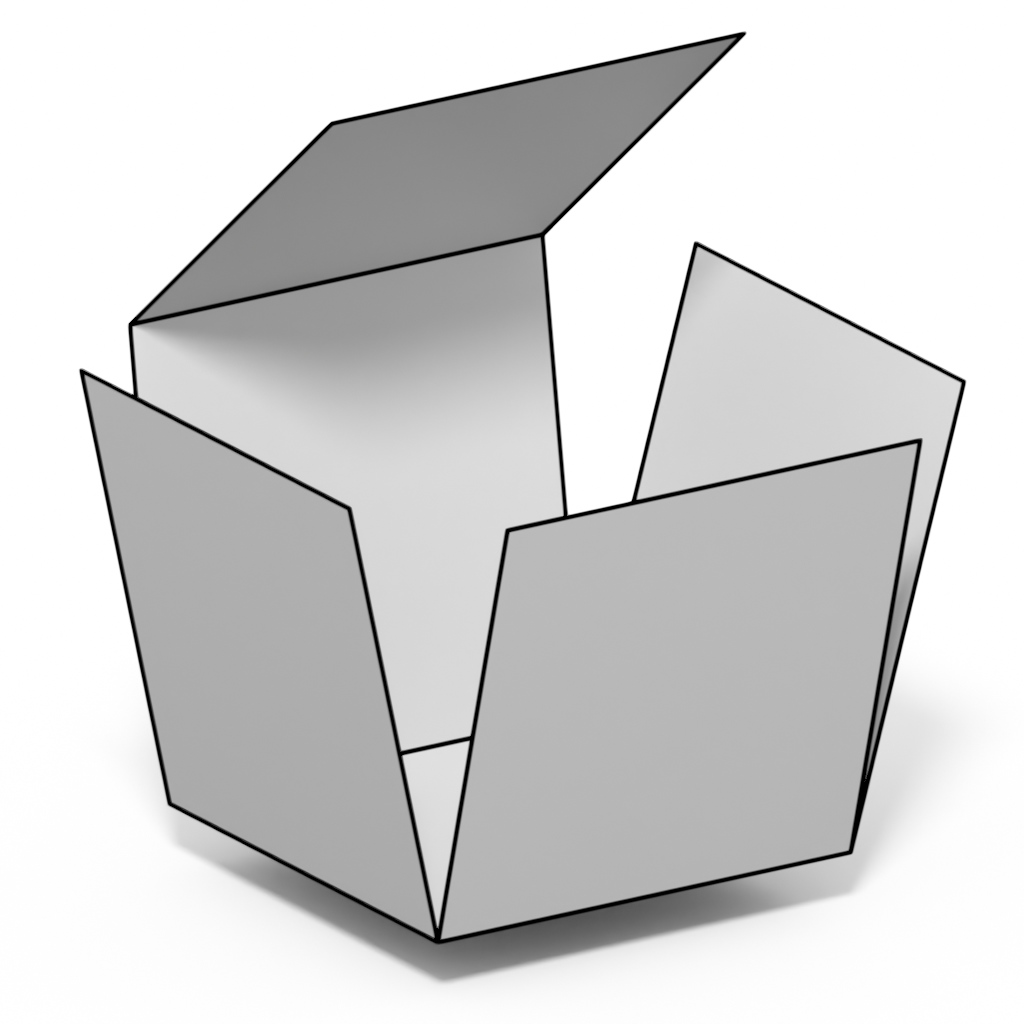}
        \includegraphics[width=\textwidth]{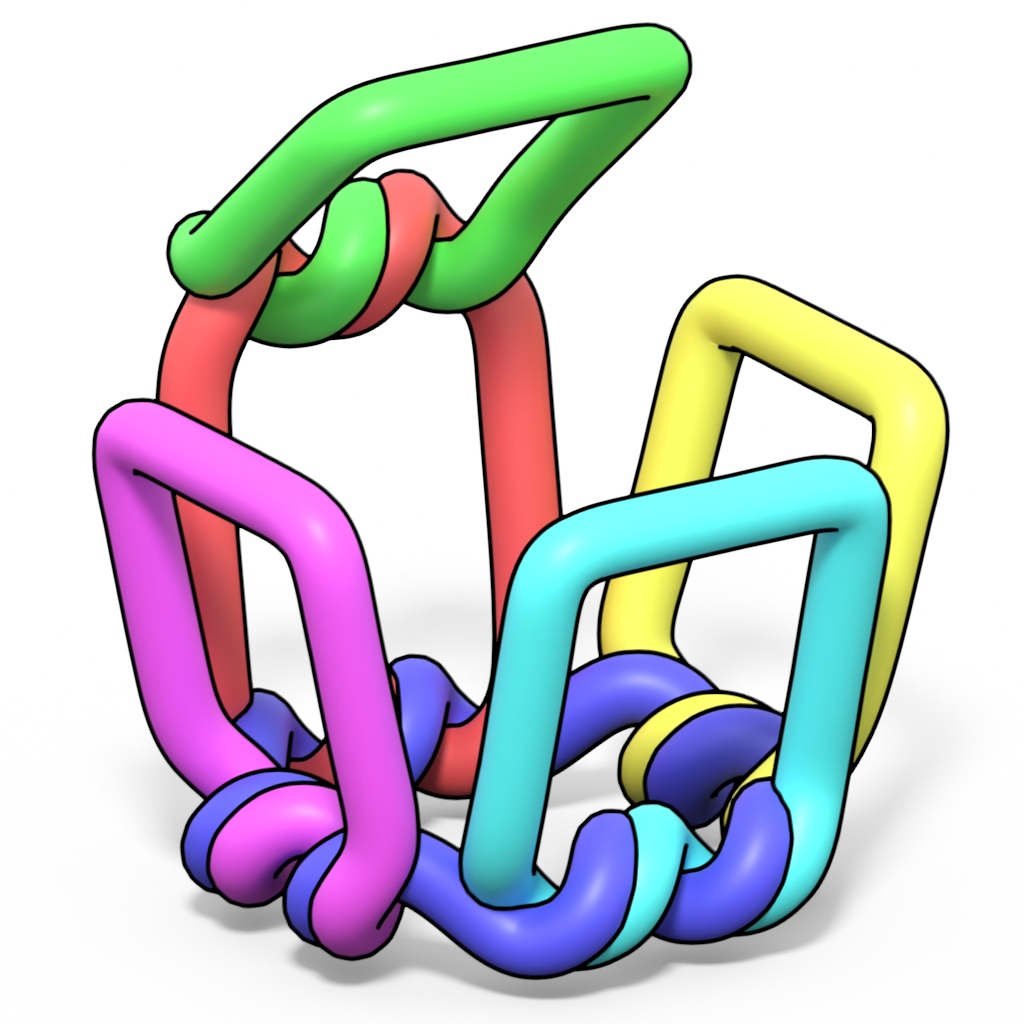}
        \caption{LK structures.}
        \label{teaser2}
    \end{subfigure}
    \hfill
    \begin{subfigure}[t]{0.16\textwidth}
        \includegraphics[width=\textwidth]{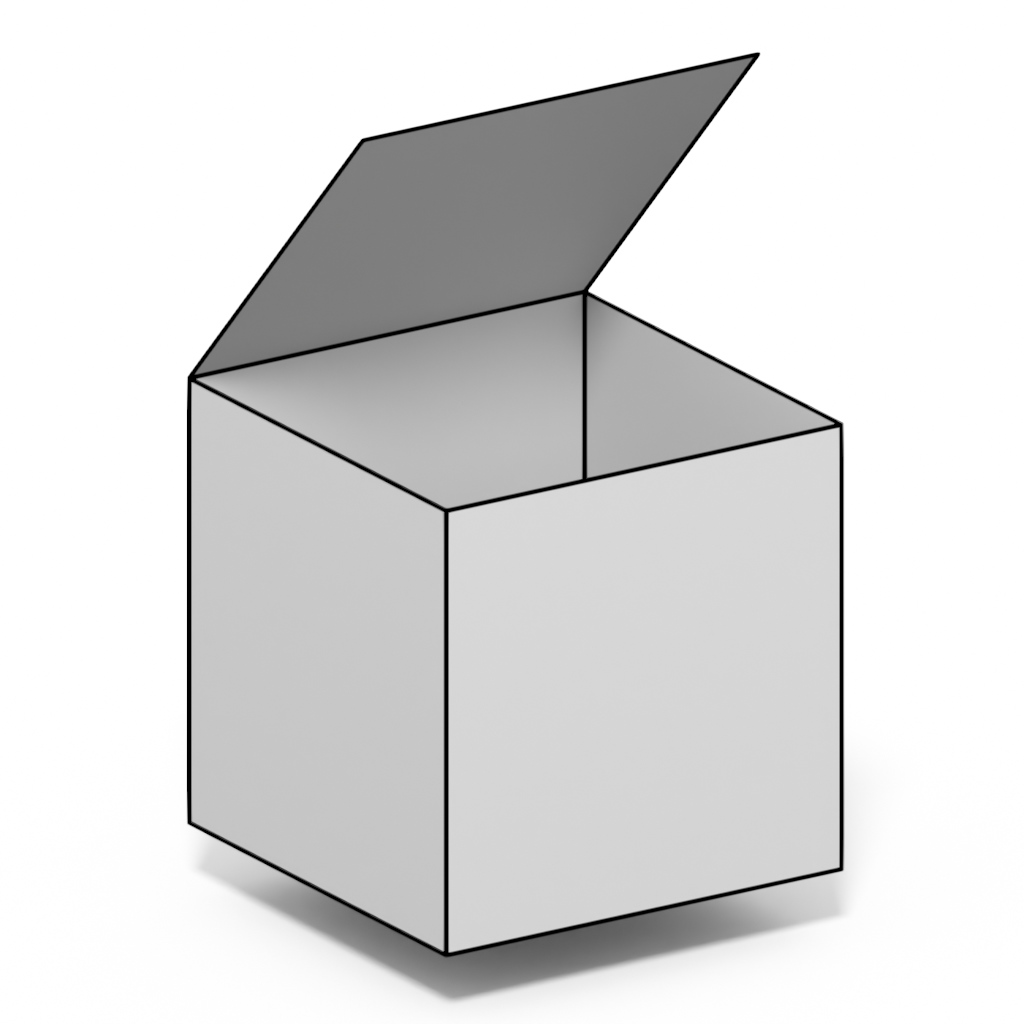}
        \includegraphics[width=\textwidth]{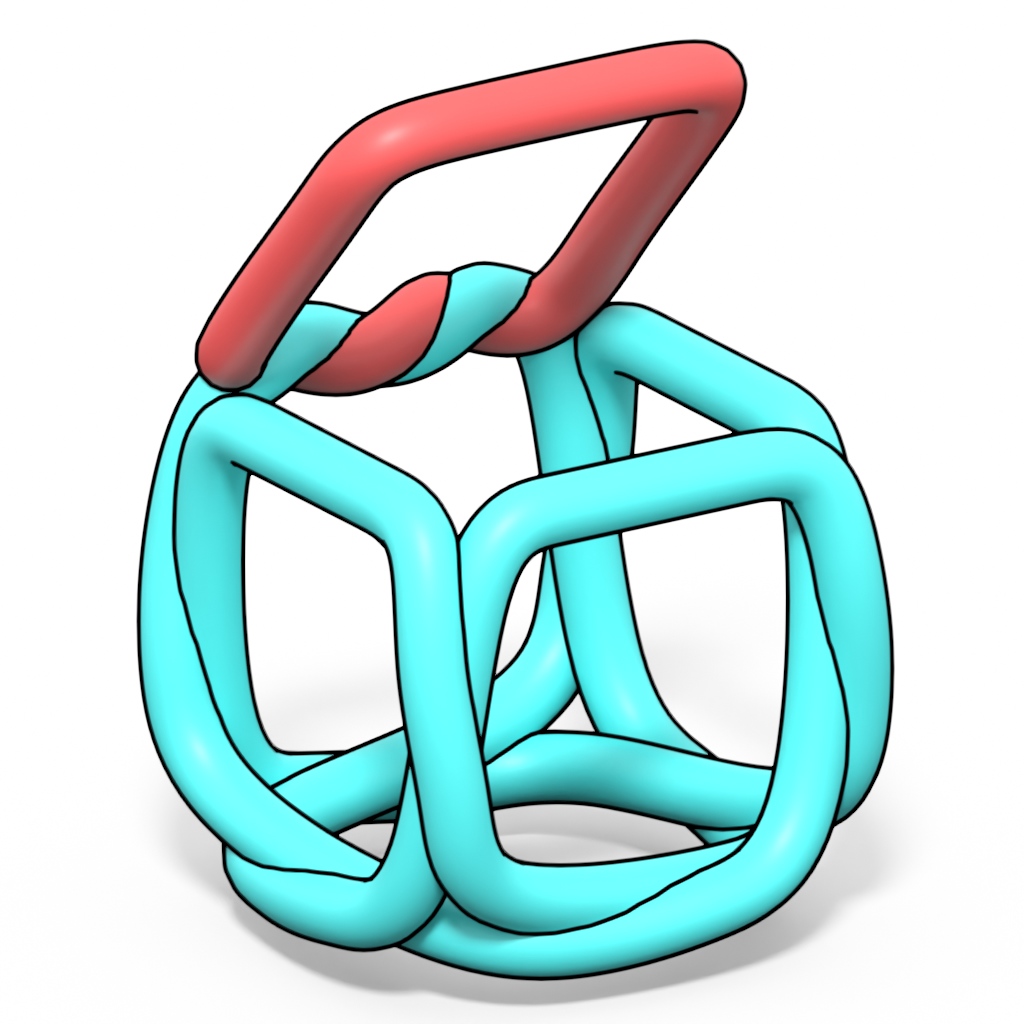}
        \caption{LK structures.}
        \label{teaser3}
    \end{subfigure}
    \hfill
    \begin{subfigure}[t]{0.16\textwidth}
        \includegraphics[width=\textwidth]{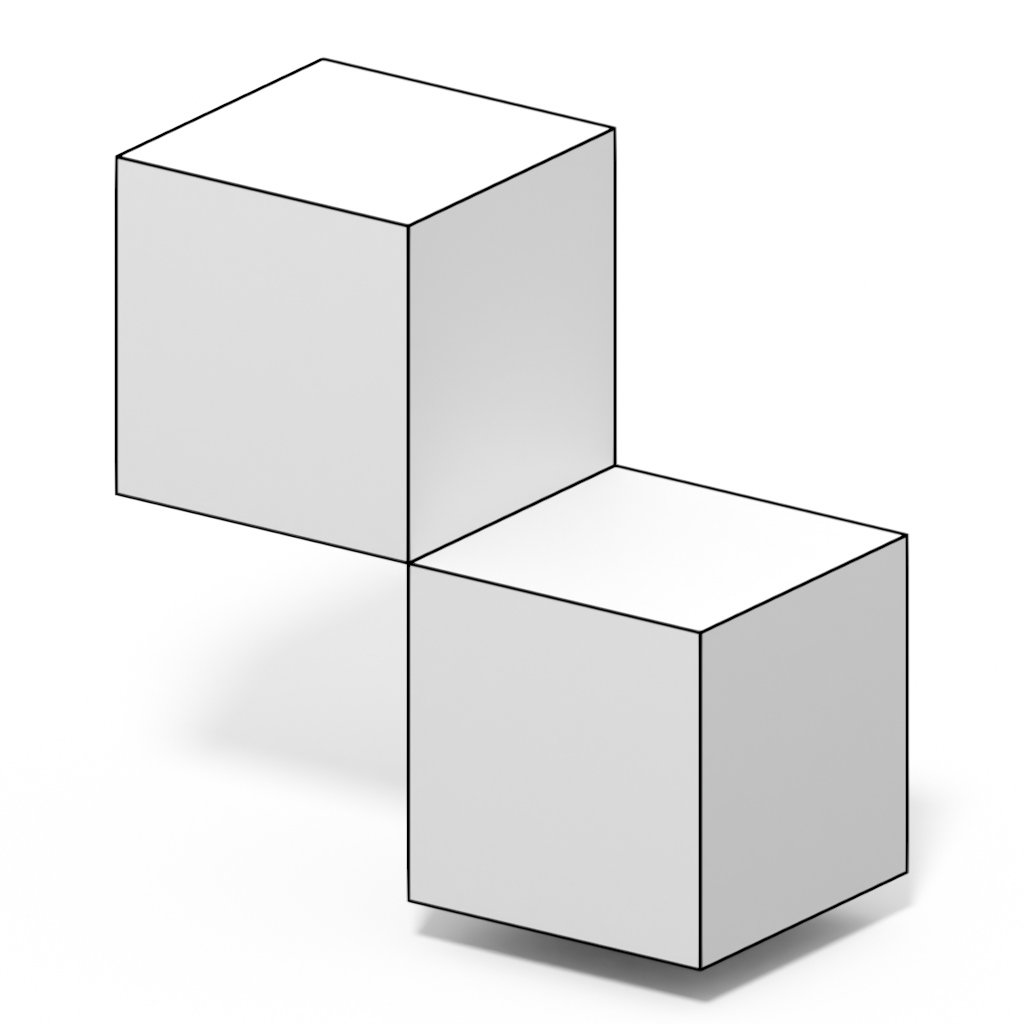}
        \includegraphics[width=\textwidth]{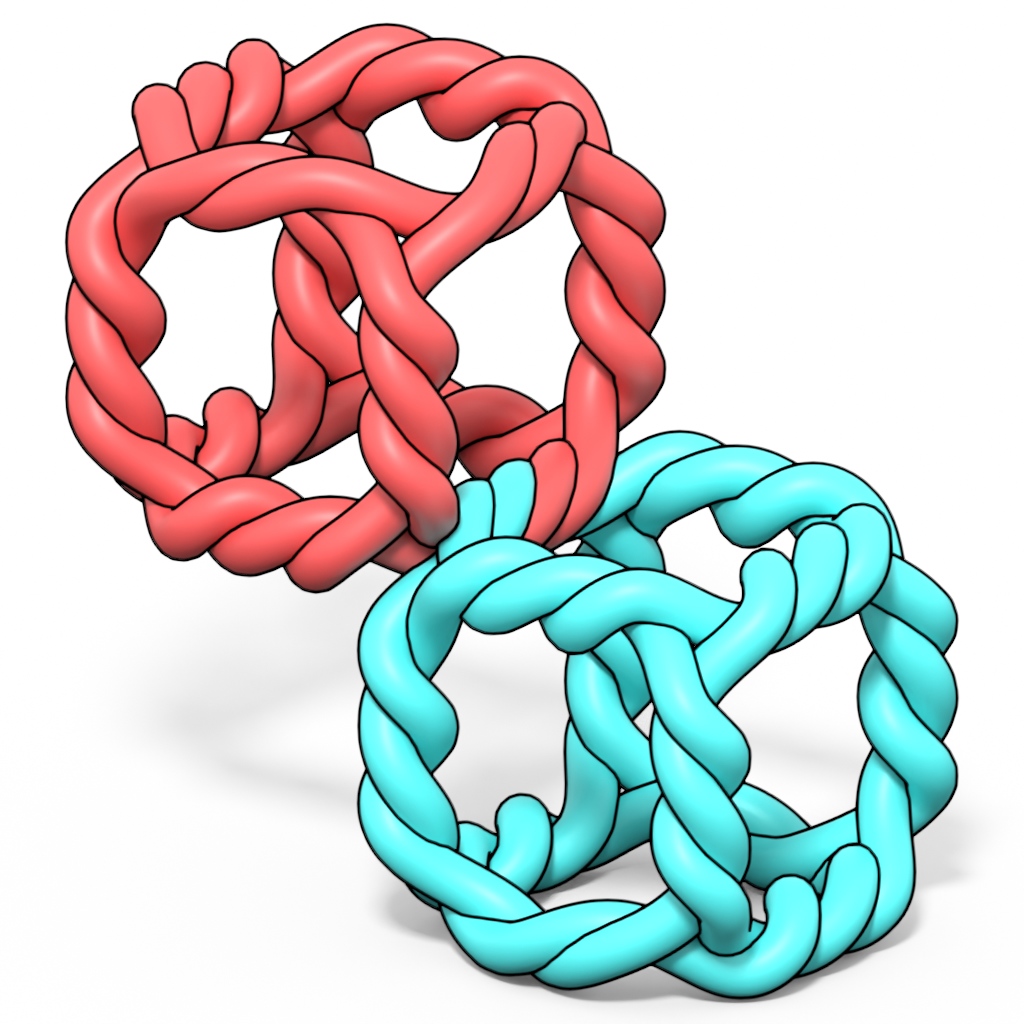}
        \caption{LK structures.}
        \label{teaser4}
    \end{subfigure}
    \hfill
    \begin{subfigure}[t]{0.16\textwidth}     
    \includegraphics[width=\textwidth]{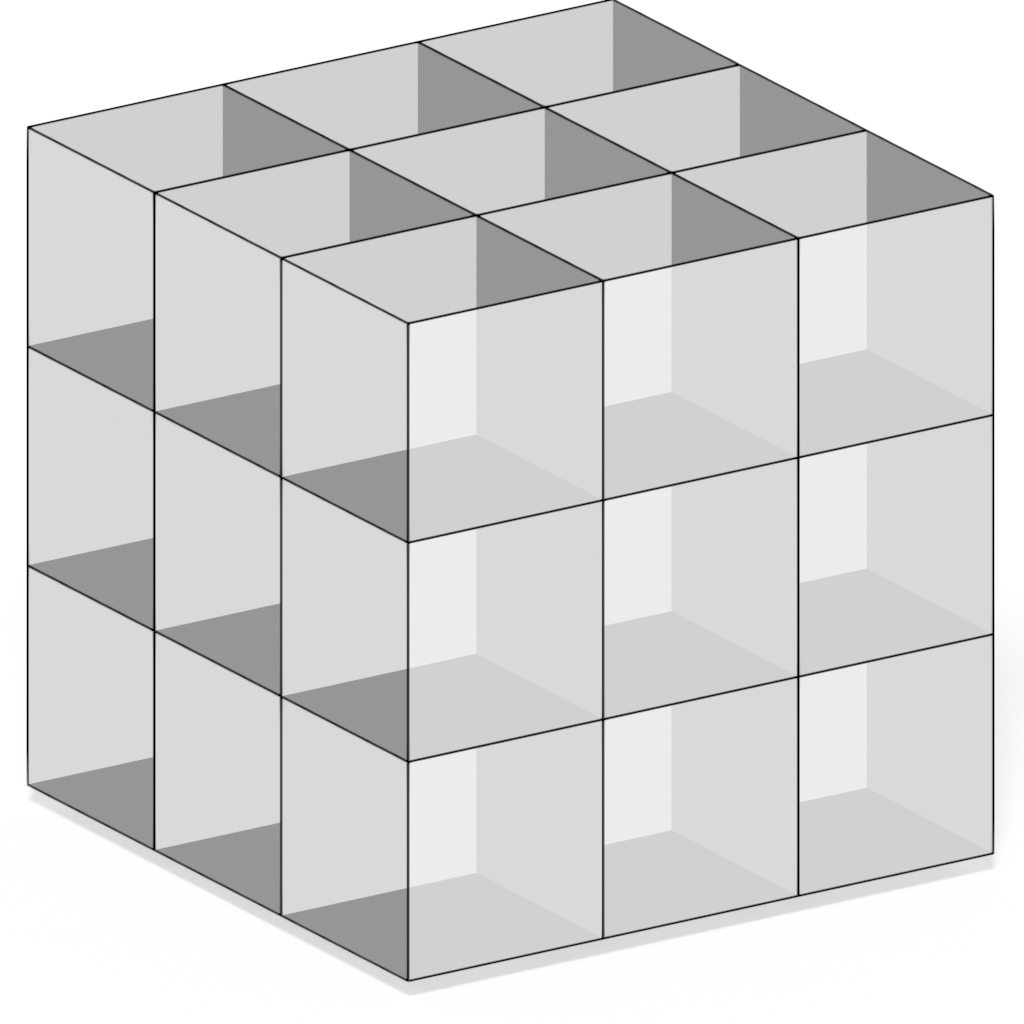} 
    \includegraphics[width=\textwidth]{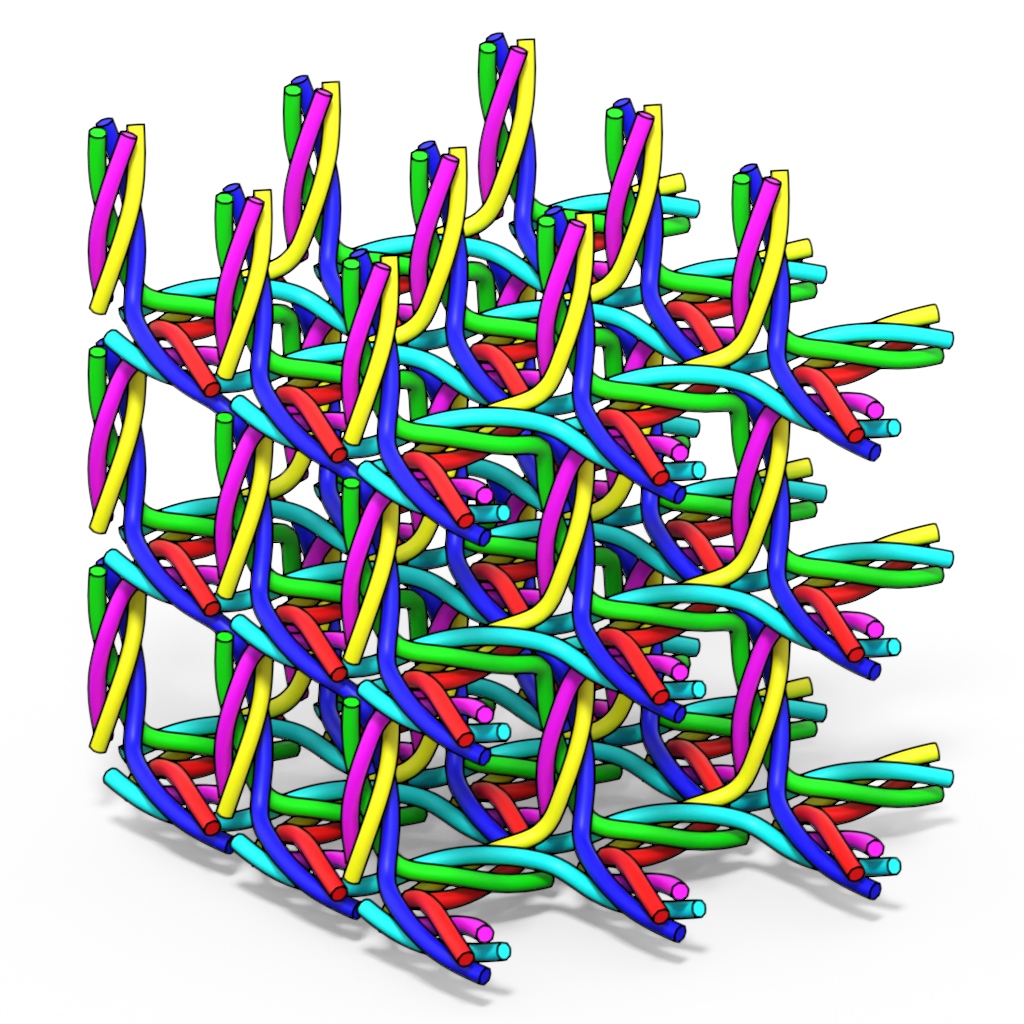}
        \caption{Periodic LKs.}
        \label{teaser5}
    \end{subfigure}
    \hfill
  \caption{
 We present a unified framework for generating \textit{Linked Knot (LK) structures} by applying integer twists to standard surface meshes. The top row shows the input control meshes, while the bottom row displays the resulting LK geometries. (a) Applying positive even integer twists to a cube produces a chainmail-like link. (b) Selective parity assignment (e.g., one even-twisted edge and all others odd) transforms the same control mesh into a single unicursal knot. (c-d) The framework naturally supports meshes with boundaries, such as unfolded cubes, yielding dynamically articulating structures. (e) Extending the approach to non-manifold inputs enables functional mechanisms, such as two cubes connected by a piano hinge. (f) The method scales to periodic, space-filling geometries, illustrated here by an LK structure derived from a cubical 3-honeycomb.
  }
  \Description{Six example images arranged in two rows. The top row shows geometric source meshes, including closed cubes, unfolded cube surfaces, hinged cube configurations, and a periodic cubical lattice. The bottom row shows the corresponding Linked Knot (LK) structures generated from these sources, rendered as thick, colored tubular strands. From left to right: a multicolored chainmail-style link; a single closed knot; two articulated LK structures derived from unfolded or hinged cube surfaces with visible openings; a pair of interconnected LK structures of different scales; and a dense periodic arrangement of many interwoven LK strands forming a space-filling lattice. Different colors indicate distinct strands or components.}
  \label{fig:teaser}
\end{teaserfigure}
\received{20 February 2007}
\received[revised]{12 March 2009}
\received[accepted]{5 June 2009}

\maketitle

\begin{figure*}[htbp!]
    \centering
    \begin{subfigure}[t]{0.16\textwidth}
        \includegraphics[width=\textwidth]{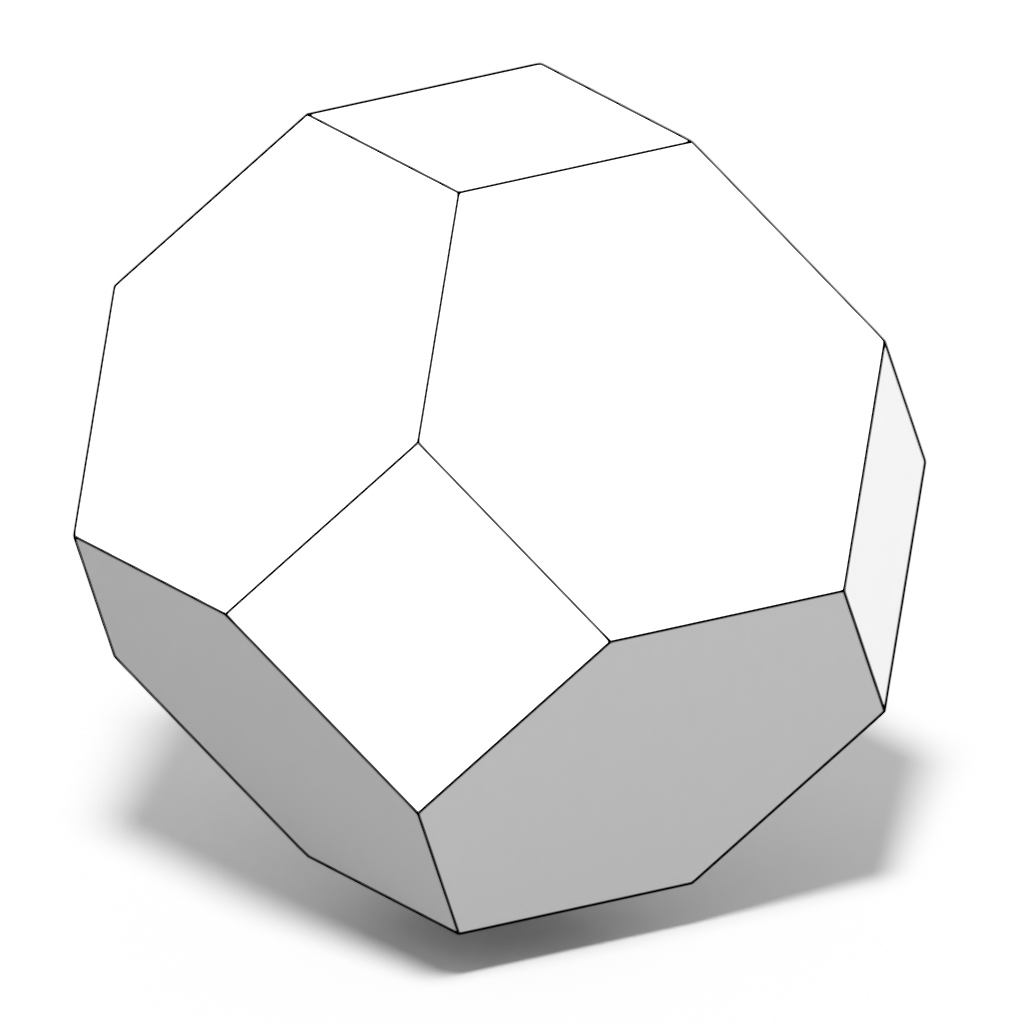}
        \includegraphics[width=\textwidth]{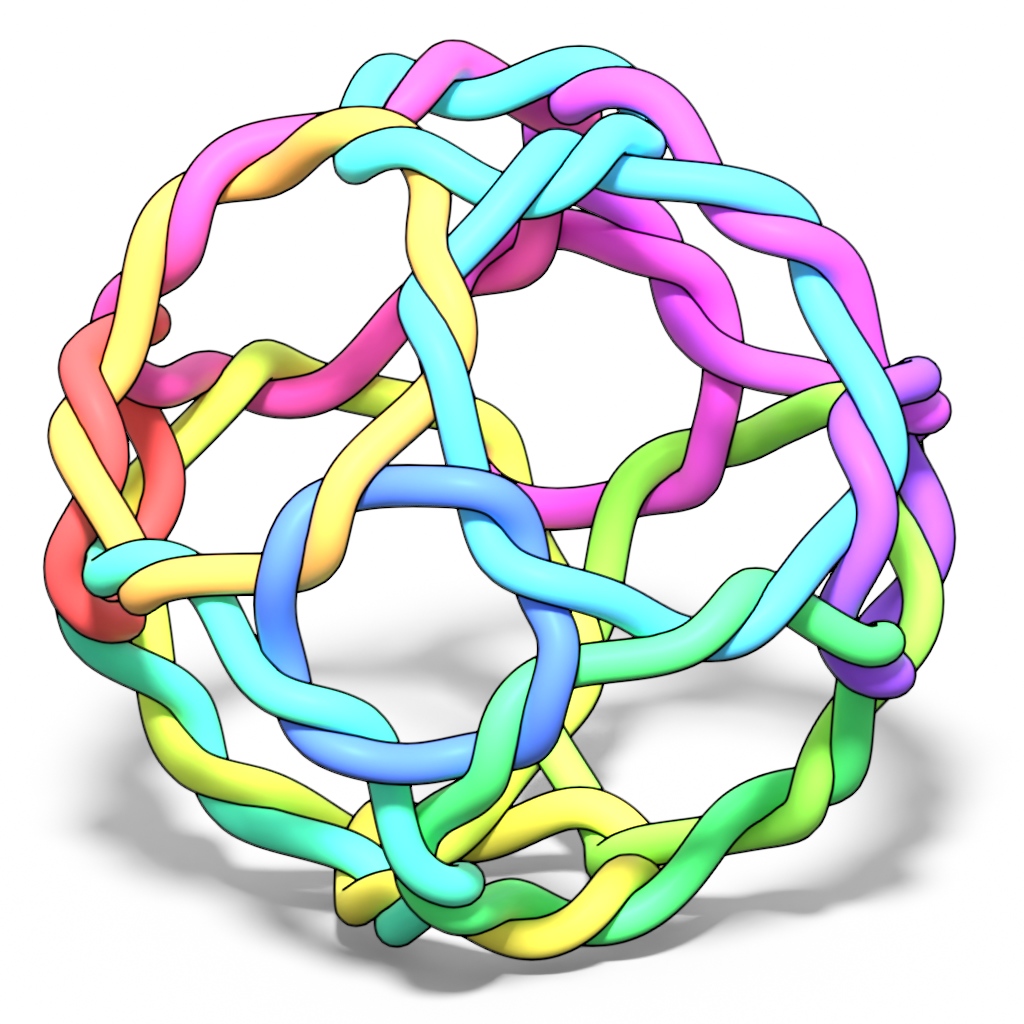}
        \caption{A link.}
        \label{teaser20}
    \end{subfigure}
    \hfill
        \begin{subfigure}[t]{0.16\textwidth}
        \includegraphics[width=\textwidth]{images/teaser/tr_oct/cf_base.jpg}
        \includegraphics[width=\textwidth]{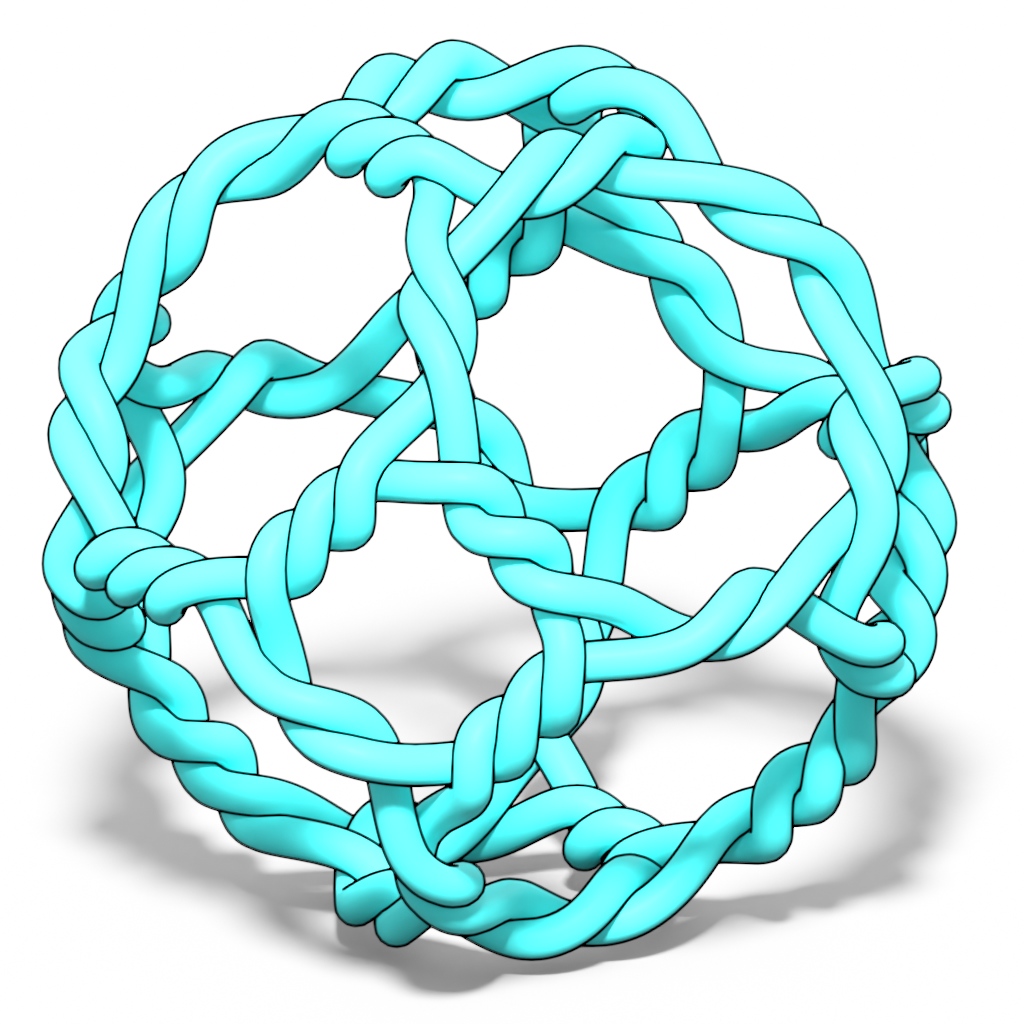}
        \caption{A knot.}
        \label{teaser21}
    \end{subfigure}
    \hfill
    \begin{subfigure}[t]{0.16\textwidth}
        \includegraphics[width=\textwidth]{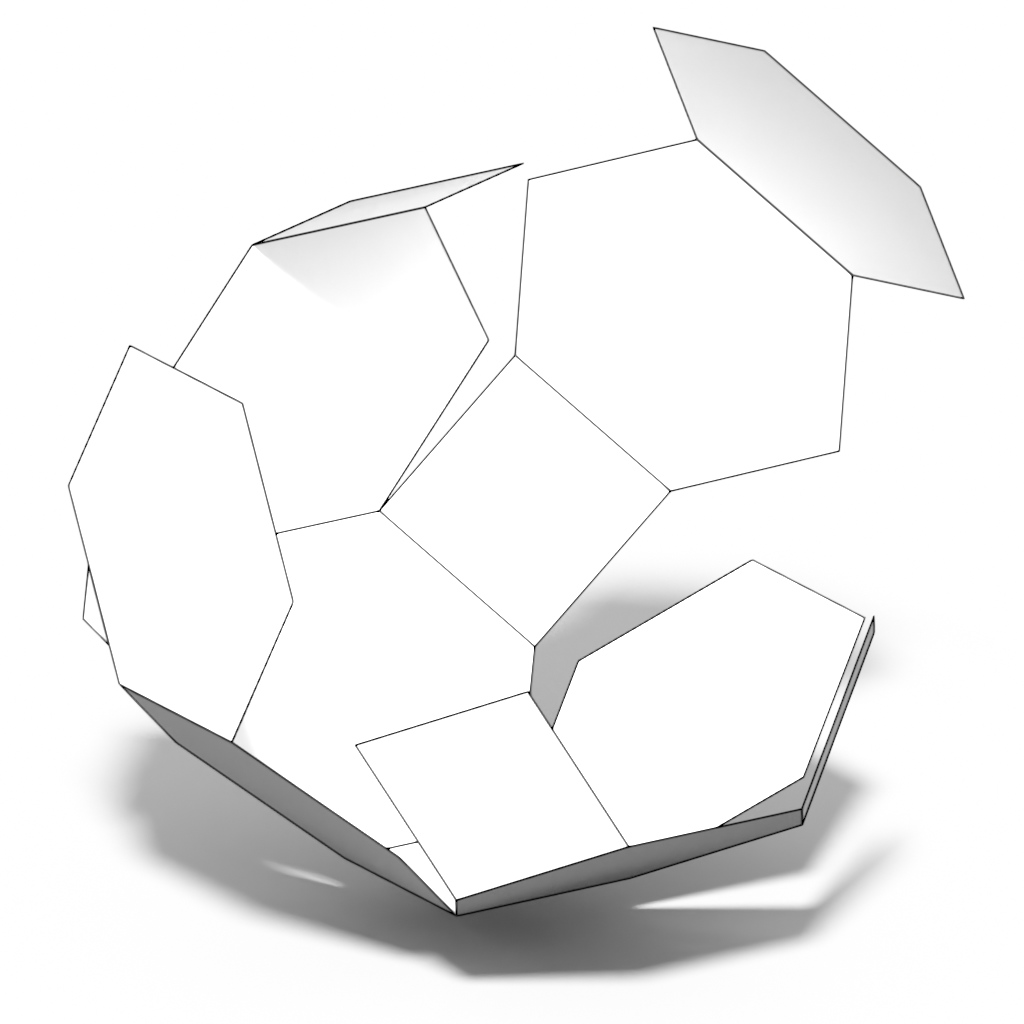}
        \includegraphics[width=\textwidth]{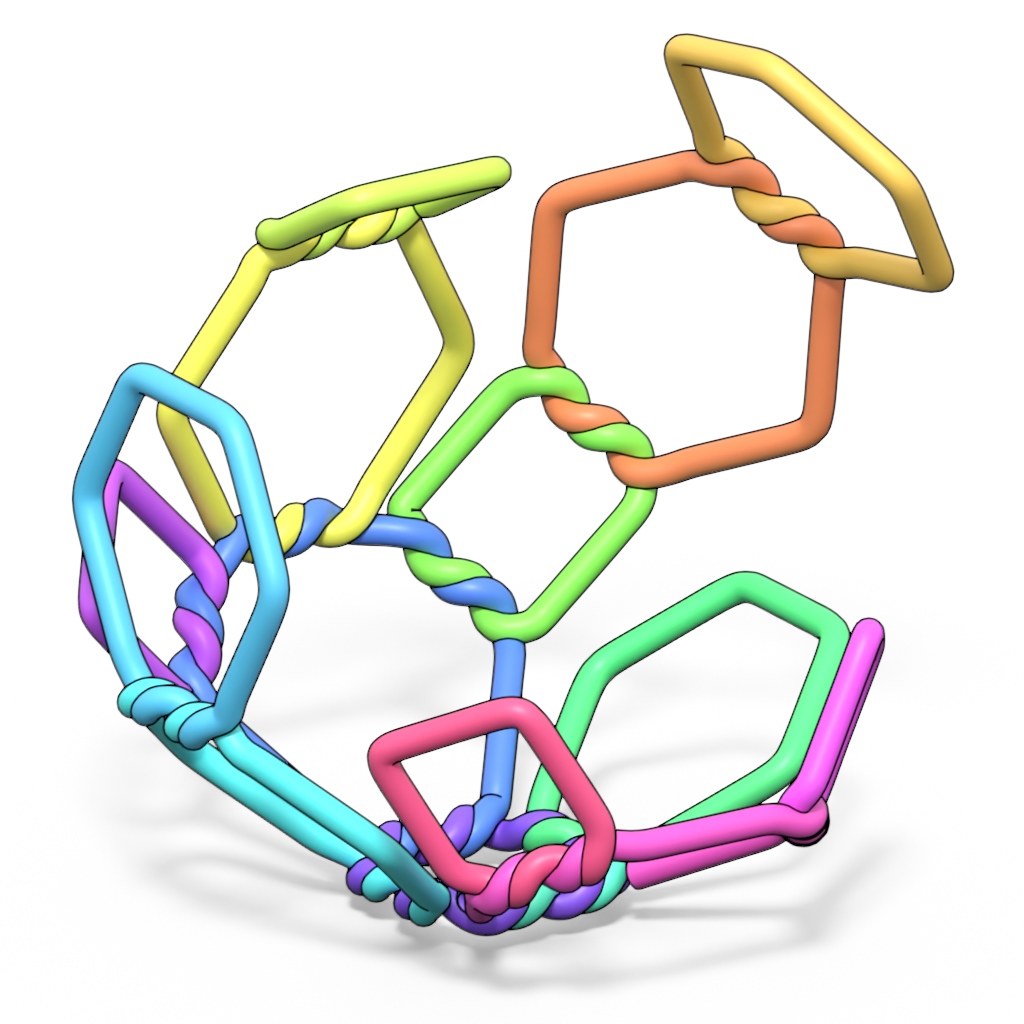}
        \caption{LK structures.}
        \label{teaser22}
    \end{subfigure}
    \hfill
    \begin{subfigure}[t]{0.16\textwidth}
        \includegraphics[width=\textwidth]{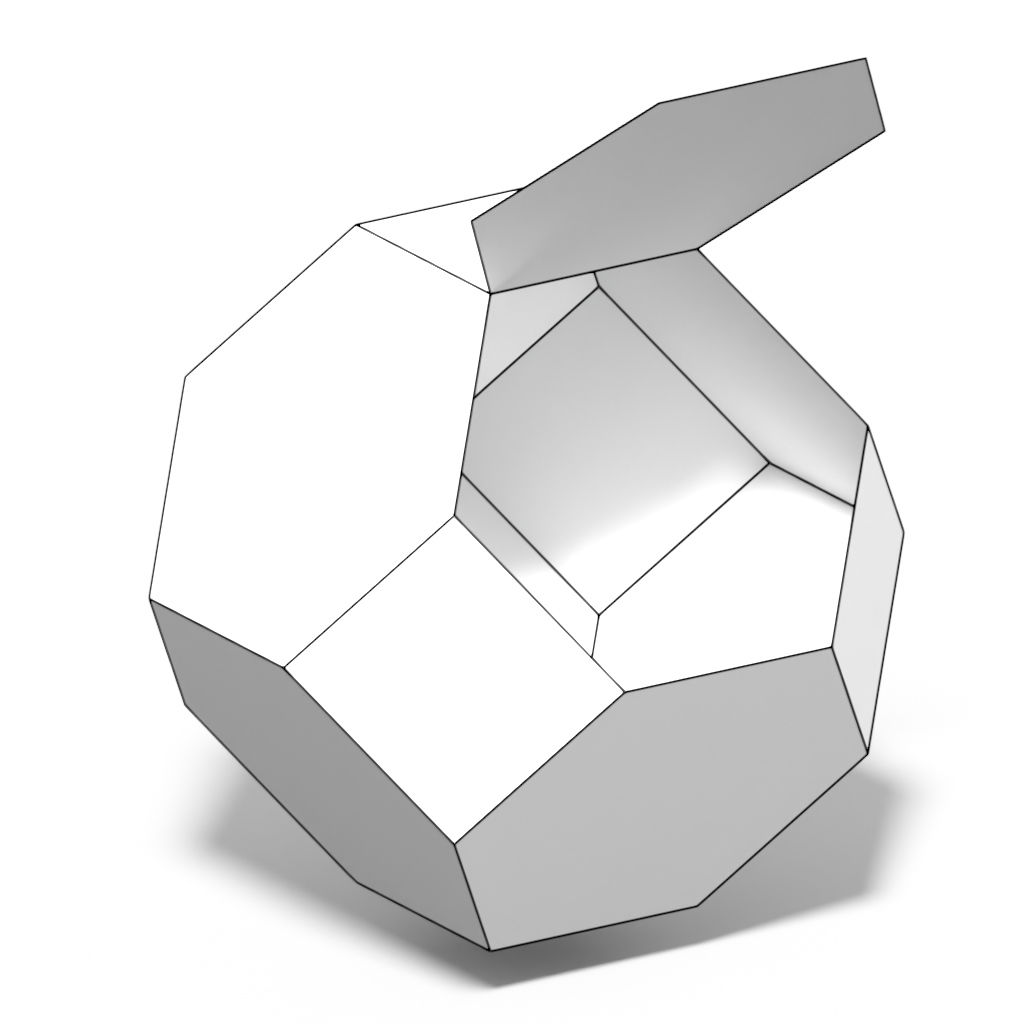}
        \includegraphics[width=\textwidth]{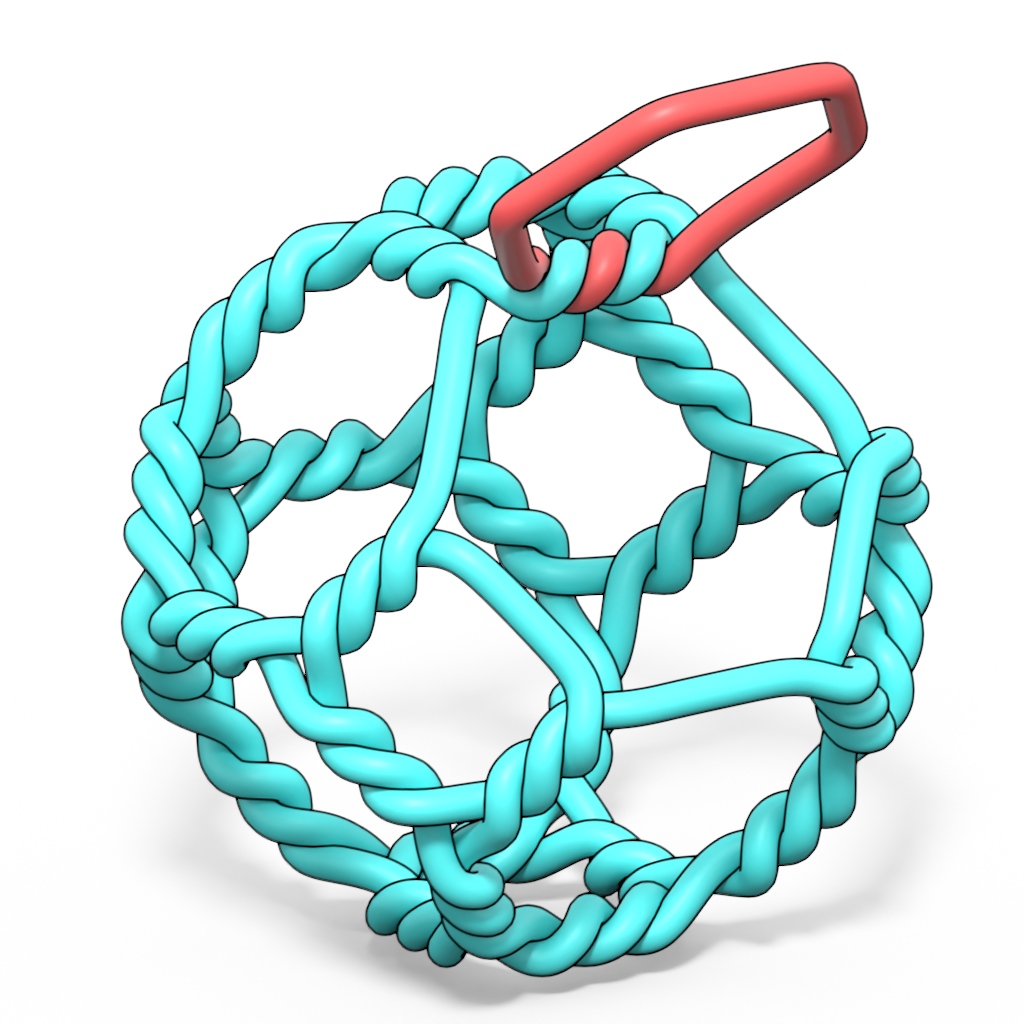}
        \caption{LK structures.}
        \label{teaser23}
    \end{subfigure}
    \hfill
    \begin{subfigure}[t]{0.16\textwidth}
        \includegraphics[width=\textwidth]{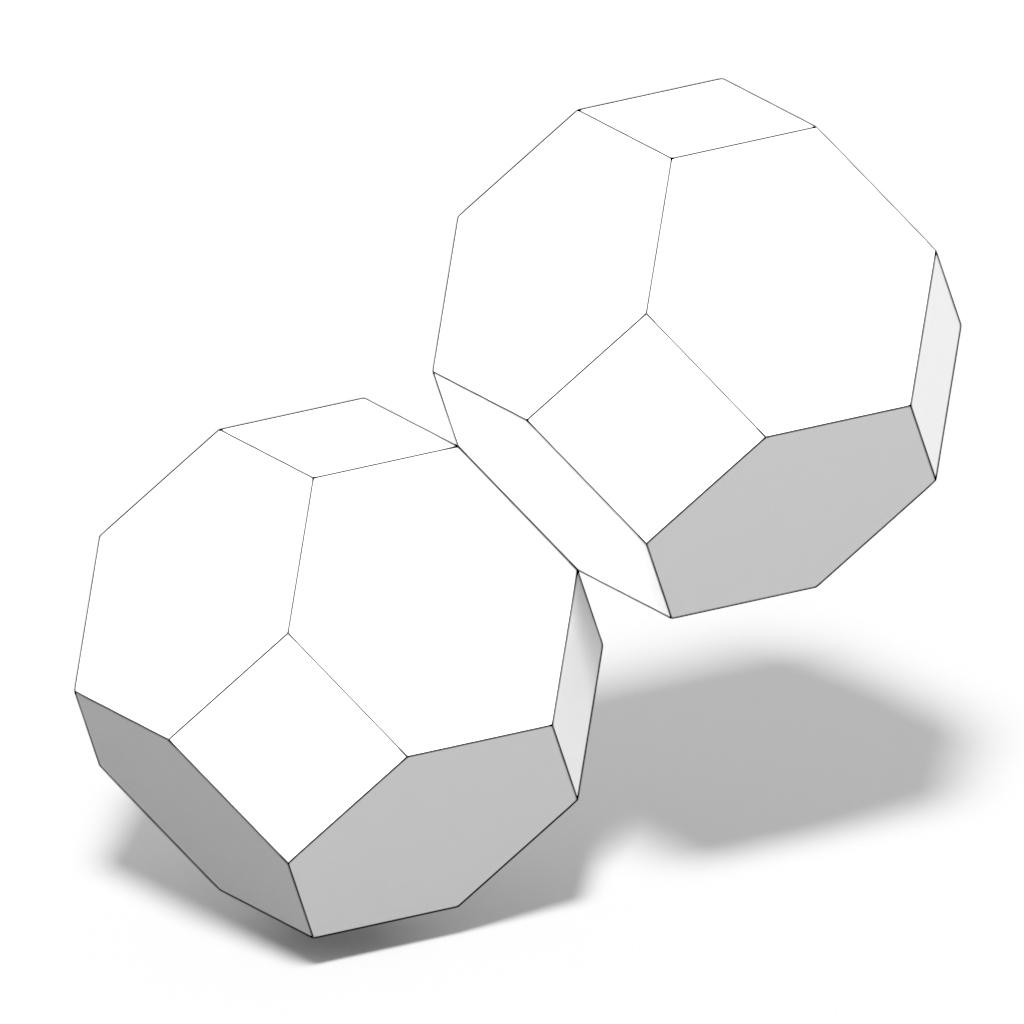}
        \includegraphics[width=\textwidth]{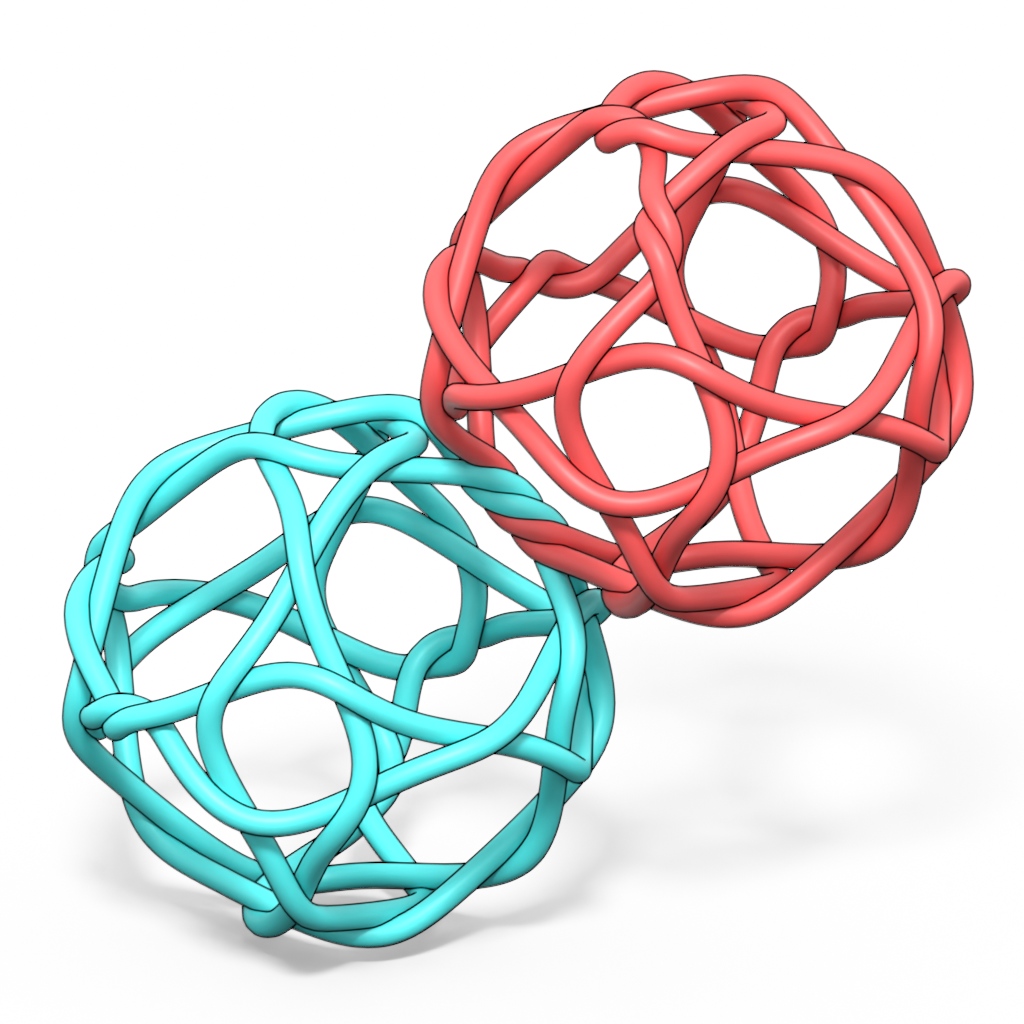}
        \caption{LK structures.}
        \label{teaser24}
    \end{subfigure}
    \hfill
    \begin{subfigure}[t]{0.16\textwidth}     
    \includegraphics[width=\textwidth]{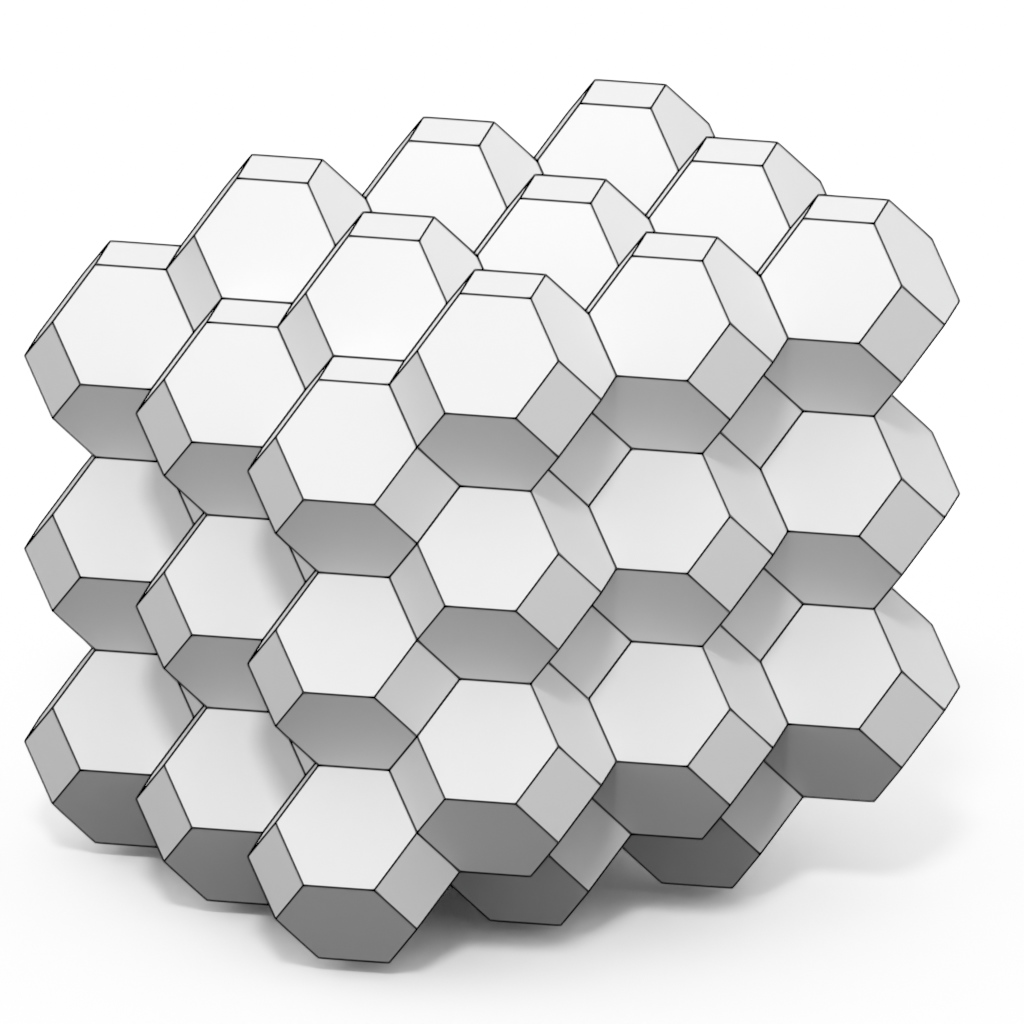}
    \includegraphics[width=\textwidth]{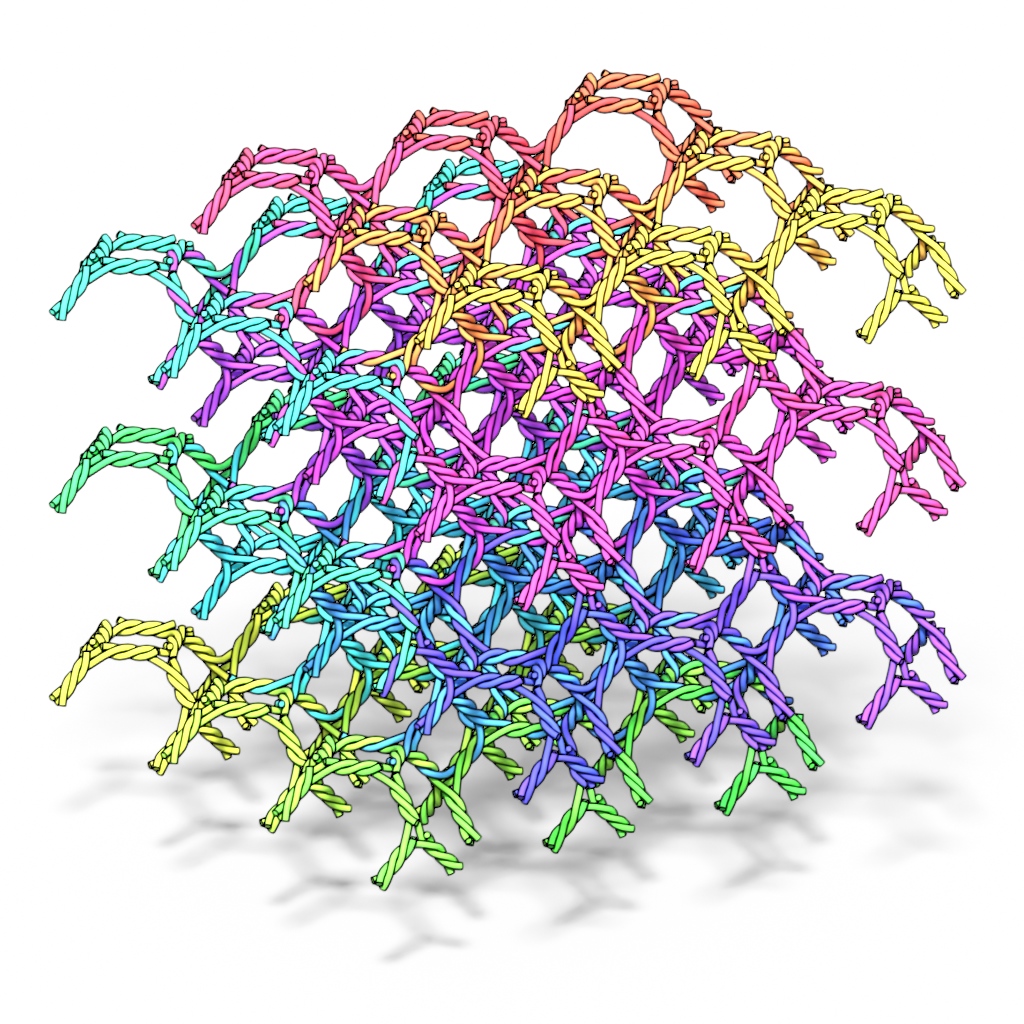}
        \caption{Periodic LKs.}
        \label{teaser25}
    \end{subfigure}
    \hfill
  \caption{Another example demonstrating the expressive range of our framework. The LK-structures in (a-e) can be generated from arbitrary manifold or non-manifold polygonal meshes. Periodic LK-structures require starting from a periodic 3-honeycomb built from space-filling polyhedra, such as the Wigner-Seitz cells. Examples include a cube, as shown in Figure~\ref{fig:teaser2}, and a truncated octahedron, as shown in this figure.
  }
  \Description{Six example images arranged in two rows. The top row shows polygonal source meshes, including a truncated octahedron in closed, partially opened, and hinged configurations, a pair of connected polyhedra, and a periodic lattice of space-filling polyhedra. The bottom row shows the corresponding Linked Knot (LK) structures rendered as thick tubular strands. From left to right, these include a multicolored interlinked structure, a single closed knot rendered in a uniform color, several articulated LK structures with visible openings, a pair of LK structures at different scales positioned side by side, and a dense periodic assembly of many interwoven LK strands. Distinct colors indicate different strands or components.}
  \label{fig:teaser2}
\end{figure*}

\section{Introduction and Motivation}

Linked, knotted, and woven structures arise naturally across scales and disciplines, typically as the outcome of physical, biological, or chemical processes and emergent material interactions, from ropes and textiles \cite{knittel2020modelling,niu2025geometric,singal2024programming} to deployable architectural gridshells \cite{yang2025topologicalwovennodes,liu2025doublelayeredegridshells}. In engineered physical materials, knotting and interlacing can likewise emerge from frictional contact, deformation, and hierarchical fiber arrangements without explicit topological abstraction \cite{moestopo2023knots}. In biology, such entanglements appear in protein knot motifs and folded peptide structures \cite{craik2001cystine}, as well as in fibrillous protein assemblies observed in butterfly wings \cite{jessop2025butterfly}. In chemistry, related phenomena arise as molecular and chemical knots, including synthetic and self-assembled molecular entanglements studied within the chemical sciences \cite{horner2016knot,fielden2017molecular,forgan2011chemical}. These ideas also manifest in DNA-based molecular assemblies, ranging from catenated DNA polyhedra \cite{zhang1994dnatroctahedron} to molecular assemblies such as DNA origami and related nanostructures \cite{glaser2021dnananostructures}. While these processes produce remarkably complex entanglements, computational methods that attempt to mimic such physical or chemical mechanisms often result in intricate, highly coupled design pipelines. In contrast, \textbf{our approach leverages topological structure directly, enabling simple, explicit, and controllable design of linked knot structures by operating at the level of discrete mesh topology rather than mimicking physical or chemical mechanisms.}

Building on this perspective, we demonstrate that linked knot (LK) structures can be constructed from any non-manifold mesh using a fundamental mesh operator, known as edge twisting. This mesh operator can be used to construct physical elements such as springs, double, and triple helices. In these physical structures, twist is not merely a visual motif but a fundamental structural mechanism that enables strength, flexibility, and controlled deformation. Despite their prevalence, the controlled synthesis of such twisted structures within computer graphics and geometric modeling remains limited in scope. Existing methods are often restricted to surface meshes, rely on global parameterizations, or target specific weaving patterns, making it difficult to generalize across dimensions, topologies, and embedding manifolds.

\textbf{A fundamental challenge in modeling knots and links is that knot topology cannot be robustly defined using geometry alone: arbitrarily small geometric perturbations can alter crossings and change knot type.} As a result, relying solely on geometric embeddings makes knot identity unstable. In contrast, if the knot structure is induced from a higher-dimensional or combinatorial scaffold, the resulting topology can be inferred uniquely and remains invariant under geometric deformation. \textbf{This observation motivates our design-level approach, in which knot and link structures arise from discrete topological decisions rather than from geometric construction.}

In this paper, we present a unified, mesh-based framework for generating twisted-edge structures from labeled non-manifold mesh surfaces. Our method applies to all 2-manifolds with or without boundary, as well as to the boundary surfaces of solids, provided they are given as discrete mesh structures. Rather than relying on continuous parameterizations or global optimization, we encode twisting behavior locally on mesh edges using an integer-valued \emph{twisting number}. These local twisting assignments define coherent spiral trajectories that propagate through mesh connectivity, producing globally consistent structures in both surface-embedded and volumetric settings. We refer to the resulting constructions as \emph{Linked Knot (LK) structures}, encompassing knots, links, and more general linked assemblies.

\begin{figure}[htbp!]
    \centering   \includegraphics[width=0.99\linewidth]{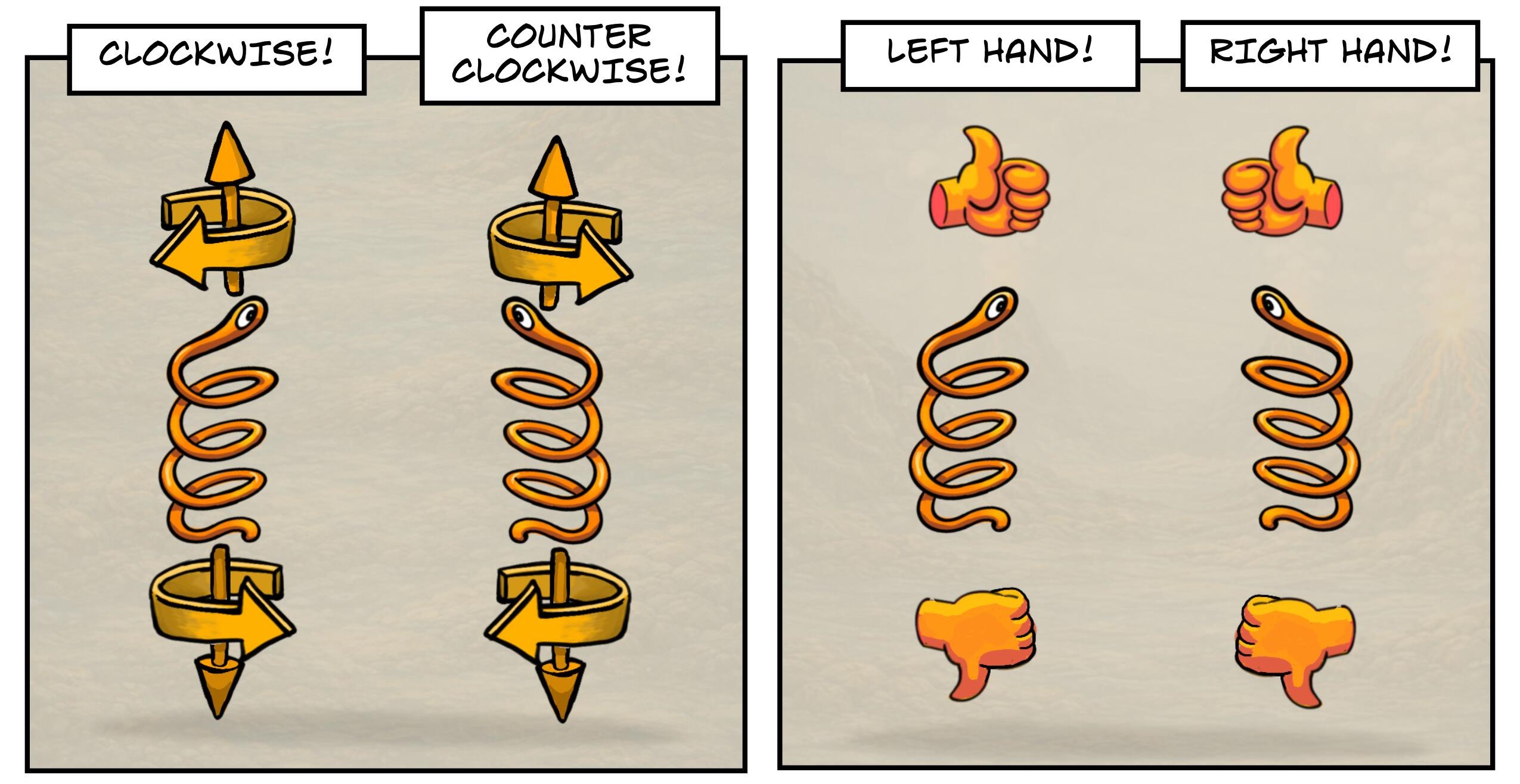}
    \caption{Our method is based on the fact that the chirality of a twist in three dimensions is uniquely defined. A twist operation generates spiral structures whose chirality can be identified precisely by fixing a direction along the spiral. When viewed along this direction, the spiral rotates either clockwise or counterclockwise. This rotation direction defines its handedness. Mirror-image spirals always rotate in opposite ways and can never be superimposed. Although this principle has been implicitly used for millennia, for example by rope-makers, a clear visual formalism is rarely presented. We therefore include this illustration to eliminate a common ambiguity that arises when formalizing twist direction in discrete mesh settings. Once one chirality is designated as a positive twist, the opposite chirality is naturally defined as negative.}
    \Description{Illustration composed of two side-by-side panels. The left panel shows two vertical spiral shapes with arrows indicating their rotation when viewed along the spiral axis, labeled “clockwise” and “counterclockwise.” Each spiral includes directional arrows at the top and bottom to fix the viewing direction. The right panel shows two similar vertical spirals accompanied by hand icons, labeled “left hand” and “right hand,” indicating the correspondence between spiral rotation direction and handedness. The spirals are visually similar but mirrored, emphasizing opposite chirality.}
    \label{fig:chiralspiral}
\end{figure}

Figures~\ref{fig:teaser} and~\ref{fig:teaser2} demonstrate the versatility of our approach. Even starting with a simple cube, we can obtain a wide variety of LK structures by changing the underlying topology and twisting numbers. In these examples of LK structures, each knot is encoded by a distinct color, revealing the underlying topological organization. These examples demonstrate how users can design individual knots and precisely control the links between them, allowing for a wide range of static and dynamic configurations within a single, coherent representation.

\begin{figure*}[htb!]
    \centering
    \includegraphics[width=0.120\textwidth]{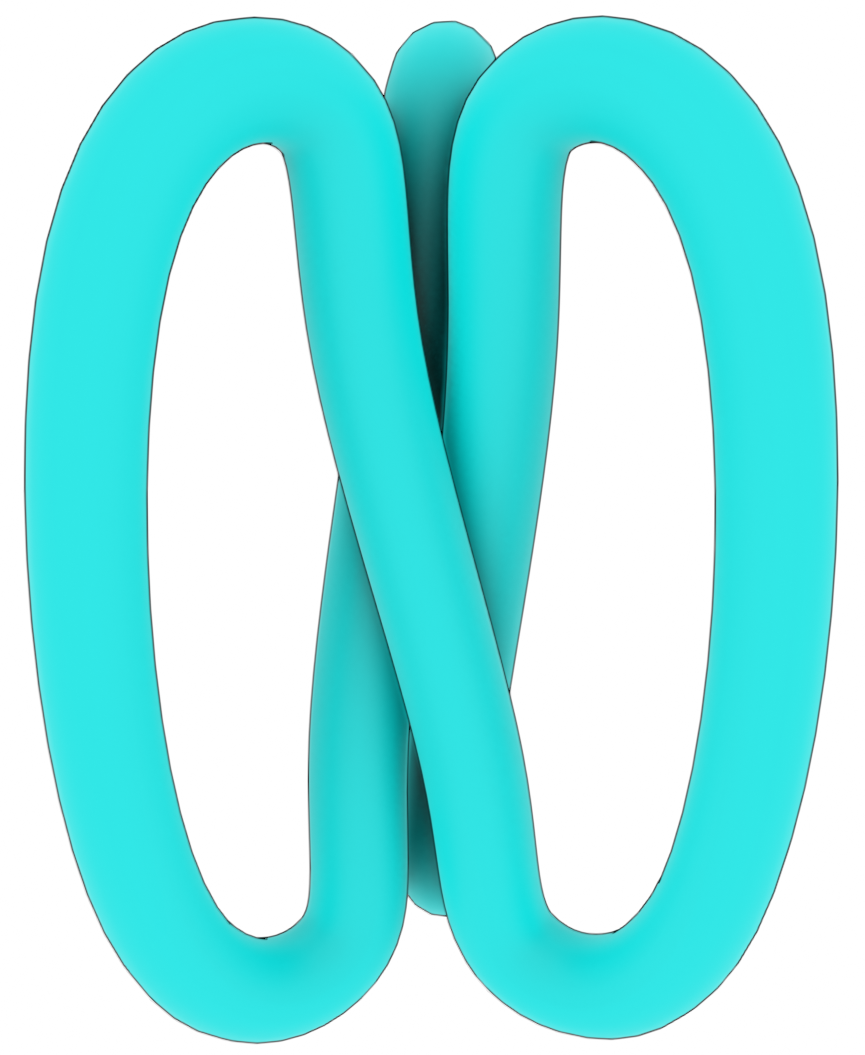}
    \includegraphics[width=0.120\textwidth]{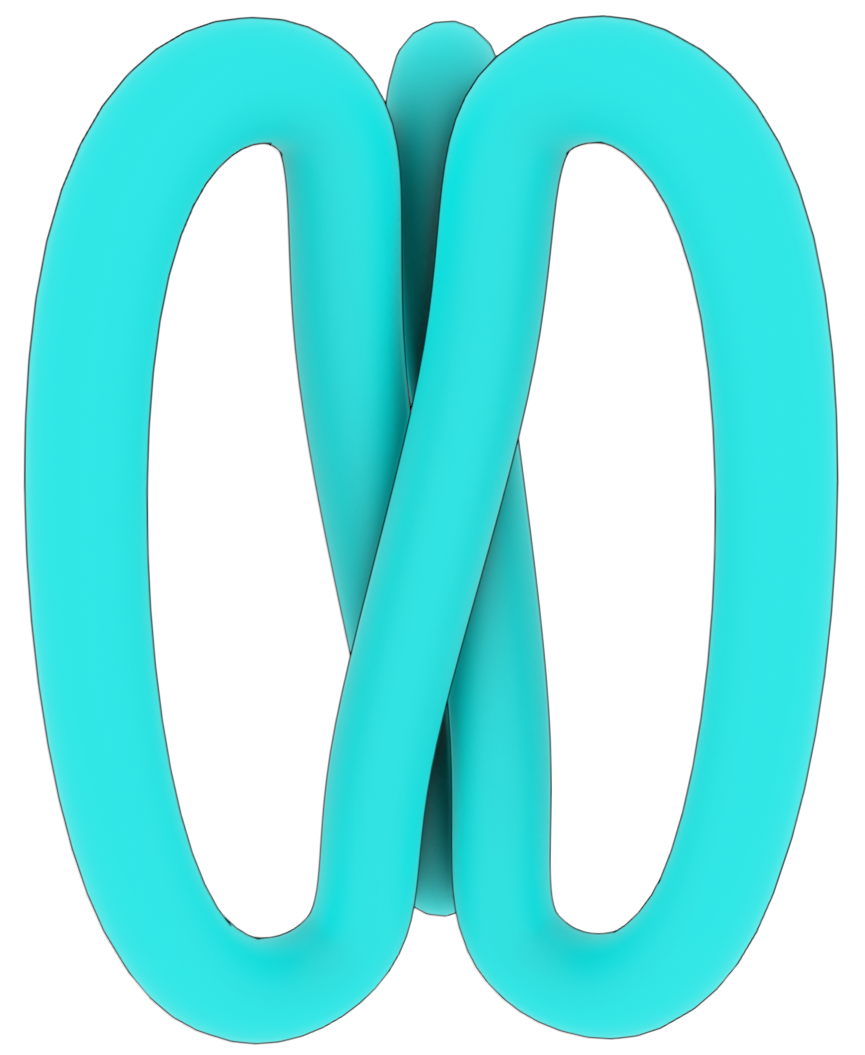}
    \includegraphics[width=0.120\textwidth]{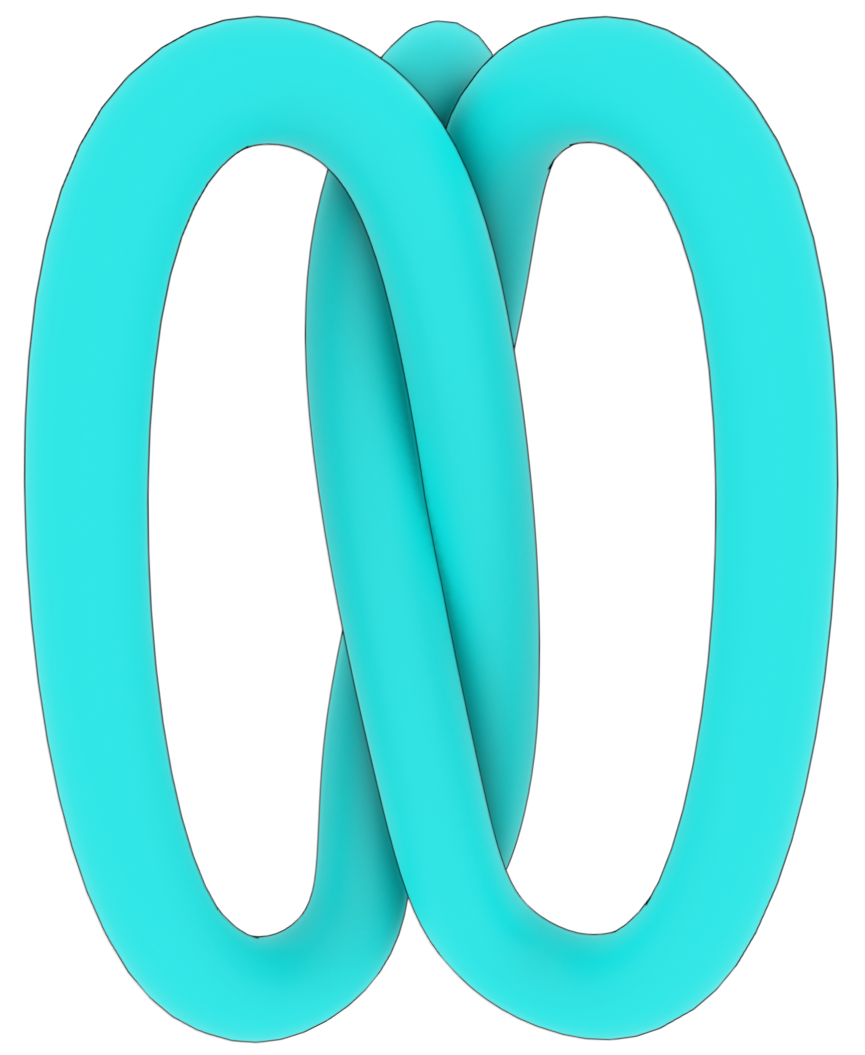}
    \includegraphics[width=0.120\textwidth]{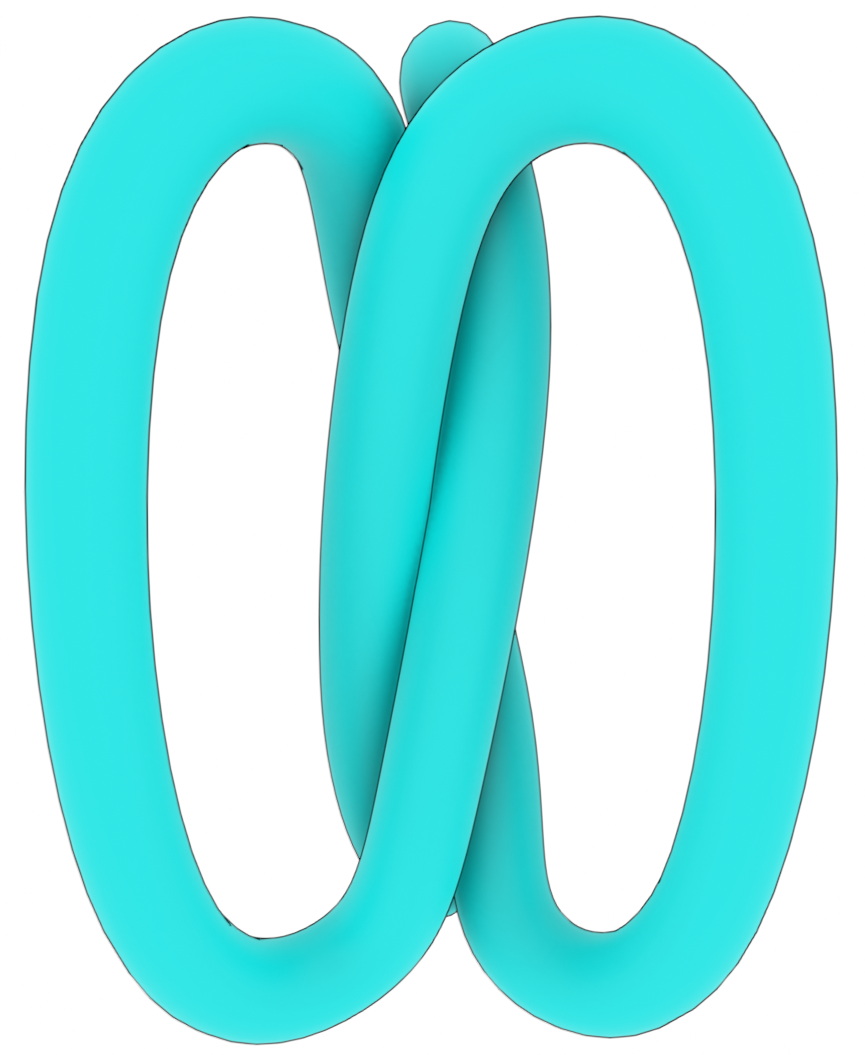}
    \includegraphics[width=0.120\textwidth]{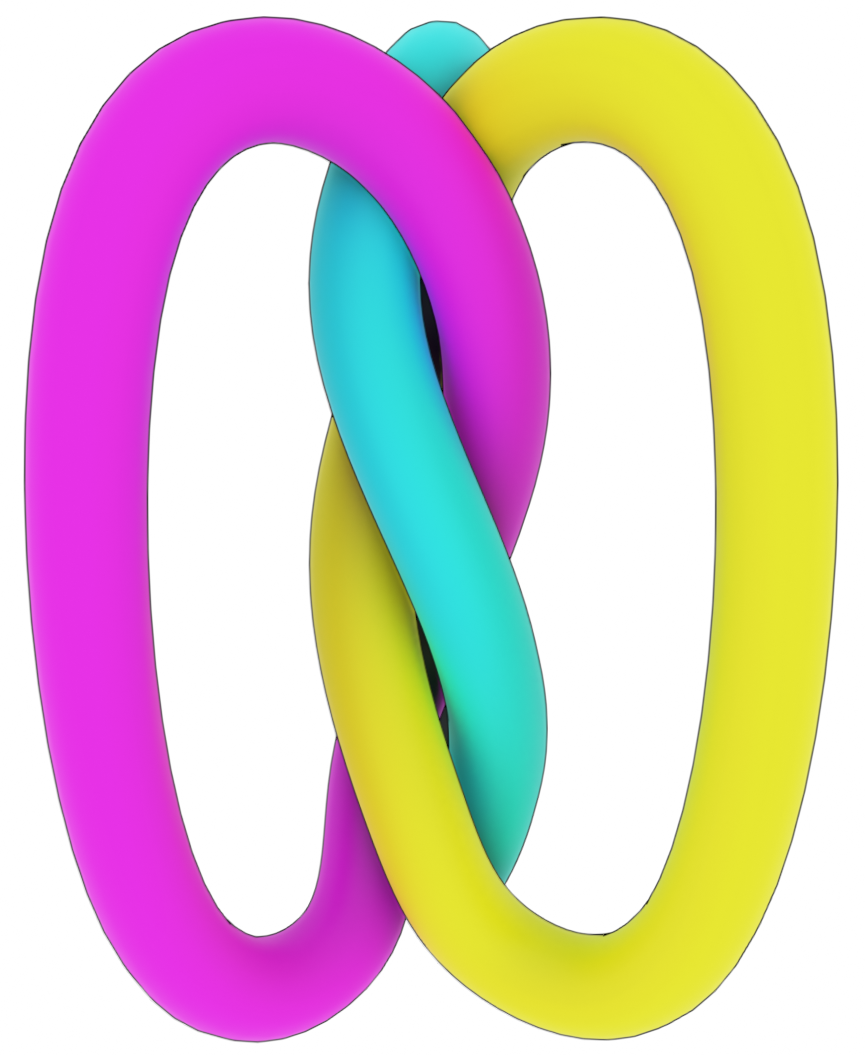}
    \includegraphics[width=0.120\textwidth]{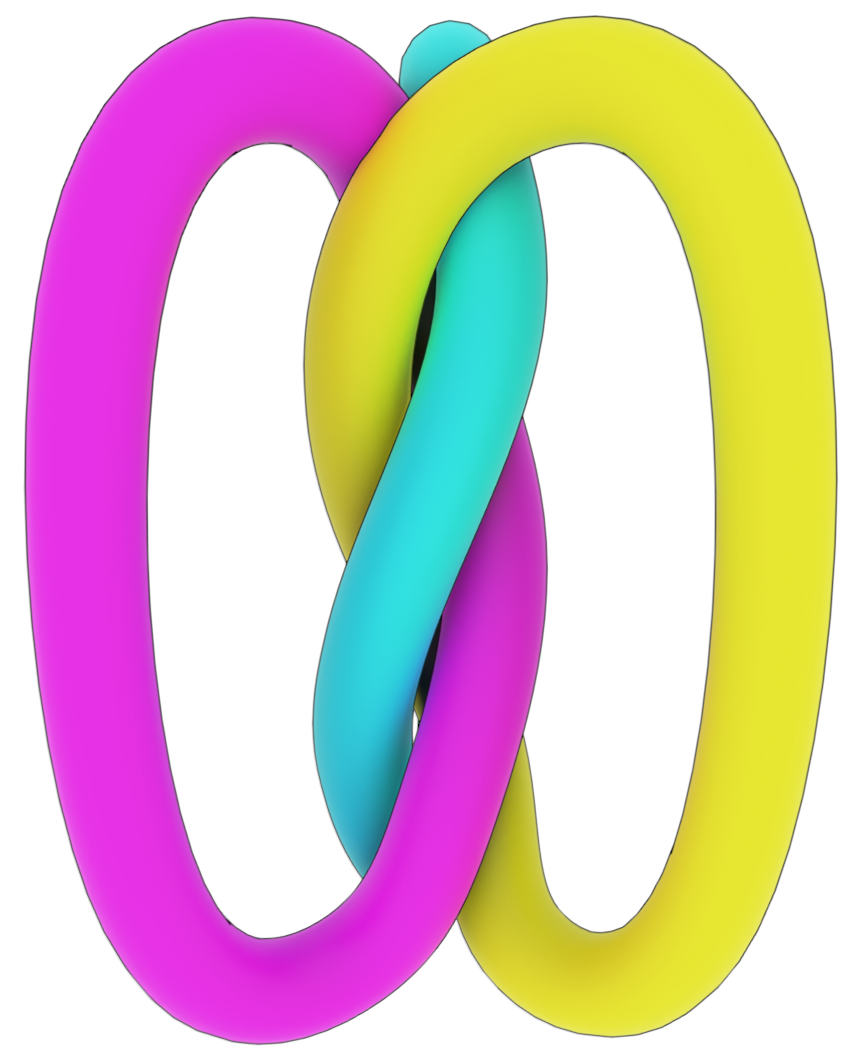}
    \includegraphics[width=0.120\textwidth]{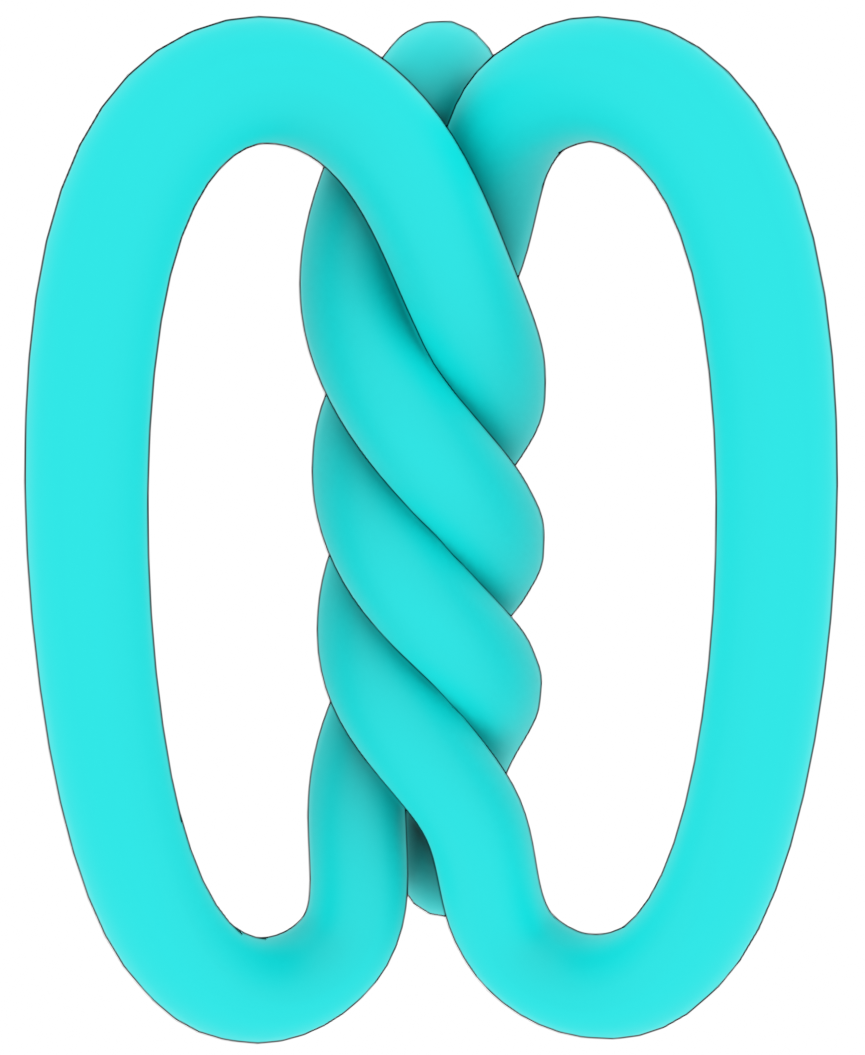}
    \includegraphics[width=0.120\textwidth]{images/chirality/4tw.png}\\
    \parbox[t]{0.120\textwidth}{\centering 1 twist}
    \parbox[t]{0.120\textwidth}{\centering -1 twist}
    \parbox[t]{0.120\textwidth}{\centering 2 twist}
    \parbox[t]{0.120\textwidth}{\centering -2 twist}
    \parbox[t]{0.120\textwidth}{\centering 3 twist}
    \parbox[t]{0.120\textwidth}{\centering -3 twist}
    \parbox[t]{0.120\textwidth}{\centering 4 twist}
    \parbox[t]{0.120\textwidth}{\centering -4 twist}
    \caption{These examples show chiral twist pairs on an edge shared by three faces. As visually shown here, the opposite signs form chiral pairs. Also note that we chose the plus sign to represent a counter-clockwise (right-handed) twist and the minus sign to represent a clockwise (left-handed) twist. These are uniquely defined in 3D space without any ambiguity (see Figure~\ref{fig:chiralspiral}).}
    \Description{A horizontal sequence of eight examples showing thick tubular loops arranged side by side. Each example depicts a pair of intertwined strands around a shared edge, rendered in cyan, with some examples including additional strands in magenta and yellow. The examples are labeled, from left to right, “1 twist,” “-1 twist,” “2 twist,” “-2 twist,” “3 twist,” “-3 twist,” “4 twist,” and “-4 twist.” Positive and negative twist examples appear as mirrored configurations, with the strands winding around the shared edge in opposite directions.}
    \label{fig:chirality}
\end{figure*}

Because the correctness of our discrete twist formalism depends on a precise notion of chirality in three dimensions, we make this assumption explicit with a visual clarification in Figures~\ref{fig:chiralspiral} and~\ref{fig:chirality}. This clarification is essential for the integer twist-label formalism introduced in Section~\ref{Sec:Twist_Labels}, where twist signs must be globally consistent across non-manifold configurations.

The key insight underlying our approach is that twisting can be treated as a combinatorial operation on mesh edges, independent of the ambient dimension. To traverse and manipulate local mesh neighborhoods, we rely on the notion of \emph{radial edge adjacency} \cite{weiler1986radialedge}, a classical concept from boundary representation modeling that provides cyclic access to faces incident on an edge. Rather than introducing a new data structure, we leverage existing radial-edge connectivity already present in modern mesh representations, specifically the corner-based mesh structure used widely in software \cite{Roosendaal1995blender}. This allows us to robustly enumerate and traverse edge-adjacent elements across manifold and non-manifold configurations.


Importantly, this work does not aim to describe a specific algorithm for computing spiral geometry or strand embeddings. Such constructions are well understood in computer graphics and can be realized using many existing techniques once the topological structure is fixed. Instead, our contribution lies in specifying the topological design space itself: how local, integer-valued decisions on mesh edges uniquely determine global linking, knotting, and articulation behavior. Geometry is treated as a downstream realization step, deliberately decoupled from the design abstraction.

Our approach is purely discrete at its core, yet yields smooth and visually rich geometric results after embedding. Because it operates uniformly on non-manifold surface meshes, it provides a general foundation for twisted structure generation across applications in computational design, digital fabrication, architectural geometry, and scientific visualization.

Beyond engineered and architectural systems and twisted structures also arise naturally in biological materials. Recent work by Jessop et al.~\cite{jessop2025butterfly} shows that the gyroid photonic structures found in butterfly wing scales develop not as smooth minimal surfaces, but as hierarchical woven fibrillar assemblies composed of helical elements. This finding challenges the common assumption that such complex geometries arise from continuous surface formation, and instead suggests that local twisting, packing, and connectivity rules play a central role in the emergence of global topology.

This biological perspective further motivates our approach. By modeling twisted structures as the result of local, discrete twisting assignments on mesh edges, our framework provides a computational abstraction for studying how complex three-dimensional knotted geometries can emerge from simple local interactions. While our method is not intended as a biological simulation, it offers a flexible modeling tool that may help explore and visualize structure–formation mechanisms observed in natural systems.

\subsection{Contributions}

Rather than introducing a new geometric construction, this paper contributes a unified topological design perspective for knotted and linked structures, grounded in non-manifold surface connectivity and integer twist labels.

The key contribution of this work is not a single geometric construction, but a design framework in which a vast space of linked and knotted structures is generated by small, local, integer labels assigned to edges. These labels act as discrete design parameters that allow systematic exploration, classification, and control of topological outcomes, rather than ad-hoc construction of individual knots.  The framework naturally produces LK structures that do not correspond to known textile patterns, architectural lattices, or previously studied link families. We intentionally separate this design layer from geometric embedding and physical realization, which allows the framework to remain general, implementation-independent, and applicable across a wide range of modeling pipelines.

To make the design principles and comparisons precise, we first introduce our conceptual framework and then position it relative to prior work. In this paper, we focus exclusively on the design and topological structure of LK structures, independent of material properties or physical simulation.

In the following section, we formalize this perspective by introducing a single local mesh operation that serves as the basis for constructing all linked knot structures considered in this work. 

\section{Conceptual Framework}

In this section, we briefly explain how non-manifold meshes, together with integer twist-labels, are required to obtain a large and expressive design space for modeling LK structures. In our framework, crossings are not prescribed geometrically; instead, they are inferred uniquely from the underlying topological scaffold defined by the labeled mesh, making the resulting knot structure independent of geometric realization.

The underlying scaffold used in this work originates from earlier block-mesh theory \cite{akleman2015block}, where face sides form cycles and 3-edge neighborhoods form cylindrical pipes; here, we reinterpret these structures as design primitives rather than merely representational constructs.

Throughout this section, we reason exclusively about connectivity, cycles, and permutations induced by twist labels. No assumptions are made about how the resulting strands are geometrically embedded as spirals; any valid embedding that respects the induced connectivity is sufficient.

Together, non-manifold connectivity and integer twist-labels form a minimal yet expressive parameterization that transforms mesh edges into discrete design handles, enabling exponential growth of realizable knot and link configurations.

For a mesh with E edges, each admitting multiple integer twist choices, the number of distinct labeled configurations grows combinatorially with E, which explains why even simple meshes such as cubes already generate a large and diverse family of linked-knot structures.

While twisting itself is ancient, our contribution lies in formalizing twist as a discrete, composable, and algorithmically controllable labeling system on arbitrary non-manifold meshes.
In Figures~\ref{fig:teaser}, and ~\ref{fig:teaser2}, we intentionally use simple polyhedra such as cubes and truncated octahedra as canonical bases, because they make the combinatorial role of twist-labels explicit; the same framework applies unchanged to any arbitrary mesh. These simple examples demonstrate that the novelty is not the act of twisting, but the realization that integer-labeled twisting on surface-based scaffolds defines a coherent, extensible, and computable design space that subsumes multiple prior representations while enabling structures previously inaccessible.

\subsection{Non-Manifold Surface Meshes}
\label{Sec:Non_Manifolds}

We adopt \emph{non-manifold surface meshes} \cite{hoffmann1989} as the starting point for generating links and knots because they provide a \textbf{natural and unambiguous combinatorial scaffold} for encoding cyclic structures and crossings directly in three dimensions. Unlike curve-based or projection-based approaches, this representation allows links and knots to be constructed and manipulated \emph{intrinsically in 3D}, without relying on planar diagrams, which often obscure spatial relationships and introduce unnecessary complexity.

In earlier work on block meshes and 3-manifold modeling \cite{akleman2015block}, higher-dimensional cells were required to guarantee topological robustness for general shape representations. In the present setting, however, modeling knots and links requires significantly less structure. Because knots arise as cycles and links arise from interactions between cycles, all necessary topological information is already encoded at the level of faces and their adjacency. As a result, non-manifold surface meshes form a minimal yet sufficient scaffold for knot and link design: volumetric cells are unnecessary, and even surface orientation is not required. Face cycles and their reconnection through twisted edges fully determine the resulting topology.

\textbf{First}, each face in a non-manifold surface mesh defines a \emph{cycle}. This cycle can be embedded directly as a closed curve in three-dimensional space, providing a natural representation for the strands that constitute links and knots. By associating strands with \emph{face cycles} rather than with abstract curves, we obtain closed loops that are inherently tied to the mesh connectivity, ensuring topological consistency and robustness under local modifications. Notably, the framework does not rely on a global orientation of the surface: twist labels encode local chirality directly in three dimensions, making surface orientation irrelevant for determining knot topology.

\textbf{Second}, non-manifold surfaces permit edges to be shared by \emph{more than two faces}, creating regions where multiple cycles meet. These shared edges correspond precisely to locations where \emph{crossings can occur}. Rather than specifying crossings through planar over-under conventions, we generate crossings \emph{directly in 3D} by twisting edges. Edge twisting locally permutes the connectivity between adjacent face cycles, producing over-under relationships as a geometric consequence of the twisting operation. In this way, crossings emerge intrinsically from the three-dimensional structure, eliminating the need to reason about projections or knot diagrams.

\begin{figure*}[htb!]
    \centering
    \includegraphics[width=0.19\textwidth]{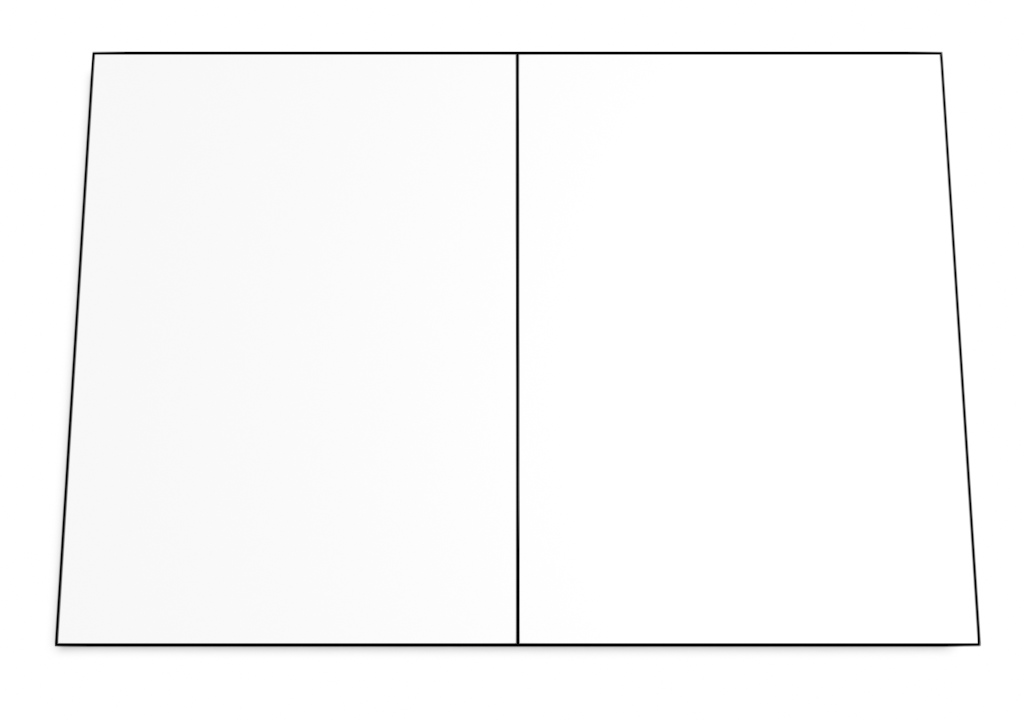}
    \includegraphics[width=0.19\textwidth]{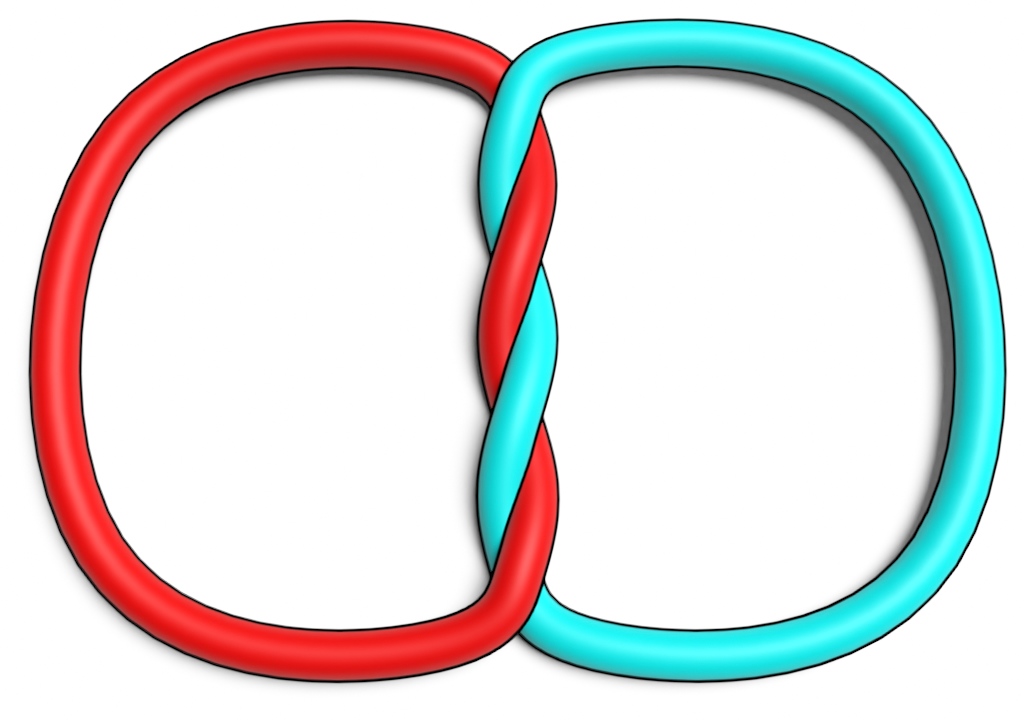}
    \includegraphics[width=0.19\textwidth]{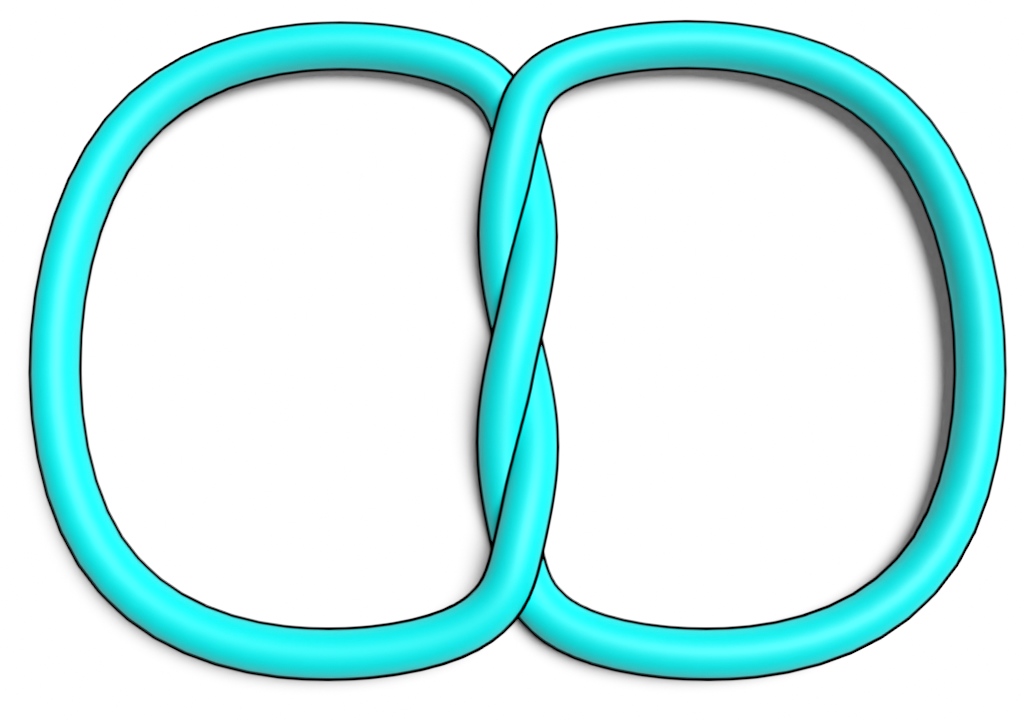}
    \includegraphics[width=0.19\textwidth]{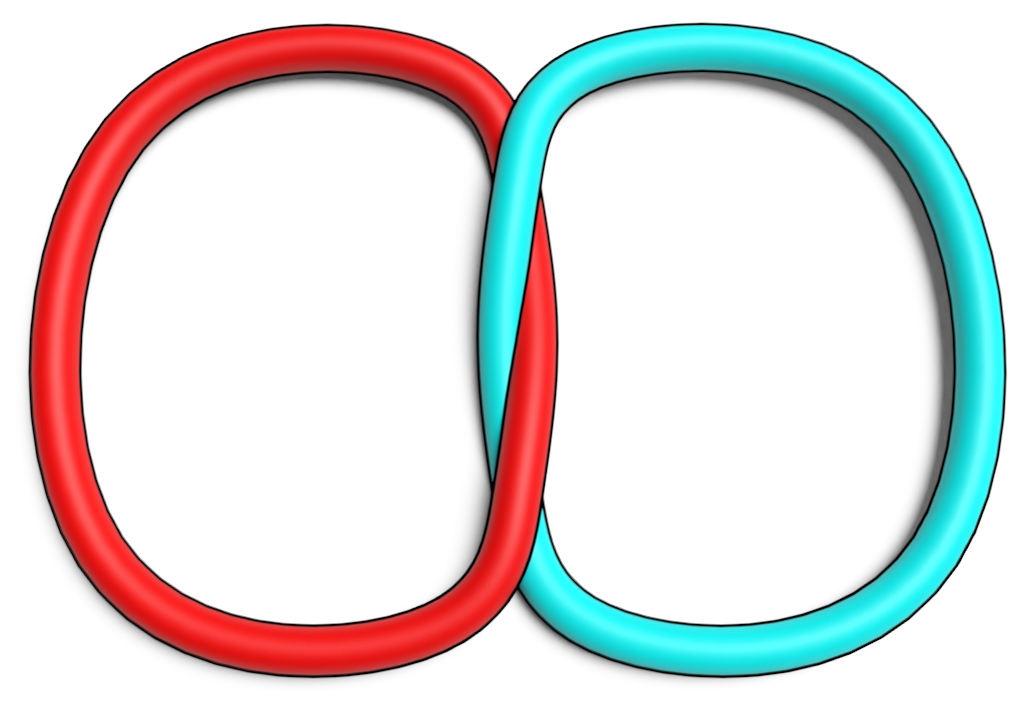}
    \includegraphics[width=0.19\textwidth]{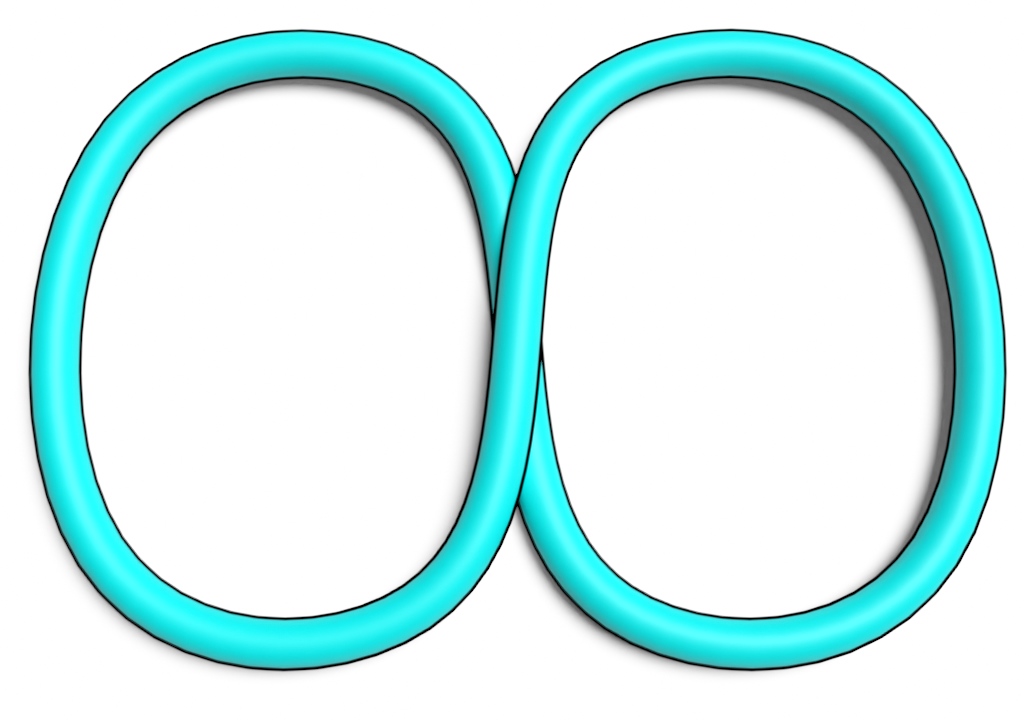}\\
    \parbox[t]{0.19\textwidth}{\centering Input mesh}
    \parbox[t]{0.19\textwidth}{\centering -4 twist}
    \parbox[t]{0.19\textwidth}{\centering -3 twist}
    \parbox[t]{0.19\textwidth}{\centering -2 twist}
    \parbox[t]{0.19\textwidth}{\centering -1 twist}\\
    \includegraphics[width=0.19\textwidth]{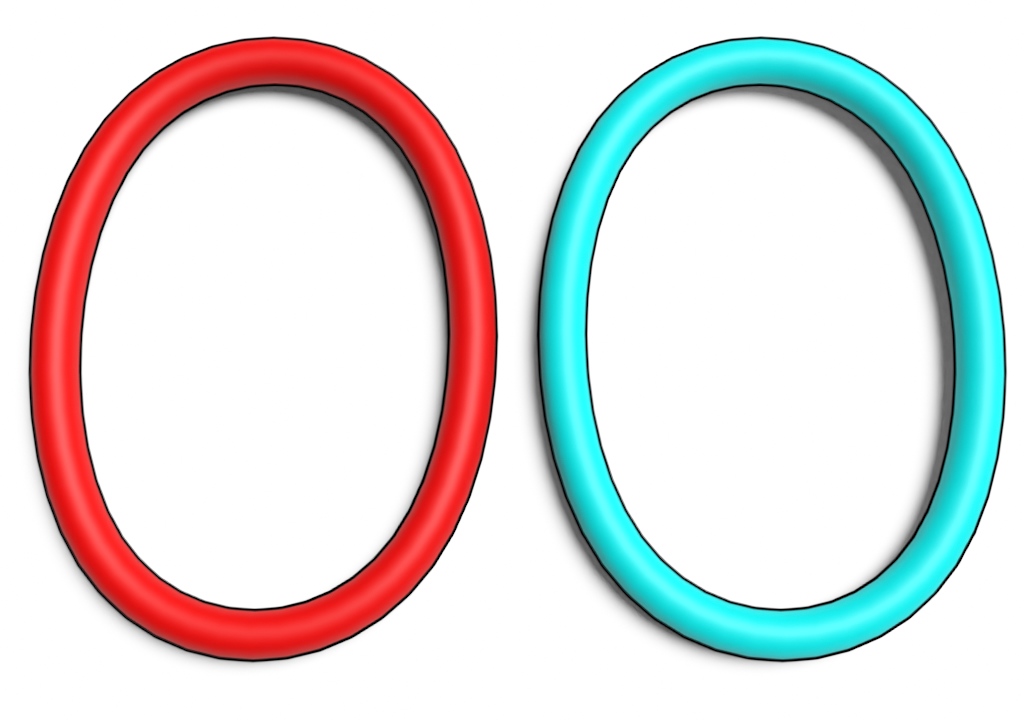}
    \includegraphics[width=0.19\textwidth]{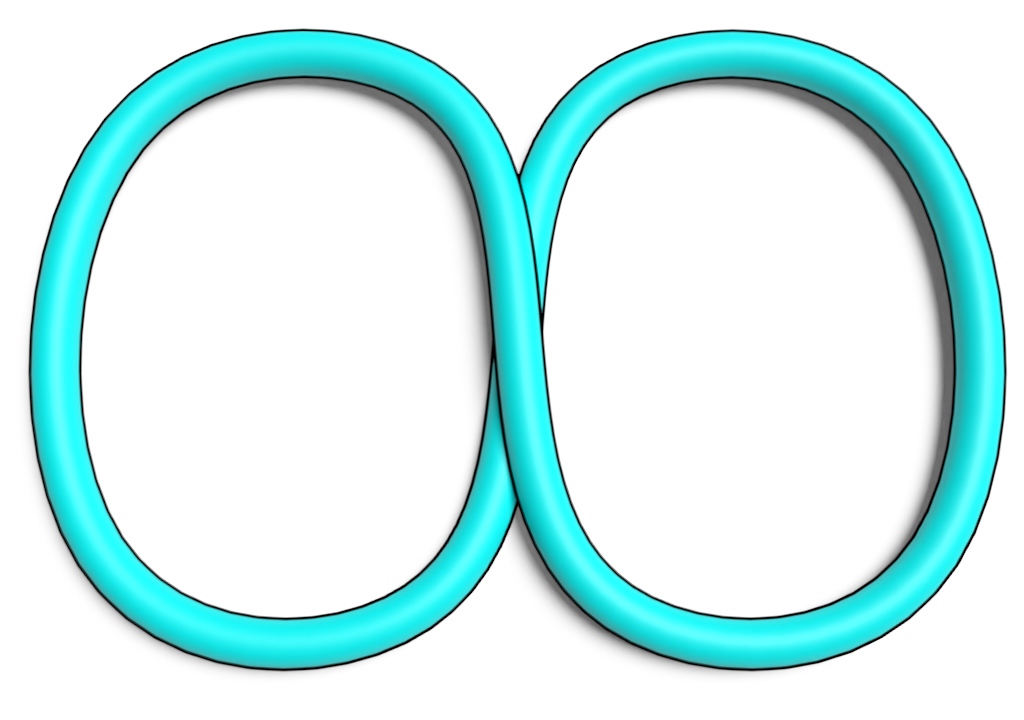}
    \includegraphics[width=0.19\textwidth]{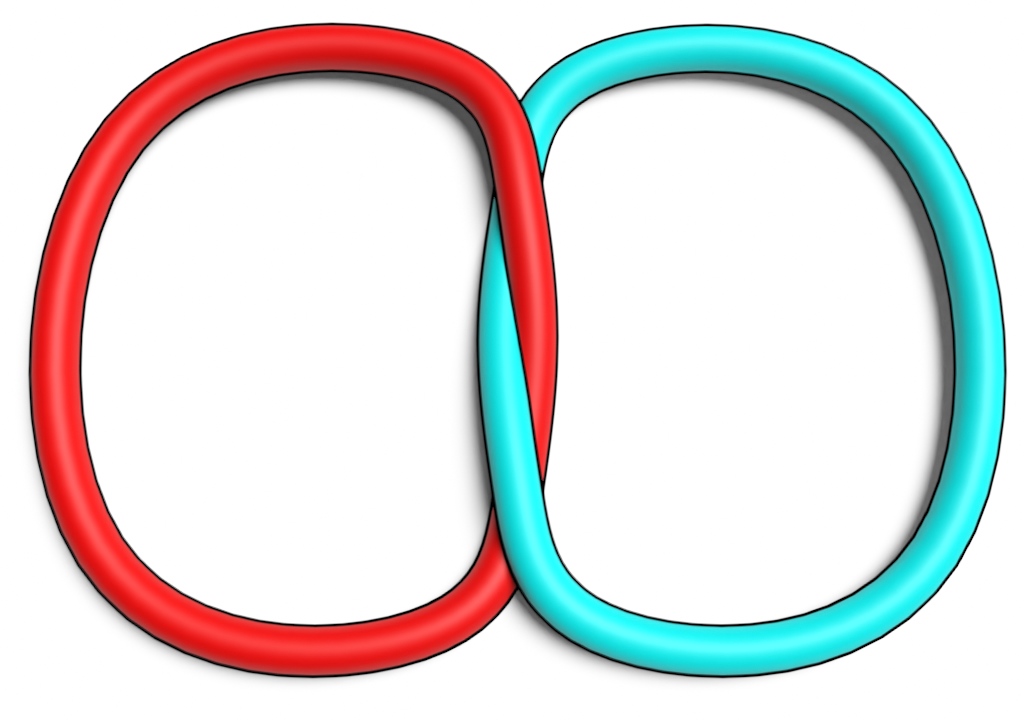}
    \includegraphics[width=0.19\textwidth]{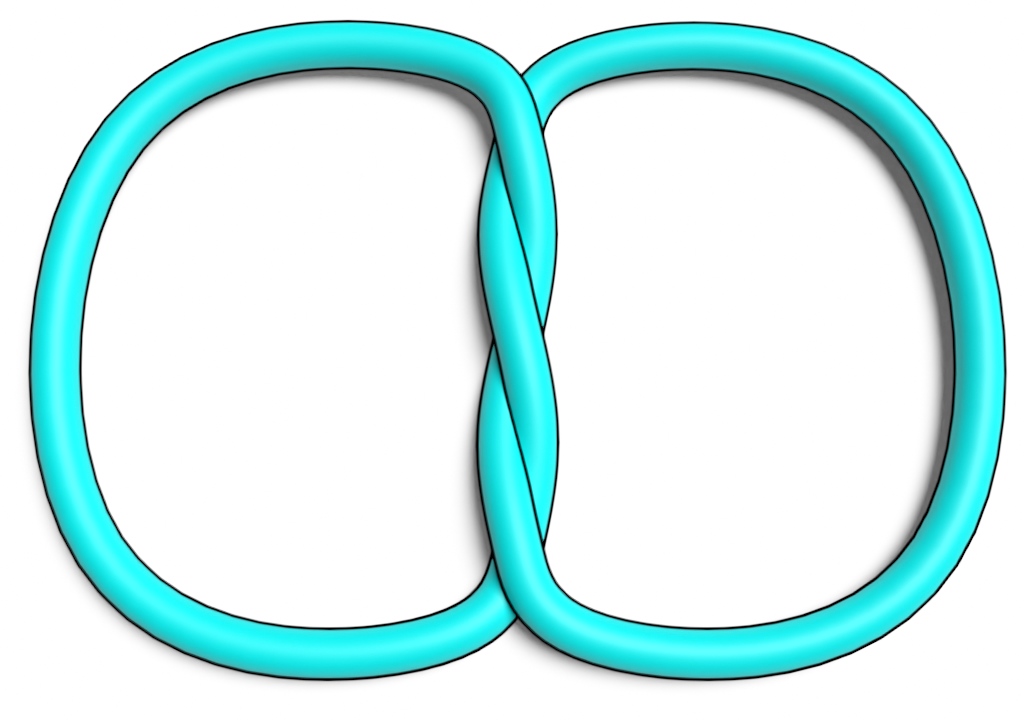}
    \includegraphics[width=0.19\textwidth]{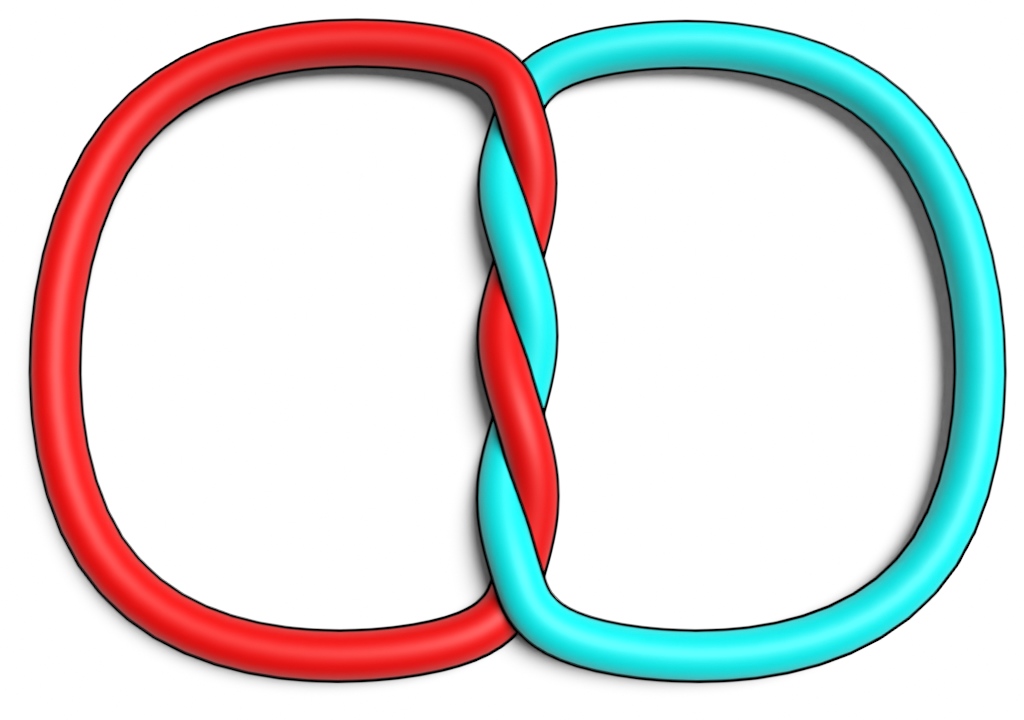}\\
    \parbox[t]{0.19\textwidth}{\centering 0 twist}
    \parbox[t]{0.19\textwidth}{\centering 1 twist}
    \parbox[t]{0.19\textwidth}{\centering 2 twist}
    \parbox[t]{0.19\textwidth}{\centering 3 twist}
    \parbox[t]{0.19\textwidth}{\centering 4 twist}\\
    \caption{This example shows the effect of twist-labels on an edge shared by two faces. Even twists result in disconnected components; non-zero even twists generate interlocked rings. An odd number of twists generates single-cycle curves, resulting in a surface knot.}
    \Description{A grid of examples illustrating the effect of different twist values on an edge shared by two faces. At the top left, a simple rectangular input mesh with a single vertical edge is shown. The remaining images show pairs of thick tubular loops rendered in red and cyan, labeled with twist values from “-4 twist” to “4 twist,” including a “0 twist” case. In the zero-twist example, the two loops appear as separate, non-intertwined ovals. In some non-zero even-twist examples, the two loops remain separate but are visibly interlocked around the shared edge. In odd-twist examples, the strands merge into a single continuous loop winding around the edge. Positive and negative twists produce mirrored winding configurations.}
    \label{fig:2_mantwist}
\end{figure*}

\textbf{Third}, the crossings induced by edge twisting are not limited to simple binary interactions. Because multiple faces may be incident on a single edge, a \emph{single twisting operation} can generate multiple, intertwined crossings locally. This enables the construction of complex linking and knotting behavior that would be difficult to design explicitly using curve-based methods. In practice, controlling such configurations without additional structure is challenging: the space of possible crossings grows rapidly, and unguided constructions easily lead to ambiguity or unintended topology.

An important consequence of formulating our framework on non-manifold surface meshes is that it naturally subsumes \emph{2-manifold surfaces with boundary} as a special case. By restricting the incidence structure so that each edge is shared by at most two faces, and allowing boundary edges to be incident to only a single face, the same representation yields standard 2-manifold surfaces with boundaries. In this setting, the number of cycles incident on an edge can be as low as one, providing a simple and intuitive configuration that aligns with classical surface modeling.

This inclusion is not merely theoretical, but provides additional practical control. Because the number of faces incident on an edge directly determines the number of cycles participating in twisting, boundaries of 2-manifold surfaces allow us to explicitly control edge degree and, consequently, the number of strands that may interact at that edge. Non-manifold edges generalize this behavior by permitting higher degrees, while boundary edges represent the minimal case.

For these reasons, non-manifold surface meshes provide a particularly effective foundation for generating links and knots. They supply \emph{explicit cycles}, \emph{natural crossing sites}, and a \emph{spatially grounded mechanism} for creating crossings, all within a unified three-dimensional framework. However, to reliably control the \emph{number}, \emph{orientation}, and \emph{complexity} of crossings produced at each edge, additional discrete guidance is required, motivating the introduction of \emph{twist labels} described next.

\begin{figure*}[htb!]
    \centering    
    \includegraphics[width=0.19\textwidth]{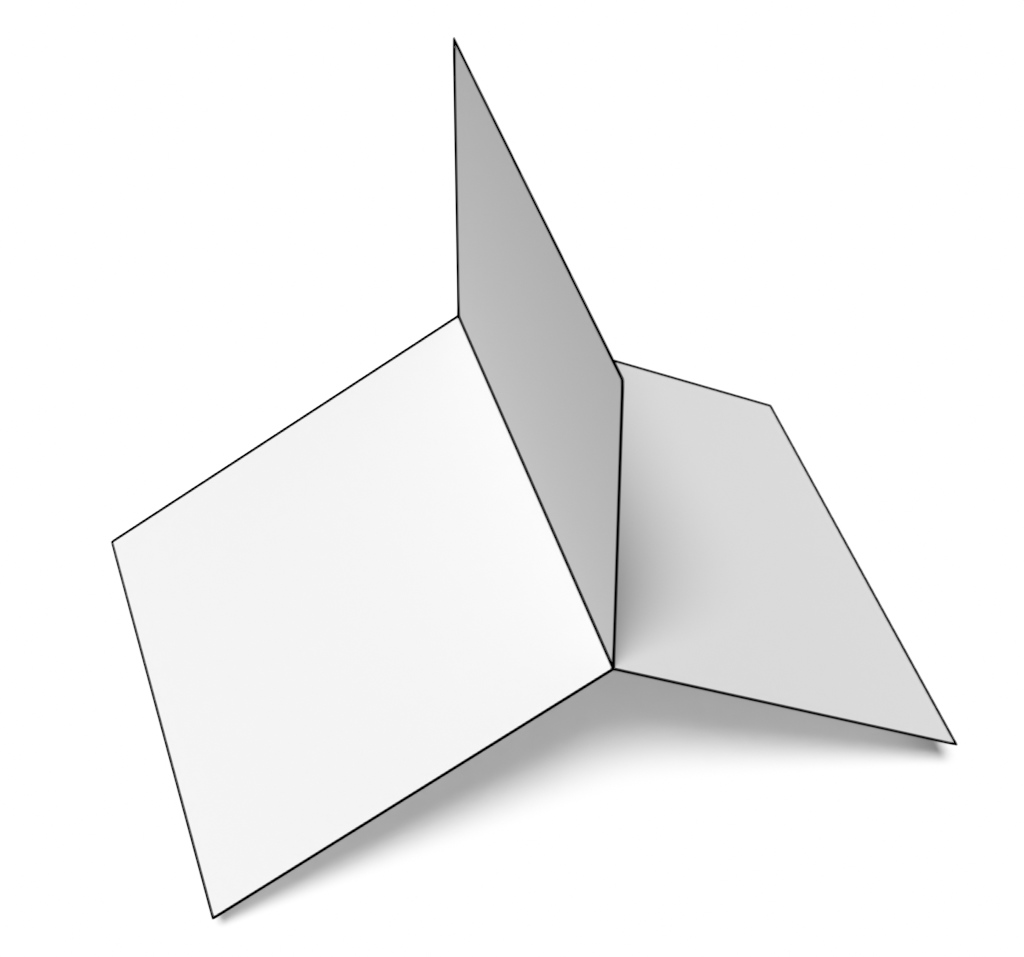}    
    \includegraphics[width=0.19\textwidth]{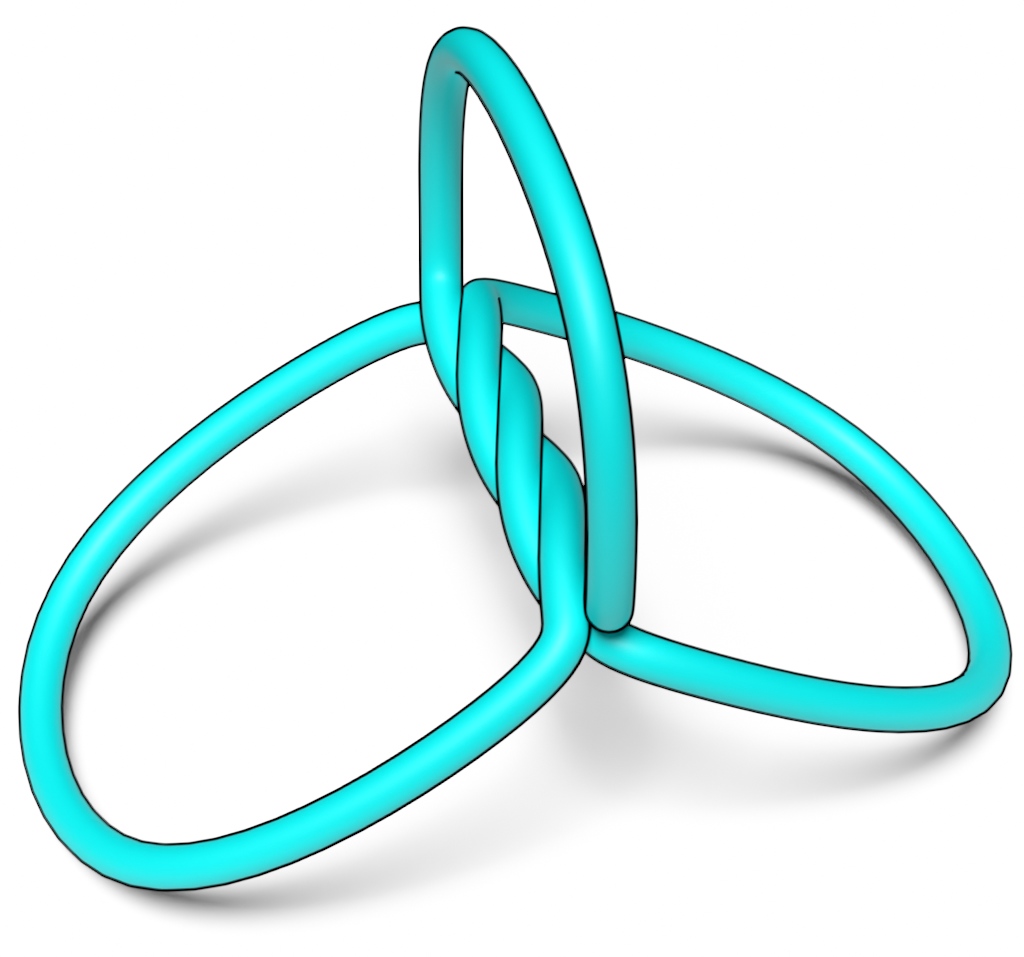}    
    \includegraphics[width=0.19\textwidth]{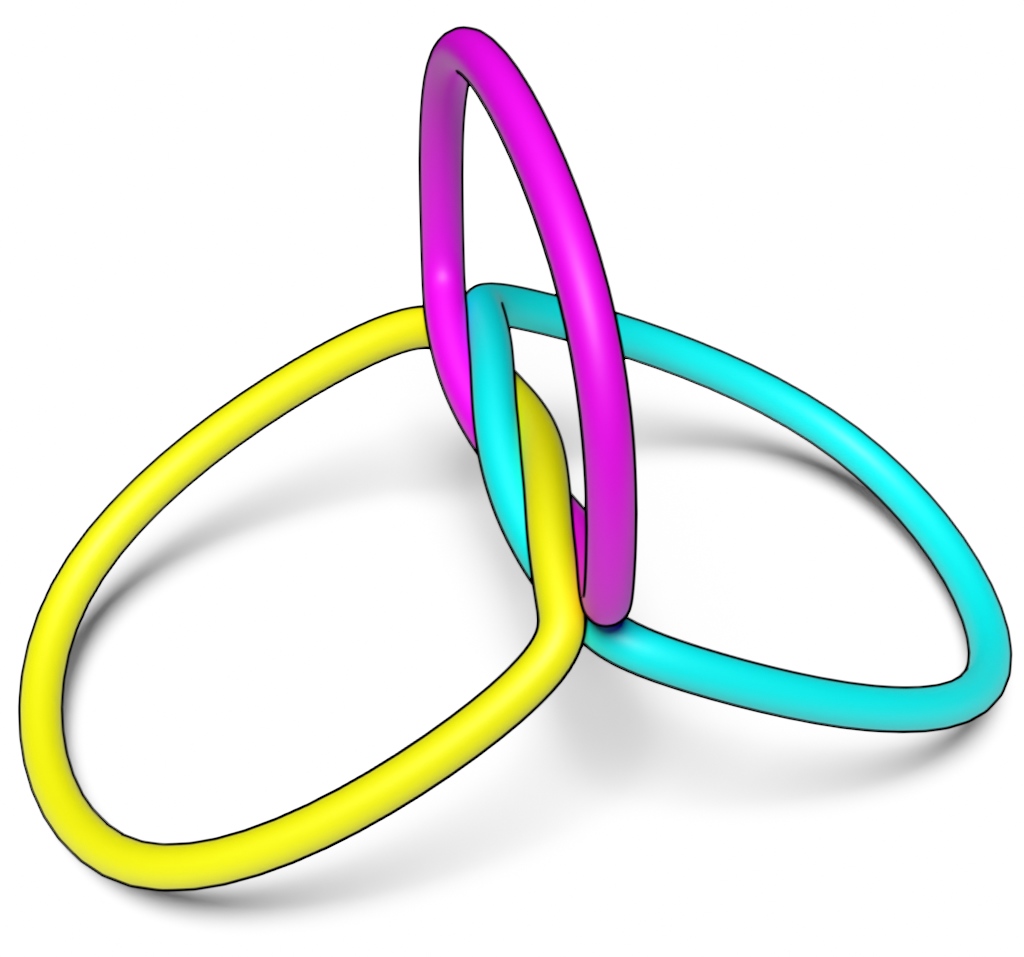}    
    \includegraphics[width=0.19\textwidth]{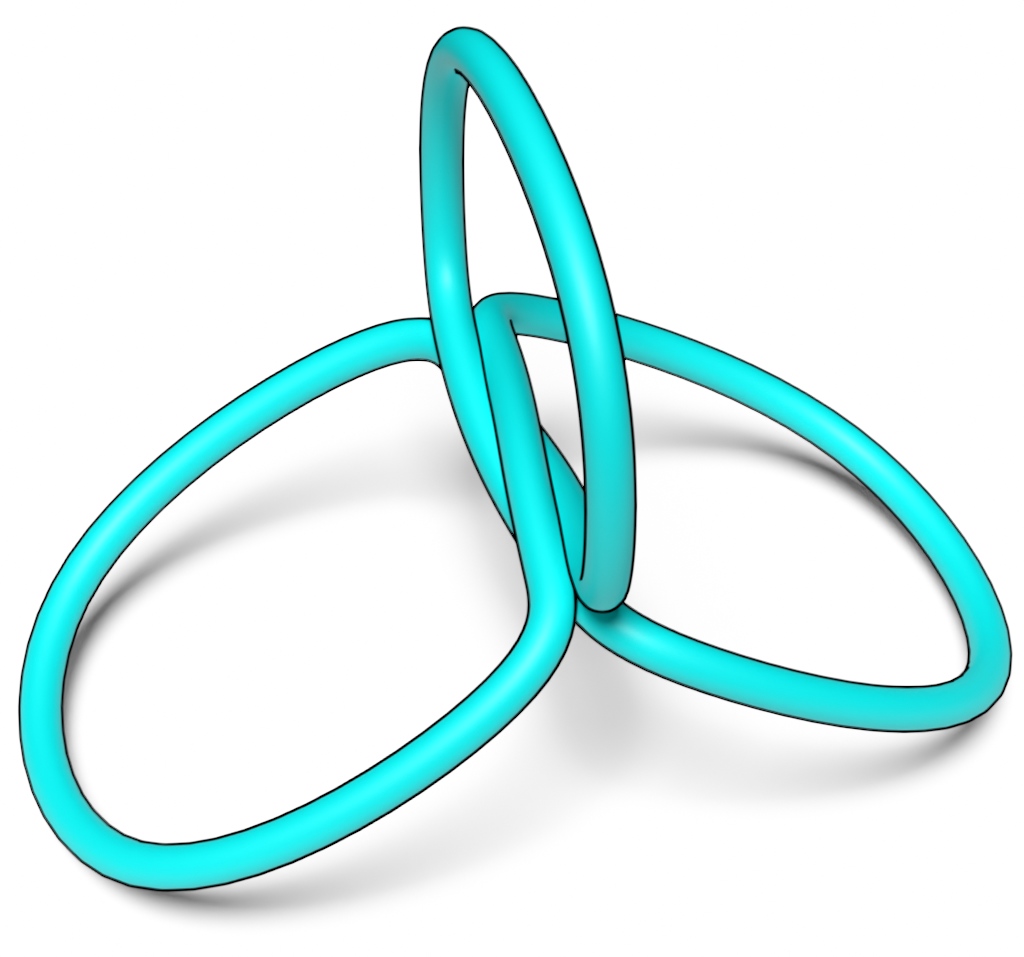}    
    \includegraphics[width=0.19\textwidth]{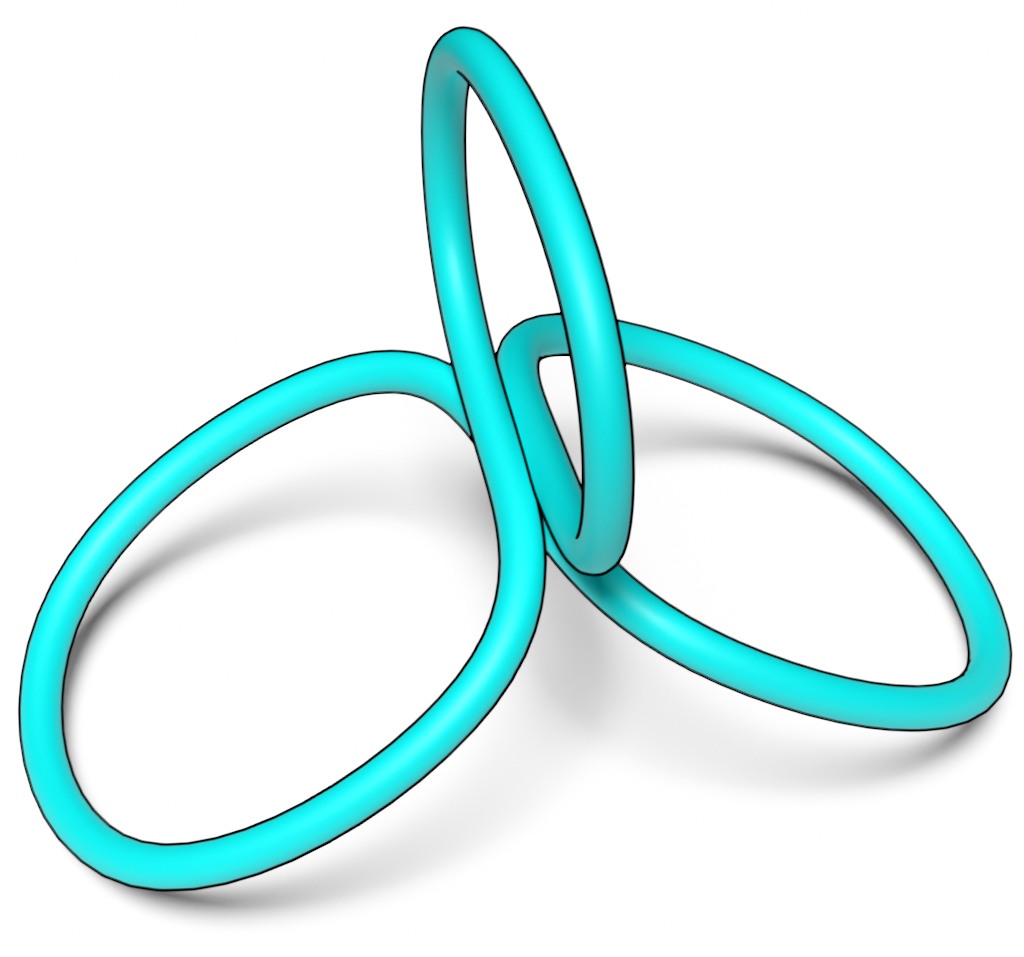}\\
    \parbox[t]{0.19\textwidth}{\centering Input mesh}
    \parbox[t]{0.19\textwidth}{\centering -4 twist}
    \parbox[t]{0.19\textwidth}{\centering -3 twist}
    \parbox[t]{0.19\textwidth}{\centering -2 twist}
    \parbox[t]{0.19\textwidth}{\centering -1 twist}\\    
    \includegraphics[width=0.19\textwidth]{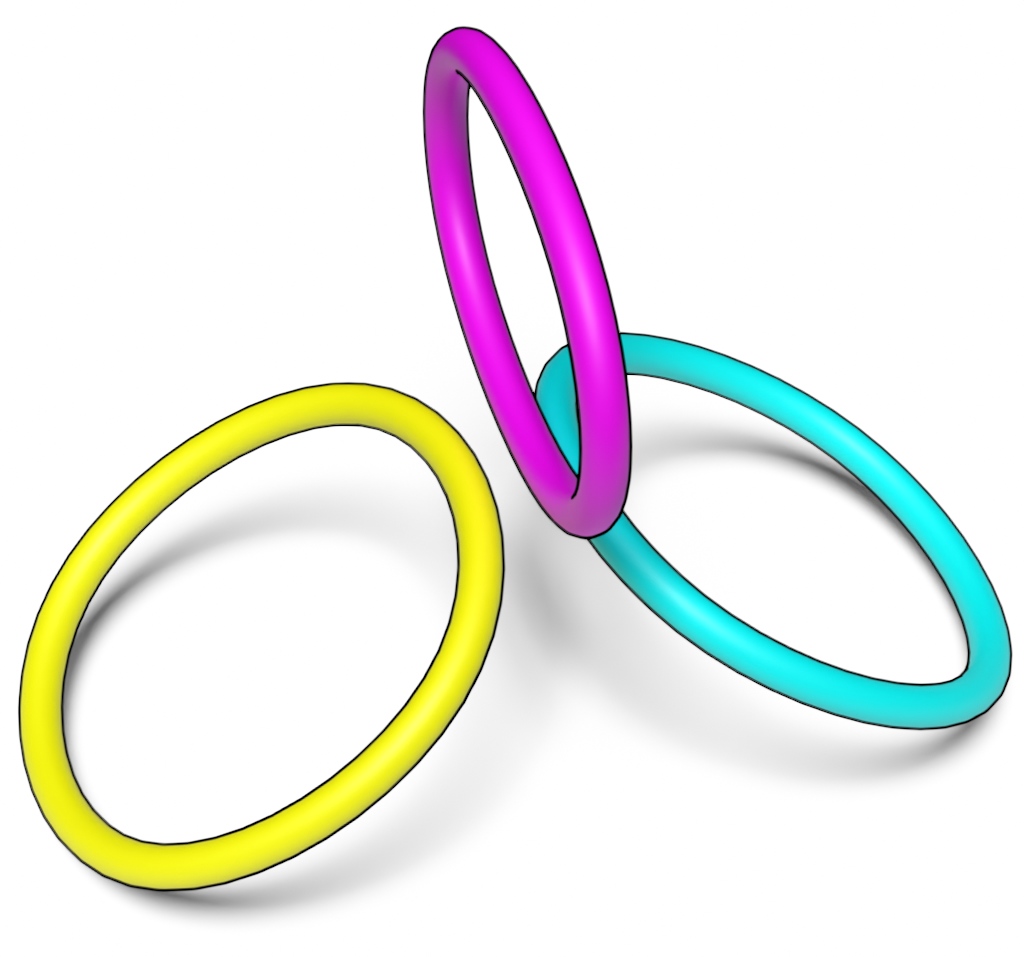}    
    \includegraphics[width=0.19\textwidth]{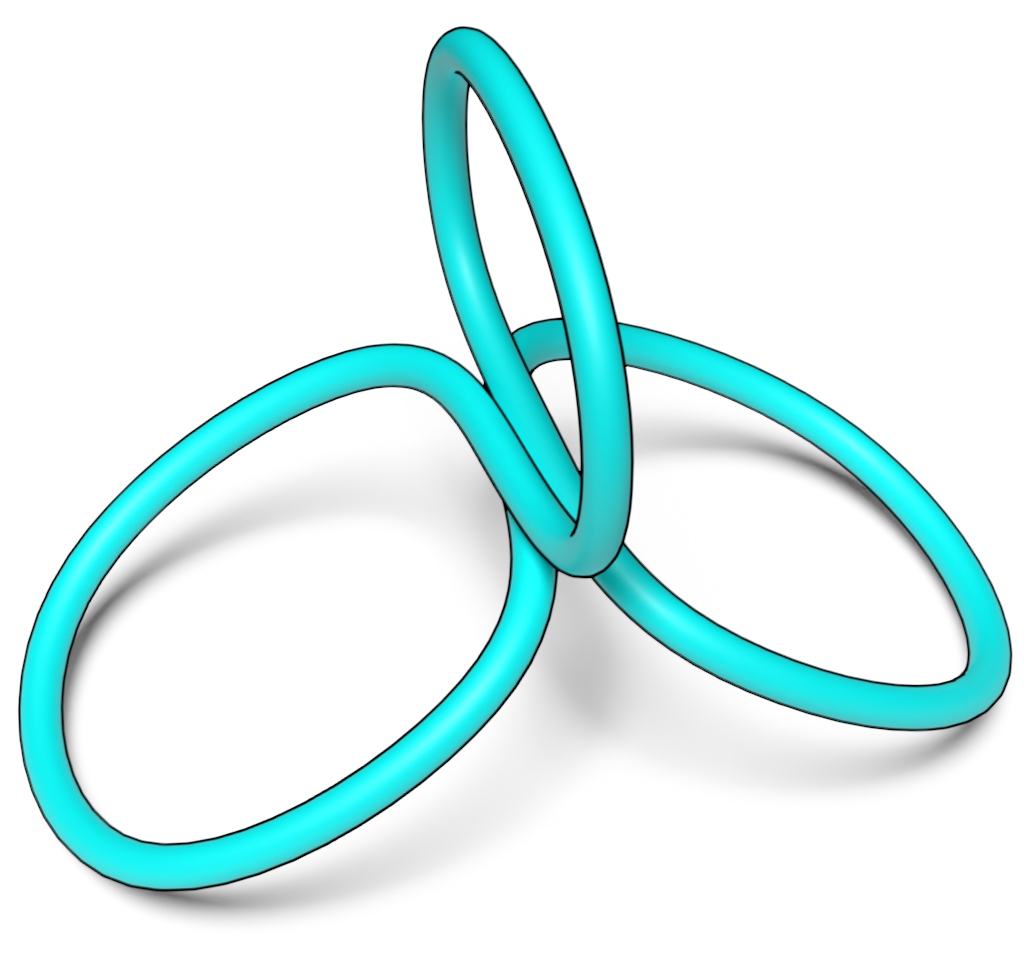}    
    \includegraphics[width=0.19\textwidth]{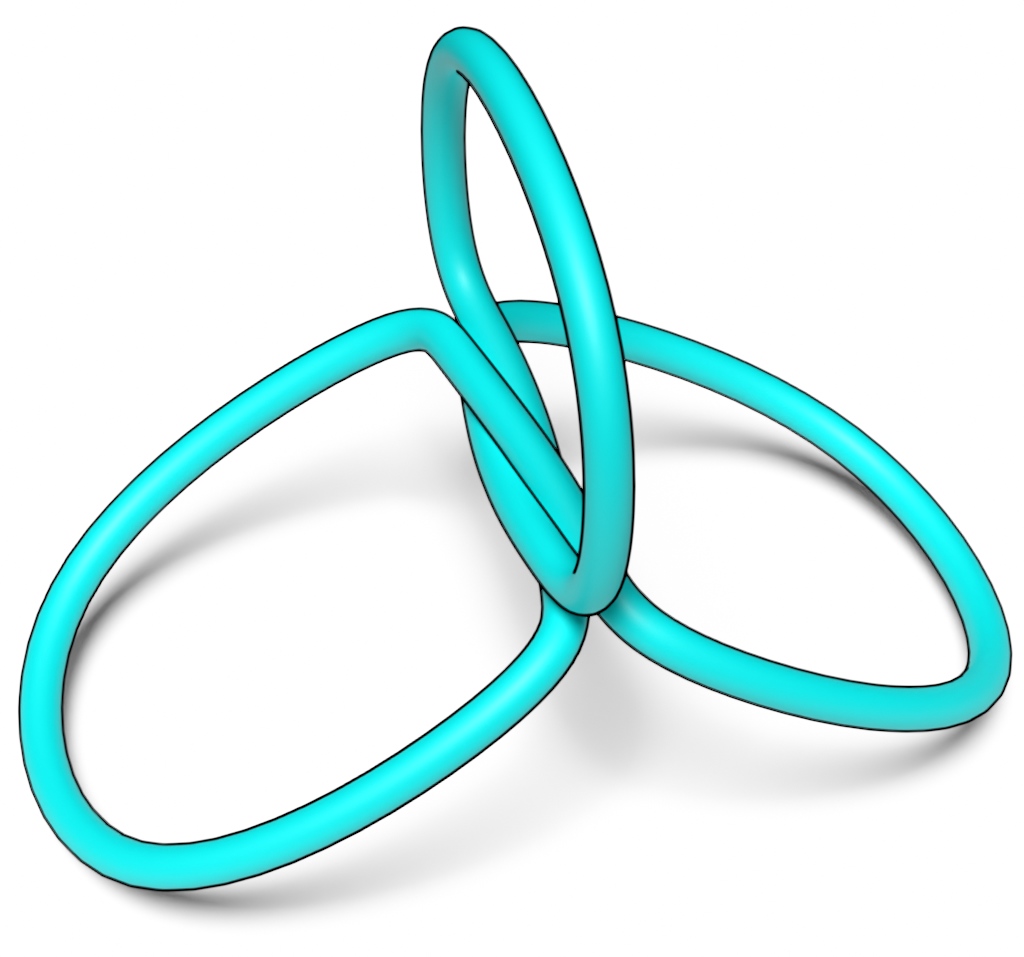}   
    \includegraphics[width=0.19\textwidth]{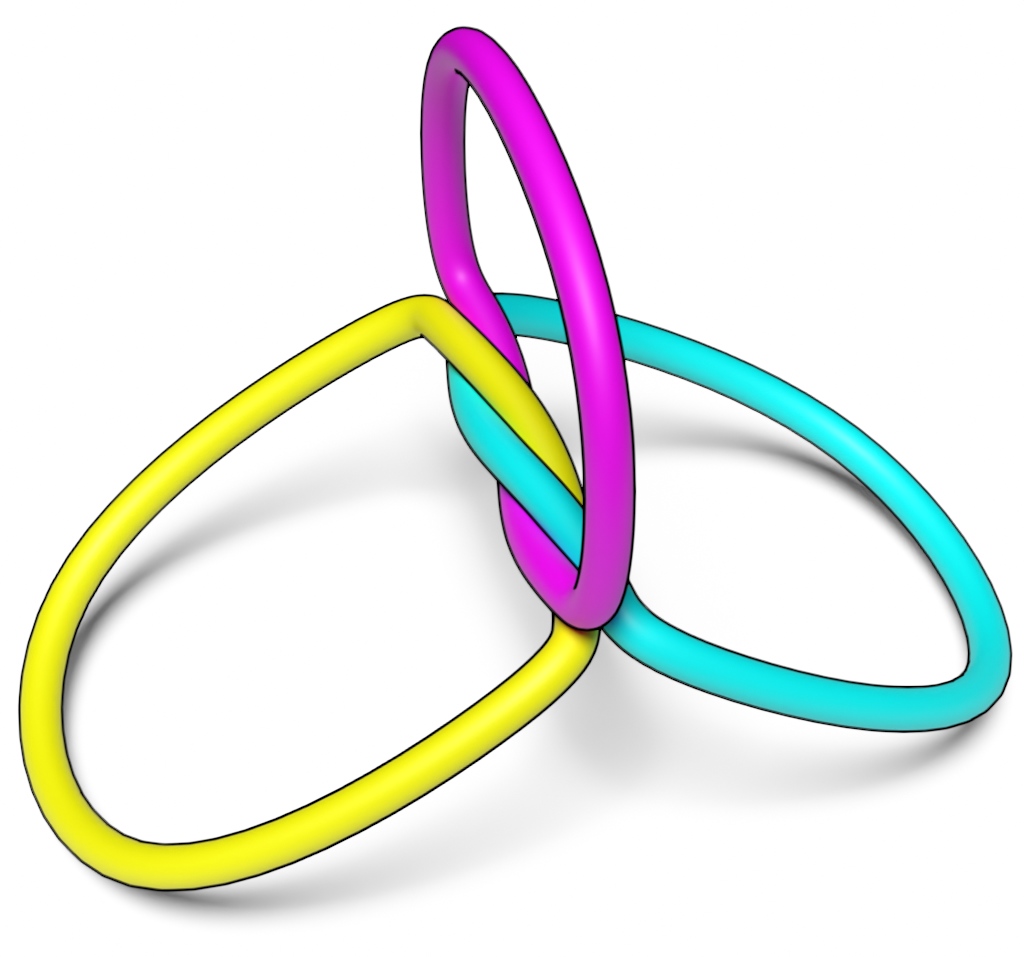}   
    \includegraphics[width=0.19\textwidth]{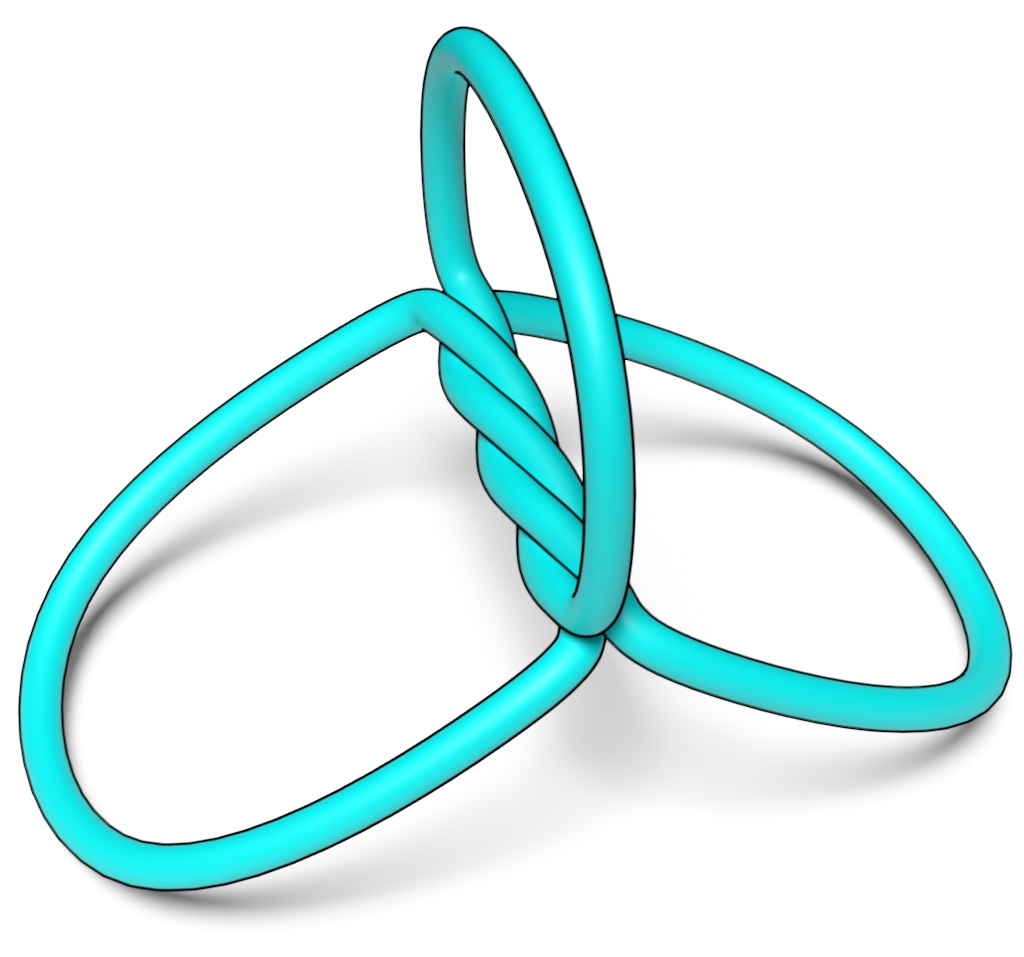}\\
    \parbox[t]{0.19\textwidth}{\centering 0 twist}
    \parbox[t]{0.19\textwidth}{\centering 1 twist}
    \parbox[t]{0.19\textwidth}{\centering 2 twist}
    \parbox[t]{0.19\textwidth}{\centering 3 twist}
    \parbox[t]{0.19\textwidth}{\centering 4 twist}\\
    \caption{These examples show the effect of twist-labels on an edge shared by three faces. The twists that are multiples of three result in disconnected but interlocked rings. Any other twist number creates a single connected cycle by combining three original cycles.}
    \Description{A grid of examples showing the effect of different twist values on an edge shared by three faces. At the top left, an input mesh with three faces meeting along a single edge is shown. The remaining images show thick tubular strands rendered primarily in cyan, with some examples including additional strands in magenta and yellow, labeled with twist values from “-4 twist” to “4 twist,” including a “0 twist” case. In the zero-twist example, three separate loops are visible without interlocking. In some other examples, multiple loops remain separate but are visibly interlocked around the shared edge. In the remaining examples, the strands merge into one continuous loop winding around the edge, forming a single connected cycle. Positive and negative twists produce mirrored winding configurations.}
    \label{fig:3_mantwist}
\end{figure*}

\subsection{Integer Twist-Labels}
\label{Sec:Twist_Labels}

While non-manifold surfaces provide natural sites for cycles and crossings, controlling how these cycles are connected is essential for constructing meaningful links and knots. Without additional guidance, local twisting operations can easily produce ambiguous or unintended connectivity. To address this, we associate each edge with an \emph{integer-valued twist label}, referred to as a \emph{twist-label}, which governs how incident face cycles are connected during traversal.

Each twist-label is an integer that encodes both \emph{orientation} and \emph{multiplicity} of twisting. Positive integers (e.g., $+1$, $+2$) denote counterclockwise twists, while negative integers (e.g., $-1$, $-2$) denote clockwise twists. Geometrically, a twist-label specifies how strands are permuted as they pass through the local neighborhood of an edge. Combinatorially, it defines a deterministic re-indexing of incident face cycles, allowing twisting to be applied uniformly across the mesh.

Importantly, positive and negative twist-labels are \emph{chiral} to each other (see Figure~\ref{fig:chirality}). A clockwise and a counterclockwise twist cannot be continuously deformed into one another without reversing orientation, and thus represent intrinsically different local configurations. This chirality is well-defined in three dimensions and does not depend on projection or viewpoint (see Figure~\ref {fig:chiralspiral}), a property that is essential for defining consistent twist labels on non-manifold meshes. In particular, the direction of a twist is invariant under the choice of endpoint: a clockwise twist remains clockwise regardless of which end of the edge is used as the reference. This guarantees that twist-labels encode an intrinsic local orientation, eliminating ambiguity even in complex three-dimensional arrangements.

Crucially, the effect of a twist-label depends on the \emph{degree} of the edge, that is, the number $K$ of faces incident on that edge. Each of these faces defines a cycle, and twisting the edge induces a permutation of these $K$ cycles. This behavior is naturally interpreted through a modular arithmetic perspective: twist-labels operate modulo $K$. When the absolute value of the twist-label is relatively prime to $K$, the induced permutation combines all $K$ cycles into a single cycle, producing a continuous strand that can be interpreted as a knot. In contrast, when the twist-label shares a common divisor with $K$, the cycles are partitioned into multiple disjoint groups, resulting in multiple strands that may be linked or remain separate. More generally, the number of independent cycles created by a twist is equal to the greatest common divisor of the twist-label and $K$, which also explains why a relatively prime relationship between the twist-label and $K$ necessarily produces exactly one cycle.

\begin{figure*}[htb!]
    \centering    \includegraphics[width=0.16\textwidth]{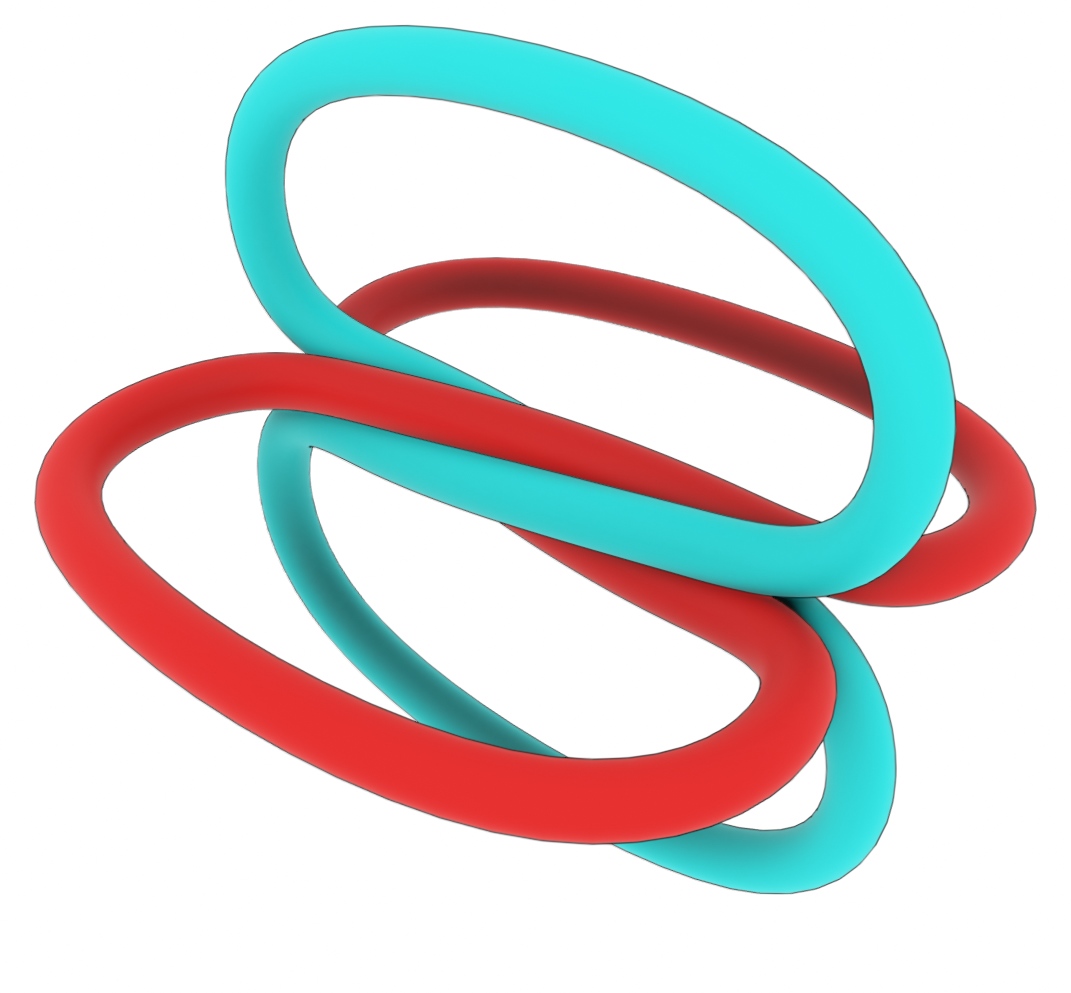}    \includegraphics[width=0.16\textwidth]{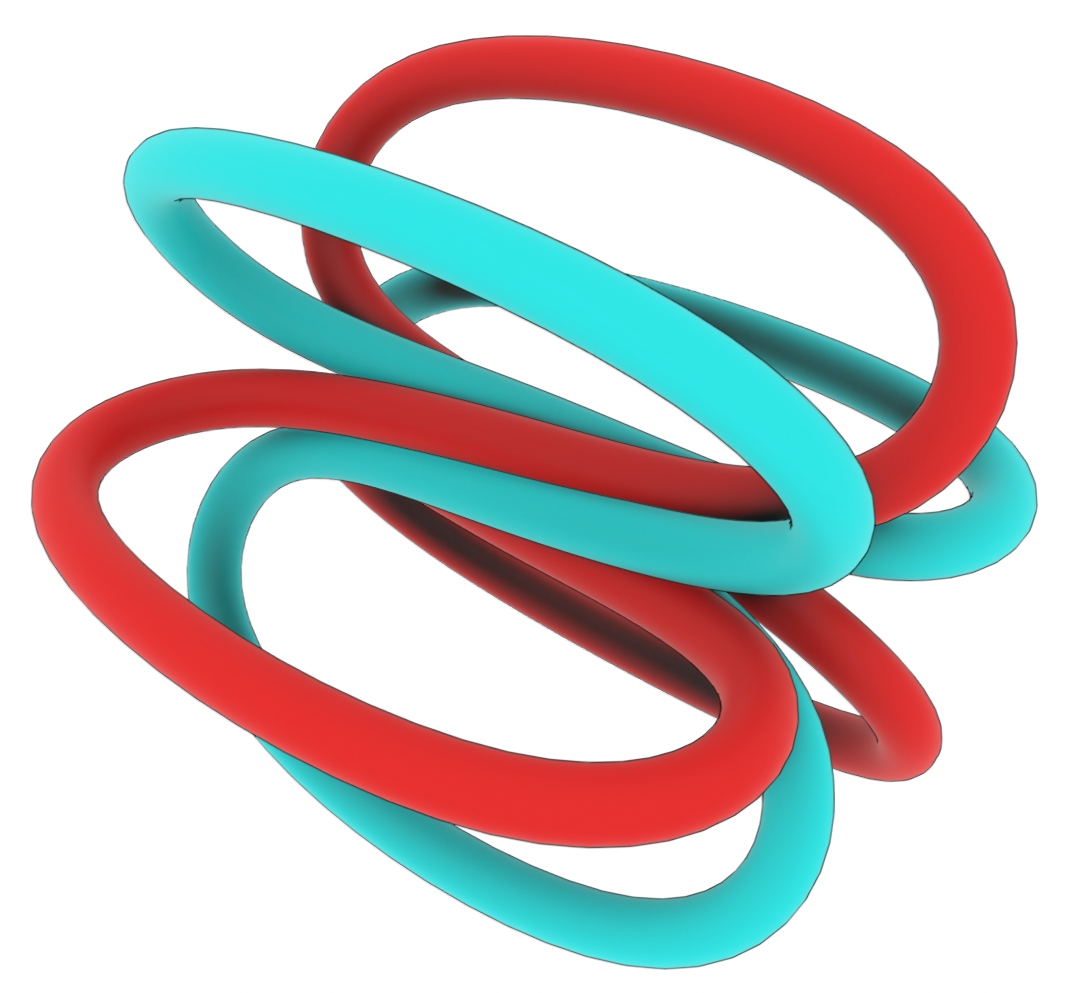}   \includegraphics[width=0.16\textwidth]{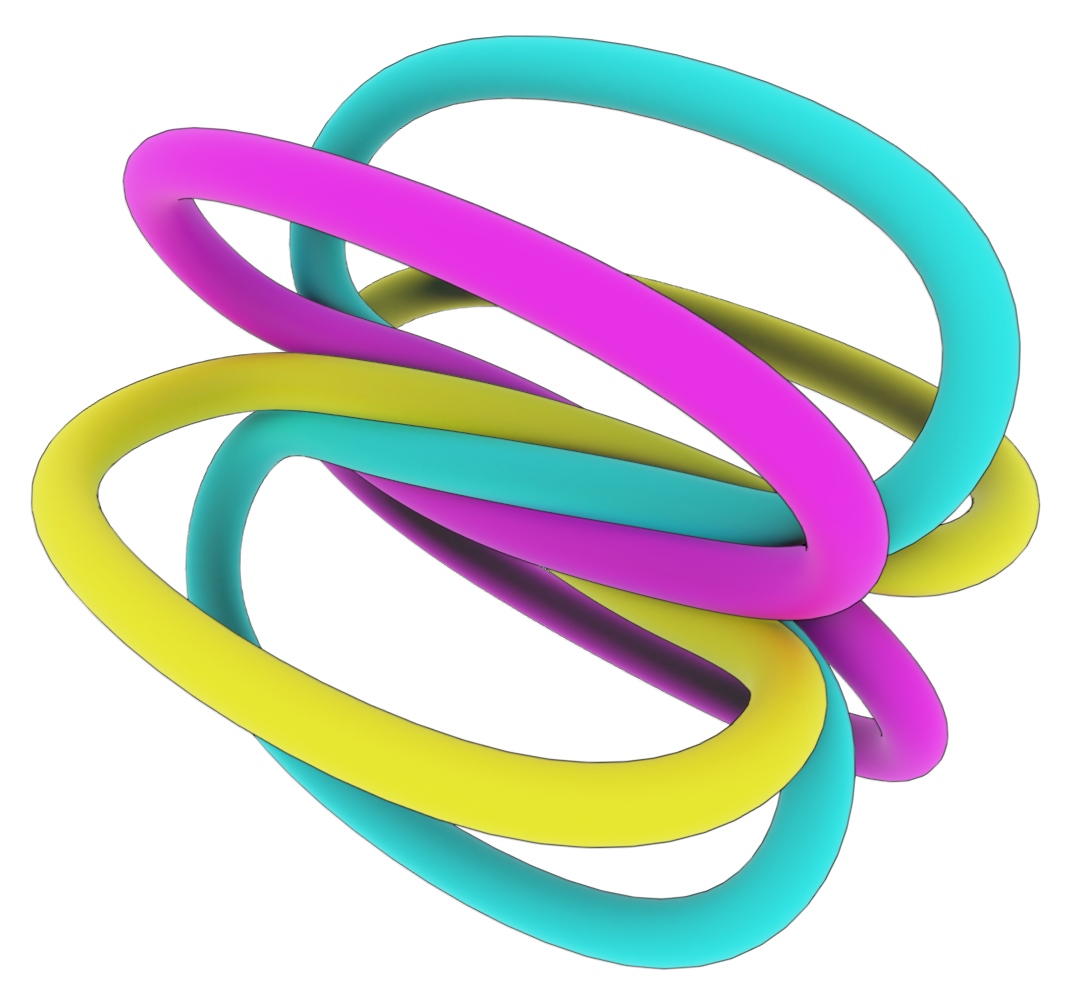}    \includegraphics[width=0.16\textwidth]{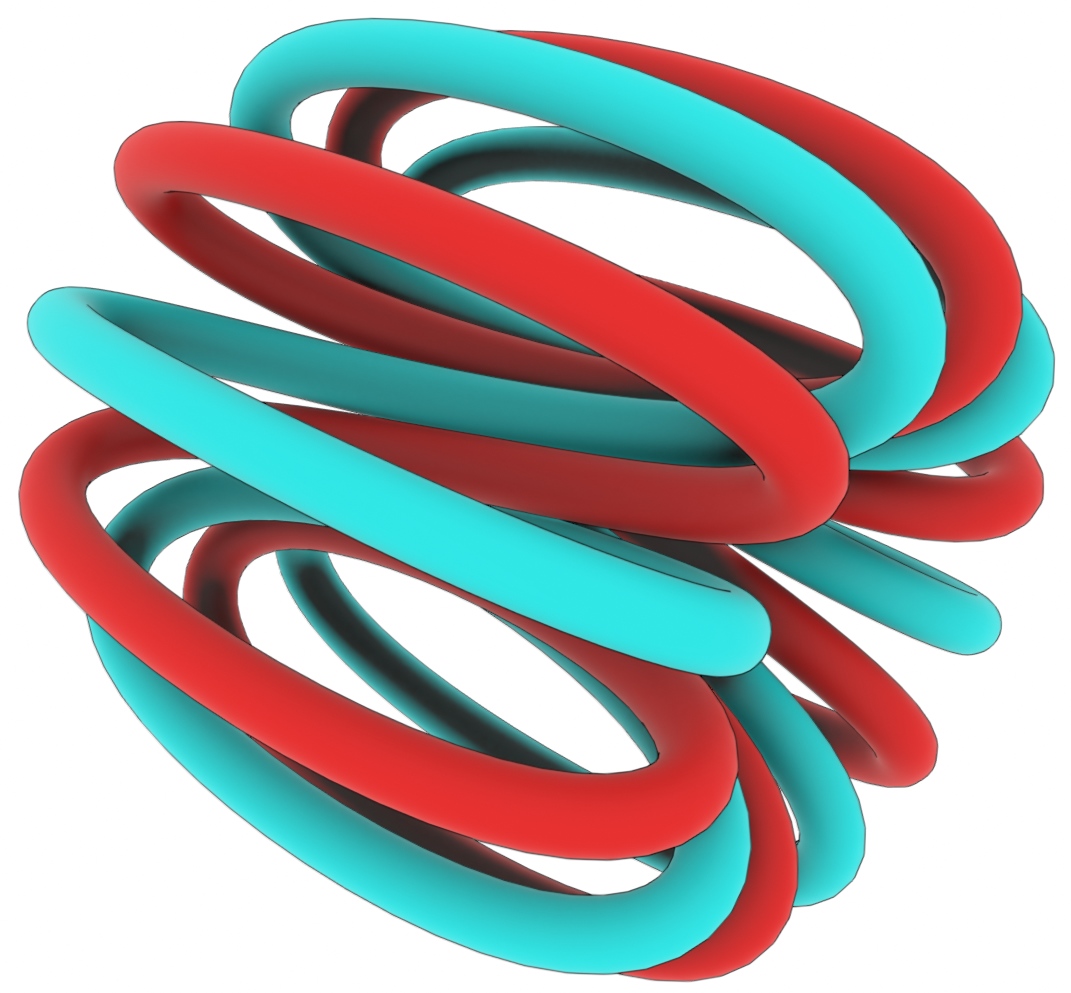}    \includegraphics[width=0.16\textwidth]{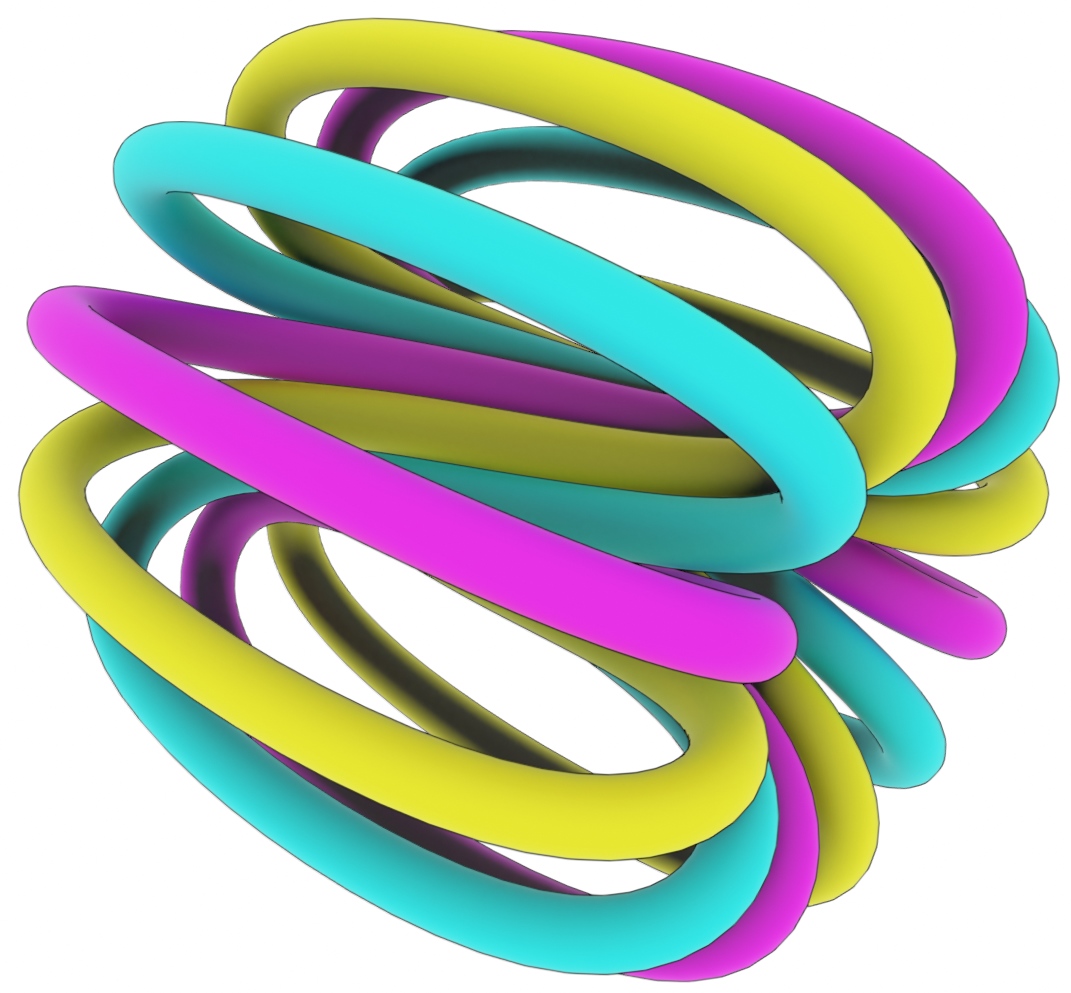}    \includegraphics[width=0.16\textwidth]{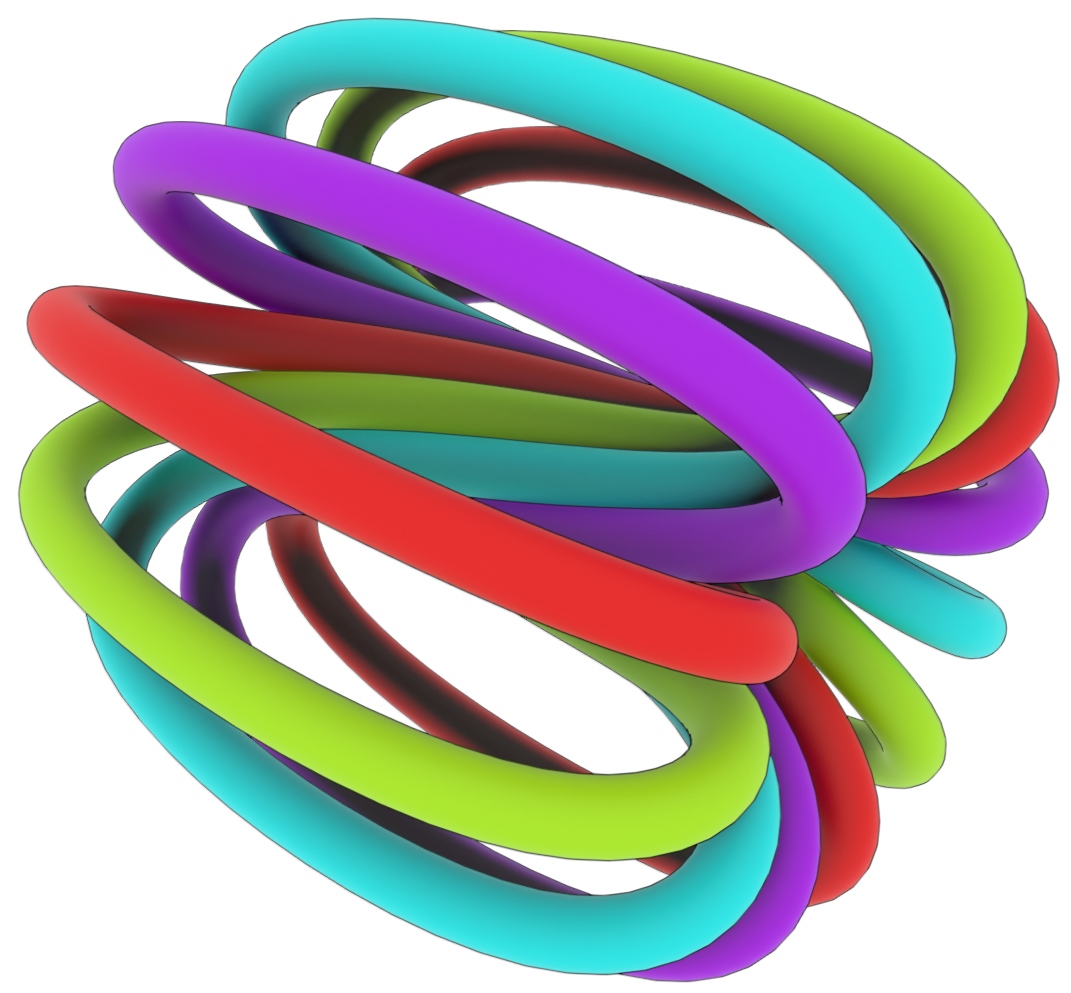}\\
    \parbox[t]{0.16\textwidth}{\centering 4 faces, 2 twist, \\2 loops}
    \parbox[t]{0.16\textwidth}{\centering 6 faces, 2 twist, \\2 loops}
    \parbox[t]{0.16\textwidth}{\centering 6 faces, 3 twist, \\3 loops}
    \parbox[t]{0.16\textwidth}{\centering 12 faces, 2 twist, \\2 loops}
    \parbox[t]{0.16\textwidth}{\centering 12 faces, 3 twist, \\3 loops}
    \parbox[t]{0.16\textwidth}{\centering 12 faces, 4 twist, \\4 loops}
    \caption{For local connectivity relations, we can change the number of twists to connect different loops and have disconnected loops at the end. Here we show examples of an edge shared by multiple faces and changing the number of twists to get various connectivity relations.}
    \Description{A horizontal sequence of six examples showing tubular loop structures generated around an edge shared by multiple faces. Each example is labeled with the number of incident faces, the twist value, and the resulting number of loops. From left to right, the images show configurations labeled “4 faces, 2 twist, 2 loops,” “6 faces, 2 twist, 2 loops,” “6 faces, 3 twist, 3 loops,” “12 faces, 2 twist, 2 loops,” “12 faces, 3 twist, 3 loops,” and “12 faces, 4 twist, 4 loops.” The structures are rendered as thick intertwined strands in cyan, red, magenta, and yellow. In each example, multiple closed loops wind around a common axis, with the number of distinct loops varying across the sequence.}
    \label{fig:twistingloops}
\end{figure*}

This behavior admits a simple algebraic interpretation. Formally, consider an edge $e$ incident to $K$ face-sides (equivalently, $K$ local cycle-elements around $e$). A twist-label $t\in\mathbb{Z}$ reconnects these elements by a cyclic shift. This action can be viewed as addition in the finite cyclic group $\mathbb{Z}_K$: indexing the $K$ elements by $i\in\mathbb{Z}_K$, a twist of $t$ maps
\[
i \;\mapsto\; i+t \pmod K .
\]
Thus, the twist induces a permutation of the $K$ local elements given by the shift operator $+t$ in $\mathbb{Z}_K$. The resulting global connectivity is determined by the orbit structure of this permutation: the elements of $\mathbb{Z}_K$ decompose into $\gcd(K,t)$ disjoint orbits, each of length $K/\gcd(K,t)$. Consequently, the reconnection around an edge of degree $K$ produces exactly $\gcd(K,t)$ distinct connected cycles (or ``strands''), which explains the $\gcd$ behavior observed in Figures~\ref{fig:2_mantwist}, ~\ref{fig:3_mantwist}, and~\ref{fig:twistingloops}.

This arithmetic structure provides a powerful and intuitive mechanism for designing topology. By selecting twist-labels appropriately, we can deliberately construct single-cycle knots, multiple knots linked together, or configurations in which each face cycle remains unknotted and independent. Importantly, this control is entirely local: global knot and link structures emerge from simple integer assignments at edges, without requiring global optimization, projection-based reasoning, or manual specification of crossings.

Twist-labels, therefore, serve as a compact and expressive interface between combinatorial topology and geometric realization. They allow complex knotting and linking behavior to be specified declaratively, using small integers, while ensuring that the resulting structures are consistent with the underlying mesh connectivity. This makes twist-labels not only a practical modeling tool, but also a conceptual bridge between discrete mesh structures and classical knot theory.

To support edge twisting uniformly across manifold and non-manifold surfaces, we adopt a labeled non-manifold mesh representation in which combinatorial adjacency is preserved even when geometric entities are absent. We maintain a complete edge–face incidence structure and use labels to control the amount of twist assigned to each edge. This design ensures that every edge admits a well-defined radial ordering of incident faces, which is essential for consistently defining twisting operations. Geometric properties such as strand thickness, spacing, or twist radius are treated as realization parameters and are intentionally decoupled from the combinatorial design framework introduced here.

In Section~\ref{sec:discussion}, we also show how this combinatorial design space subsumes many classical knots and links as emergent instances rather than explicitly constructed targets. In some cases, it may be convenient (though not necessary) to allow degenerate configurations (e.g., null edges or null faces) to represent open strands or incomplete cycles, without introducing additional topological structure, as discussed in Section~\ref{sec:discussion}. Importantly, these constructs do not introduce additional topological complexity; they merely relax closure constraints while preserving the underlying design abstraction.

\begin{figure}[htb!]
    \centering
    \begin{subfigure}[b]{0.32\columnwidth}       \includegraphics[width=\textwidth]{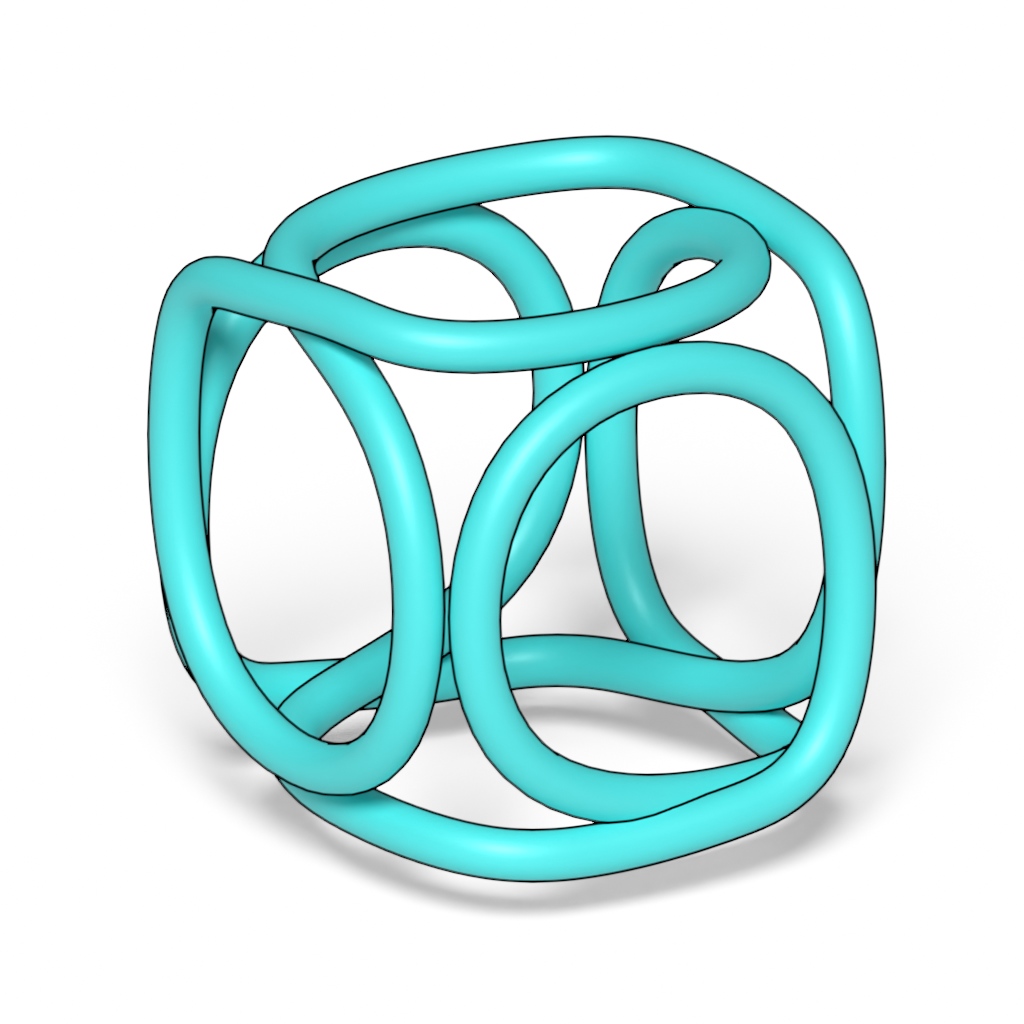}
        \caption{Only +1 and zero twist}
        \label{singlecycle0}
    \end{subfigure}
    \hfill
    \begin{subfigure}[b]{0.32\columnwidth}
        \includegraphics[width=\textwidth]{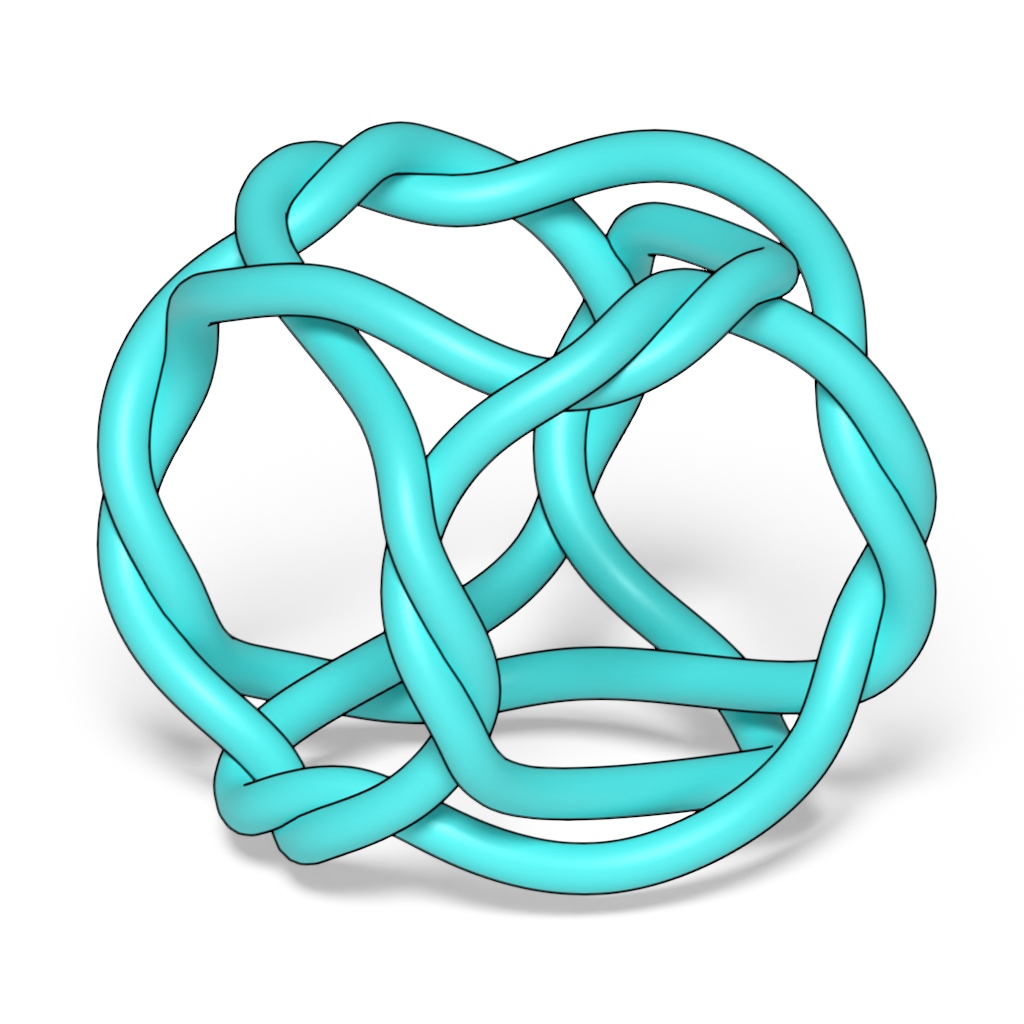}
        \caption{Replacing zero twists with 2-twists.}
        \label{singlecycle1}
    \end{subfigure}
    \hfill
        \begin{subfigure}[b]{0.32\columnwidth}
        \includegraphics[width=\textwidth]{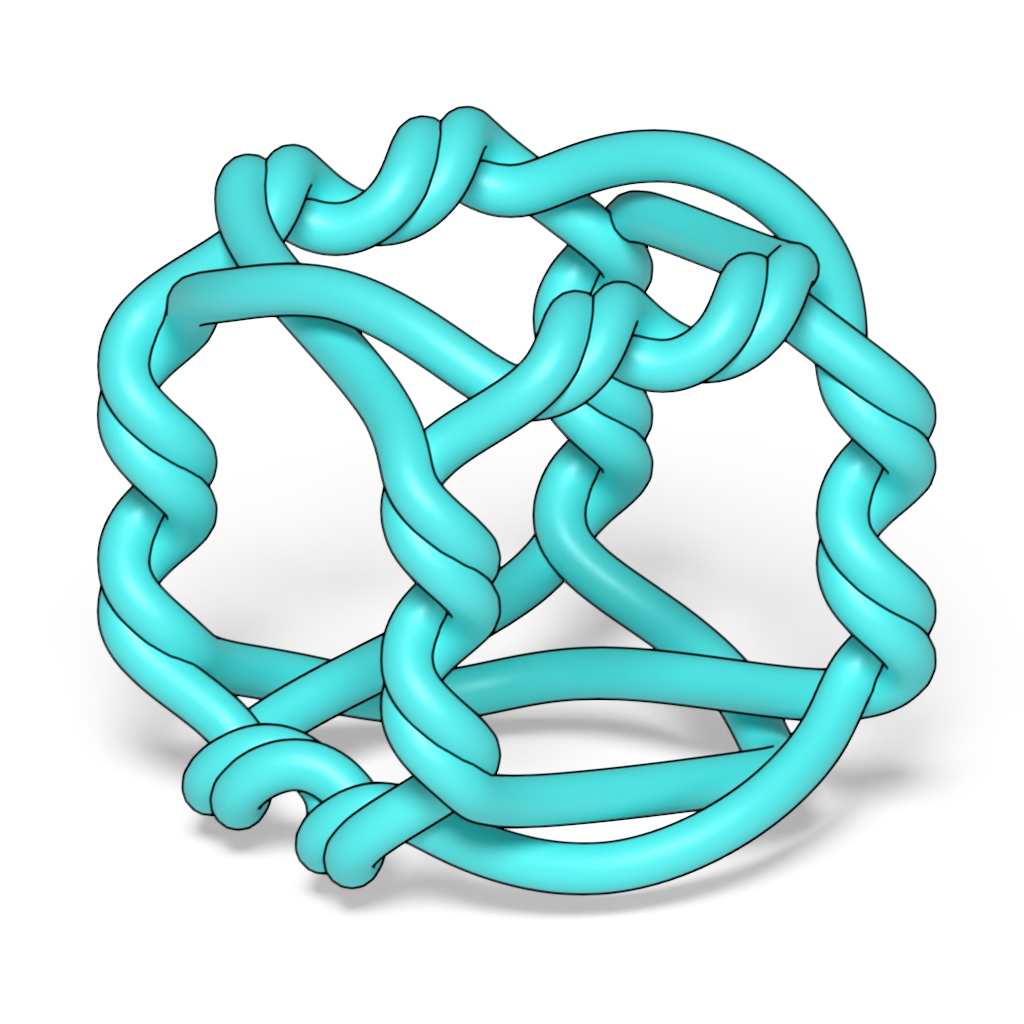}
        \caption{Replacing zero twists with 4-twists.}
        \label{singlecycle2}
    \end{subfigure}
    \hfill
  \caption{Replacing zero-twist edges with any non-zero even twist in 2-manifold meshes still results in single-cycle constructions. While zero twists produce parallel strands that are only weakly constrained geometrically, non-zero even twists preserve the zero-twist effect at the combinatorial level while introducing additional crossings that locally bind strands together. This demonstrates how multi-twist labels increase geometric coherence without altering topological guarantees.}
  \Description{Three side-by-side examples of tubular loop structures generated from the same closed surface. All three examples are rendered as thick cyan strands forming a single continuous loop. In the left image, labeled “Only +1 and zero twist,” several strands run in parallel with minimal local crossings. In the middle image, labeled “Replacing zero twists with 2-twists,” the strands follow similar global paths but exhibit additional local crossings where strands wrap around each other. In the right image, labeled “Replacing zero twists with 4-twists,” the strands show more frequent local windings and tighter interlacing, while maintaining the same overall loop structure.}
  \label{fig:single_cycle}
\end{figure}

\subsection{Guaranteeing Single Cycle with Multi-Twist Labels and Local Geometric Coupling}

Prior work \cite{xing2010single} has shown that for any 2-manifold surface mesh, assigning only zero-twist and single-twist labels to edges is sufficient to always produce single-cycle woven structures, which are topological knots, and that the number of such knots grows exponentially with the number of edges. However, while zero-twist edges play a crucial role in this combinatorial construction, they also give rise to locally parallel strands that are only weakly constrained geometrically, resulting in configurations that can visually appear loose or flap in space.

The introduction of multi-twist labels generalizes this framework by decoupling the net twist effect from the local crossing structure. In 2-manifolds, non-zero even-valued twist labels such as ±2 induce a zero-twist effect at the level of strand orientation, while simultaneously introducing additional crossings that locally bind neighboring regions together. As a result, multi-twist labels preserve the topological guarantees and exponential design space established by single- and zero-twist constructions, while providing a new degree of geometric control that ties the structure together more tightly.

This observation generalizes naturally beyond 2-manifold edges. For an edge incident to $K$ face cycles, a zero twist may be replaced by any nonzero twist that is a multiple of $K$ without altering the resulting cycle structure. Such twists act trivially at the combinatorial level—preserving the same permutation of cycles and, in particular, the single-cycle guarantee, while introducing additional local crossings that strengthen geometric interlocking. In the special case $K=2$, this reduces to the even-twist substitution shown in Figure~\ref{fig:single_cycle}; for higher-degree edges, the same principle applies with multiples of $K$. Consequently, geometric coupling can be increased arbitrarily through higher-magnitude twist labels without affecting global topology. This decoupling between topological guarantees and geometric tightness further reinforces the design-oriented nature of the framework: integer twist labels provide independent control over connectivity and local binding strength while preserving the exponential size of the design space.

\section{Scaffold Based Design for Linked Knot Construction}

To better situate this construction within existing approaches to knot and link design, we now review related work on knots and links, which have a long history in mathematics, topology, and art, and a wide range of computational approaches have been proposed for generating knotted, linked, and woven structures \cite{tait1877knots,rolfsen1976knots,adams1994knot}. Classical knot theory is primarily concerned with the classification and equivalence of embeddings, rather than with construction or design \cite{tait1877links,conway1970enumeration,neuwirth1979theory,rolfsen1976knots,adams1994knot}. Consequently, while these formulations provide deep theoretical insight, they offer limited direct control over how knots are generated from a small set of explicit design parameters.

From a design perspective, only a few ``general'' conceptual approaches have been developed for constructing knots and links \cite{tait1877knots,akleman2015extended,akleman2020topologically,akleman2015block}. Historically, these approaches can be understood as relying on different types of scaffolds—such as graphs, surfaces, or higher-dimensional cell complexes—from which cycles, crossings, and linking behavior are induced. Each scaffold imposes inherent constraints on how knots and links can be generated, and each introduces limitations that restrict the size, controllability, or expressiveness of the resulting design space. Although there exists mathematical work on periodic entanglements \cite{evans2013periodiceentanglement1,evans2013periodiceentanglement2}, volumetric weaving \cite{yildiz2025woven}, and links \cite{yildiz2025linked} that demonstrate the richness of linked knots, these efforts focus on specific families of LK structures rather than on a general, extensible design framework.

Partly because of these limitations, much of the engineering, architectural, and scientific work on knots and links does not rely explicitly on the topological properties of an underlying scaffold. Instead, interlaced or entangled forms are typically constructed using a wide variety of ad hoc physical or geometric methods. A substantial body of work explores textile-based and interlocking material systems, where knot-like and linked configurations emerge from local assembly rules, material behavior, or fabrication constraints rather than explicit topological control \cite{borhani2016material,borhani2016emergence}. Other approaches focus on elastic and transformable structures, in which linking and entanglement arise through deformation, deployment, and geometric layout of continuous elements \cite{kozlov2013structures,liu2025doublelayeredegridshells,ren2021curvedribbons}. A further line of work emphasizes discrete linking systems, including chain-mail–like assemblies and ring-based constructions built from repeated geometric units \cite{verhoeff2011chainlinkfence,roelofs2007rings}. More recently, component-based weaving strategies have been introduced, where interlacing is resolved locally at connection regions between modular elements, such as cylindrical components, without invoking a global knot or link topology \cite{yang2025topologicalwovennodes}.

In addition to these constructive approaches, there also exists a body of survey and catalog-style work that documents, systematizes, or contextualizes knot-like and linked structures across disciplines. These include surveys of DNA-based building block structures, such as DNA nanostructures and DNA origami, which catalog linked and entangled forms emerging from molecular self-assembly \cite{glaser2021dnananostructures}, as well as broader overviews of artistic and architectural knot forms that trace the use of knots as aesthetic, symbolic, and structural elements in design practice \cite{kozlov2018knots}. Related efforts also present systematic listings of mathematical link families, such as triply periodic link structures, offering comprehensive inventories of possible configurations without focusing on their physical realization or design applicability \cite{gailiunas2022tplinks}.

Motivated by this gap between classification and construction, we shift attention from knot equivalence to the scaffolds from which LK structures are generated. Scaffolds are chosen because they determine cycles, crossings, and controllability simultaneously. In this section, we provide an overview of three common scaffolds, graphs, 2-manifolds, and 3-manifolds, as alternative scaffolds for constructing LK structures, and compare them in terms of expressiveness and design efficiency. We treat these representations as design frameworks that impose different constraints on crossings, cycles, and local interactions, offering varying degrees of control and uniqueness. This scaffold-based comparison highlights the trade-offs between generality and controllability, and motivates the selection of a representation that is expressive enough for design while avoiding unnecessary complexity. This scaffold-based view allows us to compare approaches in terms of uniqueness, controllability, and the size of the resulting design space.

\subsection{Graphs as Scaffolds}

An influential line of work in classical knot theory originated with Tait, one of the first to study knots in the 19th century \cite{tait1877knots}. Tait’s classical construction is often described in terms of medial graphs, which are obtained by replacing each edge of a planar graph with a crossing, which is related to a single twist\footnote{In Tait’s original construction, crossings are induced by fixing a global cyclic ordering of edges and assigning positive or negative crossings in a planar diagram \cite{kauffman1987knots,gross2001topological}. Although often described as being related to twisting, this mechanism does not correspond to an explicit local twist operator: it introduces no notion of three-dimensional chirality, and all crossings arise from a uniform rotation convention. In contrast, modern twist-based formulations treat twisting as a local, intrinsically three-dimensional operation with well-defined sign and multiplicity.}. Although this is a historically important observation, an important subtlety creates a significant problem. An abstract graph alone does not uniquely determine the cycles used to form knots. The cycles become well-defined only after fixing a cyclic ordering of incident edges at each vertex, i.e., a rotation system, which is equivalent to embedding the graph on a surface of a given genus
\cite{edmonds1960combinatorial}. 

\begin{figure}[htb!]
  \centering
  \includegraphics[width=0.9\columnwidth]{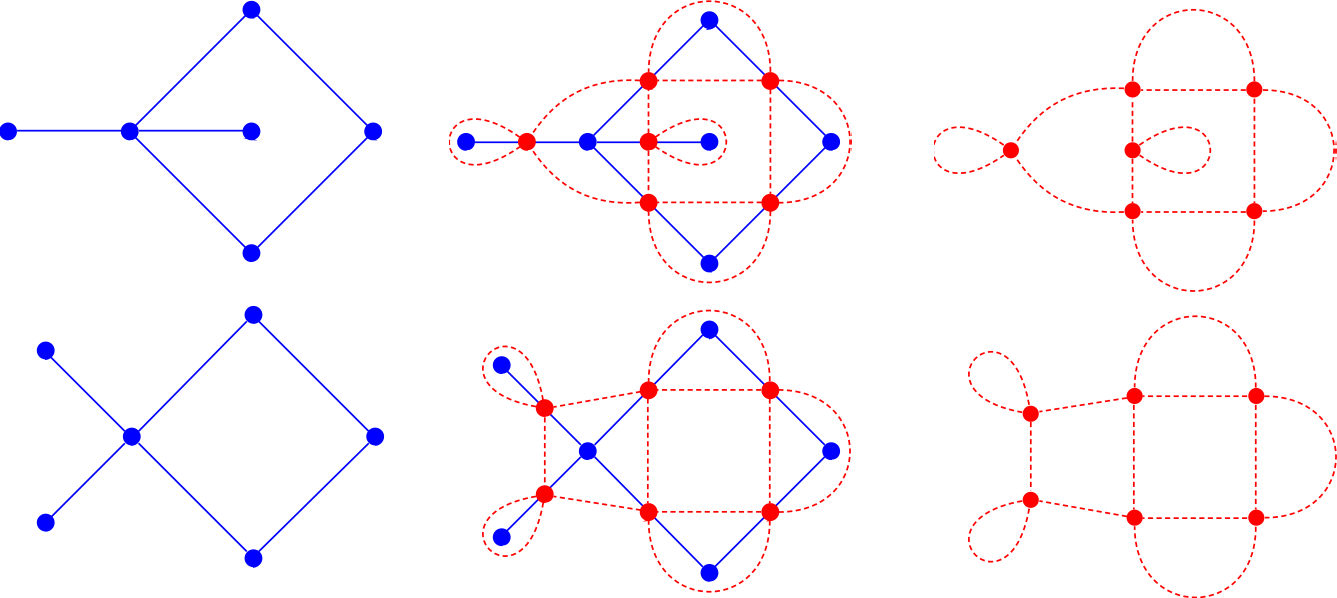}
  \caption{The two red graphs are both medial graphs of the blue graph, but they are not isomorphic. In other words, the same underlying graph can lead to different medial graphs and therefore to non-isomorphic knot structures. This figure uses a standard illustrative example adapted from Wikipedia, with minor layout adjustments~\cite{wikipediaMedialGraph}.}
  \Description{A diagram composed of blue and red graphs arranged side by side. On the left, a blue graph with vertices connected by straight line segments is shown in two layouts. On the right, two different red graphs are shown, each overlaid with curved dashed edges and circular arcs. Both red graphs correspond to the same blue graph but differ in their edge connections and overall structure. The red graphs have similar vertex positions but distinct edge routing, resulting in visually different configurations.}
  \label{fig:tait}
\end{figure}

In Tait’s original context, the graphs were implicitly planar drawings on paper, corresponding to embeddings on a genus-zero surface, where face cycles are uniquely determined. For general graphs, however, different rotation systems yield different sets of cycles and thus different knots, even when the underlying abstract graph is the same \cite{gross2001topological,akleman1999}. As illustrated in Figure~\ref{fig:tait}, different cycles can correspond to the same graph. This non-uniqueness is a fundamental limitation of graph-only formulations; graphs alone are therefore insufficient as a controllable design scaffold.

\subsection{2-Manifolds as Scaffolds}

The need for graph rotation systems~\cite{edmonds1960combinatorial} is well known in topological graph theory~\cite{gross2001topological} and underlies combinatorial surface representations such as winged-edge~\cite{baumgart1972}, half-edge~\cite{mantyla1988}, quad-edge~\cite{guibas1985}, and related data structures~\cite{akleman1999}, where rotation systems are made explicit. Graph rotation systems embed graphs onto surfaces and uniquely determine cycles, which are then used to define the faces of 2-manifold meshes. For example, using rotation systems, the two blue graphs in Figure~\ref{fig:tait}, which are drawn on the same 2D plane, can be distinguished, and their corresponding medial graphs are uniquely defined.

As a result, surface-based formulations derived from graph rotation systems emerge as an alternative 2-manifold scaffold representation, in which LK structures are constructed by augmenting surface meshes with local twisting or gyro information~\cite{akleman2009cyclic,akleman2015extended,xing2010single,lipschutz2022linkedknots,sequin2021polyhedral}. By encoding twist direction directly into a rotation-based representation, these approaches enable deterministic construction of single-cycle woven structures and avoid the non-uniqueness inherent in purely graph-based formulations.

In classical topological graph theory, twisting is formalized as a single, purely topological operator that introduces a half-twist along an edge. This operation is binary: applying the twist twice is equivalent to untwisting, and no notion of chirality is defined or needed\footnote{
This dimensional dependence explains why classical topological graph theory does not distinguish integer twist.
While chirality and twisting are essential for knotted curves in three dimensions, any one-dimensional knot becomes unknotted when embedded in four dimensions \cite{freedman1990topology}.
Nontrivial knotted surfaces do exist in four-dimensional space \cite{kamada2002surfaces}; however, when the ambient dimension exceeds four, even surface knots become topologically trivial up to isotopy \cite{guillemin1974differential,milnor1965topology}.
As a result, higher-dimensional topological frameworks typically do not encode integer twist or accumulated rotation, which are inherently three-dimensional phenomena.
 }. As a consequence, the twist does not distinguish between chiral configurations. A key property of this operator is that it converts orientable 2-manifolds into non-orientable ones by inserting a M\"obius strip, so the resulting knots can be interpreted as immersions of non-orientable surfaces into three-dimensional space. 

Because only a single type of twist is permitted, the resulting knot structures are necessarily alternating, reflecting an inherent limitation of this formulation~\cite{akleman2009cyclic,akleman2015extended}. Consequently, this formulation only allows LK structures that correspond to alternating knots and links, such as cyclic plain-woven objects~\cite{akleman2009cyclic}.

It should be noted that generating other classes of woven structures, such as twill weaves, requires the explicit use of chirality~\cite{akleman2011cyclic}. Introducing chirality leads to two distinct types of half-twists, corresponding to clockwise and counterclockwise configurations. Geometrically, this distinction can be interpreted through the existence of two chiral M\"obius strips, which are mirror images of one another. Incorporating these chiral half-twists extends the purely topological formulation and enables a more general theory of surface weaving, capable of representing a broader range of woven and linked structures~\cite{akleman2015extended,akleman2020topologically}.

These frameworks are limited to twisting edges on 2-manifold surfaces: edges are always incident to exactly two cycles, and crossings are inherently restricted to two local layers. As a result, geometric coupling and articulation are constrained. Later work on block meshes and CMM-pattern coverable meshes removed this restriction entirely by defining topology through a higher-dimensional combinatorial scaffold, namely 3-manifold meshes~\cite{akleman2015block}. 

\subsection{3-Manifolds as Scaffolds}

In this framework~\cite{akleman2015block,hyde2022tangledpolyhedra}, 3-manifolds or polytopes play the role of a scaffold. Most importantly, face sides form cycles, and the local neighborhoods around 3-edges naturally form cylindrical pipes that can accommodate any number of parallel strands. Moreover, crossings created through twist operations can produce more complex configurations, thereby providing additional design flexibility.

While block-mesh formulations were essential for defining robust topology in general volumetric modeling, the present work shows that, for knot and link design, the same robustness can be achieved using only non-manifold surface connectivity, resulting in a simpler and more direct abstraction.
In other words, although this paper builds directly on a theoretical foundation based on 3-manifolds, it shifts the focus away from general volumetric mesh representation and toward the design and construction of LK structures, which do not require the full expressive power of 3-manifold meshes.

\subsection{Non-Manifold Surfaces as Scaffolds}

The present work substantially generalizes surface-based approaches by introducing integer-valued twist labels on non-manifold surface meshes. Prior rotation-based and gyro-based methods encode a single, binary twist per edge, which fundamentally limits them to alternating structures; our framework generalizes this to oriented, integer-valued twist, enabling non-alternating, articulated, and selectively linked configurations.
Allowing arbitrary integer twists decouples net twisting effects from local crossing structure, enabling tighter geometric coupling without sacrificing topological guarantees. At the same time, non-manifold connectivity dramatically expands the underlying design space, supporting articulated, hierarchical, and space-filling linked-knot structures that cannot be represented within strictly 2-manifold frameworks. Unlike 2-manifold-based approaches that emphasize analysis, classification, or visual similarity, the proposed method is explicitly design-oriented: local, discrete labels act as controllable parameters whose number grows with mesh complexity, yielding an exponentially large and systematically explorable design space.

In conclusion, having compared graphs, surface-based, and volumetric representations as design scaffolds, we show that labeled non-manifold surface meshes provide sufficient expressive power for linked knot design without incurring unnecessary topological overhead.

In the rest of the paper, we explore the design space enabled by labeled non-manifold surface meshes through a unified perspective. Rather than presenting isolated examples, we organize LK designs into two complementary regimes: finite LK structures, which consist of a finite number of components realized on bounded meshes, and periodic LK structures, which generate space-filling or conceptually infinite structures through repetition and symmetry.

Both regimes are constructed using the same edge-twist formalism introduced earlier; the distinction arises from the topological and combinatorial properties of the underlying surface mesh. By systematically varying these properties, we demonstrate how a single framework supports a wide range of knotted, linked, articulated, and space-filling designs. Having reviewed existing scaffolds and their limitations, we now describe the design space enabled by edge twisting on non-manifold meshes. We distinguish between finite and periodic LK structures because they lead to fundamentally different topological constraints and design affordances.

\section{Design Space of Finite LK structures}

Having established the conceptual framework and positioned it within prior work, we now explore the design space enabled by bounded non-manifold surface meshes with a finite number of cells, which give rise to a finite number of knots and links. These constructions demonstrate how local, integer-valued twist labels provide direct and systematic control over global knotting and linking behavior. Rather than targeting individual classical knots, we show how families of knots, links, chainmail structures, and articulated assemblies emerge naturally as points within a unified design space defined by mesh topology, boundary structure, and twist assignments. While familiar patterns such as classical weaves or chainmail arise as special cases, the majority of LK structures enabled by this design space have no direct counterpart in existing textile, architectural, or knot-theoretic constructions.

We begin by examining labeled 2-manifold surface meshes, which form a conceptually simple yet expressive subclass of non-manifold surfaces. Although non-manifold connectivity is not required in this initial setting, introducing integer-valued twist labels already leads to a dramatic expansion of the design space. As demonstrated in the following subsections and accompanying figures, even the simplest symmetric meshes give rise to a large number of distinct knots, links, and chainmail configurations under different twist assignments. We then progressively extend this construction to manifolds with boundary and to non-manifold surfaces, showing how additional connectivity enables linked assemblies, articulation, and controlled motion, while preserving the same underlying twist-based design principles.

\subsection{Single-Cycle Knots from Labeled 2-Manifolds}
\label{subsec:designingknots}

We begin by considering the simplest setting in which labeled surface meshes give rise to single-cycle LK structures, corresponding to knots. Although non-manifold connectivity is not required in this case, introducing integer-valued twist labels already leads to a surprisingly rich design space. This subsection focuses on bounded 2-manifold surface meshes and demonstrates how local twist assignments control the emergence and diversity of single-cycle structures.

For a bounded surface mesh with no nontrivial symmetries, the number of distinct single-cycle LK structures that can be generated using unit twists is closely related to the number of spanning trees of the dual graph. Each spanning tree specifies a minimal set of edges whose twisting merges all face cycles into a single cycle. As a result, different spanning trees generically lead to different single-cycle knot configurations. Since the number of spanning trees grows exponentially with mesh complexity, the number of possible single-cycle LK structures exhibits the same exponential growth behavior.

\begin{figure}[htb!]
    \centering        \includegraphics[width=0.23\columnwidth]{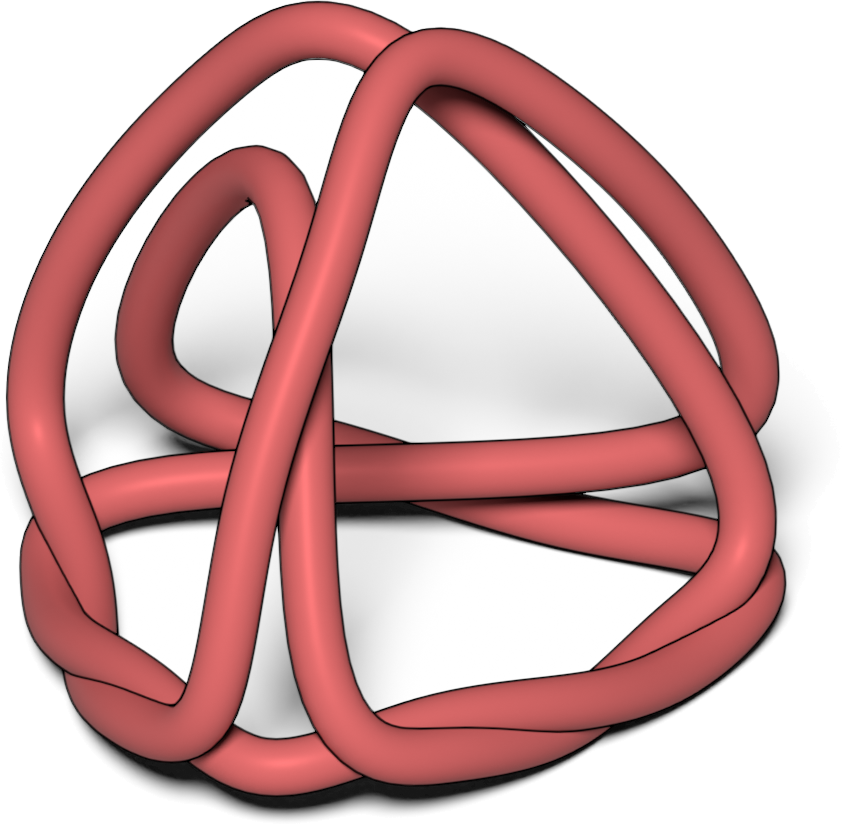}
\includegraphics[width=0.23\columnwidth]{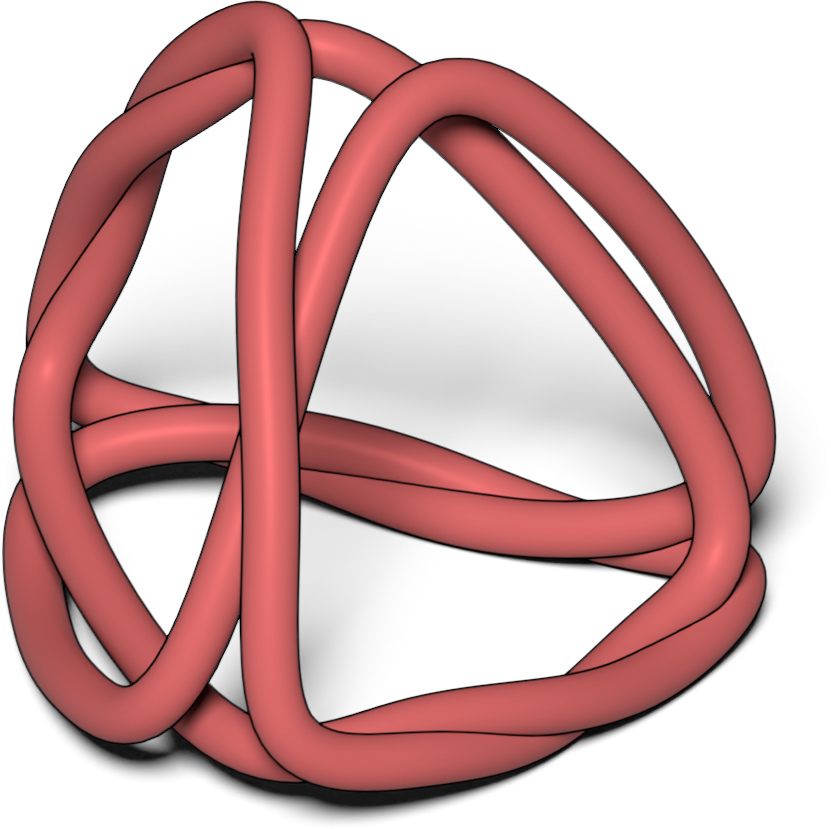}
\includegraphics[width=0.23\columnwidth]{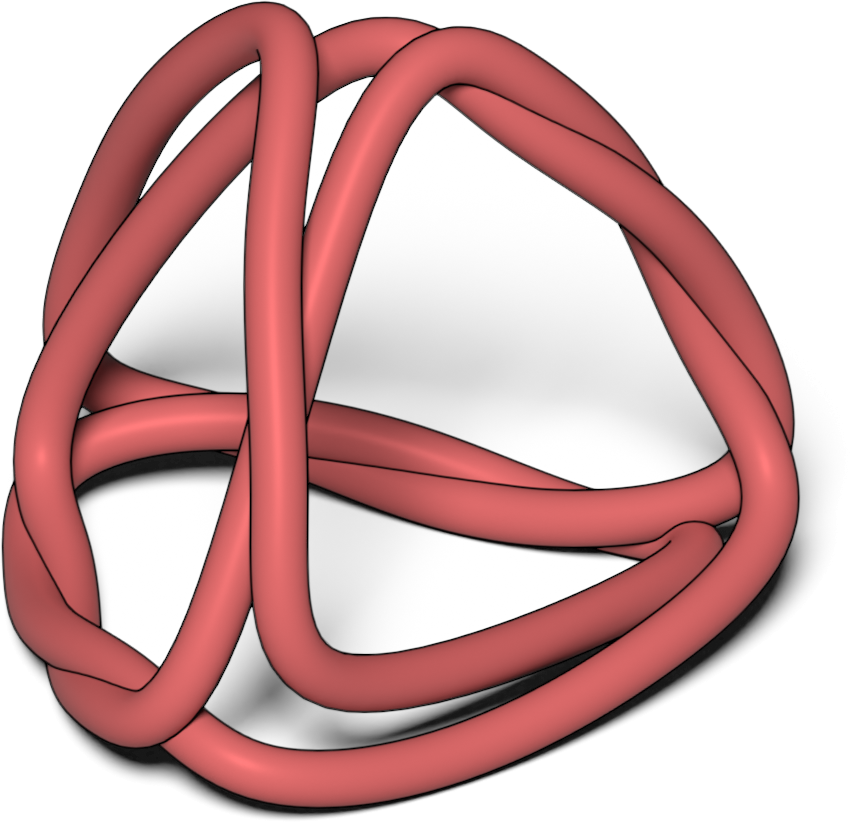}
\includegraphics[width=0.23\columnwidth]{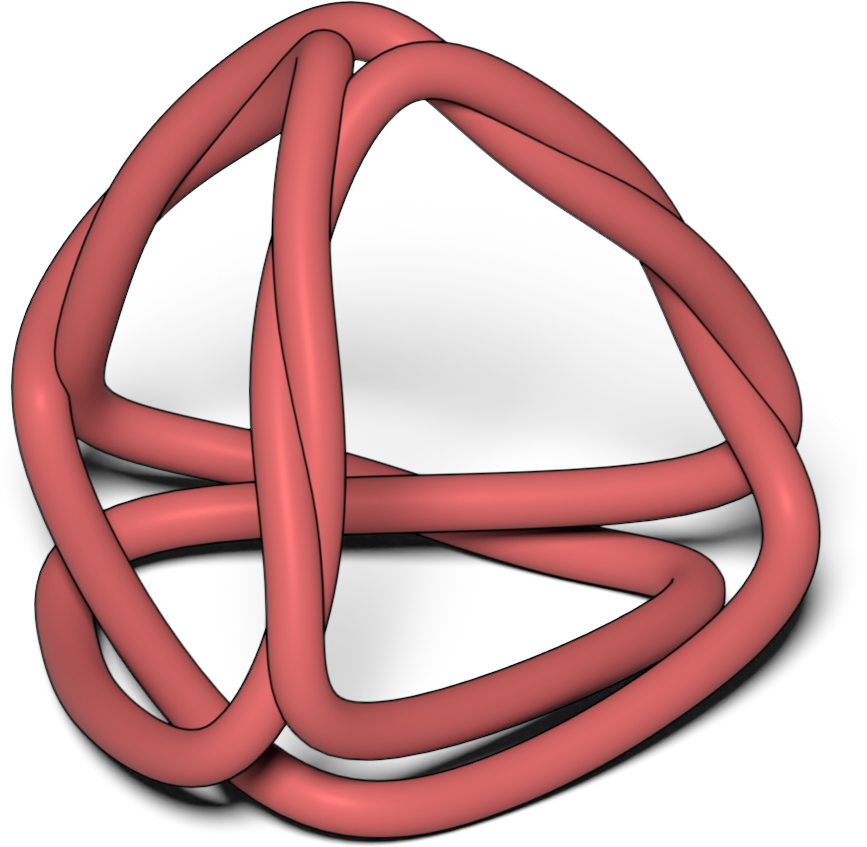}
  \caption{Four distinct knots obtained from a regular tetrahedron by twisting edges +1, -1, +2, and -2 to achieve a different single cycle in each case. In total, even this simplest mesh yields 32 distinct cases.}
  \Description{Four side-by-side images showing closed tubular knot structures rendered in a uniform red color. Each knot has a triangular overall shape corresponding to a tetrahedral arrangement, with the strand looping multiple times through the interior before closing. Although similar in size and general form, the four knots differ in how the strand winds and crosses itself, producing visibly distinct knot patterns. The knots are presented against a white background and viewed from a similar perspective for direct comparison.}
  \label{fig:tetra_knot}
\end{figure}

For highly symmetric meshes, such as the regular tetrahedron, counting distinct configurations is more subtle, since different spanning trees may be equivalent under symmetry. In such cases, multiple twist assignments can lead to geometrically identical LK structures. However, this reduction is purely a consequence of symmetry. Once symmetry is broken, for example, by perturbing edge lengths or face geometry, the same connectivity supports a significantly larger number of distinct single-cycle designs. Notably, even for tetrahedral connectivity, a rich family of distinct knots can be generated using only small integer twist values drawn from $\{+1,-1,+2,-2\}$, as illustrated in Figures~\ref{teaser1},  \ref{teaser21}, and~\ref{fig:tetra_knot}.

Another important observation is that the single-cycle property depends only on the parity of twists assigned along edges of a spanning tree. Any odd twist ($\pm1, \pm3, \ldots$) introduces a local M\"obius-type connection and contributes equivalently to cycle merging, while any nonzero even twist plays the same topological role as $\pm2$, differing only in geometric tightness. Consequently, half-twists may be freely replaced by higher odd twists, and double twists by arbitrary nonzero even twists, without affecting the single-cycle nature of the resulting structure. Even under these strong restrictions, the resulting design space is already exponentially large, highlighting how rapidly complexity emerges from simple integer-labeled surface scaffolds.

\begin{figure}[htb!]
    \centering        \includegraphics[width=0.23\columnwidth]{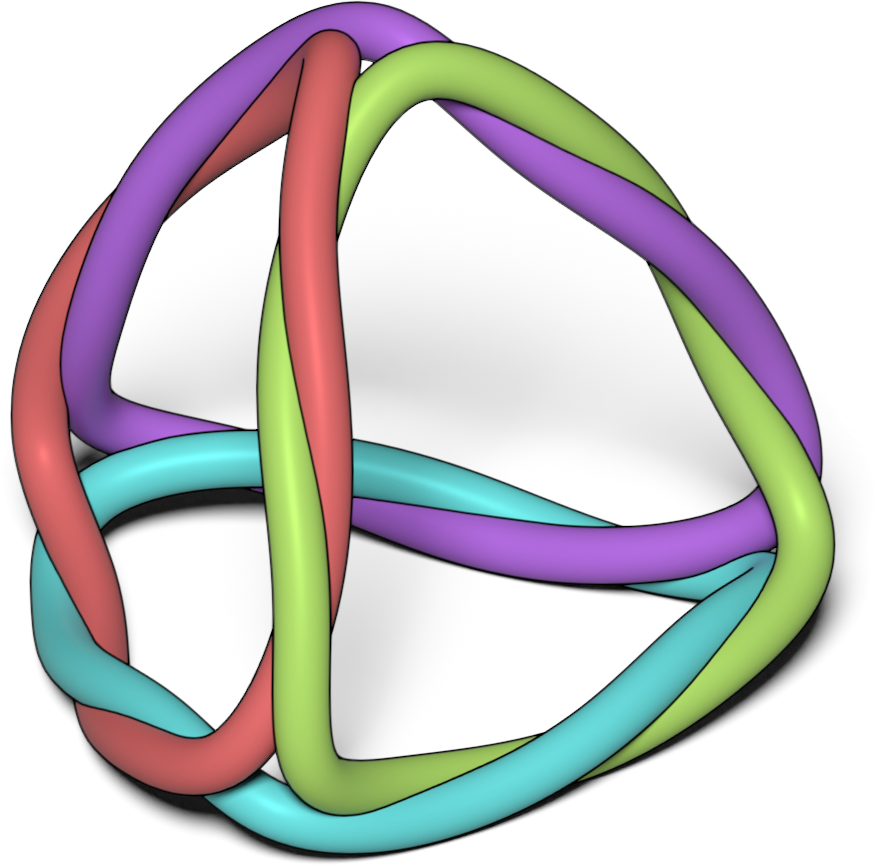}
\includegraphics[width=0.23\columnwidth]{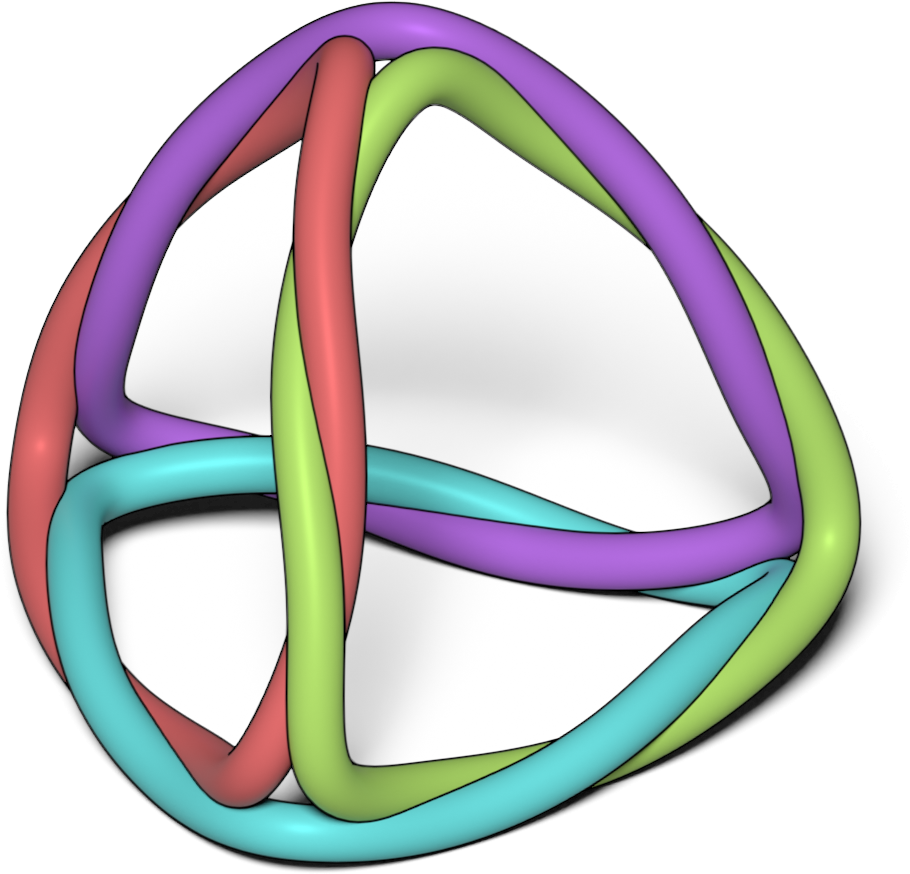}
\includegraphics[width=0.23\columnwidth]{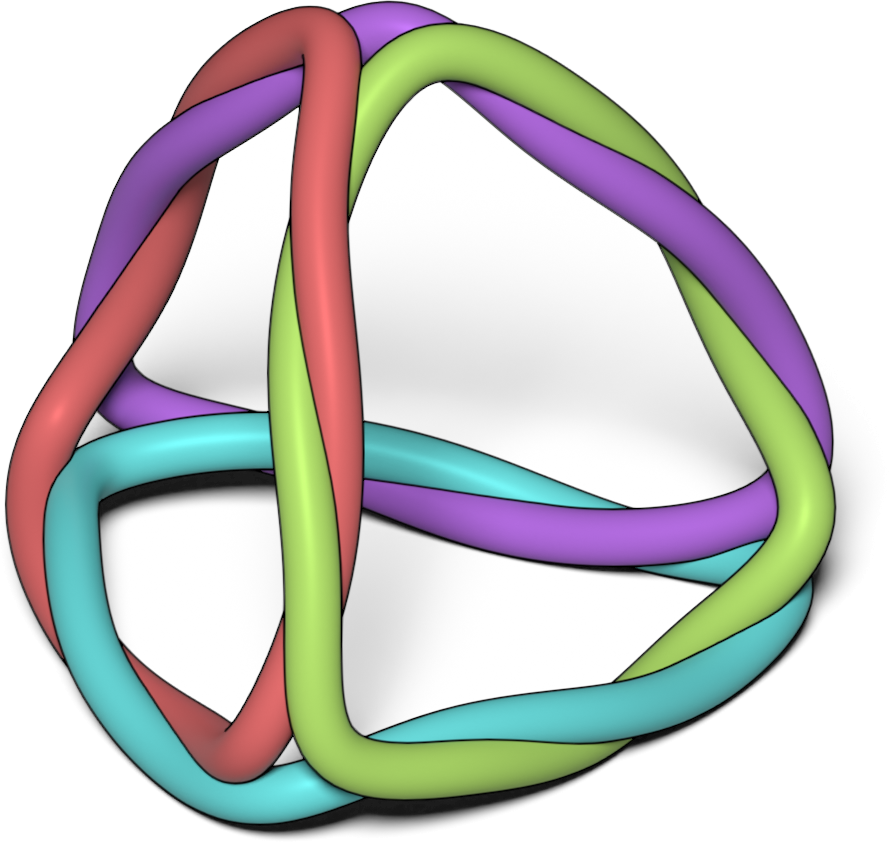}
\includegraphics[width=0.23\columnwidth]{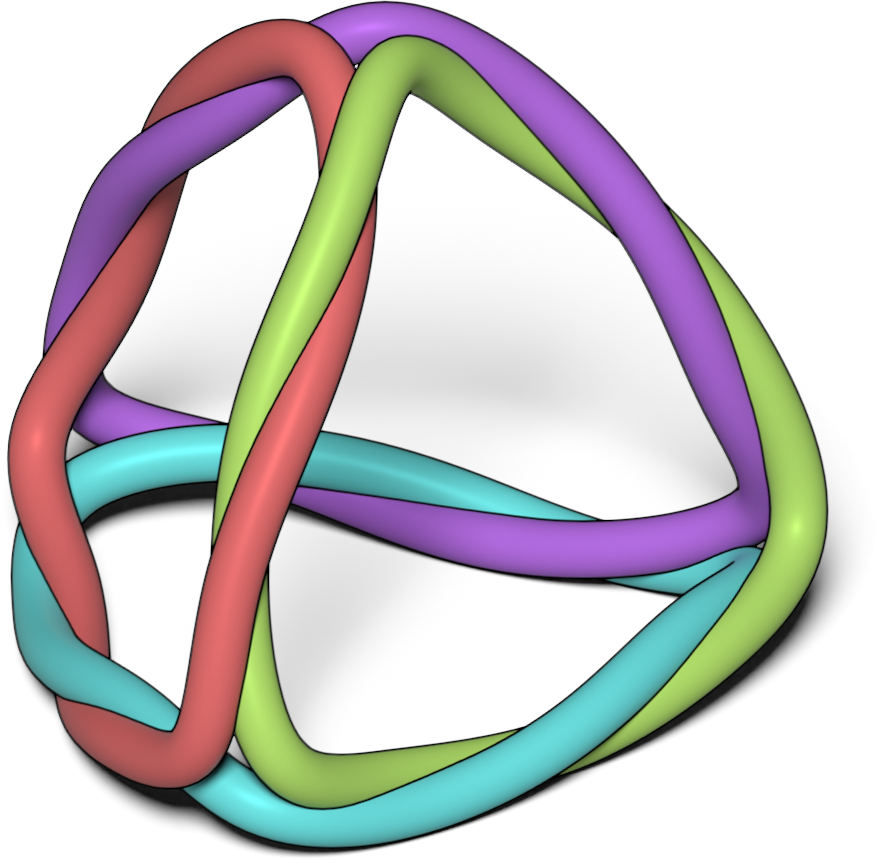}
  \caption{Four distinct chainmail structures obtained from a regular tetrahedron by assigning twist labels of +2 and -2 to its edges. Each choice produces a different chainmail configuration. In total, even this simplest mesh yields 12 distinct cases. }
  \Description{Four side-by-side images showing chainmail-style structures rendered as multiple interlinked tubular loops in different colors, including cyan, red, green, and purple. Each structure has an overall triangular form corresponding to a tetrahedral arrangement, with several closed loops passing through one another. While the four examples share the same general shape and scale, the ordering and interweaving of the colored loops differ, producing visibly distinct chainmail configurations. All examples are shown from a similar viewpoint against a white background.}
  \label{fig:tetra_link}
\end{figure}

\subsection{Chainmails (Links) from 2-Manifolds}
\label{subsec:designingchainmails}

Chainmail structures represent the simplest class of linked LK structures and can be generated trivially within our framework from essentially any surface mesh. In contrast to single-cycle constructions, chainmails arise when face cycles are preserved as distinct loops but are systematically interlinked with their neighbors \cite{klotz2024chirality}. This behavior is obtained by assigning nonzero even twist labels to all edges, ensuring that no cycles merge while guaranteeing consistent linking throughout the mesh.

From a topological perspective, this construction is remarkably simple: assigning any even twist to an edge preserves the separation of adjacent face cycles, while nonzero values ensure that neighboring cycles are interlocked rather than merely adjacent. As a result, twisting all edges of a surface mesh by even integers transforms the mesh into a fully connected chainmail-like structure in which each face becomes a closed loop linked to its neighbors. Even when restricting the allowable twist labels to the two chiral choices $\{+2,-2\}$, the resulting design space is already large, since different sign assignments produce distinct global linking configurations, as illustrated in Figures~\ref{teaser0},  \ref{teaser20}, and~\ref{fig:tetra_link}.

\begin{figure}[htb!]
    \centering    \includegraphics[width=0.46\columnwidth]{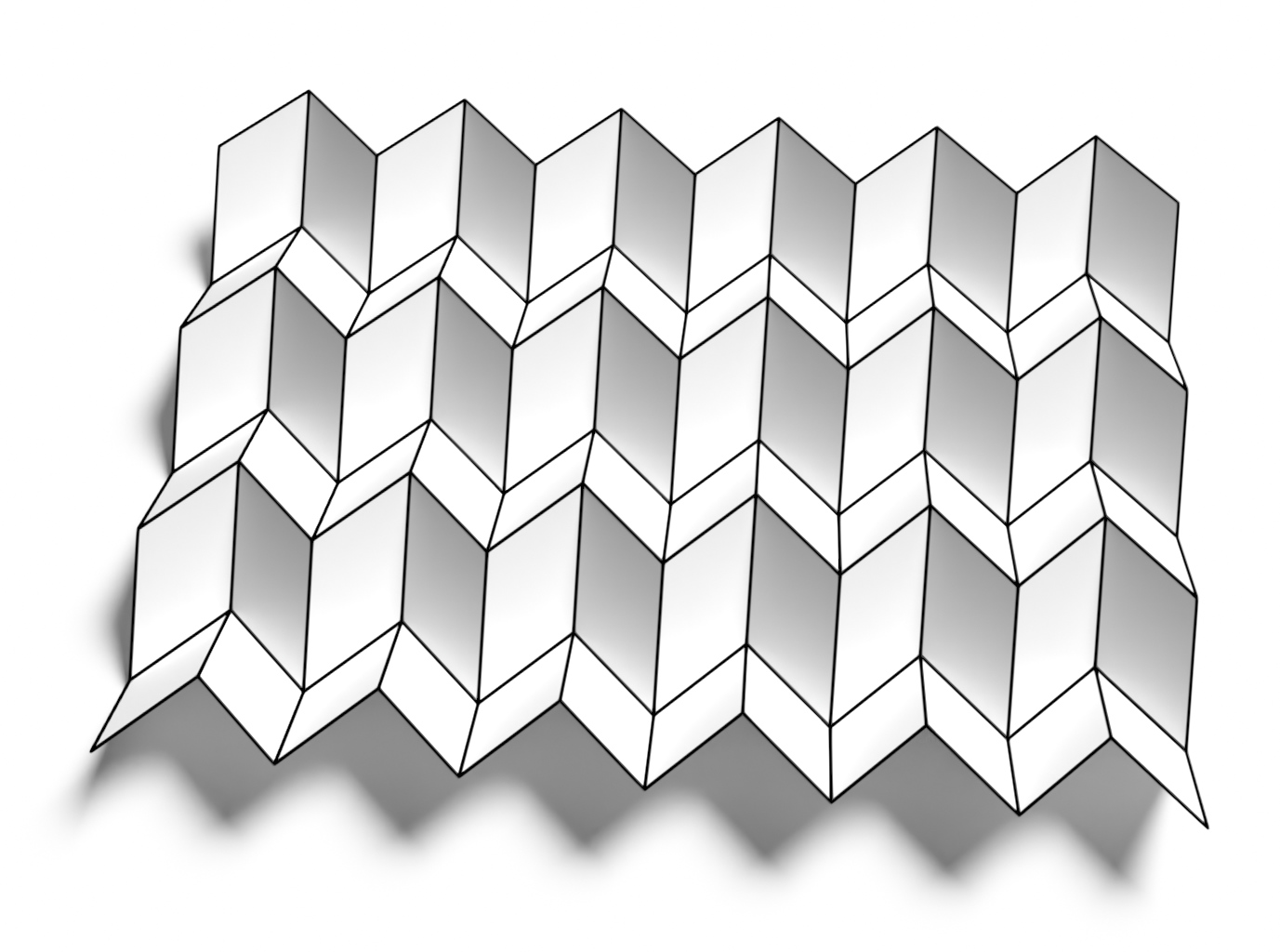}    \includegraphics[width=0.46\columnwidth]{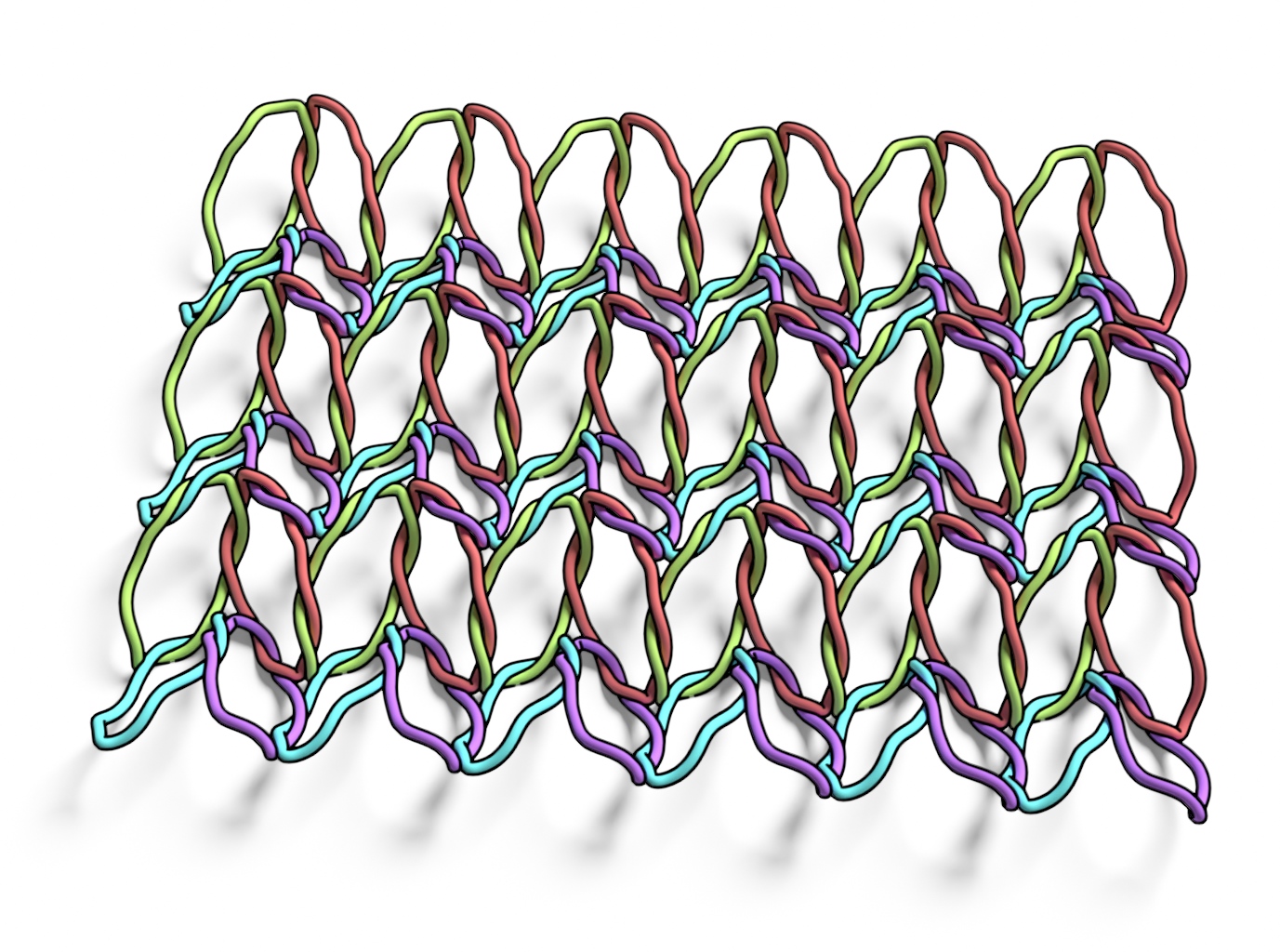}
    \caption{Herringbone origami as an LK structure. Left: A planar polygonal mesh representing a herringbone origami pattern, modeled as a 2-manifold surface with boundary. Right: The corresponding LK structure obtained by assigning twist labels to the mesh edges, transforming the surface pattern into a linked-knot configuration.}
    \Description{Two images shown side by side. On the left, a planar polygonal mesh arranged in a repeating zigzag pattern is shown, forming a herringbone-like tiling with folded facets and visible edges. On the right, a corresponding three-dimensional structure is shown, composed of many thick tubular strands rendered in multiple colors. The strands follow the same repeating pattern as the mesh, forming a dense array of interlinked loops arranged in rows and columns. Both images are shown from a similar viewpoint for comparison.}
    \label{fig:herringbone}
\end{figure}

As in the case of knots, symmetry plays an important role: for highly symmetric meshes, such as regular polyhedra, multiple twist assignments may lead to equivalent configurations. Although symmetry influences how many twist assignments lead to geometrically distinct outcomes, it does not fundamentally limit the size of the design space. The total number of possible chainmail configurations grows exponentially with the number of edges, since each edge admits independent twist choices. Symmetries merely identify subsets of these configurations as equivalent, introducing at most a linear or polynomial reduction relative to the exponential growth. Consequently, even highly symmetric meshes support a vast number of distinct chainmail structures, and the full exponential diversity is immediately apparent.

Labeled 2-manifolds with boundary also support folding and cutting behaviors reminiscent of origami and kirigami. By introducing boundaries through cuts or unfoldings, and by selectively controlling which edges participate in twisting, it is possible to design chainmail structures that open, fold, or articulate while preserving well-defined linking relationships between loops. In these cases, geometric flexibility arises not from mechanical hinges, but from topological linking combined with boundary placement. The example in Figure~\ref{fig:herringbone} illustrates how planar origami patterns can be systematically lifted into LK structures while preserving their underlying combinatorial organization. This demonstrates that the framework naturally extends from static chainmail patterns to reconfigurable and deployable linked structures, using the same discrete twist-label formalism.

\begin{figure}[htb!]
    \centering
    \includegraphics[width=0.24\linewidth]{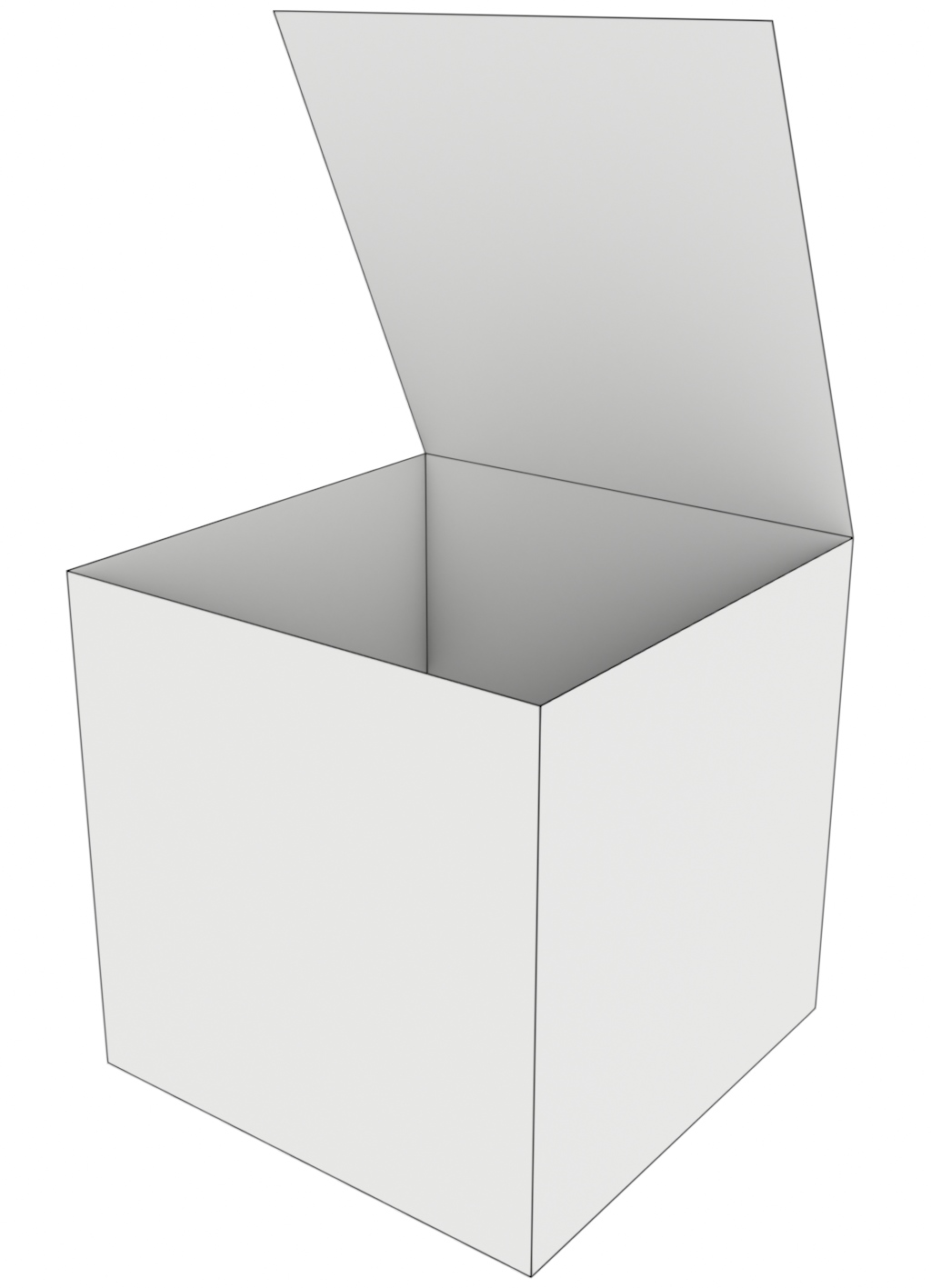}
    \includegraphics[width=0.24\linewidth]{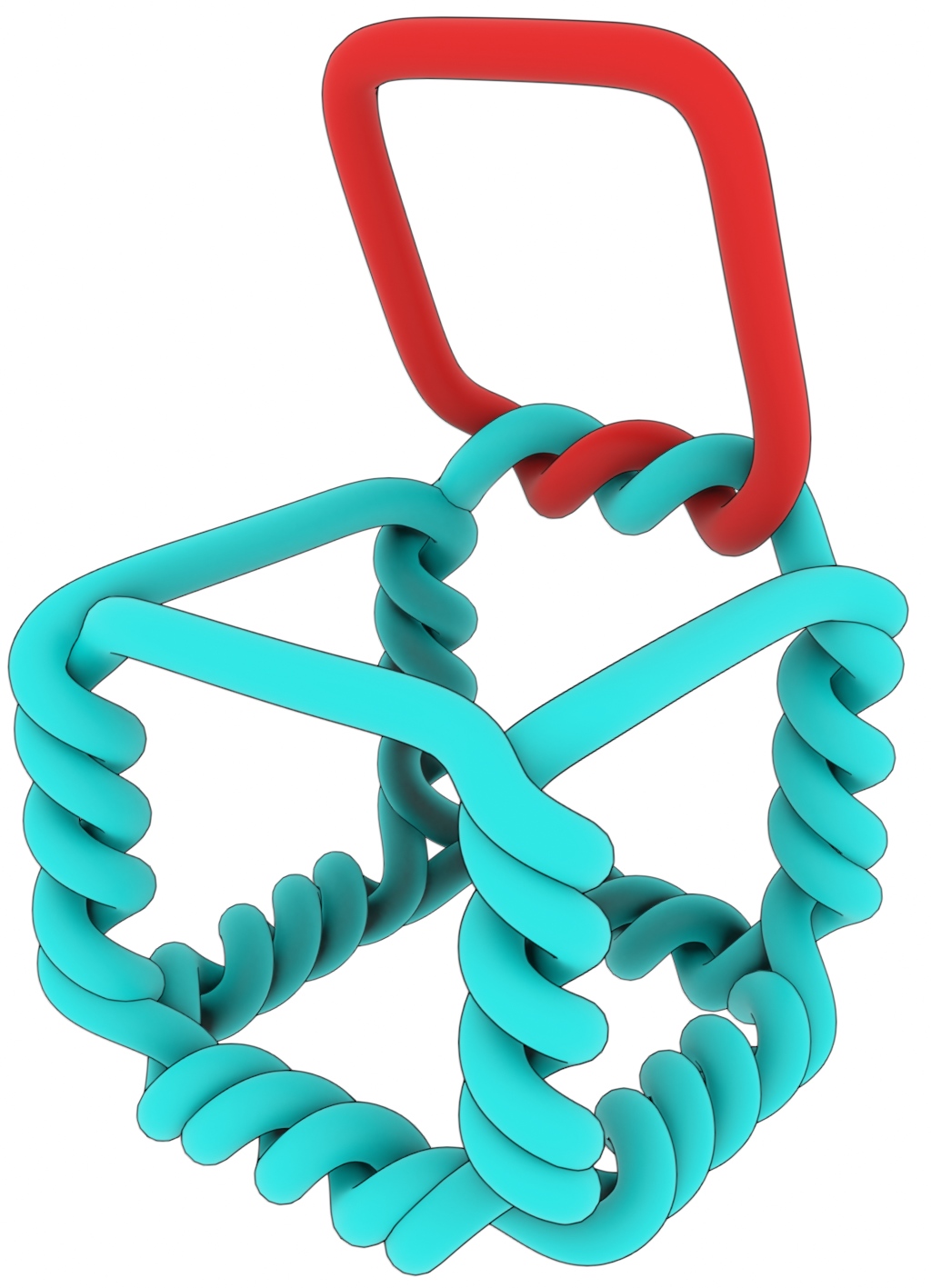}
    \includegraphics[width=0.24\linewidth]{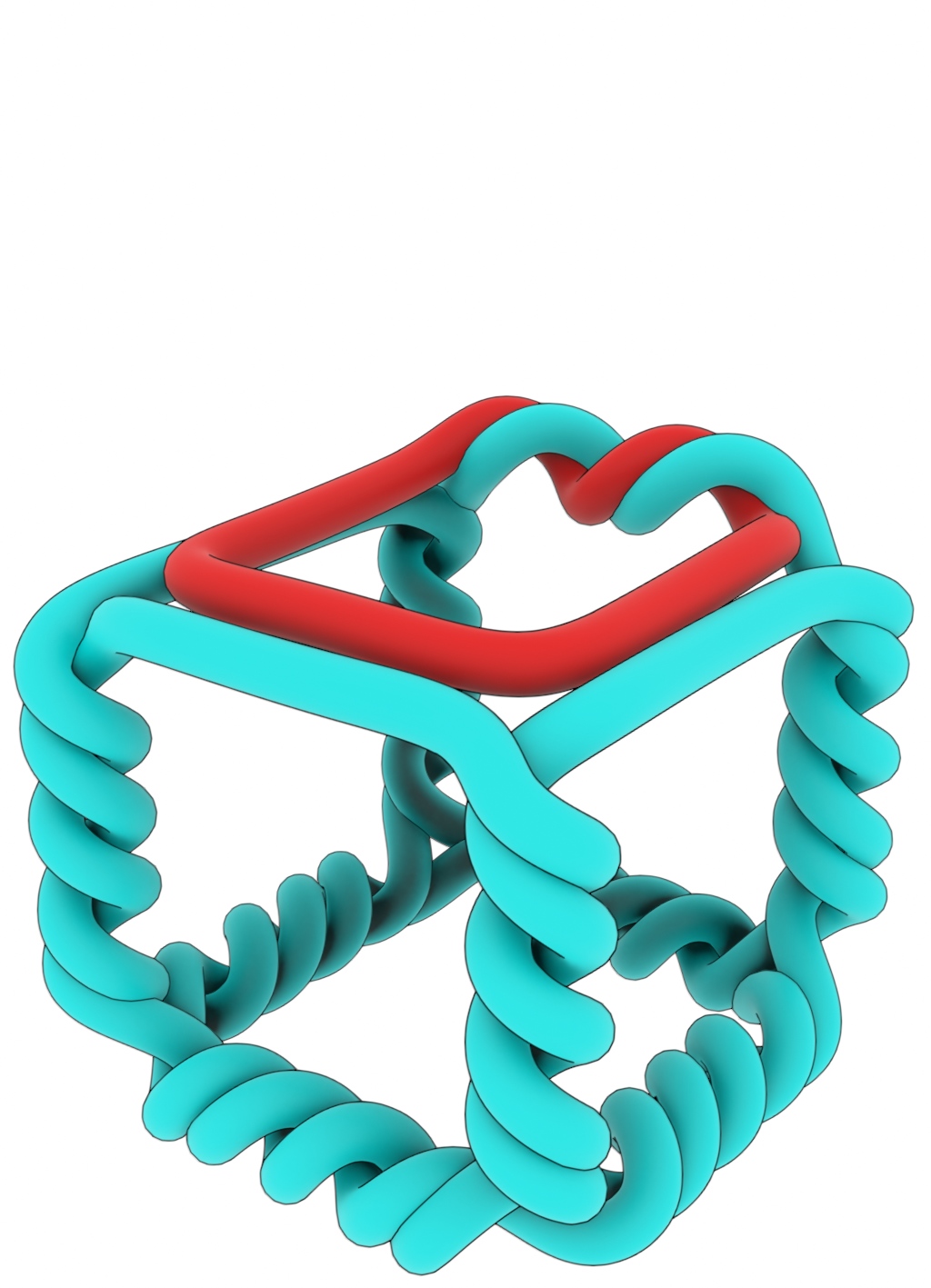}
    \includegraphics[width=0.24\linewidth]{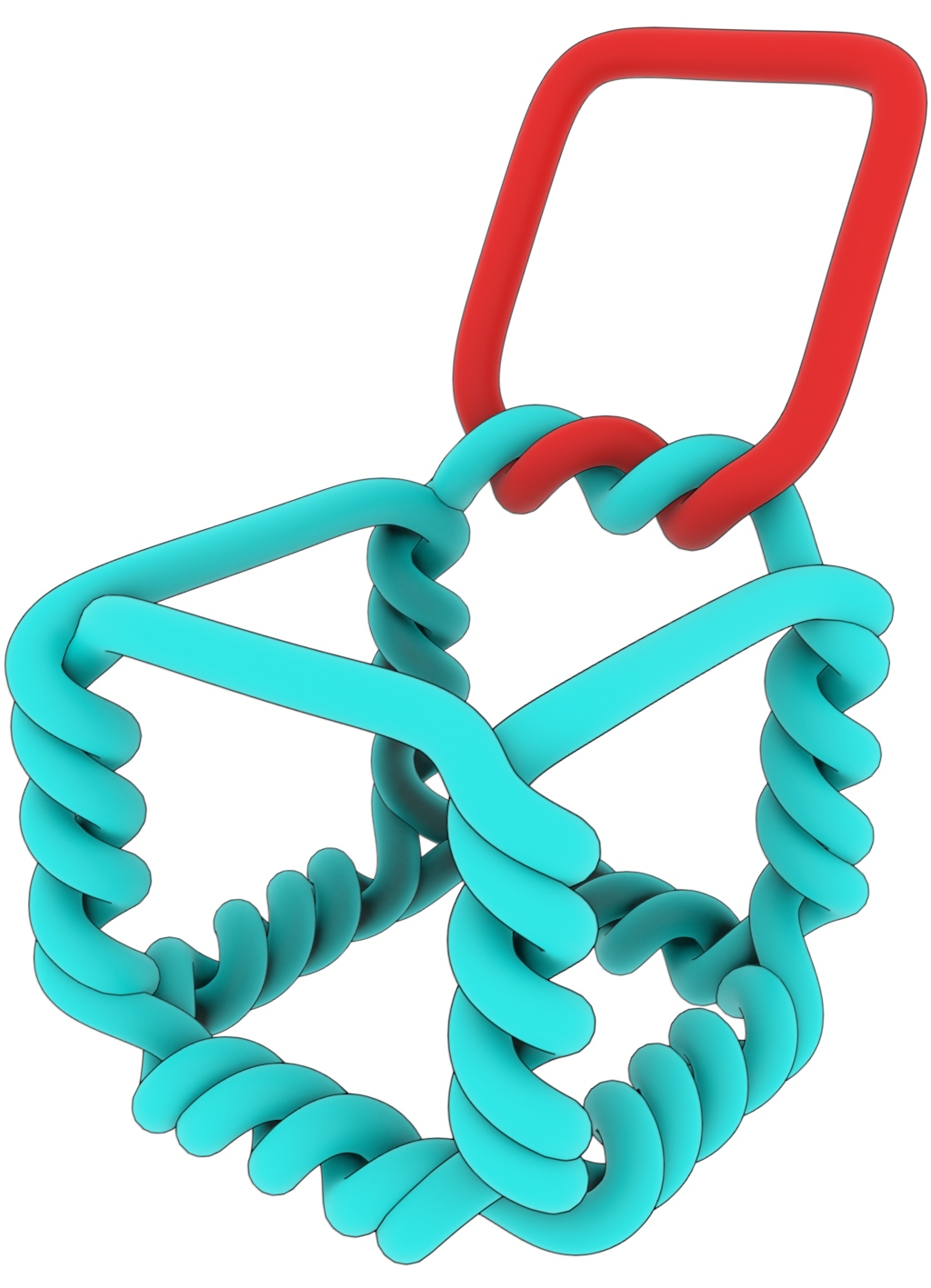}\\
    \parbox[t]{0.24\linewidth}{\centering Input Mesh}
    \parbox[t]{0.24\linewidth}{\centering Open Lid}
    \parbox[t]{0.24\linewidth}{\centering Closed Lid}
    \parbox[t]{0.24\linewidth}{\centering Open Lid}
    \caption{Using unfolded polyhedra, we can design joints that can rotate around an edge. In this example, we are using an unfolded face of a cube to design the lid.}
    \Description{Four images arranged from left to right. The first image shows a simple cube with one face unfolded upward, labeled as the input mesh. The remaining three images show corresponding linked tubular structures rendered mainly in cyan, with a red segment indicating a movable lid. These three examples are labeled “Open Lid,” “Closed Lid,” and “Open Lid,” showing the lid in different rotational positions around a shared edge. The linked structure maintains a cubic overall shape while the red loop changes orientation relative to the rest of the structure.}
    \label{fig:lidexample}
\end{figure}

\subsection{Articulated LK Structures from 2-Manifolds with Boundary}
\label{subsec:designingunfolding}

In this section, we extend the framework to design multiple knots that are intentionally linked to one another. By introducing boundaries into surface meshes and combining them with carefully chosen twist labels, we can control when face cycles merge into individual knots and when they remain distinct but linked. This enables the deliberate construction of linked-knot assemblies in which both the number of components and their interconnections are specified through local edge labels.

Unfolded polyhedra provide a particularly intuitive setting for designing articulated linked-knot structures (see Figures~\ref{teaser2} and~\ref{teaser22}). By introducing boundaries through unfolding, certain edges no longer participate symmetrically in cycle merging, creating hinge-like connections whose behavior is governed entirely by the twist labels assigned along shared edges. These boundaries act as design handles that localize articulation while preserving topological linking.

More complex boundary configurations, such as surfaces with flaps or lids (Figures~\ref{teaser3},~\ref{teaser23}, and~\ref{fig:lidexample}), extend this idea further. In these examples, attached faces function as topological appendages whose motion is constrained by their attachment edges. Although these structures remain ordinary 2-manifolds with boundary, the placement of boundaries selectively prevents cycle merging and introduces asymmetric articulation patterns. From a design perspective, flaps and lids allow parts of the structure to rotate, open, or close relative to the rest of the assembly, while remaining consistently linked to neighboring components.

We emphasize that these connections are not mechanical hinges in the classical sense. Their apparent articulation arises from topological linking combined with boundary placement, and can be further shaped through geometric realization parameters such as strand thickness, twist magnitude, and local spacing. This distinction reinforces the design-oriented nature of the framework: articulation is specified discretely at the level of surface connectivity and twist labels, while geometry serves as a means of realization rather than control.

While manifolds with boundary enable localized articulation through selective cycle separation, the next section shows how introducing non-manifold connectivity further generalizes this idea by allowing articulation and coupling to be embedded directly into the surface topology itself.

\subsection{Articulated LK structures from Piano-Hinged 2-manifolds}
\label{subsec:designingpeanohinged}

In this subsection, we consider a broad class of LK structures derived from \emph{edge-hinged non-manifold surface assemblies}, in which multiple 2-manifold surface components are connected along shared edges that function as topological hinges. A wide range of classical reconfigurable constructions fall into this category, including the Yoshimoto cube~\cite{yoshimoto1971cube}, Peano-hinged dissections~\cite{frederickson2002hinged}, flexagons~\cite{stone1942flexagons,sherman1994flexagons}, kaleidocycles~\cite{schattschneider1987escher}, and related hinged surface assemblies explored in both recreational mathematics and design. These examples differ in the structure of their hinge connectivity—some forming tree-structured (acyclic) assemblies, others forming cycle-structured assemblies that support intrinsic state transitions—but they share a common underlying principle: articulation and reconfiguration arise from non-manifold edge connectivity rather than from mechanical joints. Our framework unifies these constructions by treating hinge edges as labeled non-manifold edges, enabling them to be systematically transformed into articulated linked-knot (LK) structures within a single design paradigm.

\begin{figure}[htb!]
    \centering
    \begin{overpic}[width=0.32\columnwidth]{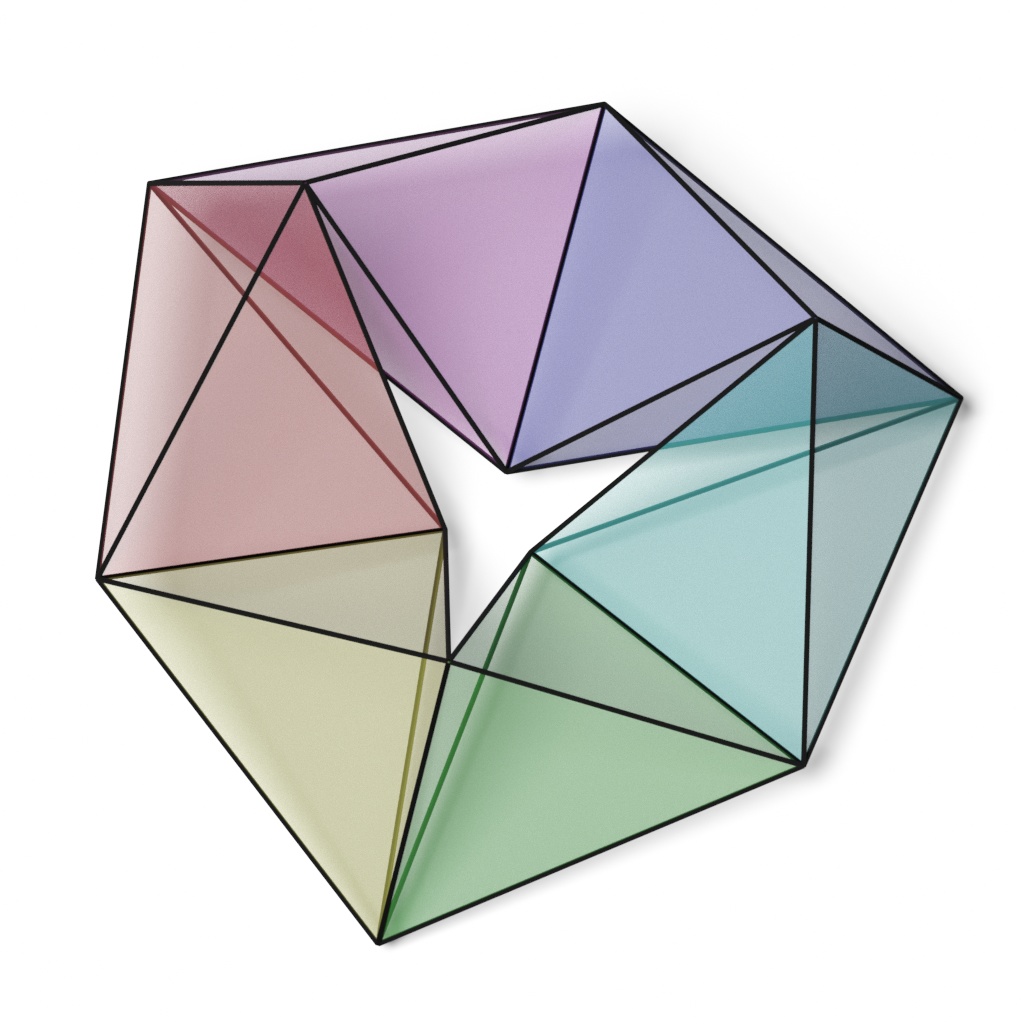}
        \put(-5,85){1}
    \end{overpic}
    \begin{overpic}[width=0.32\columnwidth]{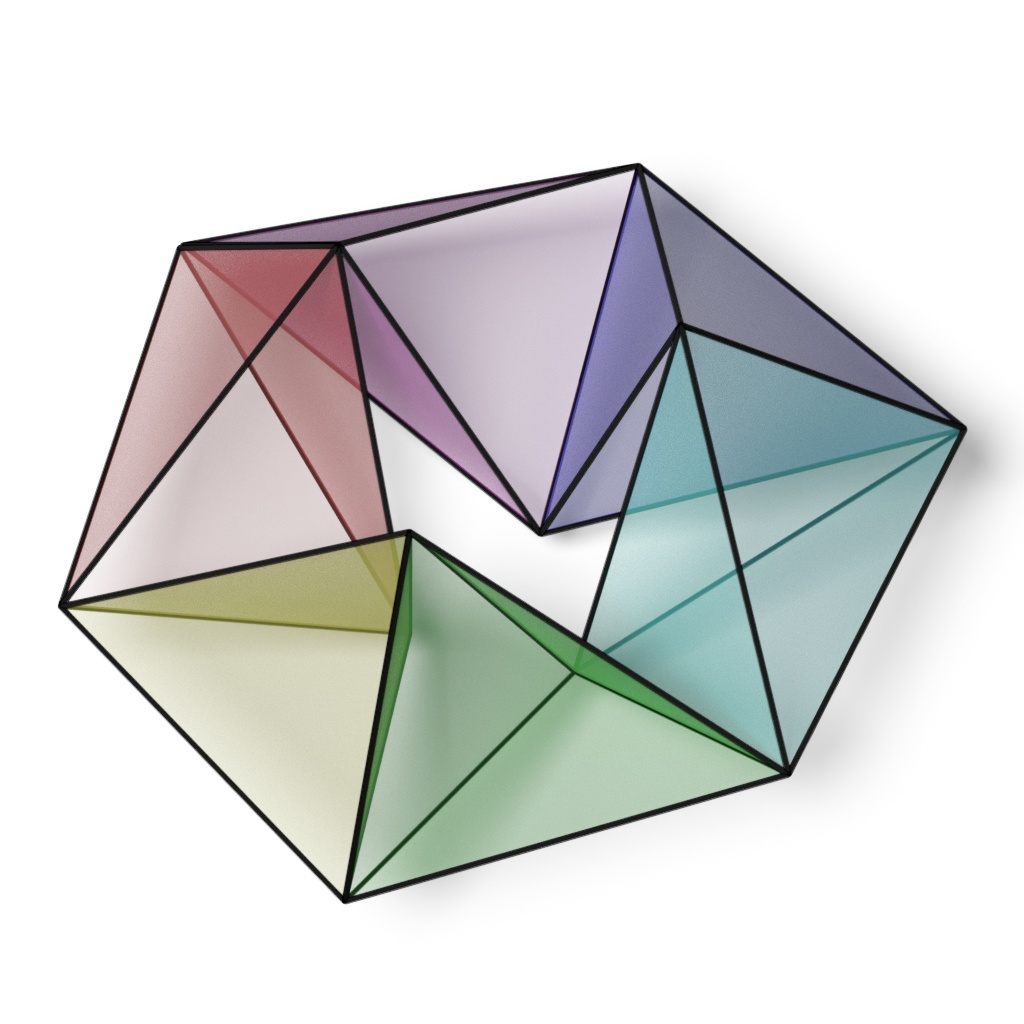}
        \put(-5,85){2}
    \end{overpic}
    \begin{overpic}[width=0.32\columnwidth]{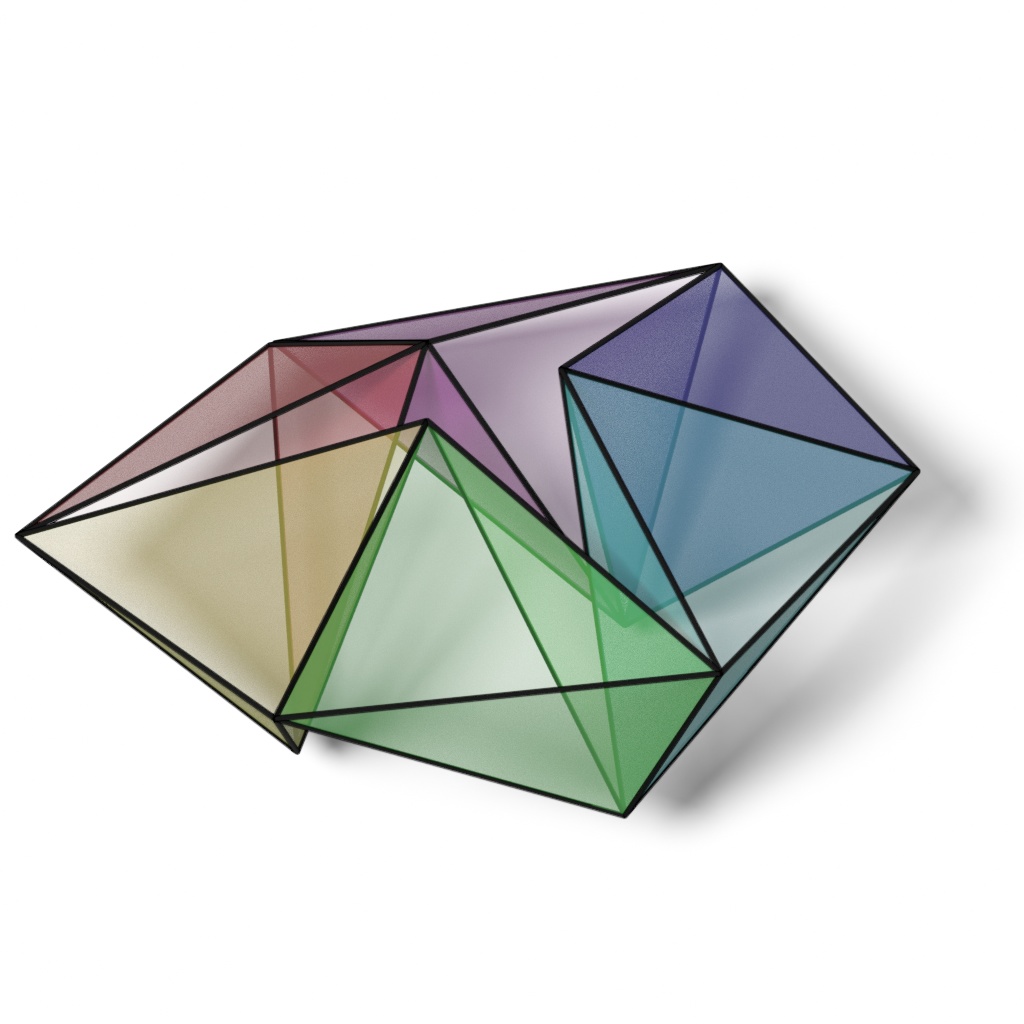}
        \put(-5,85){3}
    \end{overpic}    \includegraphics[width=0.32\columnwidth]{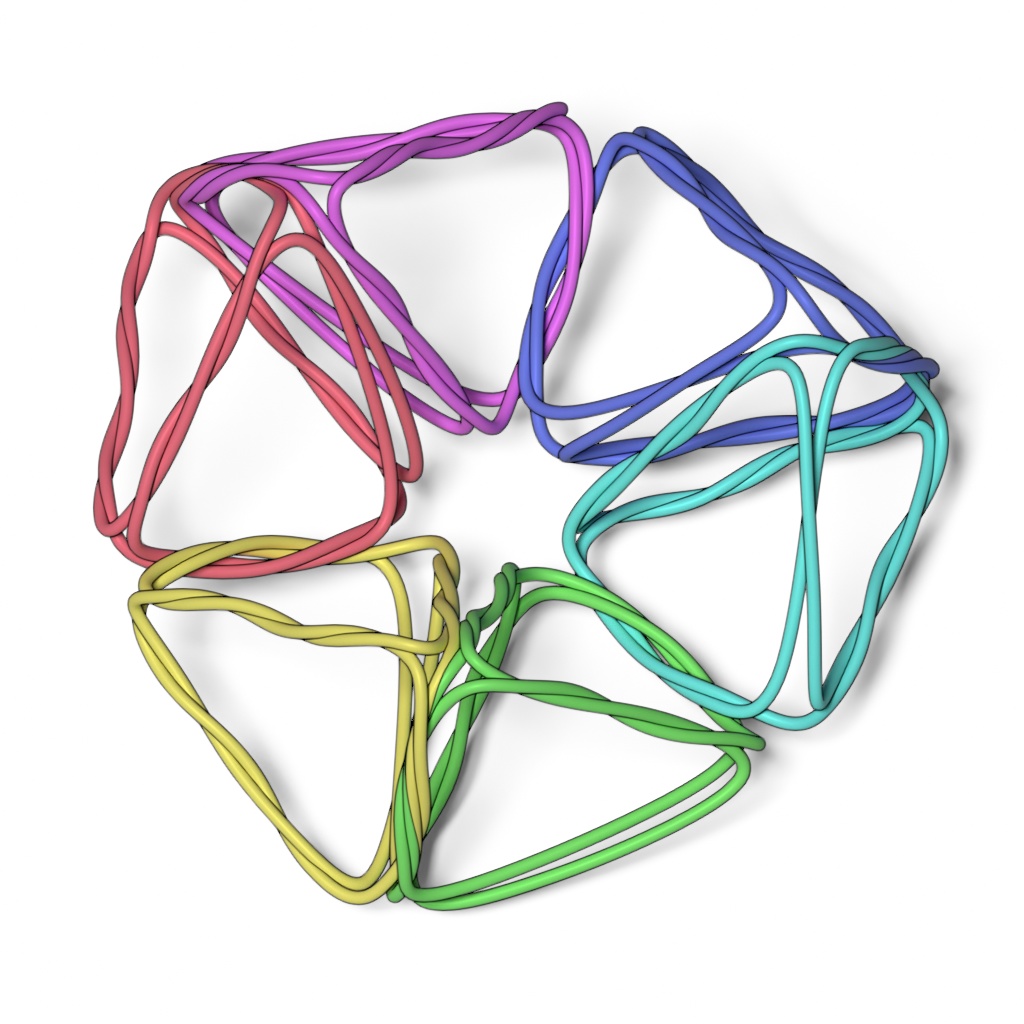}    \includegraphics[width=0.32\columnwidth]{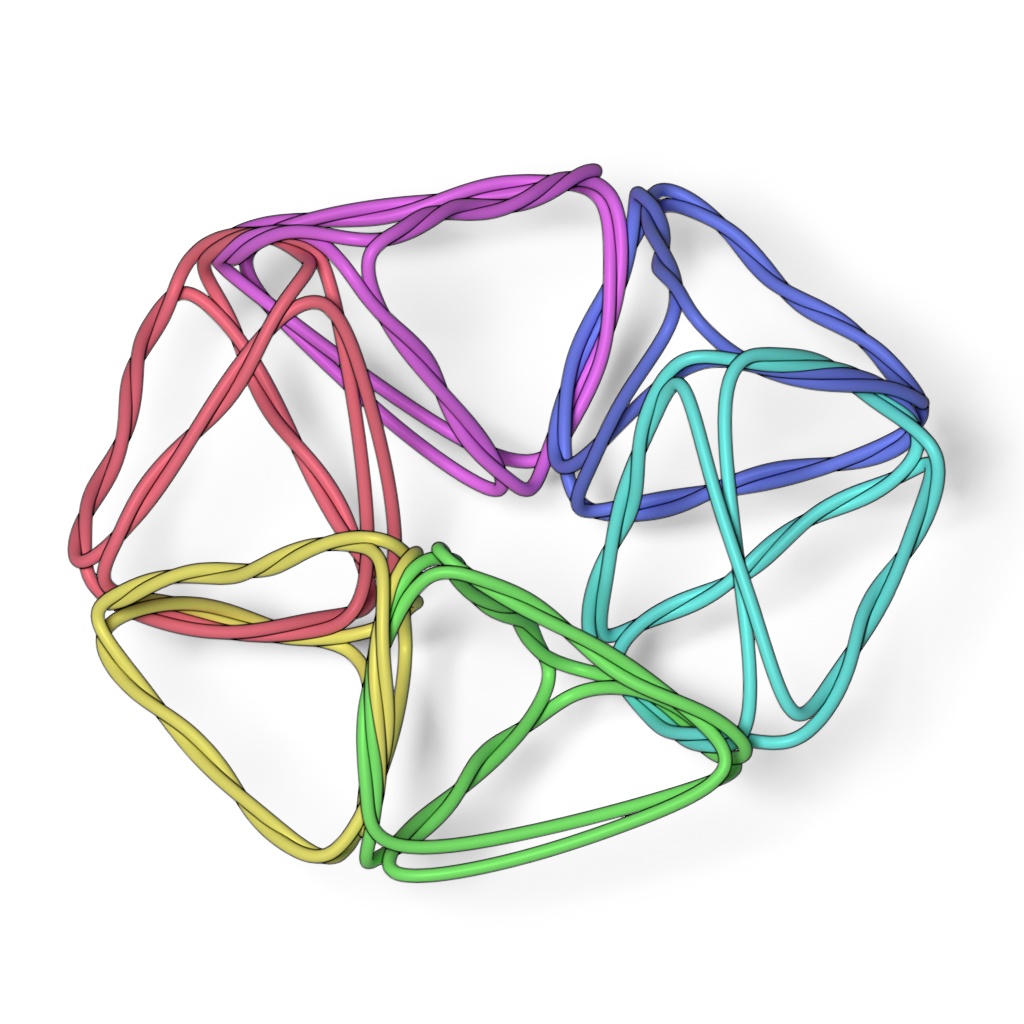}
   \includegraphics[width=0.32\columnwidth]{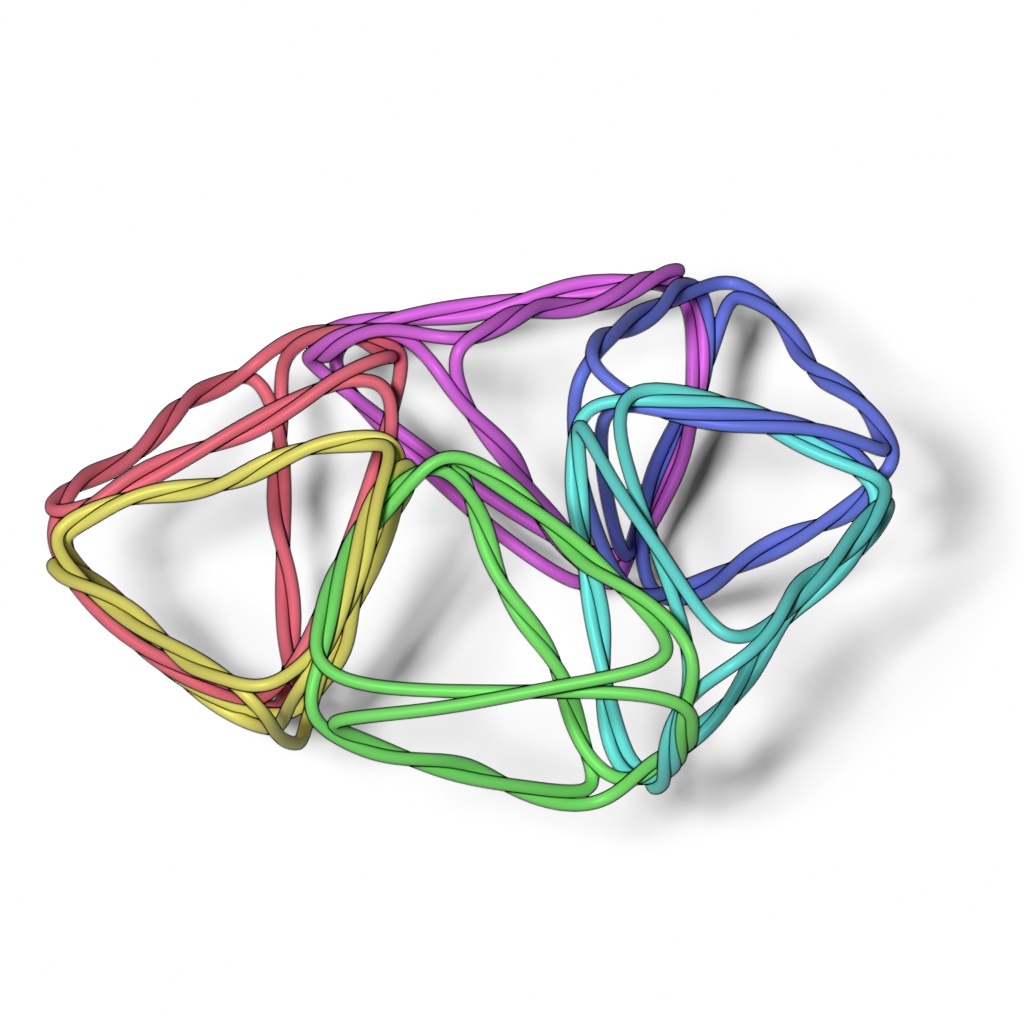}
    \begin{overpic}[width=0.32\columnwidth]{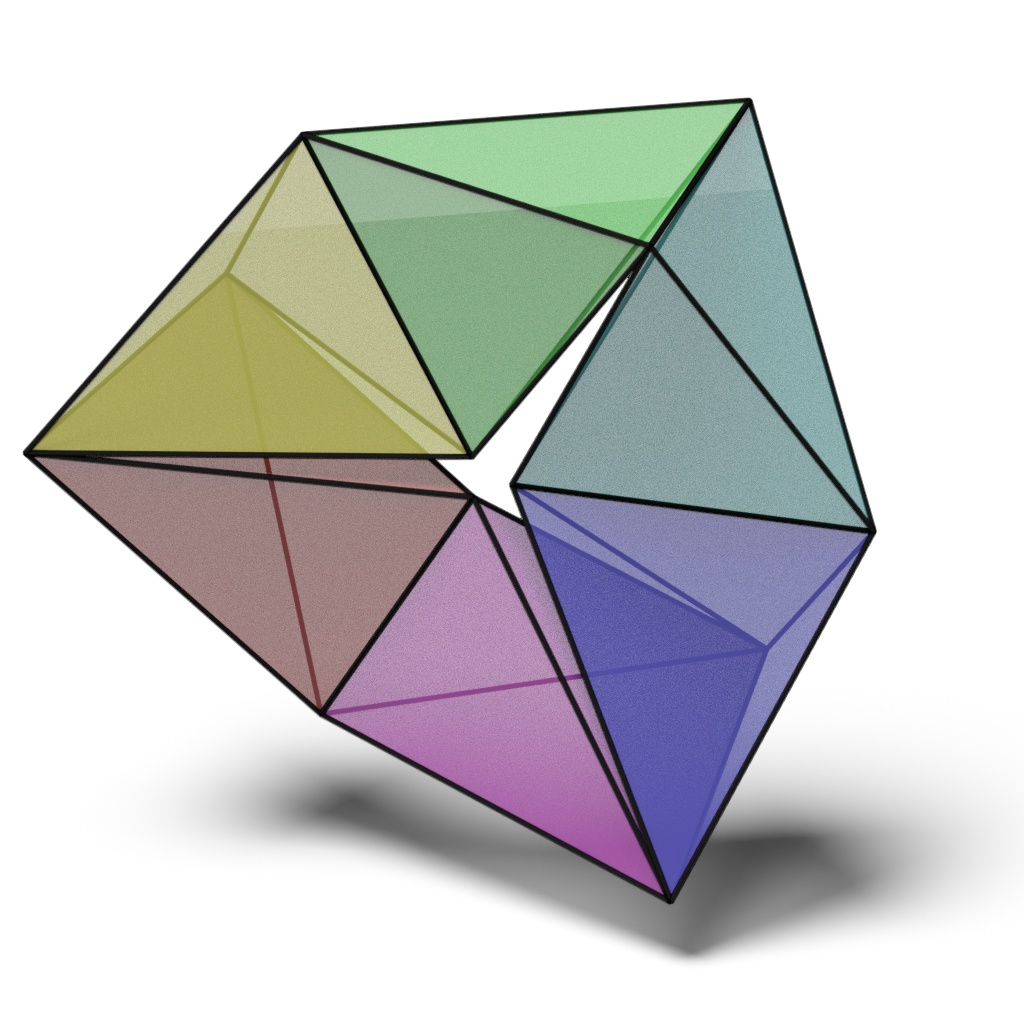}
        \put(-5,85){4}
    \end{overpic}
    \begin{overpic}[width=0.32\columnwidth]{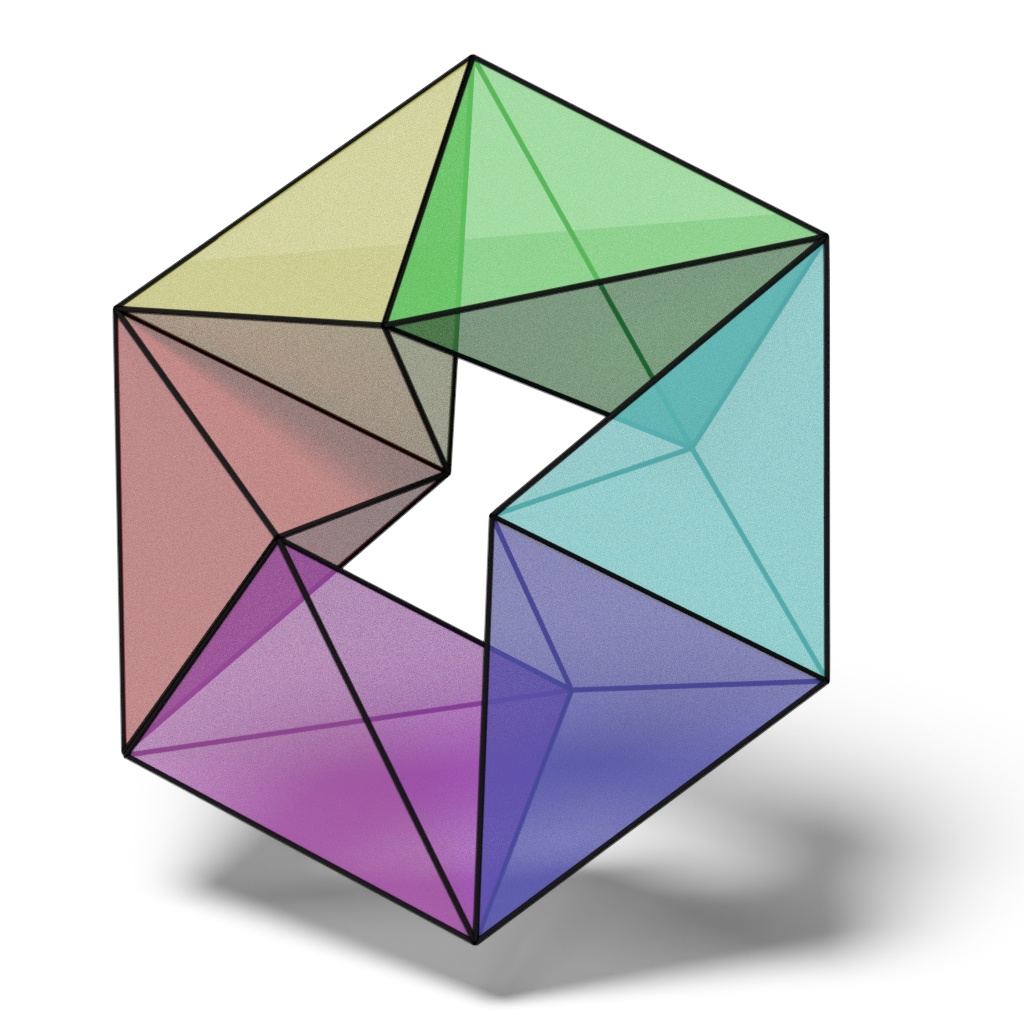}
        \put(-5,85){5}
    \end{overpic}
    \begin{overpic}[width=0.32\columnwidth]{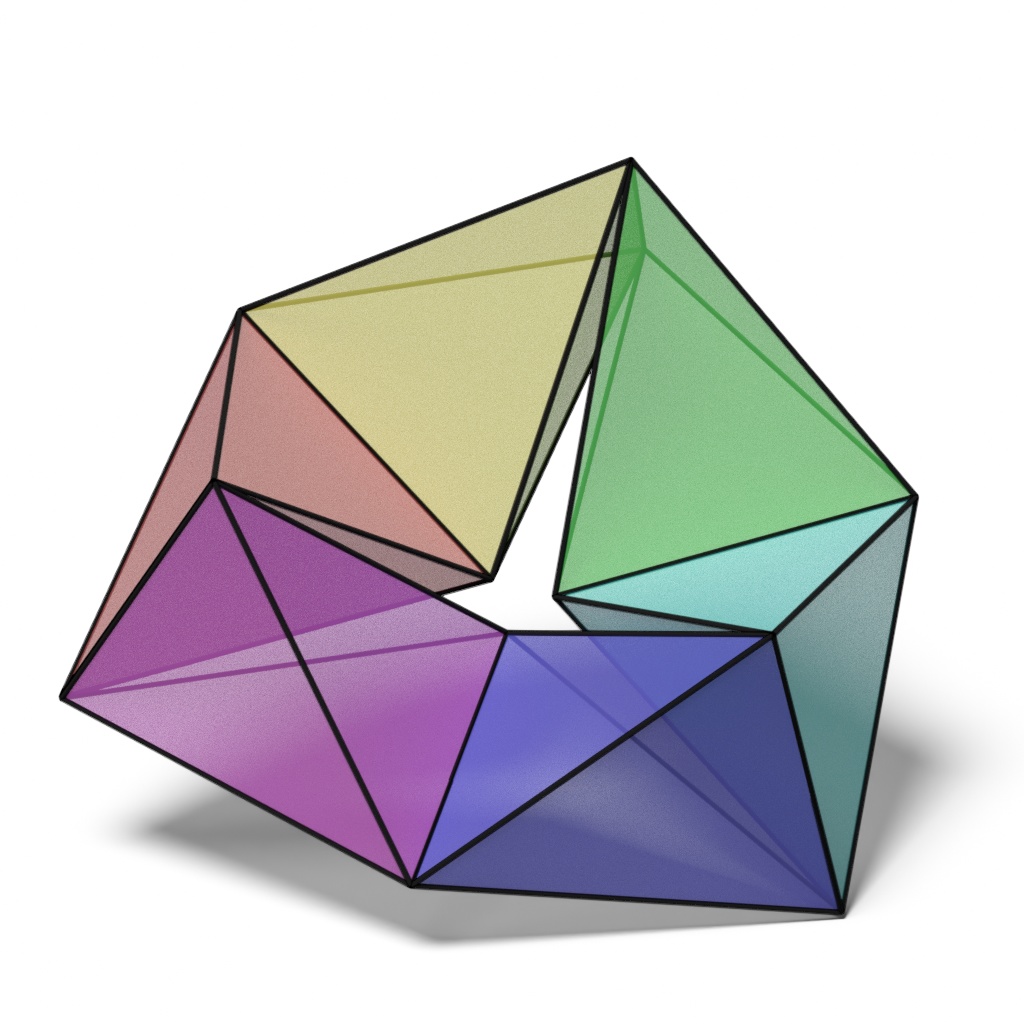}
        \put(-5,85){6}
    \end{overpic}
   \includegraphics[width=0.32\columnwidth]{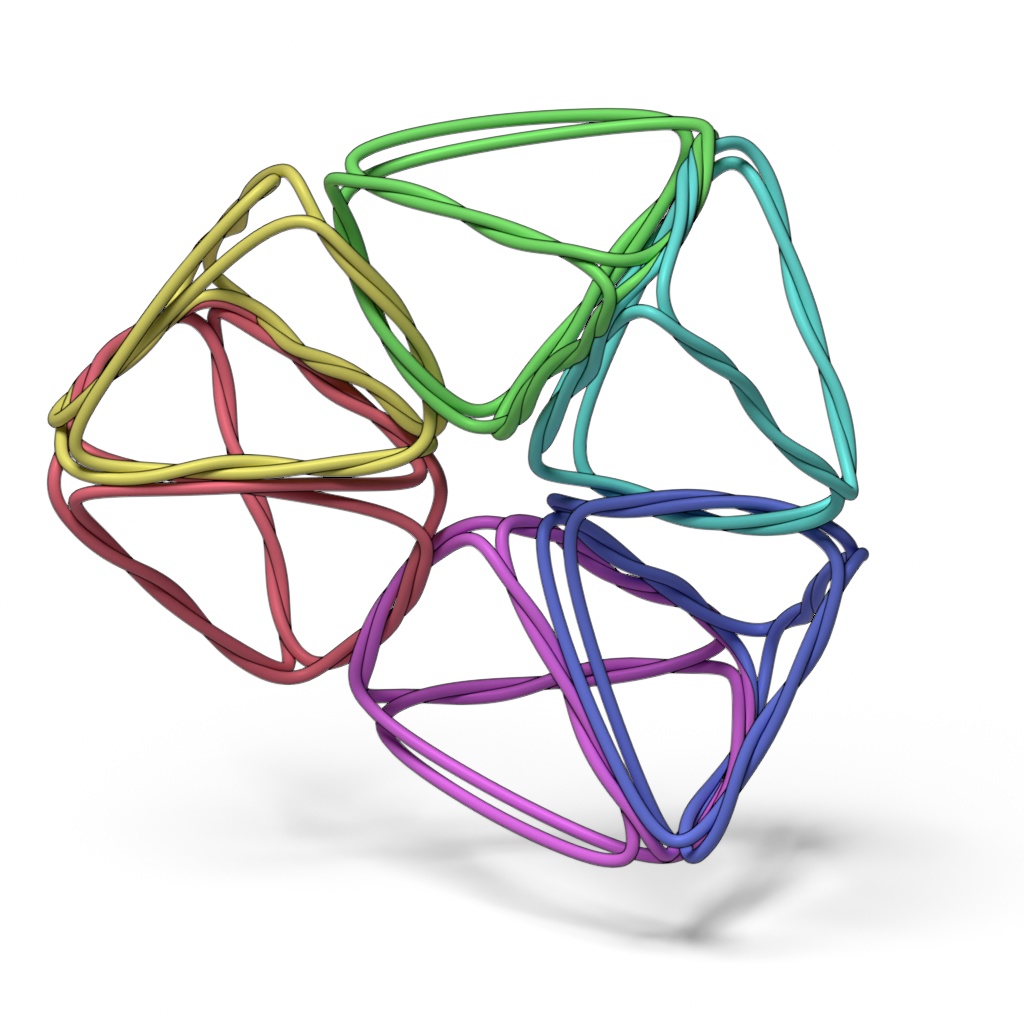}
 \includegraphics[width=0.32\columnwidth]{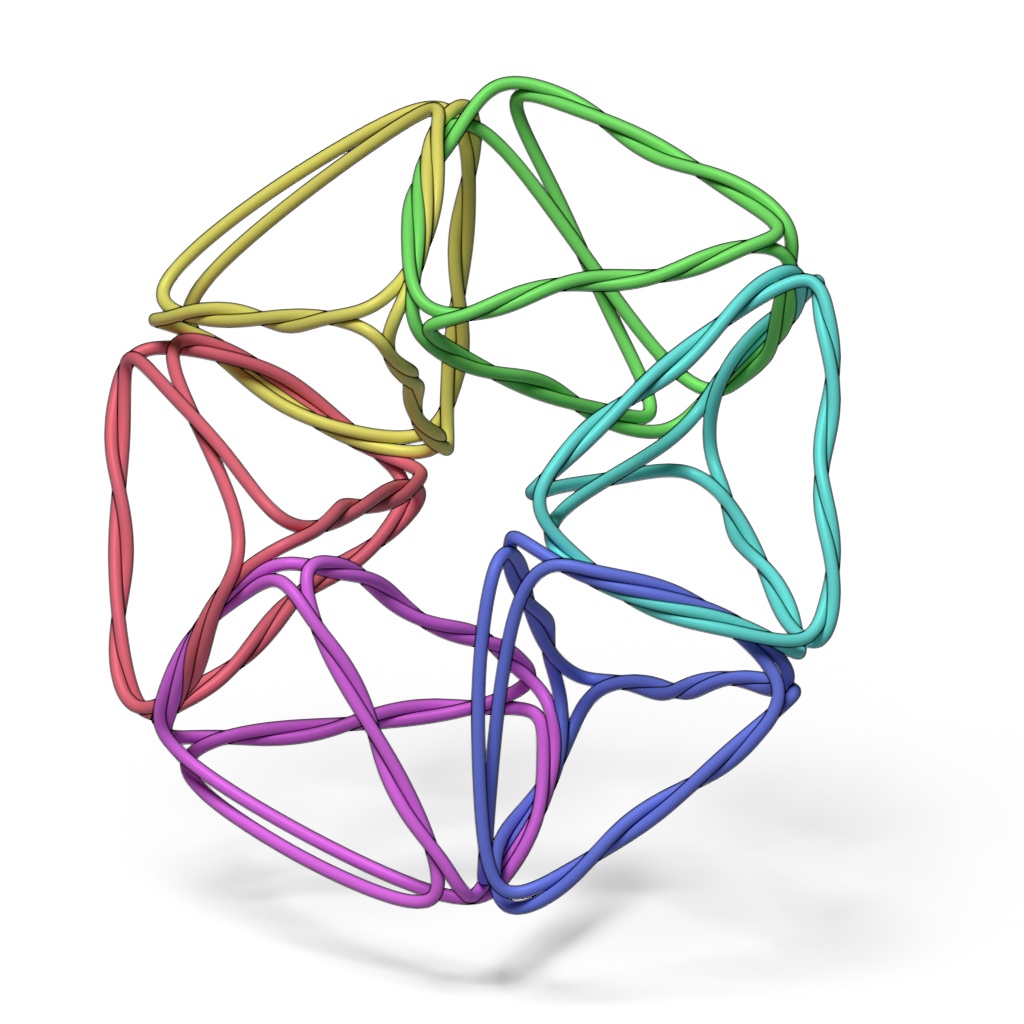}    \includegraphics[width=0.32\columnwidth]{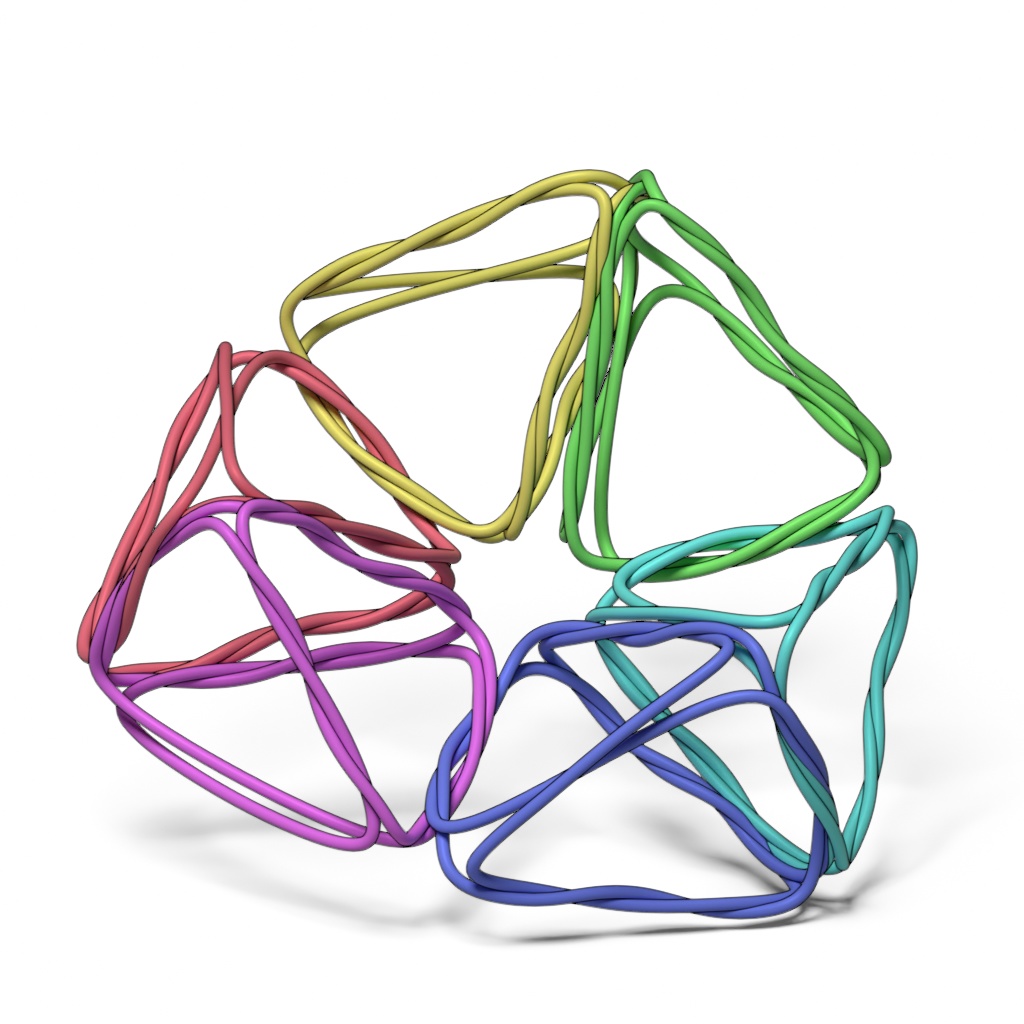}
    \caption{Six reconfiguration states of a flexagon-based LK structure. Each image shows a distinct configuration produced by continuous deformation of the same labeled non-manifold surface scaffold, with topology preserved. The structure is defined by integer twist labels on edges, merging tetrahedral components into knots while allowing controlled relative motion. This example demonstrates how classical flexagon reconfiguration emerges naturally within the LK framework.}
    \Description{A grid of six numbered examples arranged in two columns and three rows. Each example consists of two vertically stacked images. The upper image shows a semi-transparent polyhedral structure composed of colored triangular faces, forming a compact three-dimensional shape with visible edges and internal openings. The lower image shows the corresponding linked tubular structure rendered as intertwined strands in matching colors. Across the six examples, the overall connectivity and coloring remain consistent, while the shapes vary through bending, folding, and relative rotation of components. All configurations are shown from comparable viewpoints to emphasize the gradual geometric reconfiguration.}
    \label{fig:flexagon}
\end{figure}

All of the constructions discussed above can be understood as assemblies of ordinary 2-manifold surfaces connected along shared edges in a non-manifold manner. Within our framework, each such 2-manifold component may be independently transformed into a knot by applying twist labels that merge its face cycles into a single cycle. These knotted components are not articulated internally; rather, they behave as flexible but topologically fixed knots. The non-manifold edges that connect them then act as higher-degree edges incident to multiple faces, and assigning twist labels that are nonzero multiples of the local edge degree $K$ preserves the individual knot structure while introducing controlled linking between components. In this way, classical edge-hinged surface assemblies can be lifted systematically into linked-knot (LK) structures, yielding linked but non-articulated configurations whose connectivity and coupling are specified entirely through labeled non-manifold edges.

Taken together, edge-hinged non-manifold surface assemblies provide a powerful mechanism for constructing complex LK structures that combine reconfiguration principles from classical hinged designs with topological linking. By lifting individual surface components into knots and coupling them through labeled non-manifold edges, the framework enables linked structures that inherit the organizational logic of familiar articulated objects while extending them into the domain of knots and links. 

Figure~\ref{fig:flexagon} demonstrates how classical flexagon-style reconfiguration emerges naturally within the LK framework, yielding multiple stable configurations from a single labeled non-manifold surface scaffold. This perspective reveals that many celebrated reconfigurable surfaces represent specific points within a broader design space of non-manifold LK structures, and that their characteristic motions and couplings can be systematically reproduced, modified, or generalized through discrete control of surface connectivity and twist labels.

Having established how bounded non-manifold surface scaffolds give rise to finite LK structures, we now turn to infinite constructions, where periodic and unbounded scaffolds generate families of LK structures that extend indefinitely in space.

\section{Design Space of Periodic LK structures}

Many knotted and linked structures encountered in practice, such as woven fabrics, knitted textiles, and chainmail, are inherently \emph{periodic}. These structures arise from the repetition of local interlacing patterns and can be naturally modeled using \emph{periodic non-manifold structures} in two or three dimensions. Within this framework, familiar textile patterns such as plain weave, twill, and satin appear as specific low-complexity instances of a much broader class of periodic linked knot (LK) structures.

While these canonical examples provide useful intuition, the true design potential of periodic LK structures lies well beyond such familiar cases. By varying both the underlying periodic non-manifold scaffold and the twist labels assigned to its repeating units, one can generate a vast family of previously unexplored woven and knitted forms. This includes volumetric weaves and three-dimensional knits that have no direct analogue in traditional textile practice and that exhibit fundamentally different modes of interlacing and connectivity. 

At the same time, this generality introduces an enormous design space. Periodic LK structures can be parameterized in multiple, largely independent ways: by the choice of periodic non-manifold connectivity, by the assignment of twist labels within a periodic unit, and by the selection of the periodic unit itself. In principle, the space of possible periodic units is unbounded, encompassing families of infinite polyhedra, regular maps, and cell-transitive two- and three-dimensional honeycombs.

In this paper, we do not attempt to exhaust this space. Instead, we focus on a structured and representative subset of periodic units: cell-transitive two- and three-dimensional honeycombs generated by \emph{Wigner-Seitz cells}. This choice provides a principled balance between expressive power and conceptual clarity, enabling systematic exploration of periodic LK structures while keeping the underlying design parameters explicit and interpretable. While some Wigner-Seitz constructions correspond to well-known textile patterns, many volumetric weaves and knits generated by these constructions appear to lie outside established classes of woven or knitted structures. In the following subsections, we illustrate this design space using progressively richer periodic examples, beginning with uniform edge-twist assignments.

\subsection{Periodic LK Structures from Wigner-Seitz-Based Cell-Transitive Honeycombs}

Our choice of Wigner-Seitz cells is not arbitrary. Much of the historical work on planar weaves can be understood implicitly as \textit{``labeled cell-transitive 2-honeycombs''} based on Wigner-Seitz cells \cite{Grunbaum80,Grunbaum85,Grunbaum86,grunbaum1988isonemal}, in which labels are assigned to faces, as discussed in \cite{yildiz2025woven}. That work \cite{yildiz2025woven}, in turn, extends naturally to volumetric weaves through labeled Wigner-Seitz-based 3-cell-transitive honeycombs, where labels are associated with volumetric cells. While these approaches reveal a rich class of woven and interlaced structures, their design expressiveness is constrained by the choice of labeling domain.

In contrast, labeling edges provides a significantly more powerful design paradigm. Edges naturally support the notion of twist, enabling direct control over local linking behavior and leading to a stronger and more general theoretical framework \cite{tait1877knots,akleman2009cyclic,akleman2015extended}. The approach presented in this paper belongs to this lineage and extends these earlier ideas by systematically exploiting edge-based labels within periodic non-manifold scaffolds.
Although the underlying operation is simple, the contribution of this work lies in identifying integer-valued edge twisting on surface-based scaffolds as a unifying representation that exposes a large, structured design space of linked and knotted forms, rather than isolated constructions.

\subsection{Wigner-Seitz Cells}

We begin by reviewing Wigner-Seitz constructions and then illustrate their use as periodic scaffolds for LK structures. Wigner-Seitz constructions are based on Bravais lattices, which provide a standard formalism for describing spatial periodicity in crystalline and material systems. A Bravais lattice is defined as the set of points generated by all integer linear combinations of a set of linearly independent basis vectors. Let
\[
B := \{\vec{v}_0, \vec{v}_1, \ldots, \vec{v}_{N-1}\} \subset \mathbb{R}^N
\]
be a basis of $\mathbb{R}^N$, and define the associated lattice
\[
\mathcal{L}(B) = \left\{ \sum_{i=0}^{N-1} a_i \vec{v}_i \,\middle|\, a_i \in \mathbb{Z} \right\}.
\]
The set $\mathcal{L}(B)$ forms a full-rank Bravais lattice in $\mathbb{R}^N$.

Given a lattice $\mathcal{L}(B)$, a fundamental domain containing exactly one lattice point can be defined via the \emph{Wigner-Seitz cell}, also known as the \emph{Voronoi cell}. This cell consists of all points in $\mathbb{R}^N$ that are closer to a given lattice point than to any other. Translated copies of the Wigner-Seitz cell tile $\mathbb{R}^N$ without overlap, providing a natural decomposition of space into congruent periodic units.

Up to topological equivalence,  there exist two distinct Wigner-Seitz cell types in $\mathbb{R}^2$, and five distinct Wigner-Seitz cell types in $\mathbb{R}^3$. These cells serve as a finite set of topological scaffolds for constructing doubly and triply periodic LK structures. By restricting LK structure generation to a single Wigner-Seitz cell and enforcing consistency across opposite cell faces, we obtain globally periodic LK structure configurations.

\subsection{Labeling Wigner-Seitz Cells to Obtain Labeling Units}

To make the implications of edge-based labeling explicit, we first consider a simple two-dimensional periodic case.
One significant advantage of this lattice-based framework is that lattice points always form an $N$-dimensional regular array. As a result, arbitrarily large repeating units in 2D and 3D can be defined by labeling rectangular or rectangular-prism arrays that serve as generic repeating units. In fact, the original work by Gr\"unbaum and Shephard exploited this property to develop an extremely simple face-labeling approach \cite{Grunbaum80}. A similar strategy has been used to label 3D lattice points in order to describe repeating units for volumetric woven structures \cite{yildiz2025woven}.

Note that, in our case, we must label edges rather than Voronoi domains (faces in 2D and volumes in 3D). This may appear to be a difficulty; however, an elegant solution already exists. A repeating Voronoi unit cannot be a compact set that includes all of its boundaries, as doing so would introduce overlaps. Instead, some boundary elements must be excluded. When this is done, the repeating unit effectively becomes a subset of the boundary. We can then use lattice positions to label these boundary elements—in our case, edges.

This approach, namely labeling boundary edges, has previously been employed in the construction of polycatenated structures \cite{yildiz2025linked}. In this work, we build on the edge classification developed in \cite{yildiz2025linked} to identify a finite set of distinct edges within each periodic cell. These edges define equivalence classes under lattice translations and symmetries, enabling an enumerable and controllable design space for triply periodic links.

\begin{figure}[htb!]
    \centering
    \rotatebox{90}{\parbox[t]{0.120\textwidth}{\centering cp}}
    \includegraphics[width=0.30\linewidth]{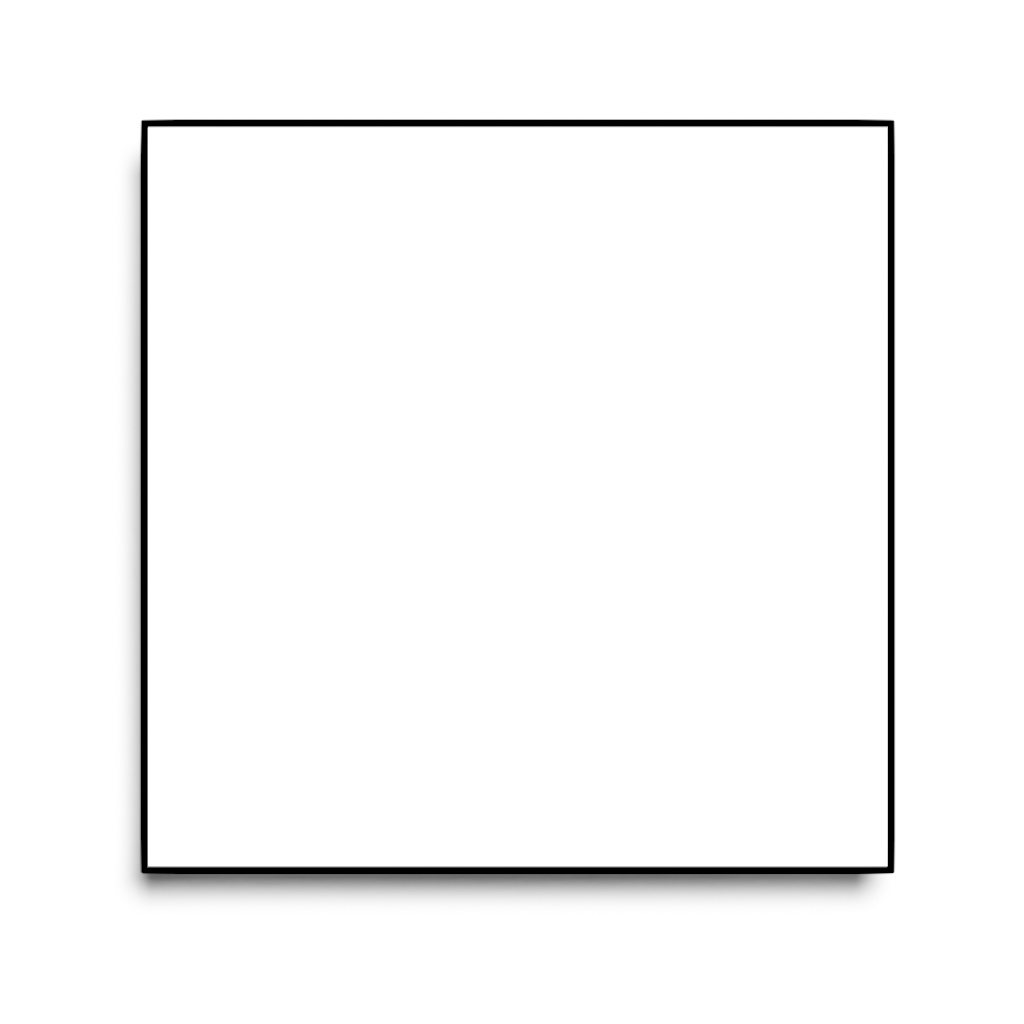}
    \includegraphics[width=0.30\linewidth]{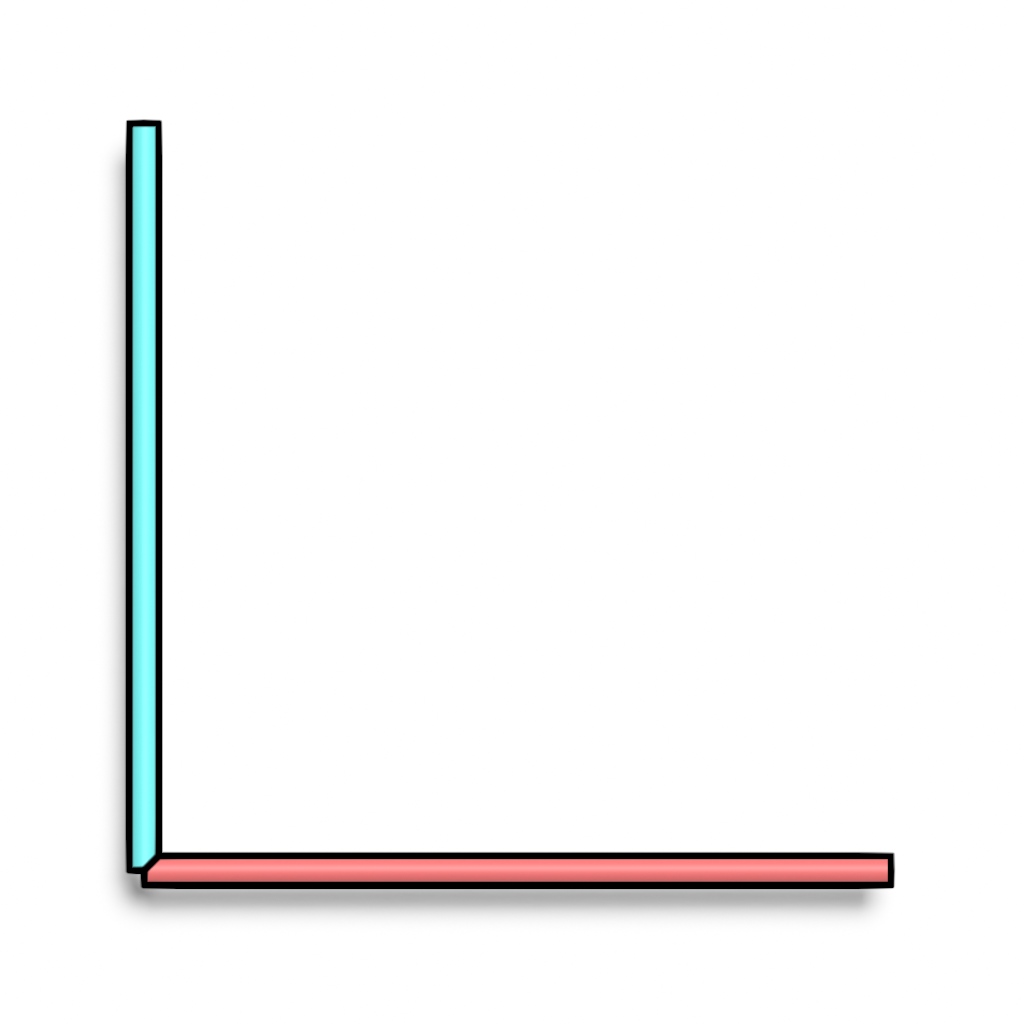}
    \includegraphics[width=0.30\linewidth]{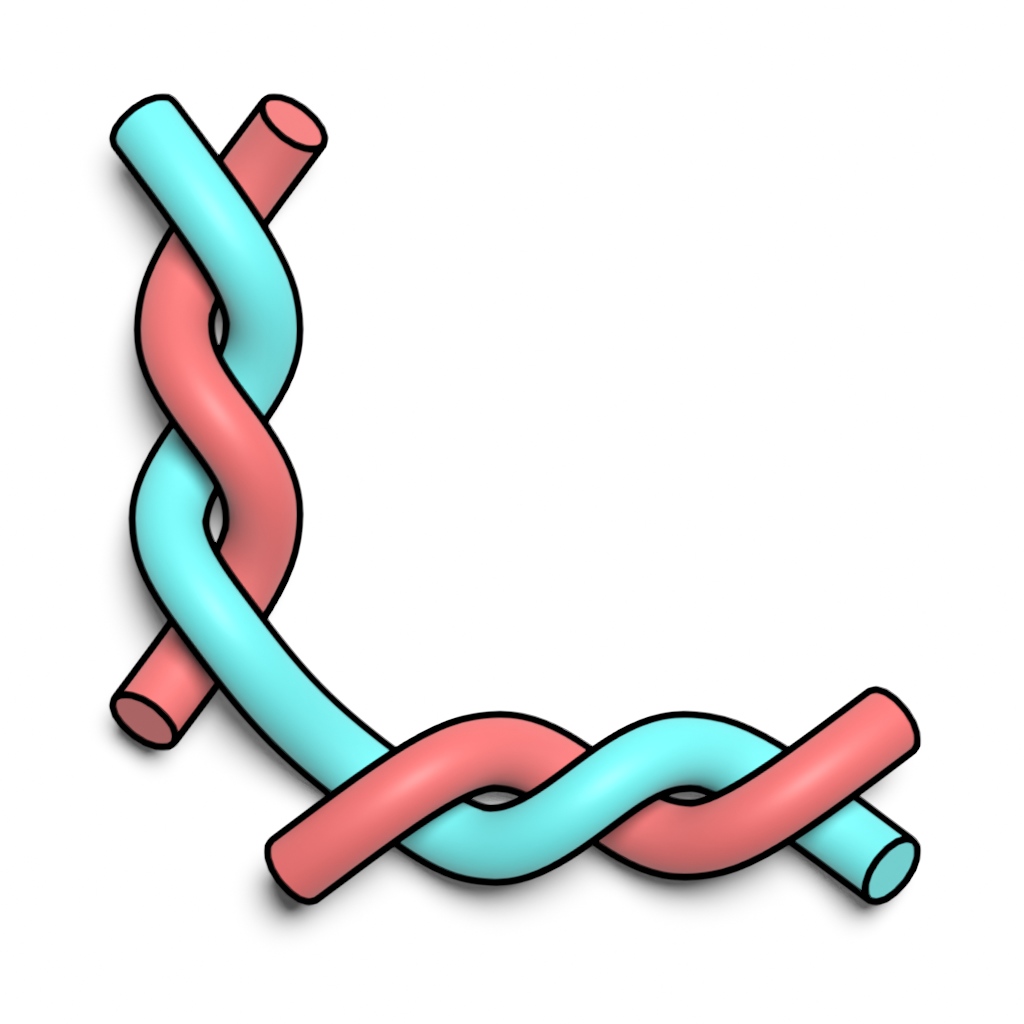}\\
    \rotatebox{90}{\parbox[t]{0.120\textwidth}{\centering hp}}
    \includegraphics[width=0.30\linewidth]{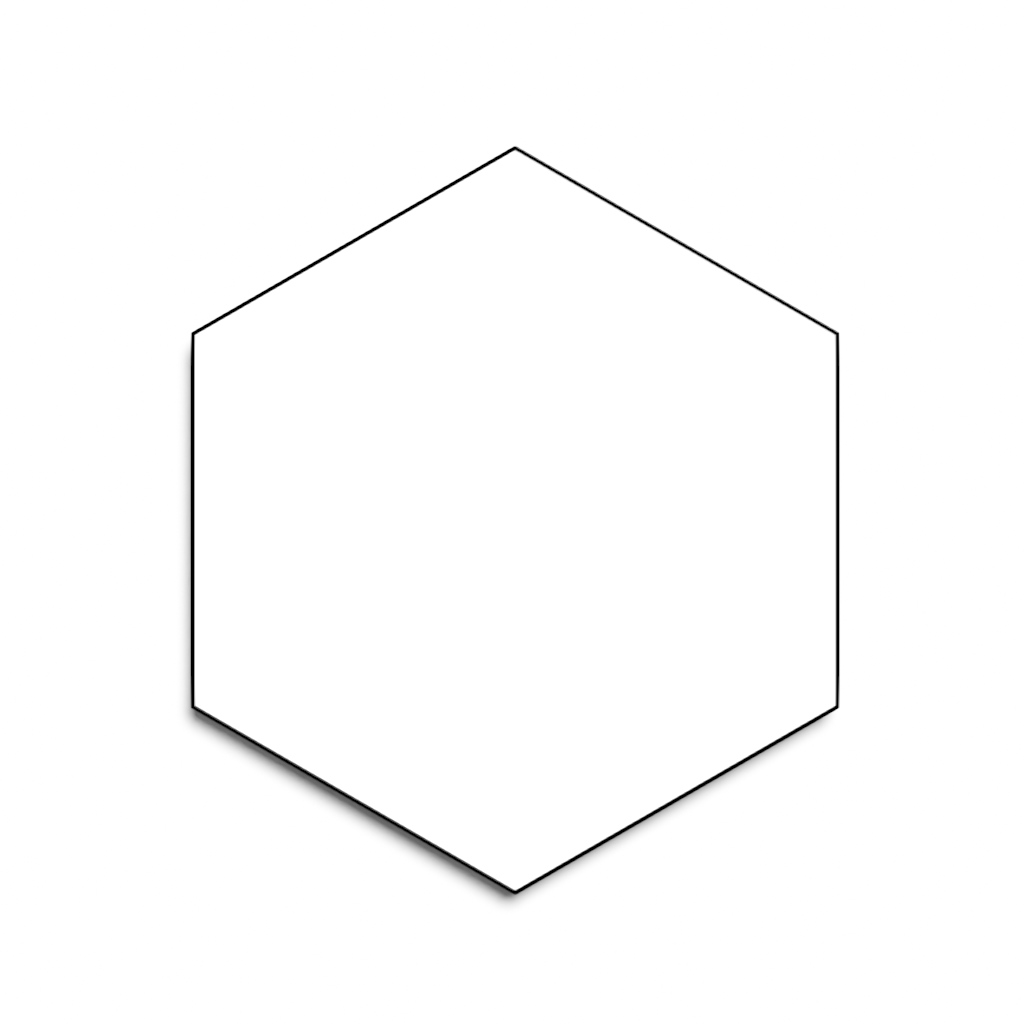}
    \includegraphics[width=0.30\linewidth]{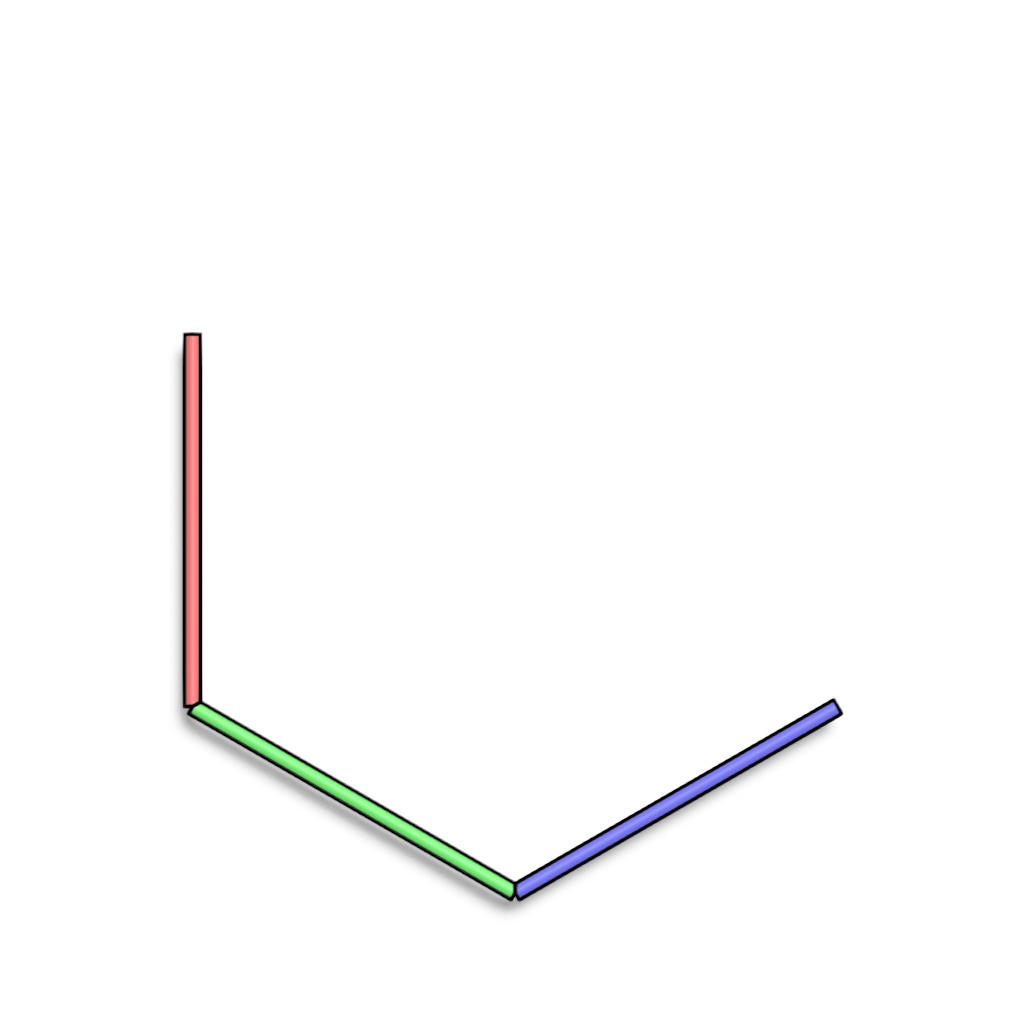}
    \includegraphics[width=0.30\linewidth]{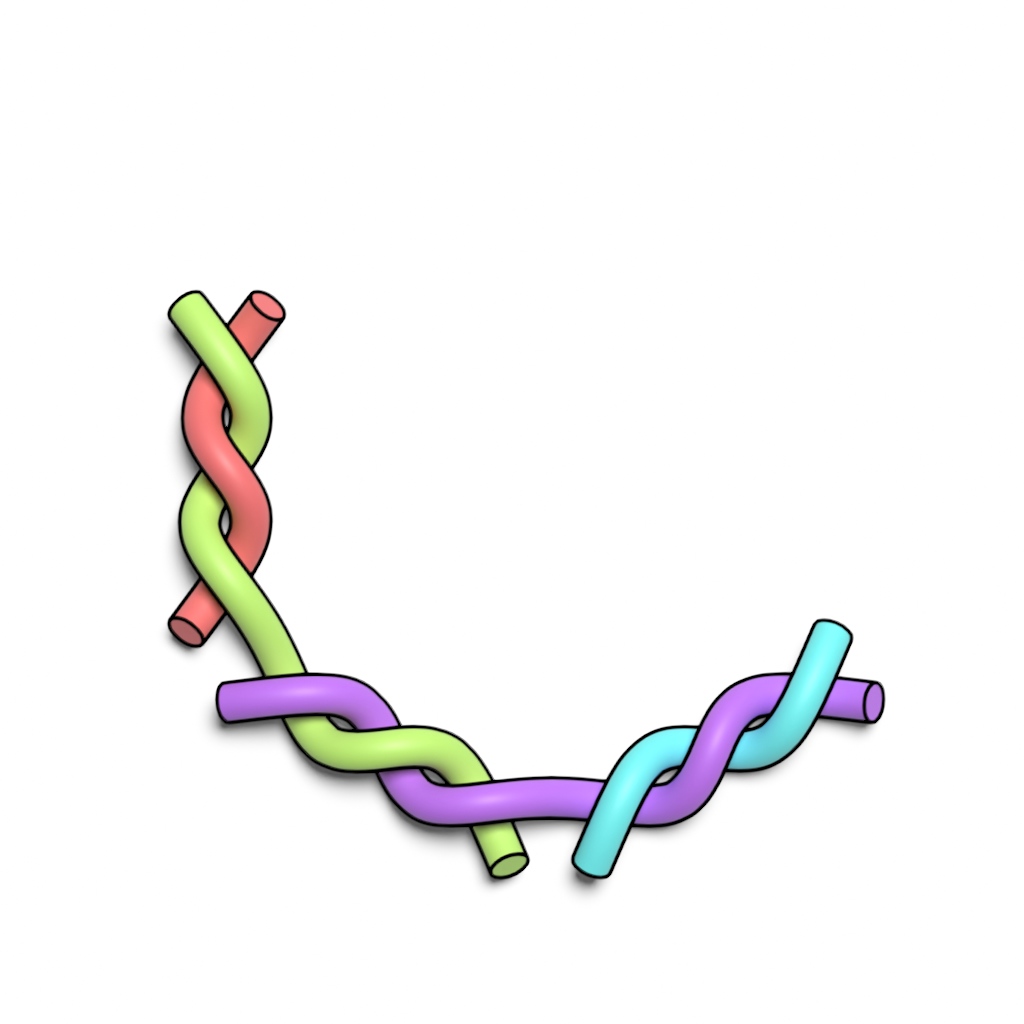}\\
    \rotatebox{90}{\parbox[t]{0.01\textwidth}{\centering }}
    \parbox[t]{0.30\linewidth}{\centering Unit Cells}
    \parbox[t]{0.30\linewidth}{\centering Unique Edges}
    \parbox[t]{0.30\linewidth}{\centering Twisting Elements}
    \caption{In 2D, there are only two topologically distinct Wigner–Seitz cells. We illustrate the unique edges arising in the tessellations of these cells and their associated twisting elements.}
    \Description{A three-column figure arranged in two rows. The left column, labeled “Unit Cells,” shows a square in the top row and a regular hexagon in the bottom row. The middle column, labeled “Unique Edges,” shows simplified edge representatives for each unit cell, drawn as straight line segments meeting at angles, with different colors indicating distinct edge types. The right column, labeled “Twisting Elements,” shows small three-dimensional twisting strand configurations corresponding to each unique edge, rendered as intertwined tubular segments in multiple colors. The top row corresponds to the square unit cell, and the bottom row corresponds to the hexagonal unit cell.}
    \label{fig:bravais2D}
\end{figure}

To explain the idea clearly, we begin with the two-dimensional case. A two-dimensional lattice always forms a regular array of points, and the Wigner-Seitz shapes induced by lattice positions fall into two classes: hexagonal and quadrilateral. For this reason, the regular hexagon and the square serve as canonical Wigner-Seitz shapes in 2D, as shown in Figure~\ref{fig:bravais2D}. 

For the square Wigner-Seitz shape, there are two distinct edge classes, whereas for the hexagonal Wigner-Seitz shape there are three. In both cases, lattice-based labeling assigns integer coordinates along the independent lattice directions, resulting in a two-dimensional labeling scheme for both square and hexagonal arrangements. Assuming that the periodic unit consists of a single Wigner-Seitz cell, the entire space can then be generated by repeating and twisting the corresponding edge elements, as illustrated in Figure~\ref{fig:bravais2D}.

\begin{figure}[htb!]
    \centering
    \rotatebox{90}{\parbox[t]{0.12\textwidth}{\centering Unit Cells}}
    \includegraphics[width=0.18\linewidth]{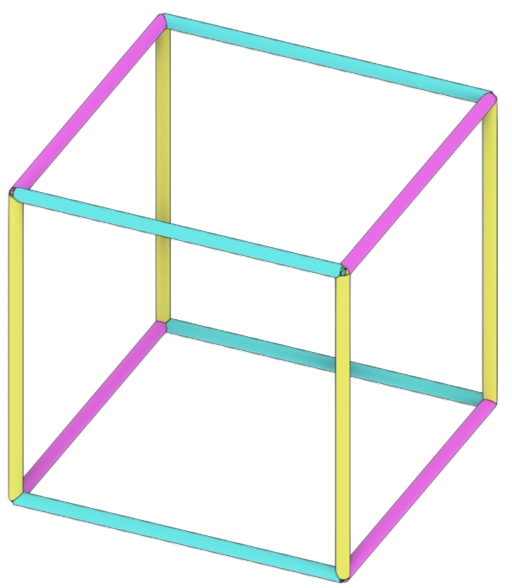}
    \includegraphics[width=0.18\linewidth]{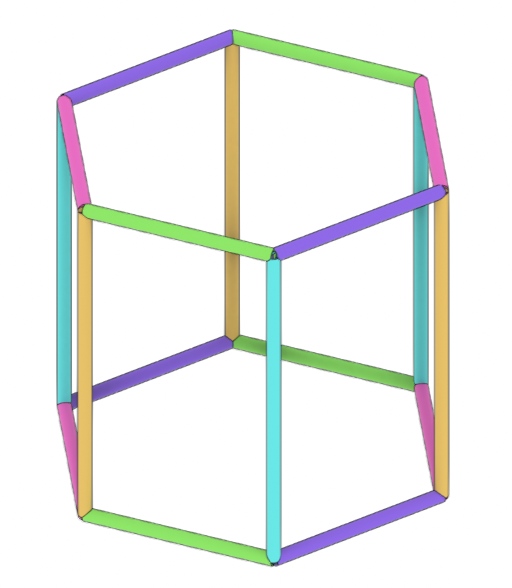}
    \includegraphics[width=0.18\linewidth]{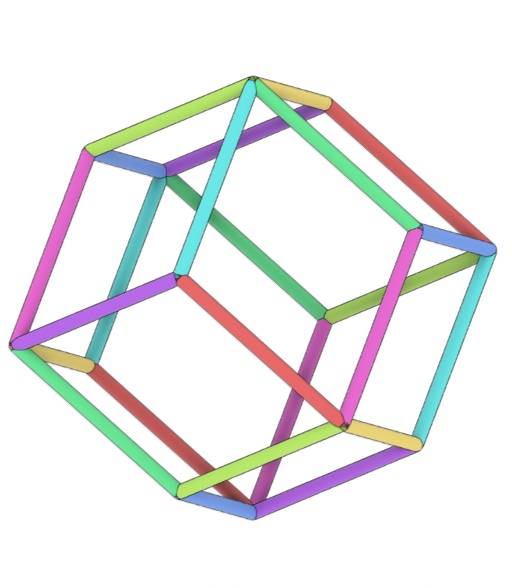}
    \includegraphics[width=0.18\linewidth]{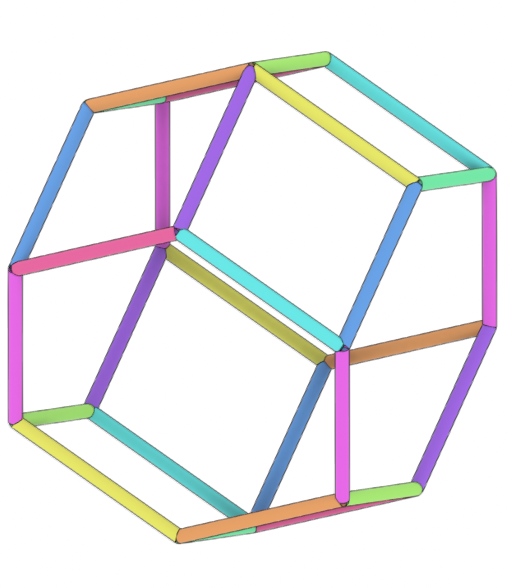}
    \includegraphics[width=0.18\linewidth]{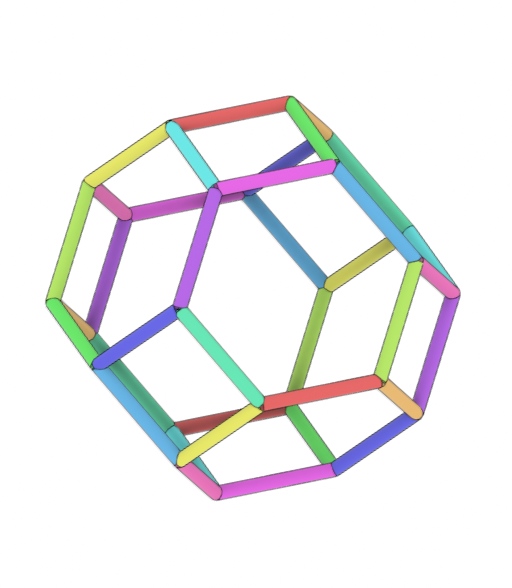}\\
    \rotatebox{90}{\parbox[t]{0.12\textwidth}{\centering Unique Edges}}
    \includegraphics[width=0.18\linewidth]{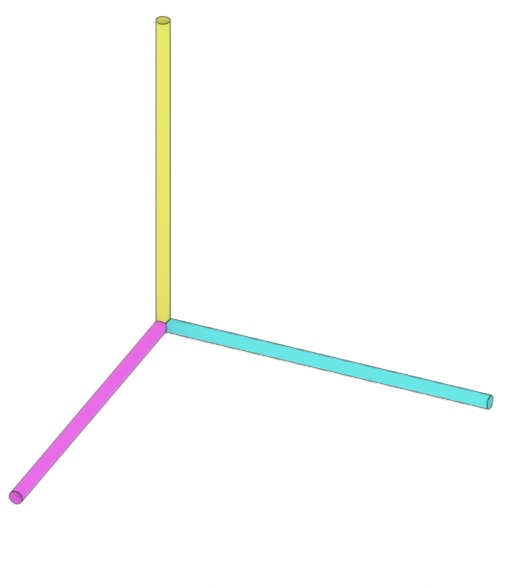}
    \includegraphics[width=0.18\linewidth]{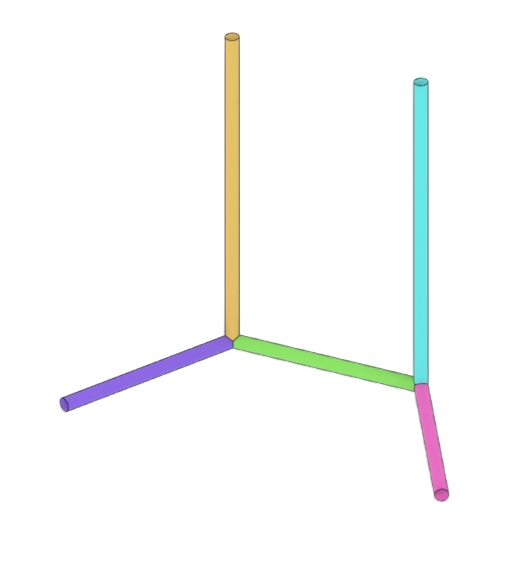}
    \includegraphics[width=0.18\linewidth]{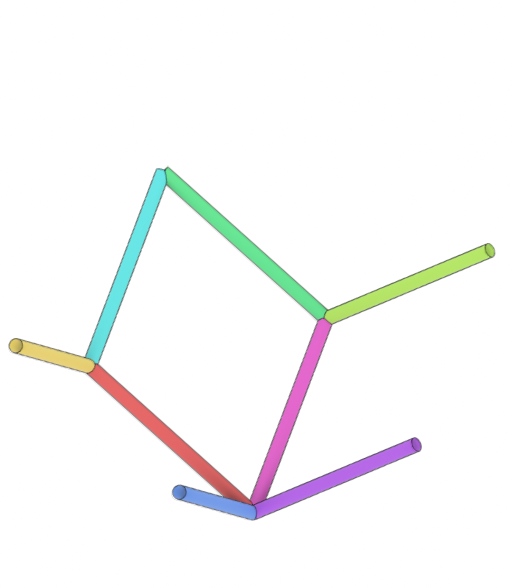}
    \includegraphics[width=0.18\linewidth]{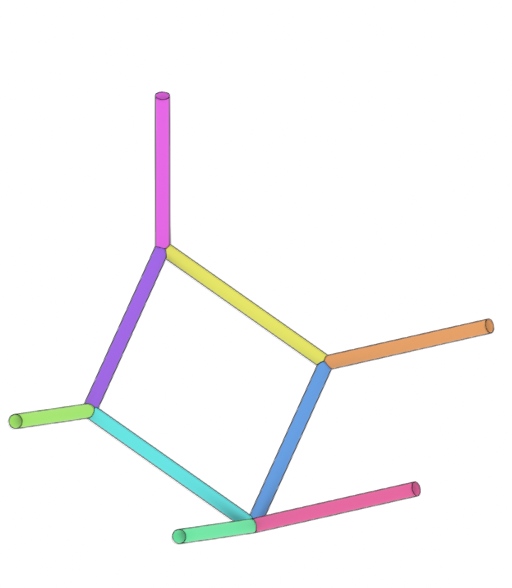}
    \includegraphics[width=0.18\linewidth]{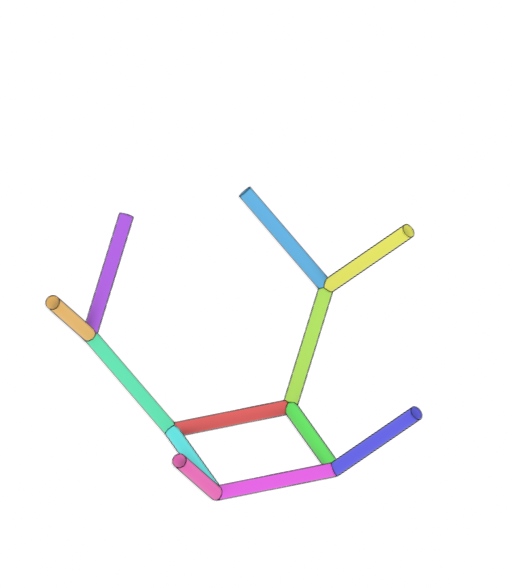}\\
    \rotatebox{90}{\parbox[t]{0.12\textwidth}{\centering Twist Units}}
    \includegraphics[width=0.18\linewidth]{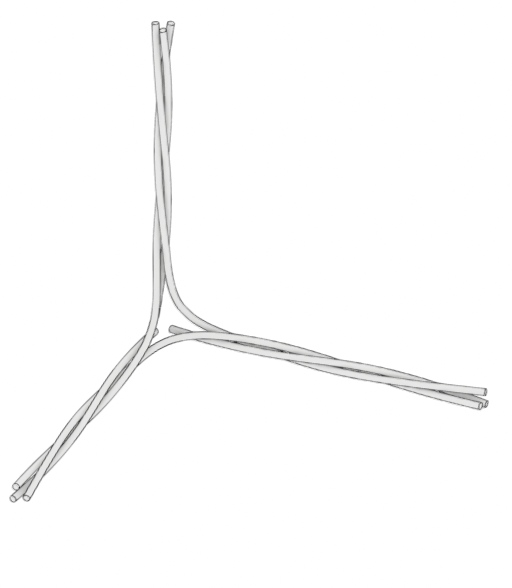}
    \includegraphics[width=0.18\linewidth]{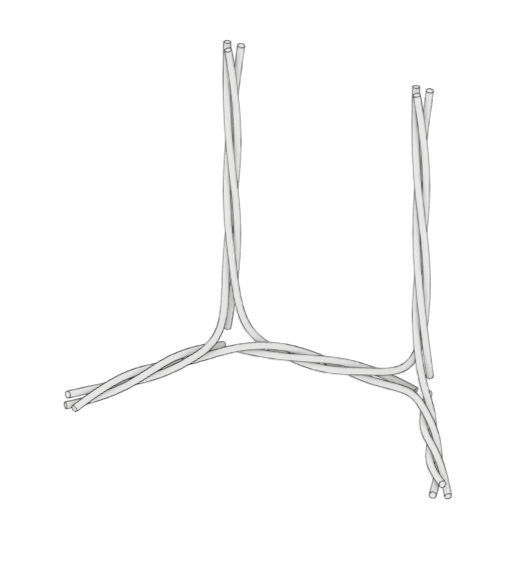}
    \includegraphics[width=0.18\linewidth]{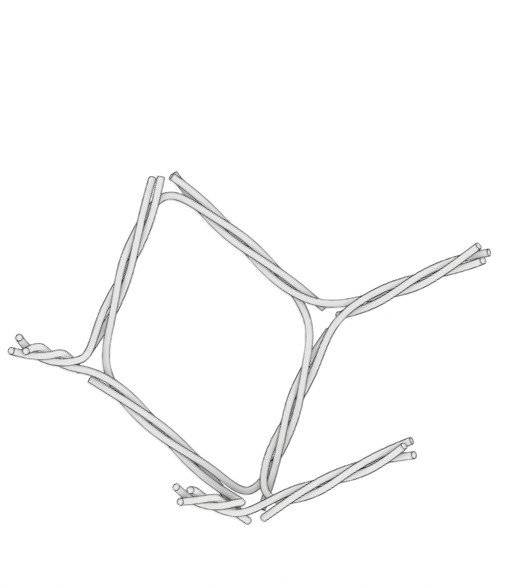}
    \includegraphics[width=0.18\linewidth]{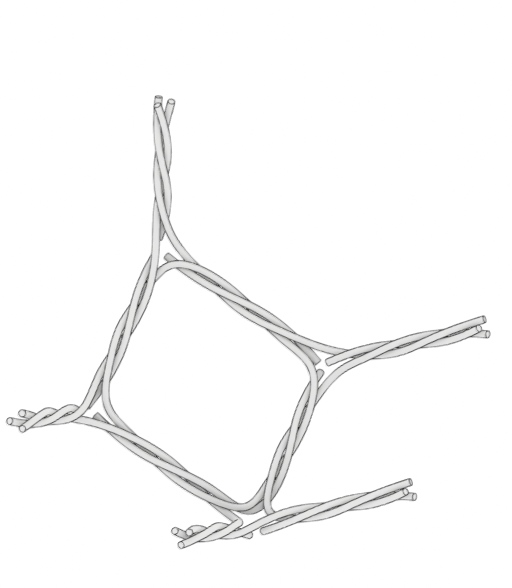}
    \includegraphics[width=0.18\linewidth]{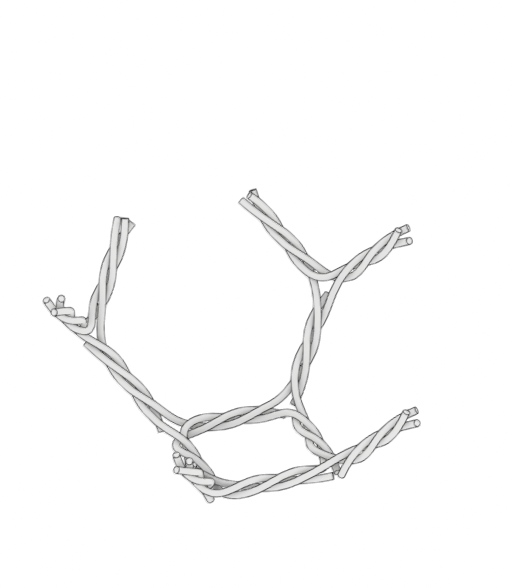}\\
    \rotatebox{90}{\parbox[t]{0.01\textwidth}{\centering }}
    \parbox[t]{0.18\linewidth}{\centering cP}
    \parbox[t]{0.18\linewidth}{\centering hP}
    \parbox[t]{0.18\linewidth}{\centering oF}
    \parbox[t]{0.18\linewidth}{\centering cF}
    \parbox[t]{0.18\linewidth}{\centering cI}
    \caption{In 3D, there are only five topologically distinct Wigner–Seitz cells. For each type of Wigner-Seitz cell in 3D, we can find the unique edges that are required to define a repeating structure in 3D. In the first two rows, we indicate the edge class with different colors. The last row shows the unit twisting element for a single cell, which can be repeated to form complex patterns.}
    \Description{A three-row figure arranged in five columns, with column labels shown along the bottom. The top row, labeled “Unit Cells,” shows five different three-dimensional polyhedral cells rendered as wireframe shapes with edges colored in multiple hues. The second row, labeled “Unique Edges,” shows simplified edge representatives for each cell, drawn as short line segments meeting at vertices, with colors matching the edge classes in the unit cells above. The third row, labeled “Twist Units,” shows small three-dimensional twisting strand elements corresponding to each cell type, rendered as thin intertwined shapes. From left to right, the columns are labeled “cP,” “hP,” “oF,” “cF,” and “cI,” with each column showing a consistent correspondence between the unit cell, its unique edges, and its twisting unit.}
    \label{fig:bravais3D}
\end{figure}

This idea generalizes naturally to higher dimensions. In 3D, there are five canonical Wigner-Seitz cells, as shown in Figure~\ref{fig:bravais3D}. For each case, the distinct edge classes of the unit cell can be identified in a manner analogous to the 2D case, as illustrated in Figure~\ref{fig:bravais3D}. Again, by assuming that the periodic unit consists of a single Wigner-Seitz cell, the entire space can be generated by repeating and twisting the corresponding edge elements, as shown in Figure~\ref{fig:bravais3D}.

\subsection{Design Space of Single Repeating Unit}

\begin{figure}[htb!]
    \centering
    \begin{subfigure}[t]{0.49\textwidth}    \includegraphics[width=0.30\linewidth]{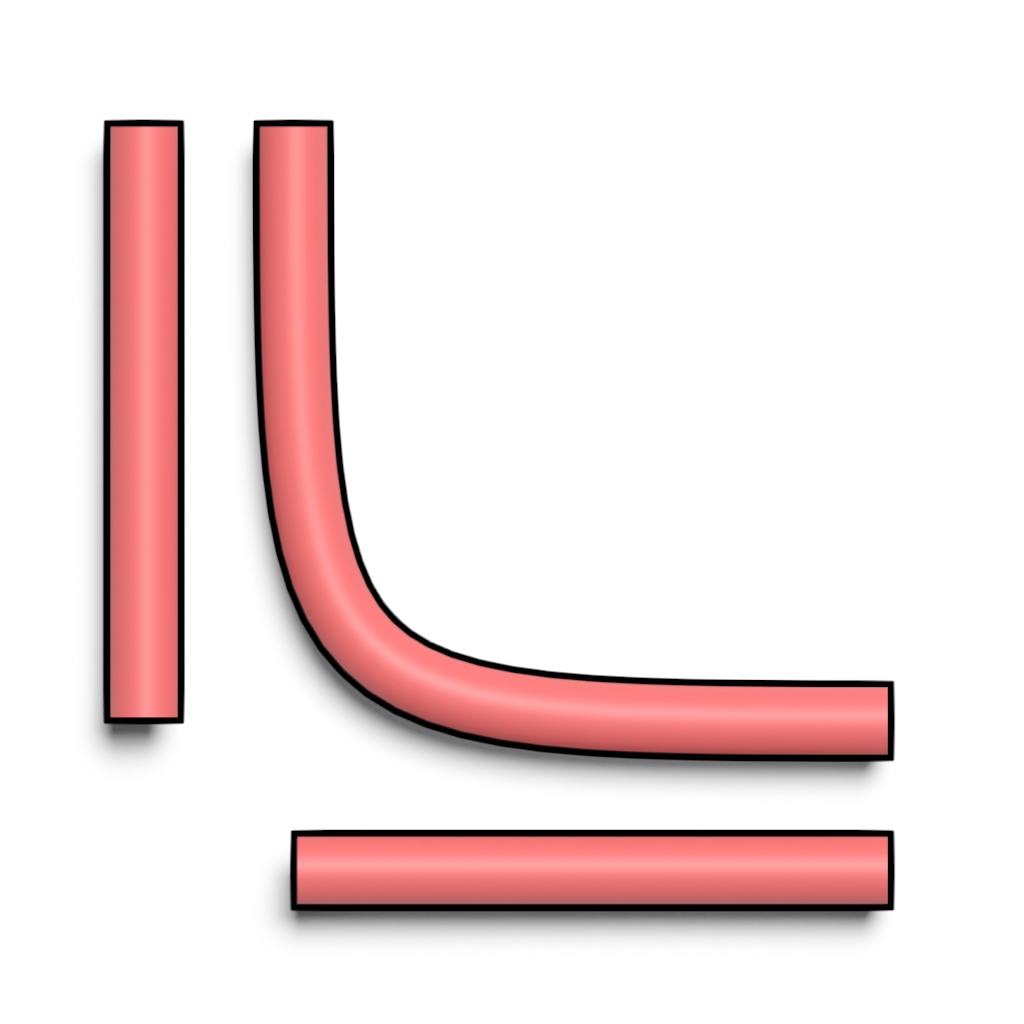}    \includegraphics[width=0.30\linewidth]{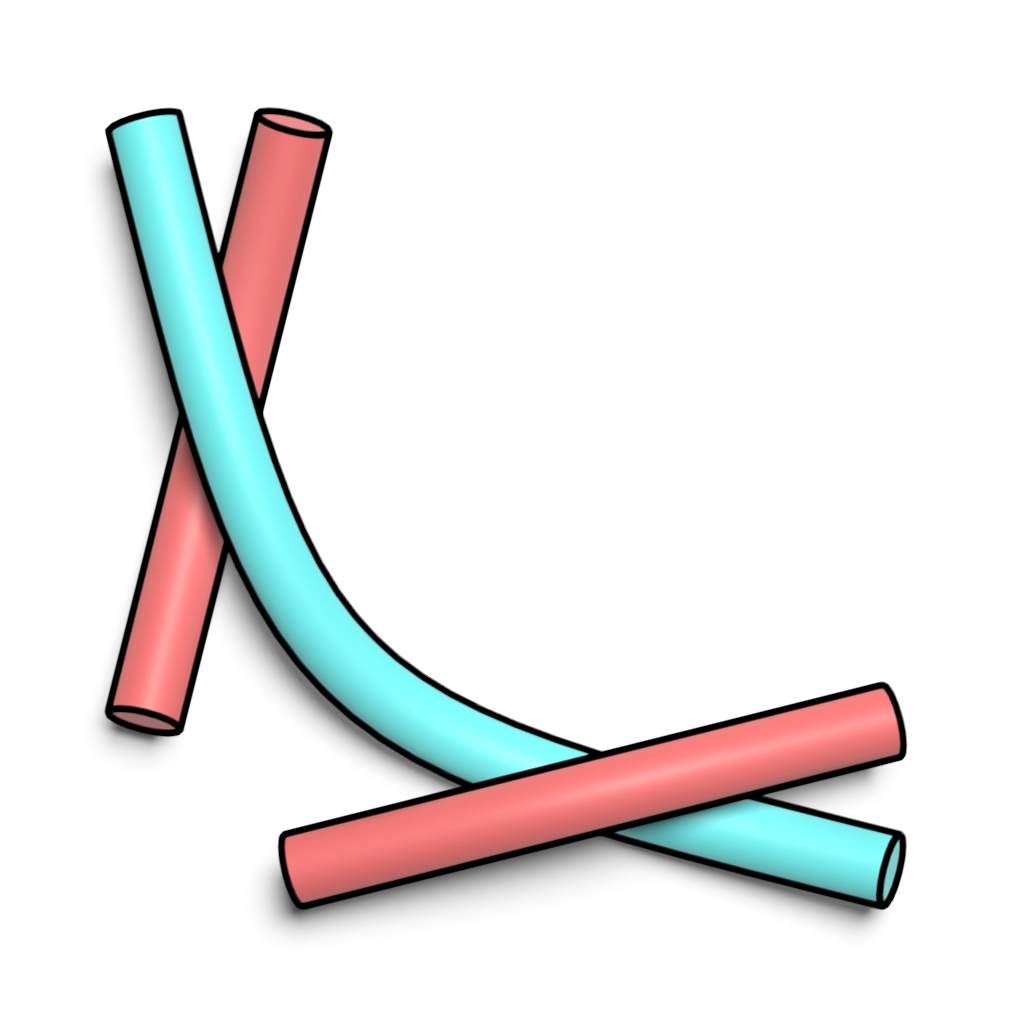}    \includegraphics[width=0.30\linewidth]{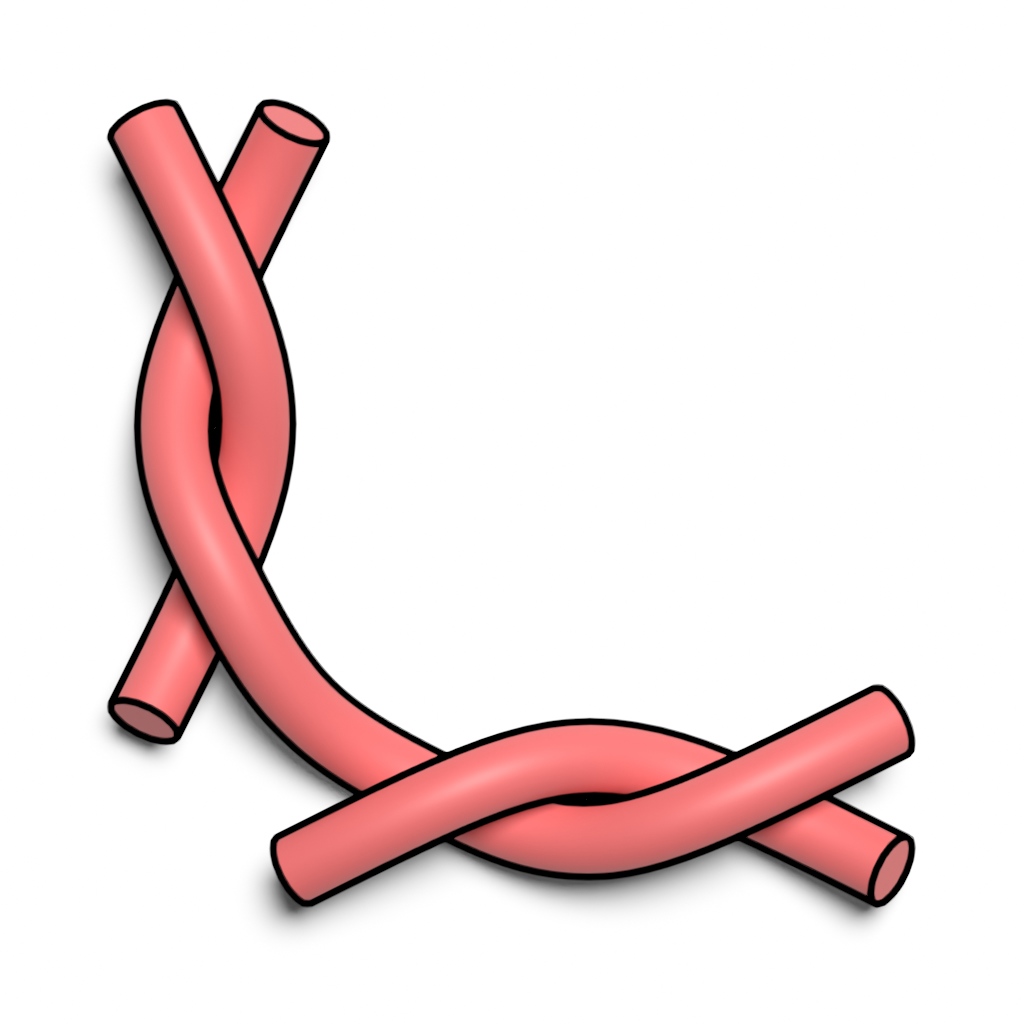}\\    \includegraphics[width=0.30\linewidth]{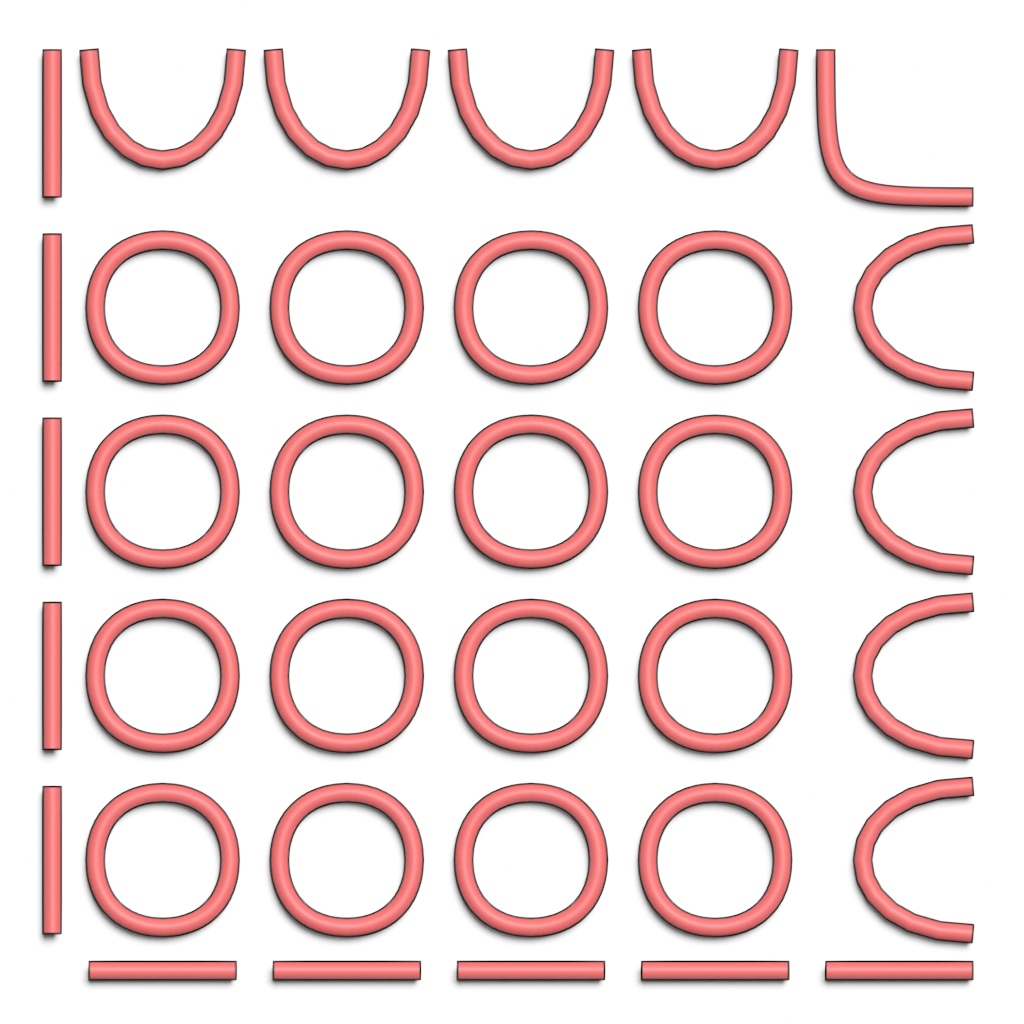}    \includegraphics[width=0.30\linewidth]{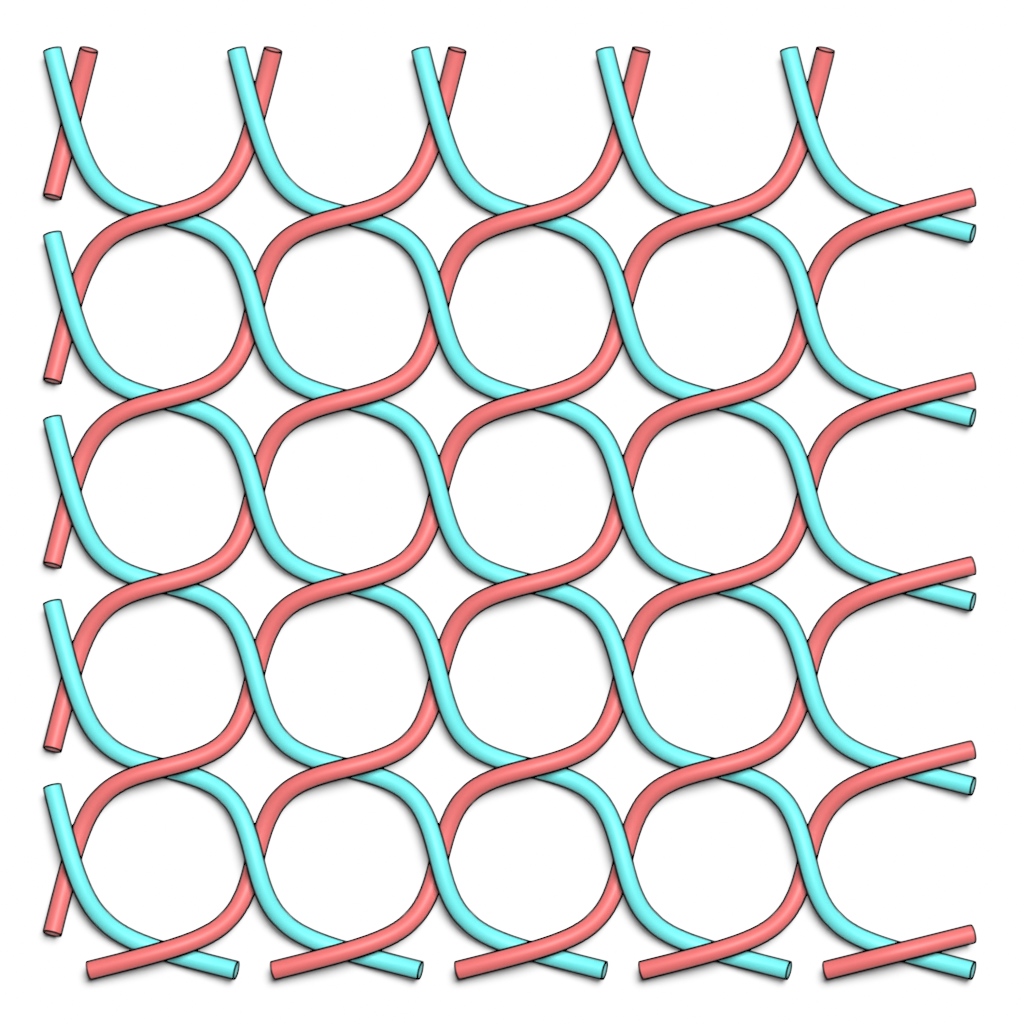}    \includegraphics[width=0.30\linewidth]{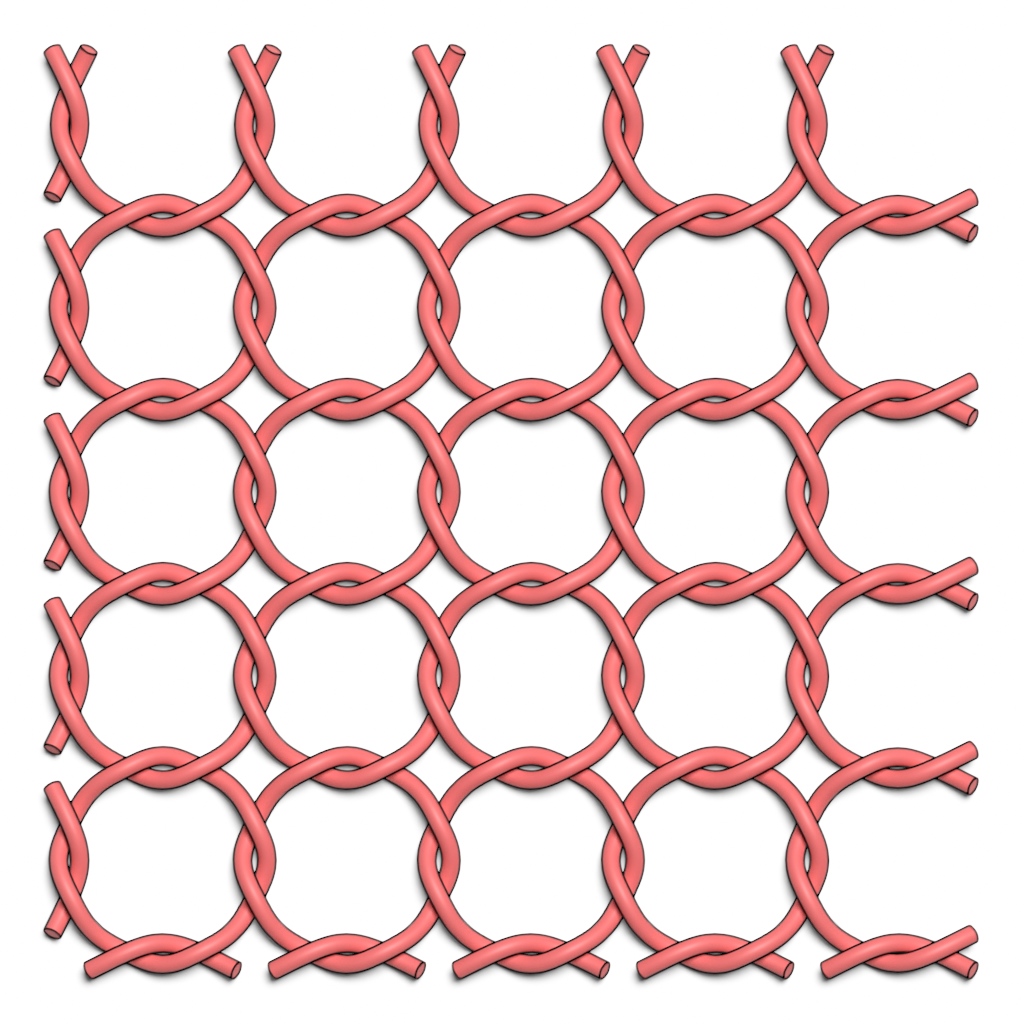}\\
    \parbox[t]{0.30\textwidth}{\centering 0 twist}
    \parbox[t]{0.30\textwidth}{\centering 1 twist}
    \parbox[t]{0.30\textwidth}{\centering 2 twist}
    \caption{Examples of uniform twist applied to square tessellation.}
    \label{fig:ws2d_quad}
    \end{subfigure}
    \begin{subfigure}[t]{0.49\textwidth}    \includegraphics[width=0.30\linewidth]{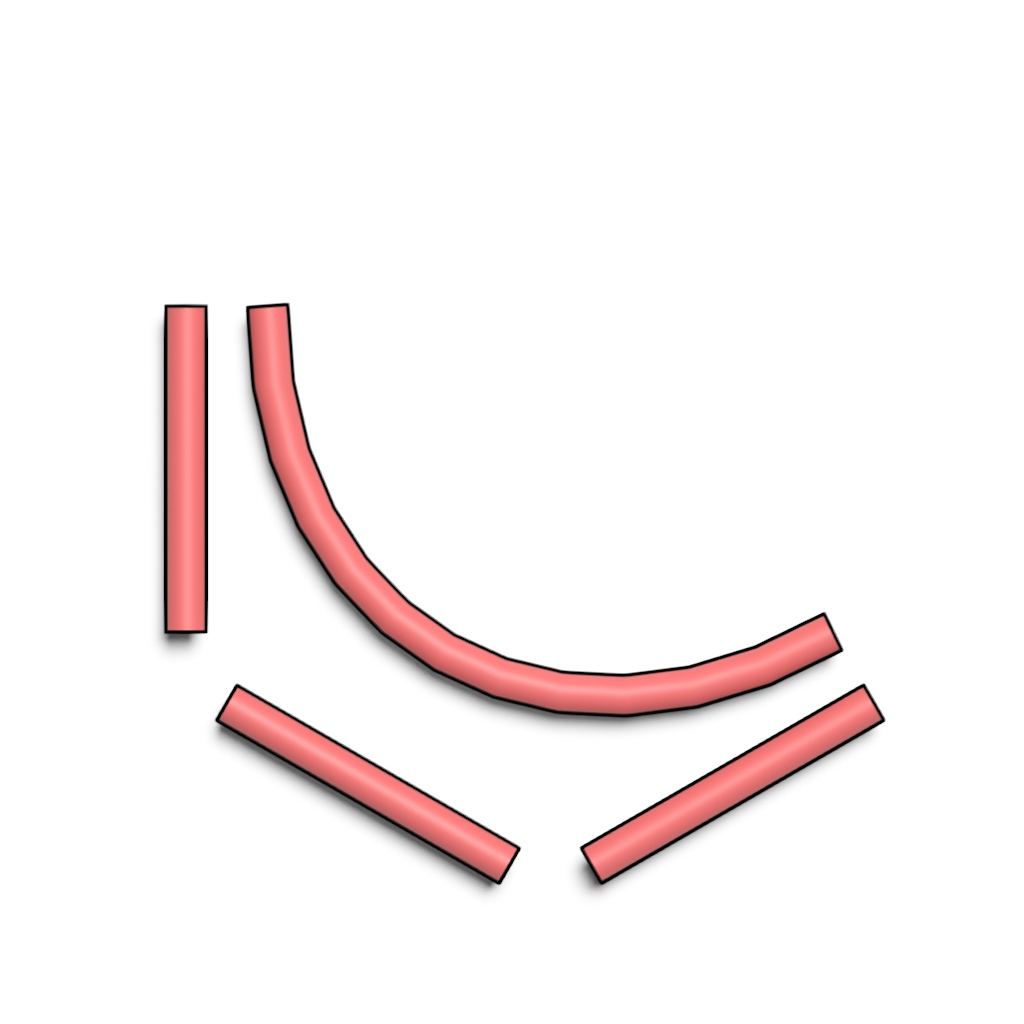}   \includegraphics[width=0.30\linewidth]{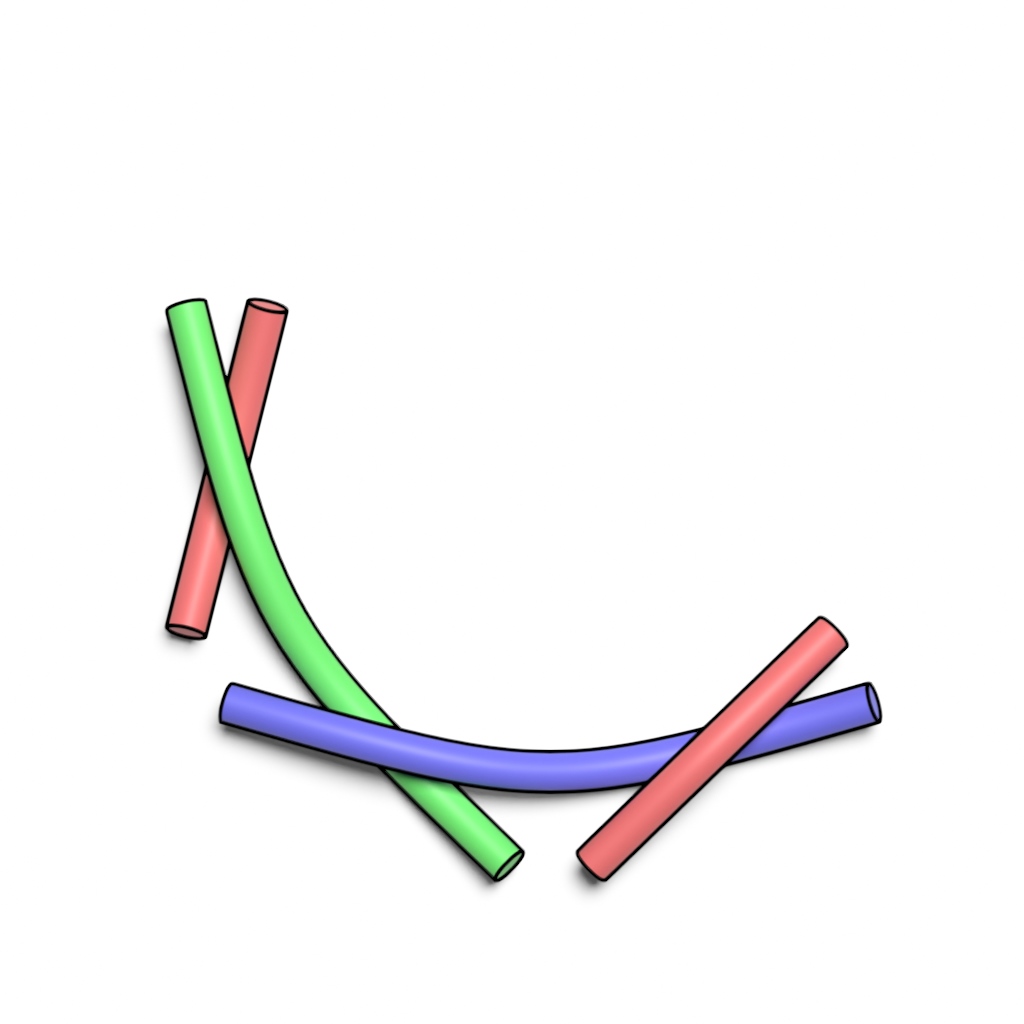}        \includegraphics[width=0.30\linewidth]{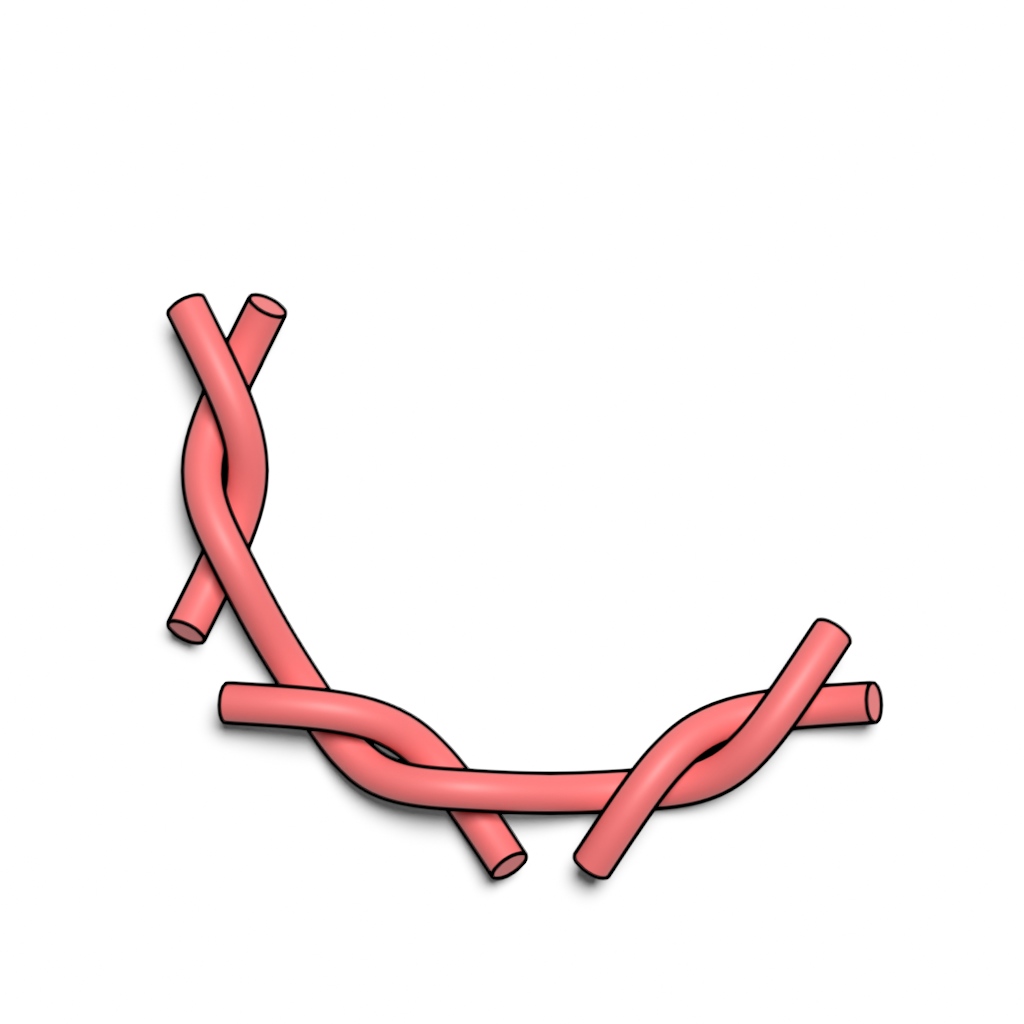}\\        \includegraphics[width=0.30\linewidth]{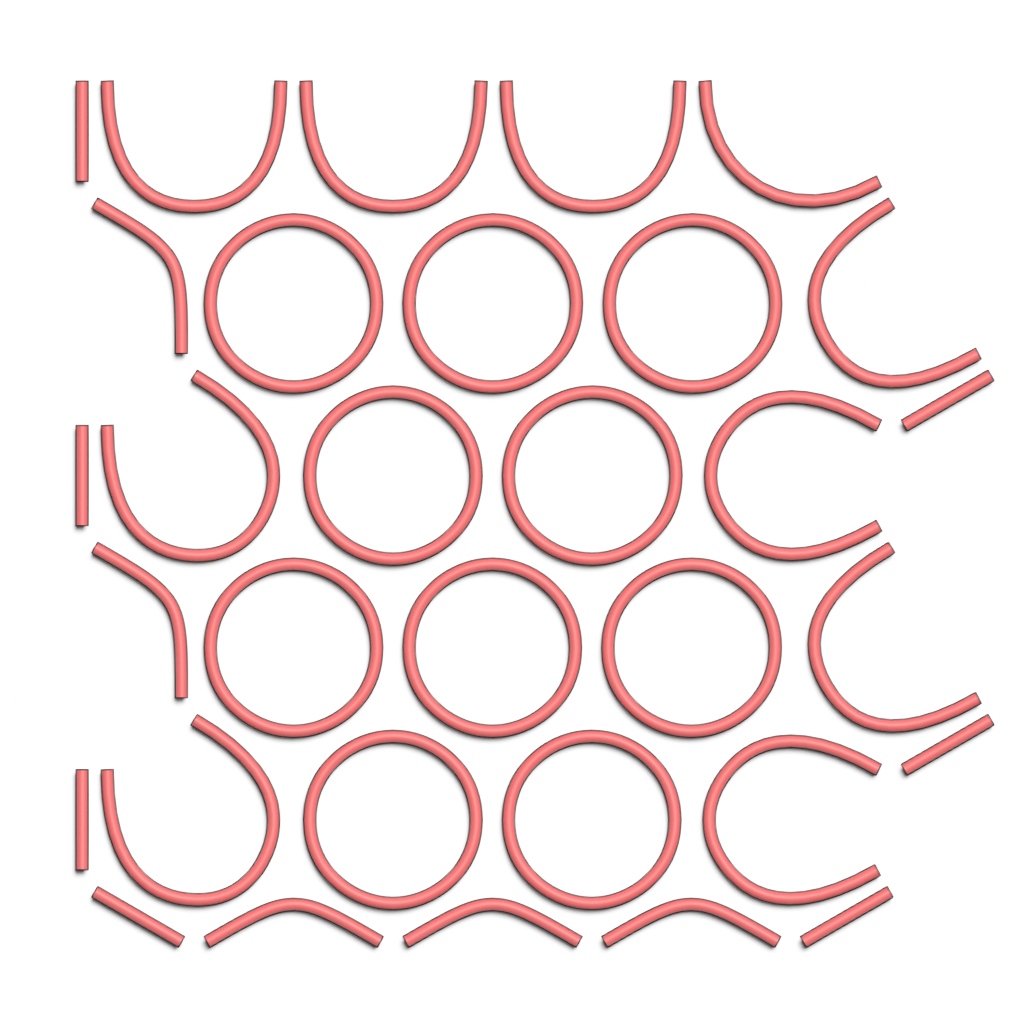}   \includegraphics[width=0.30\linewidth]{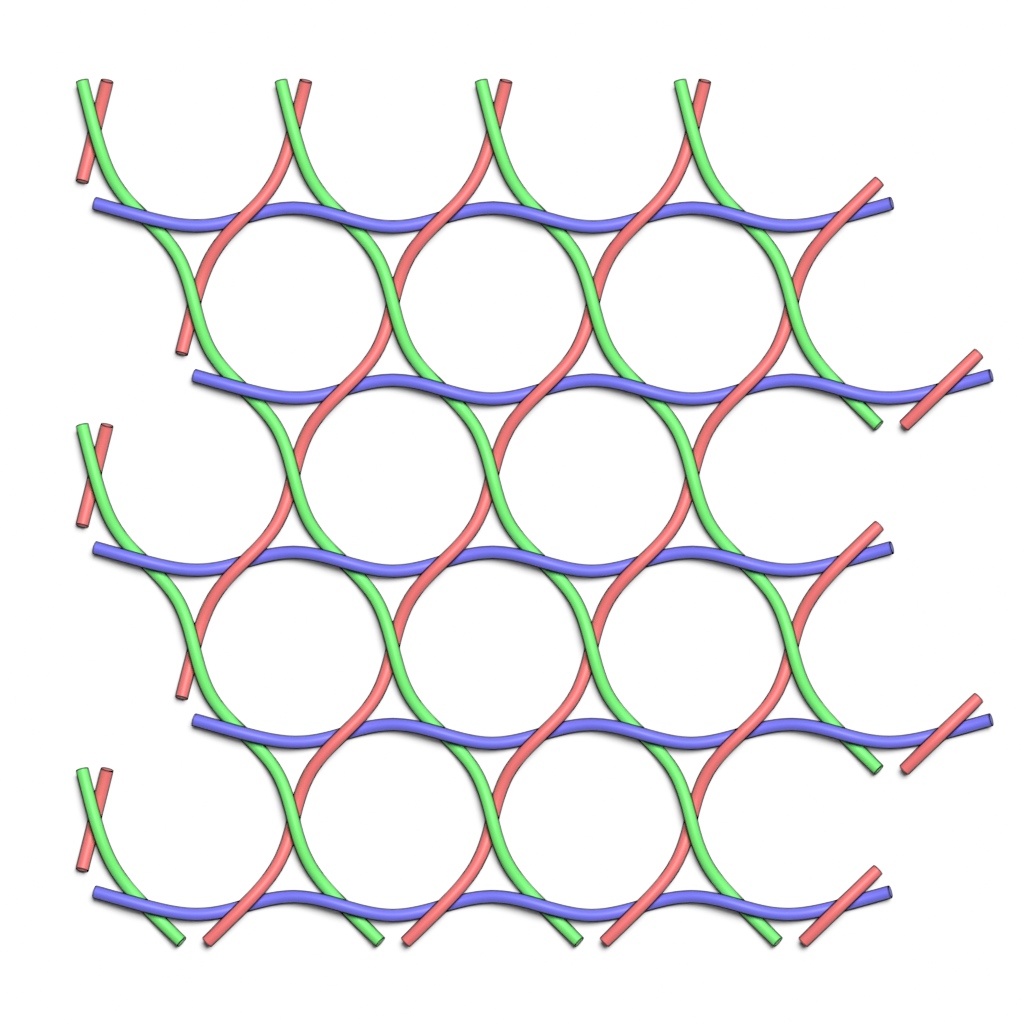}    \includegraphics[width=0.30\linewidth]{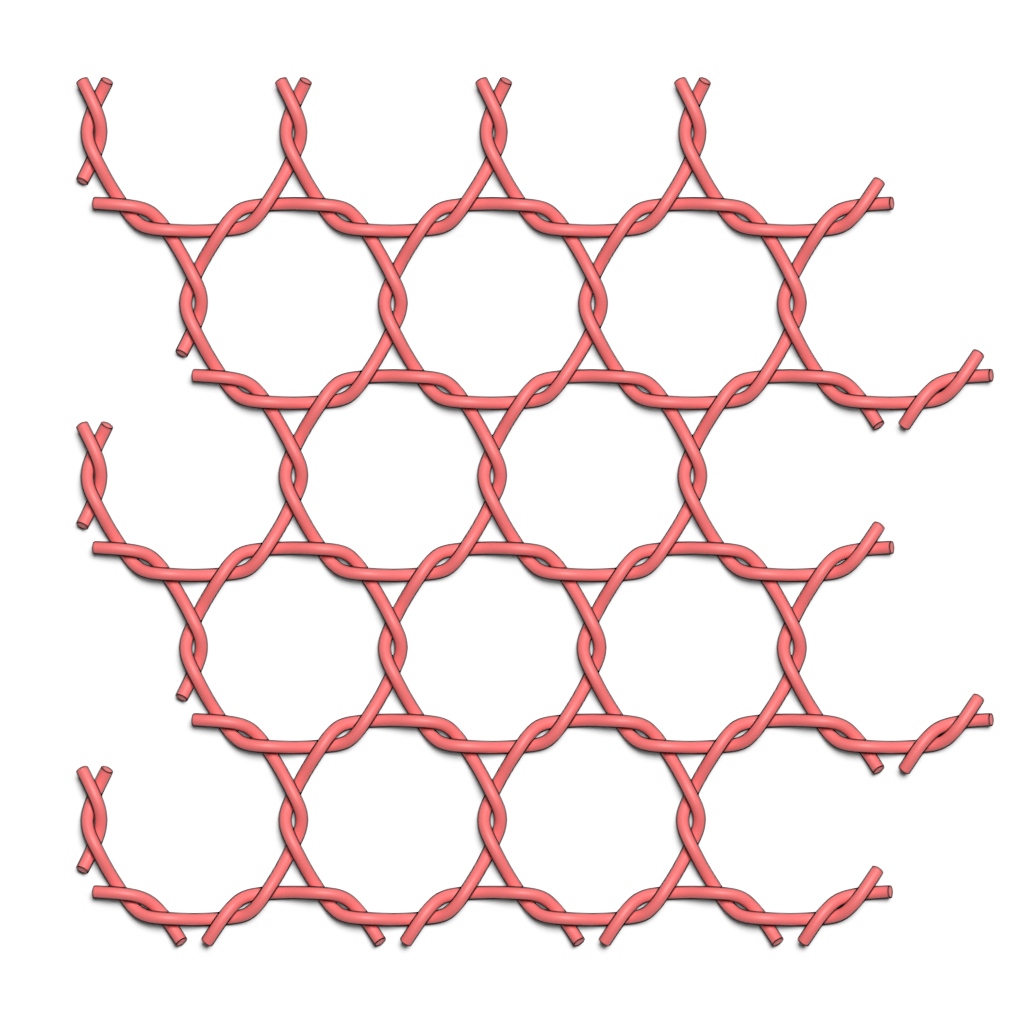}\\    
    \parbox[t]{0.30\textwidth}{\centering 0 twist}
    \parbox[t]{0.30\textwidth}{\centering 1 twist}
    \parbox[t]{0.30\textwidth}{\centering 2 twist}
    \caption{Examples of uniform twist applied to hexagonal tessellation.}
    \label{fig:ws2d_hex}
    \end{subfigure}
    \caption{Effect of uniform edge twisting in a 2D tessellation. With zero twist, the structure consists of disconnected links. Applying a single twist generates (a) biaxial and (b) triaxial weaving patterns. Applying two twists yields two distinct chainmail-like configurations, shown in (a) and (b).}
    \Description{Two panels labeled (a) and (b) showing the effect of uniform twisting in two-dimensional tessellations. In each panel, the top row shows schematic twisting elements rendered as short tubular segments, illustrating increasing twist values labeled “0 twist”, “1 twist”, and “2 twist”. The bottom row in each panel shows the resulting repeating structures. In panel (a), corresponding to a square tessellation, the zero-twist case shows many separate circular loops arranged in a grid, the one-twist case shows loops intersecting in a regular woven pattern, and the two-twist case shows a dense interlinked arrangement of loops, rendered primarily in red and cyan. In panel (b), corresponding to a hexagonal tessellation, the zero-twist case again shows separate circular loops, the one-twist case shows a multi-directional woven pattern with strands rendered in three distinct colors, and the two-twist case shows a tightly interlinked lattice of loops. Colors are used to distinguish different strands and emphasize crossings and interconnections.}
    \label{fig:bravais2D_repeated}
\end{figure}

In Figure~\ref{fig:bravais2D_repeated}, all edges are periodically assigned the same twist label. The resulting structures show that uniform edge twisting is sufficient to generate a range of woven and chainmail-like LK configurations, without introducing any local variation or additional design parameters. These outcomes arise solely from the uniform twist value and do not require any local variation or additional parameters.

\begin{figure*}[htp!]
    \centering
        \begin{overpic}[width=0.23\linewidth]{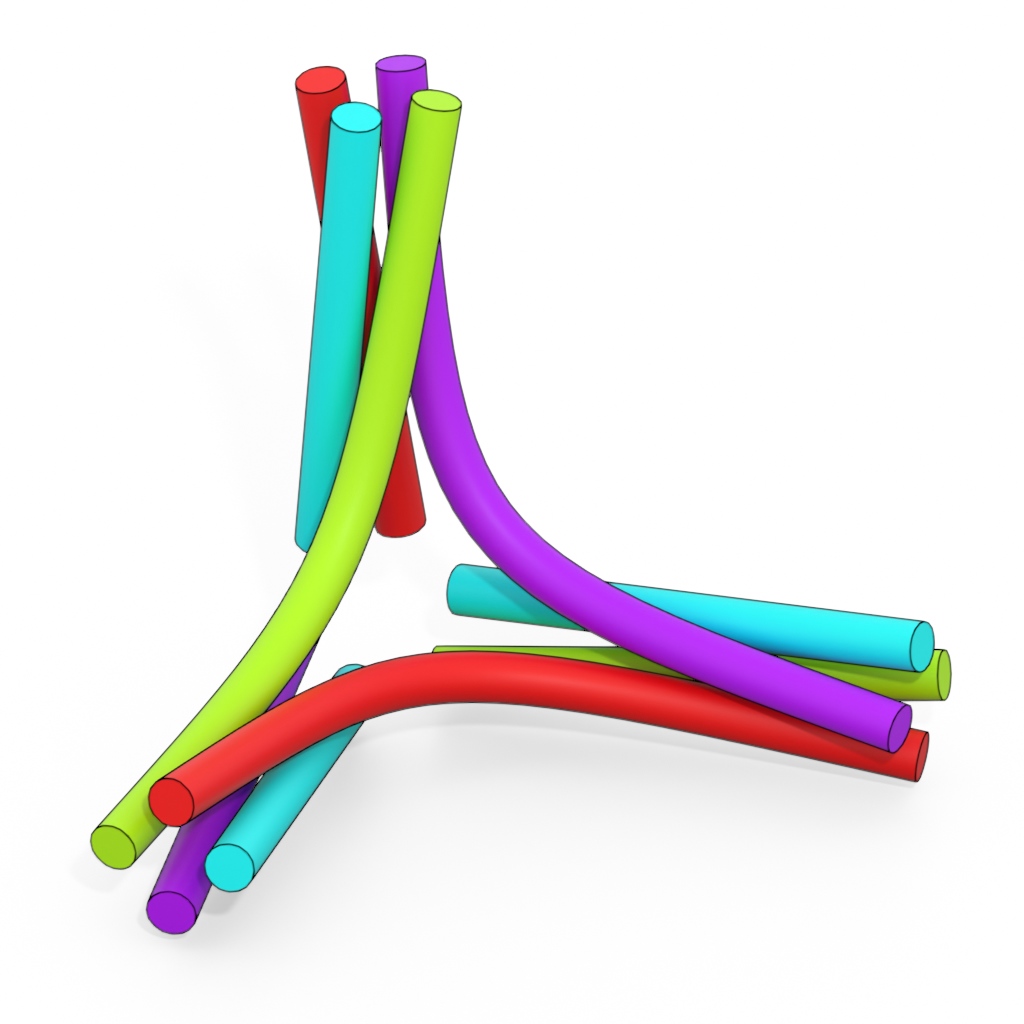}
            \put(-5,85){\centering 1}
        \end{overpic}
        \begin{overpic}[width=0.23\linewidth]{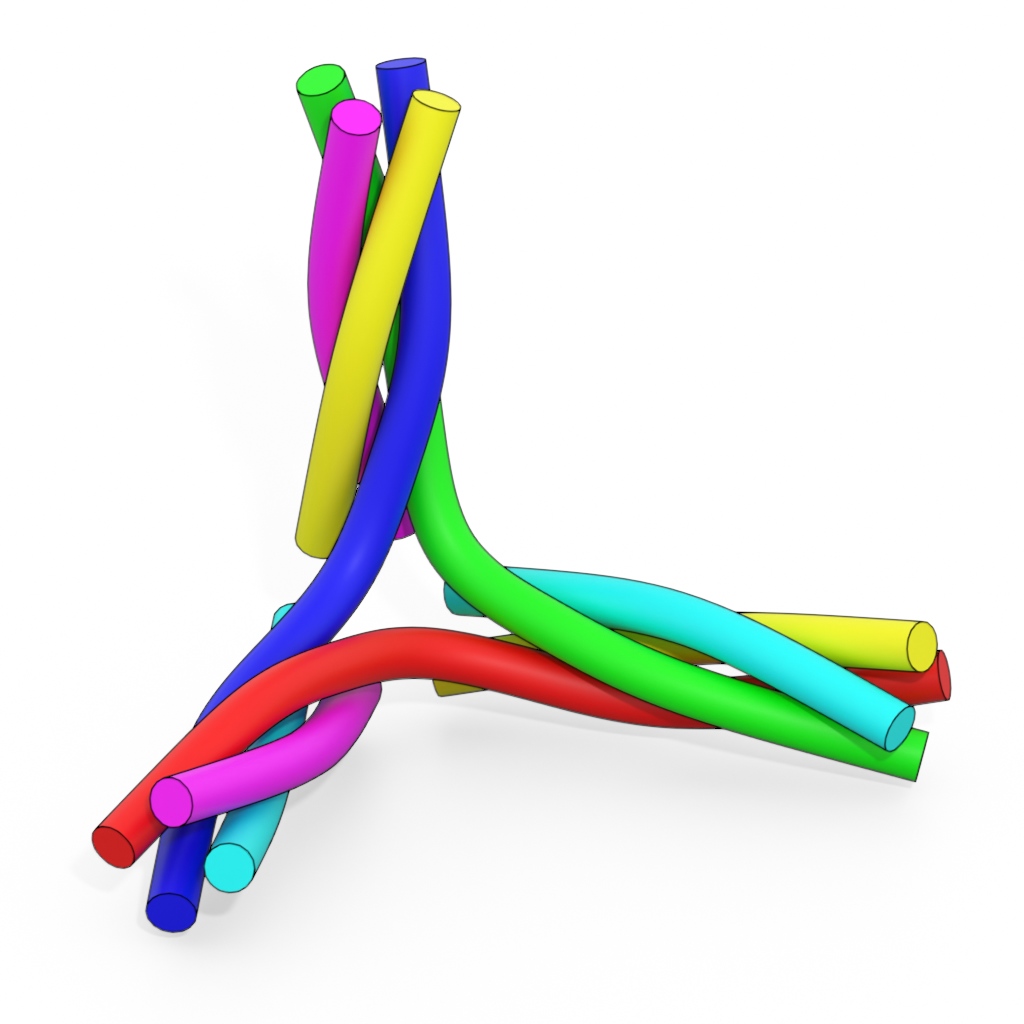}
            \put(-5,85){\centering 2}
        \end{overpic}
        \begin{overpic}[width=0.23\linewidth]{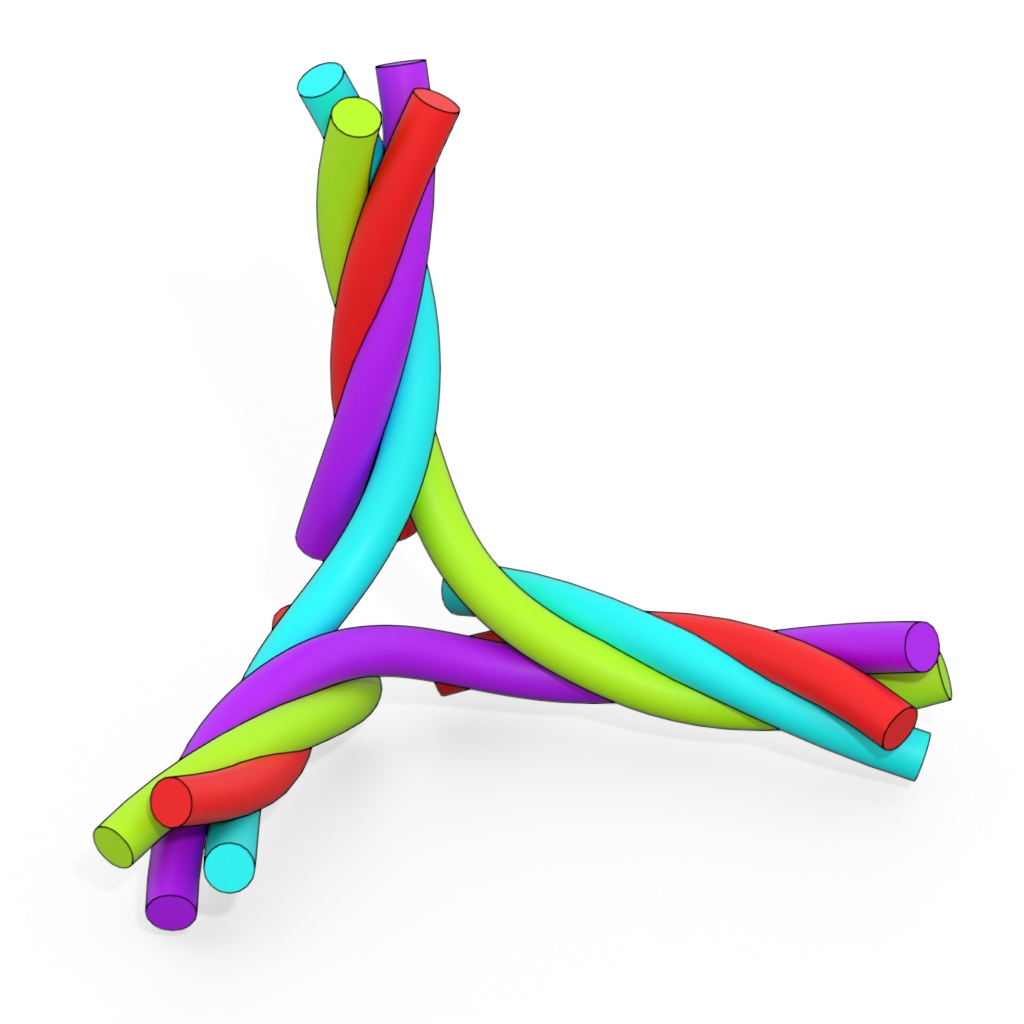}
            \put(-5,85){\centering 3}
        \end{overpic}
        \begin{overpic}[width=0.23\linewidth]{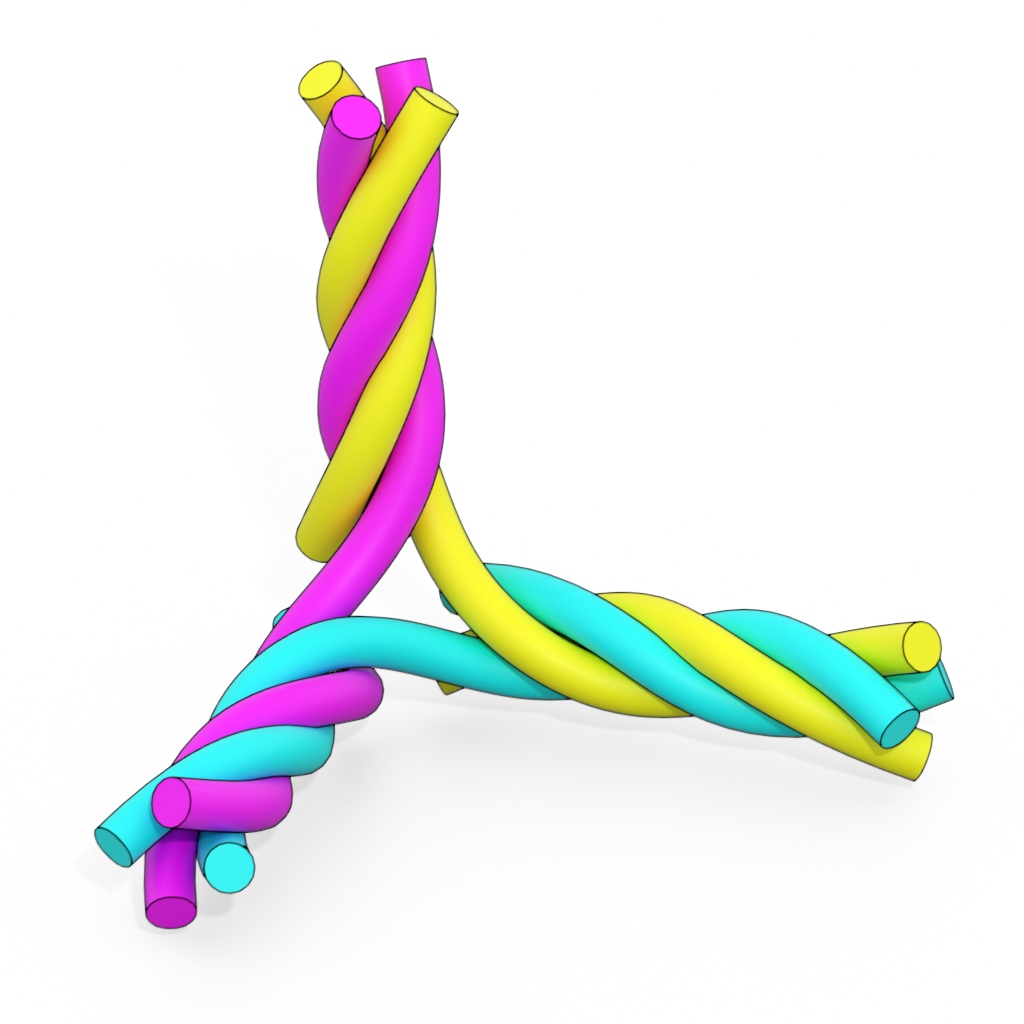}
            \put(-5,85){\centering 4}
        \end{overpic}
        \includegraphics[width=0.23\linewidth]{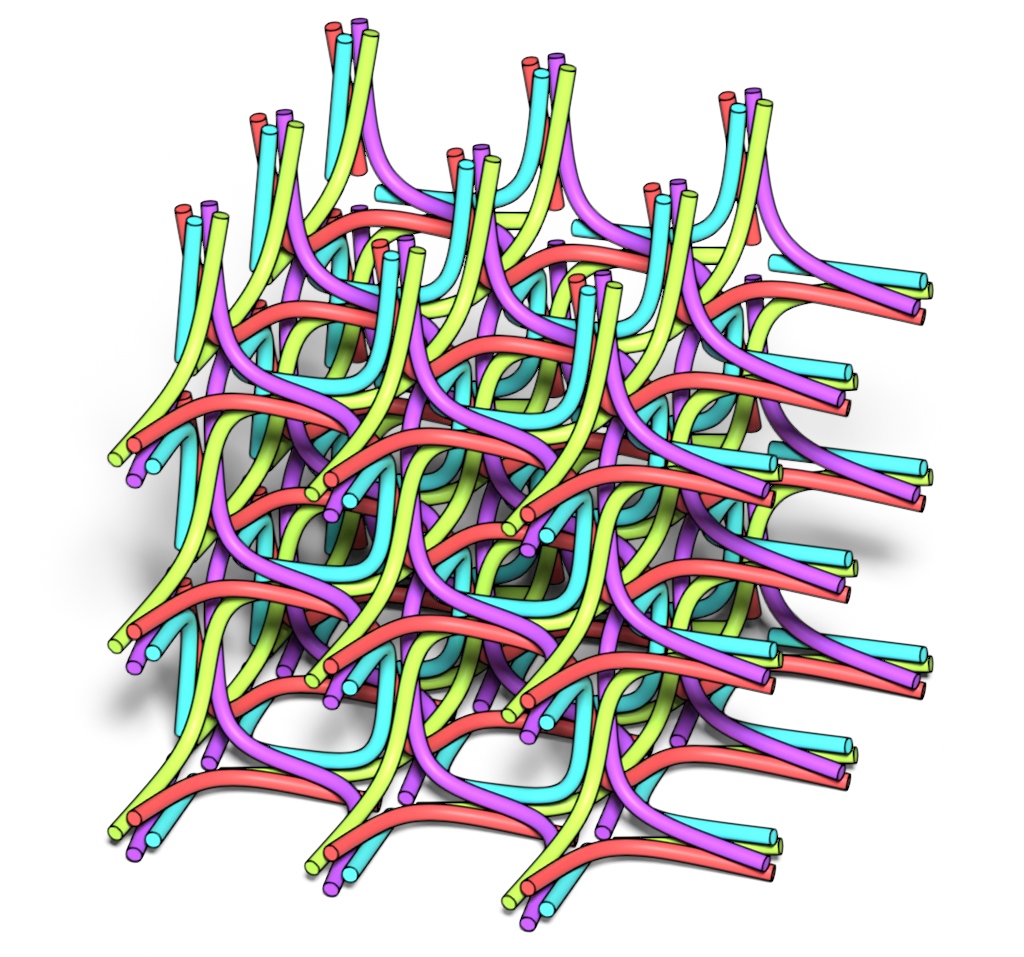}        \includegraphics[width=0.23\linewidth]{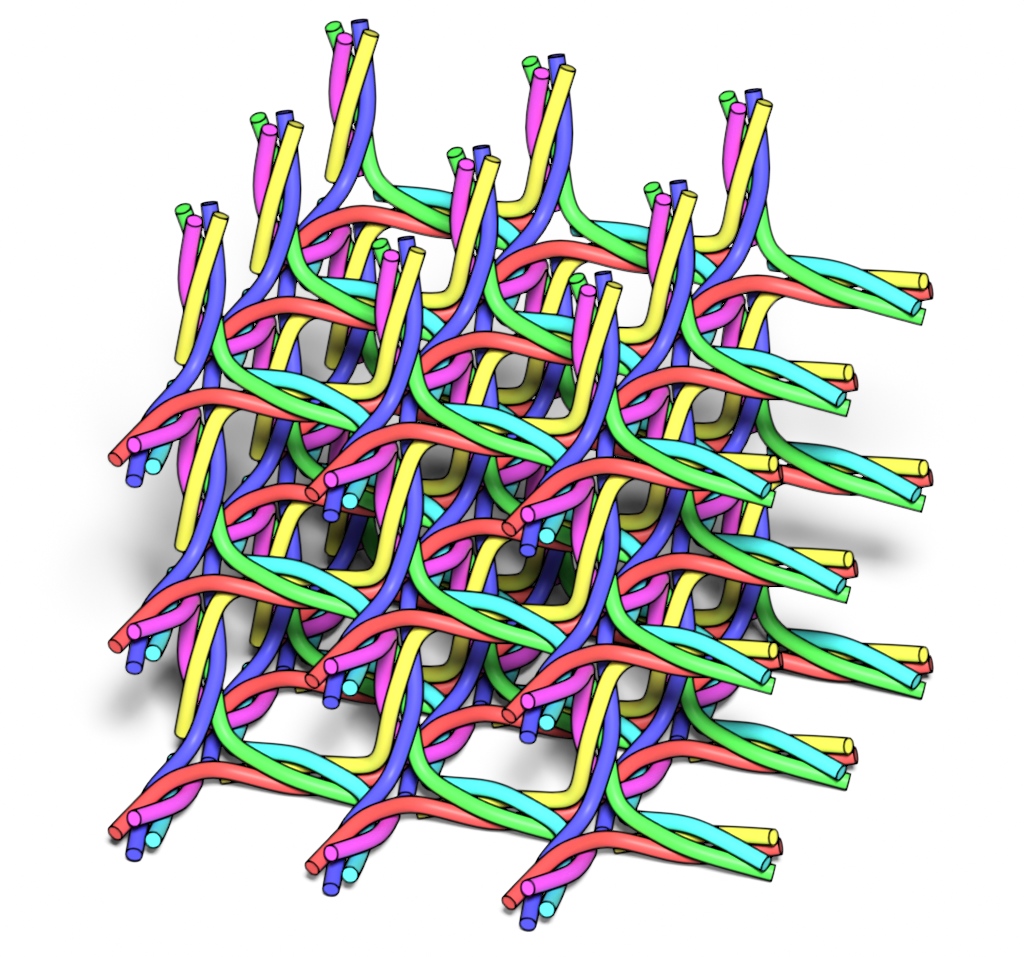}        \includegraphics[width=0.23\linewidth]{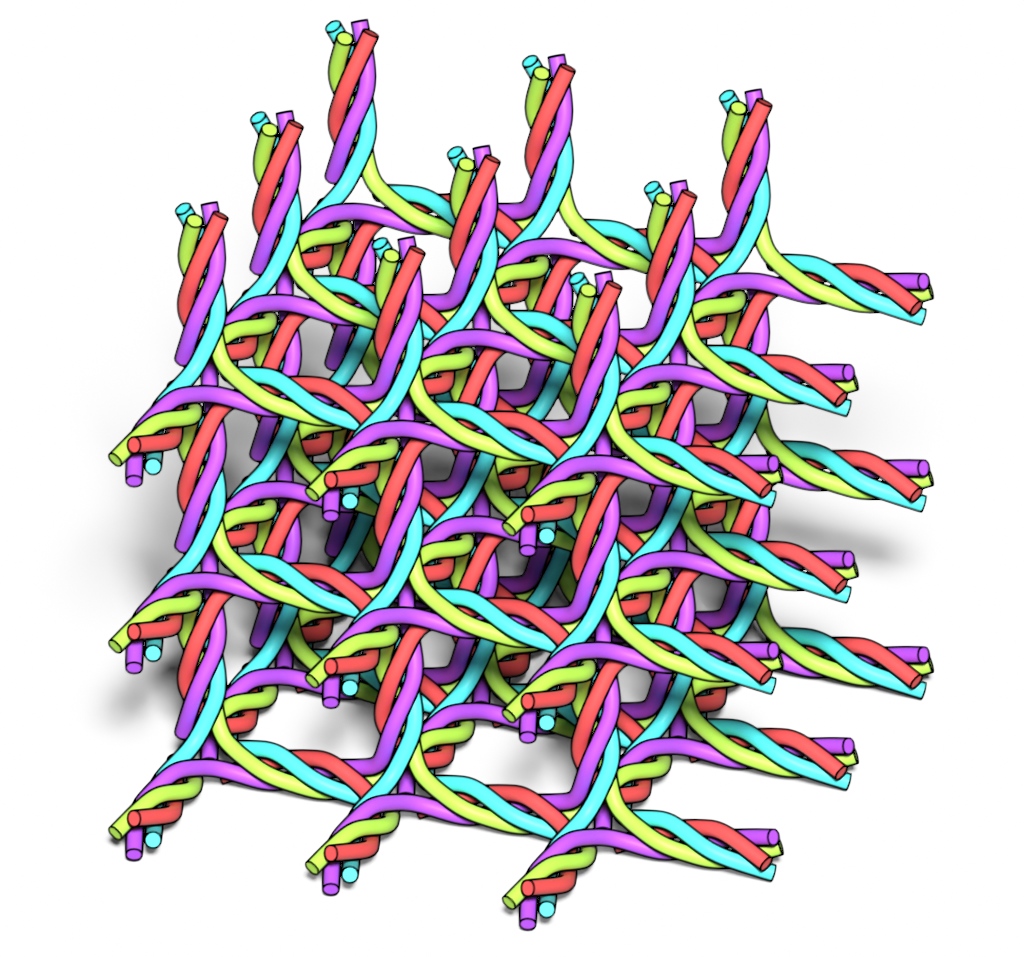}
        \includegraphics[width=0.23\linewidth]{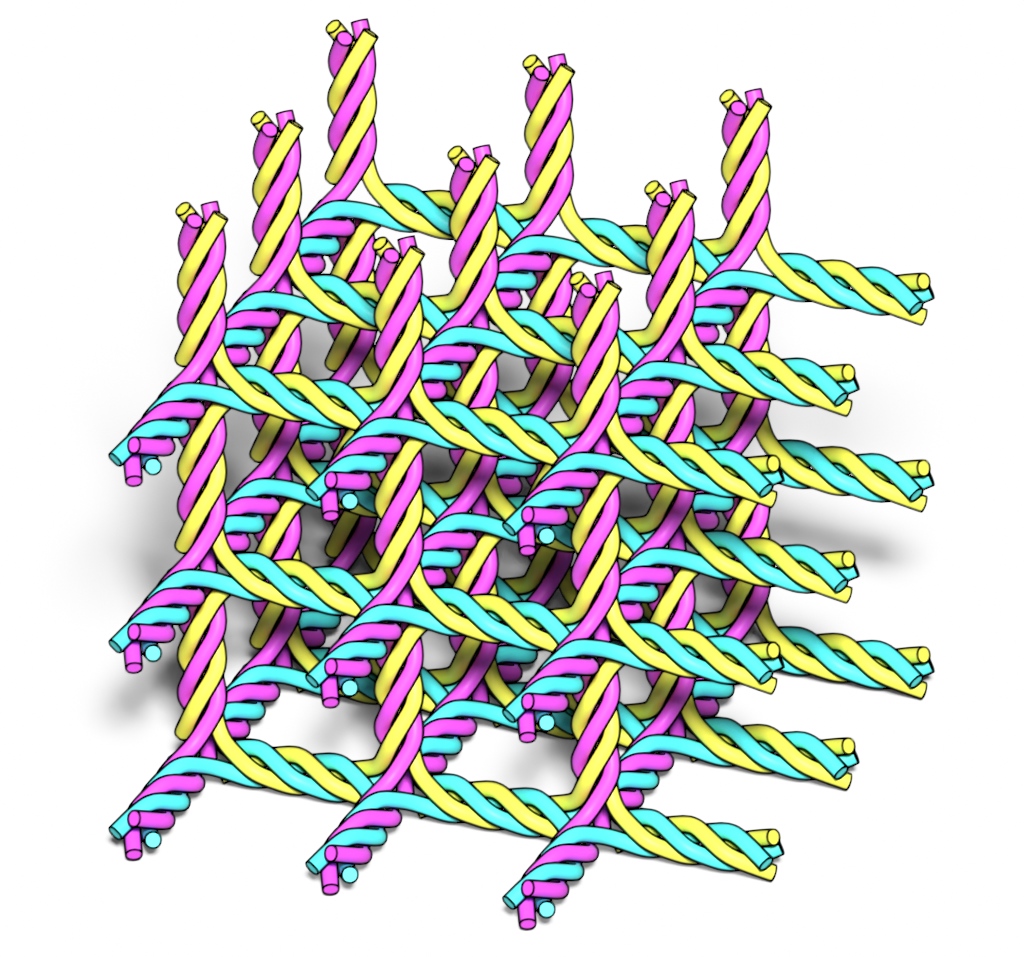}
    \caption{Using a tileable twisting module, we can use the distinct edges of the cube as a tile. Adjacency of the cubic tessellation gives 4 threads per edge, and there are 3 distinct edges leading to 12 threads per module. Assuming all thread components have the same twisting operations, depending on the twisting number, the connected components change. In the case of 0 and 4 twists (modulo 4 twist), there are 3 distinct components; 1 and 3 twists give 4 distinct components, and 2 twists give 6 distinct components for cubic tesselation.}
    \Description{A figure arranged in four numbered columns labeled 1 through 4, each column containing two images stacked vertically. In the top row, each column shows a single tileable twisting module composed of multiple thick tubular strands in different colors, including red, cyan, green, yellow, purple, and blue. The strands meet at a central junction and extend outward along three orthogonal directions, with varying degrees of twisting between the strands across the four columns. In the bottom row, each column shows a larger repeating structure formed by tiling the corresponding module, resulting in a dense three-dimensional lattice of intertwined strands. The overall arrangement and strand colors remain consistent across columns, while the local twisting patterns and resulting global connectivity visibly differ from left to right.}
    \label{fig:bravais3D_repeated}
\end{figure*}

\begin{figure*}[htp!]
    \centering
    \begin{overpic}[width=0.23\linewidth]{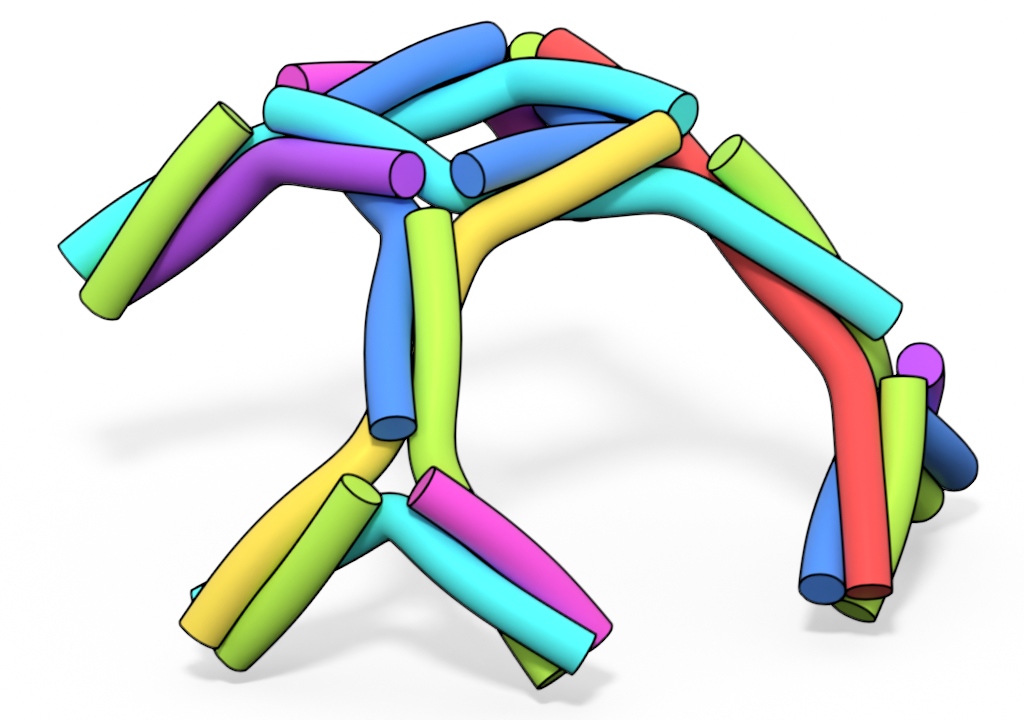}
        \put(-5,65){\centering 1}
    \end{overpic}
    \begin{overpic}[width=0.23\linewidth]{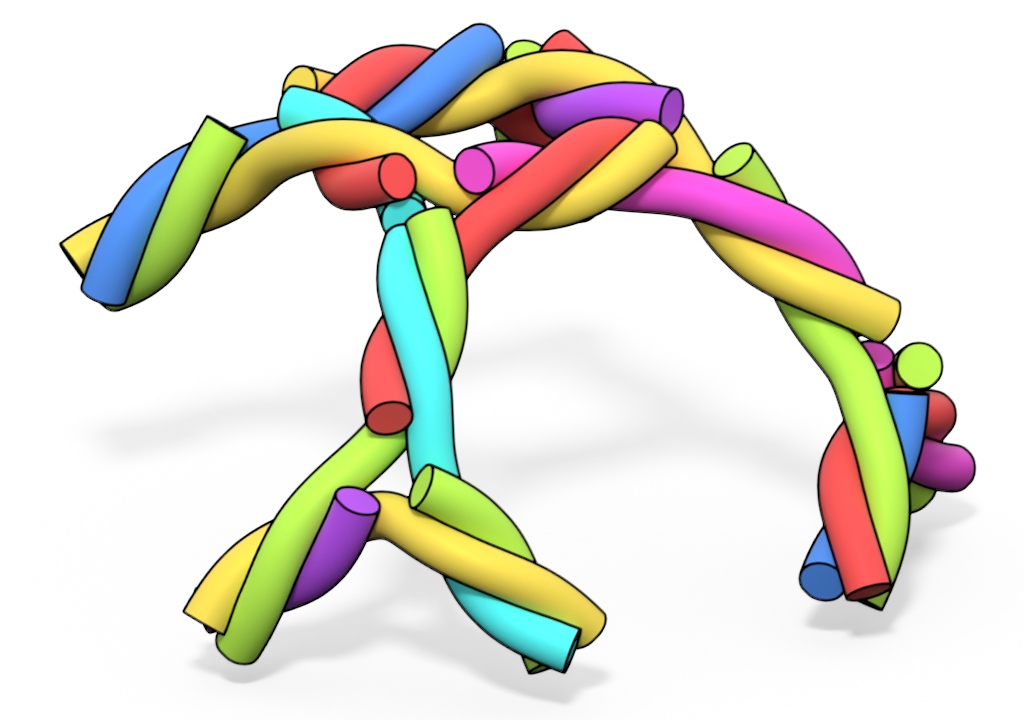}
        \put(-5,65){\centering 2}
    \end{overpic}
    \begin{overpic}[width=0.23\linewidth]{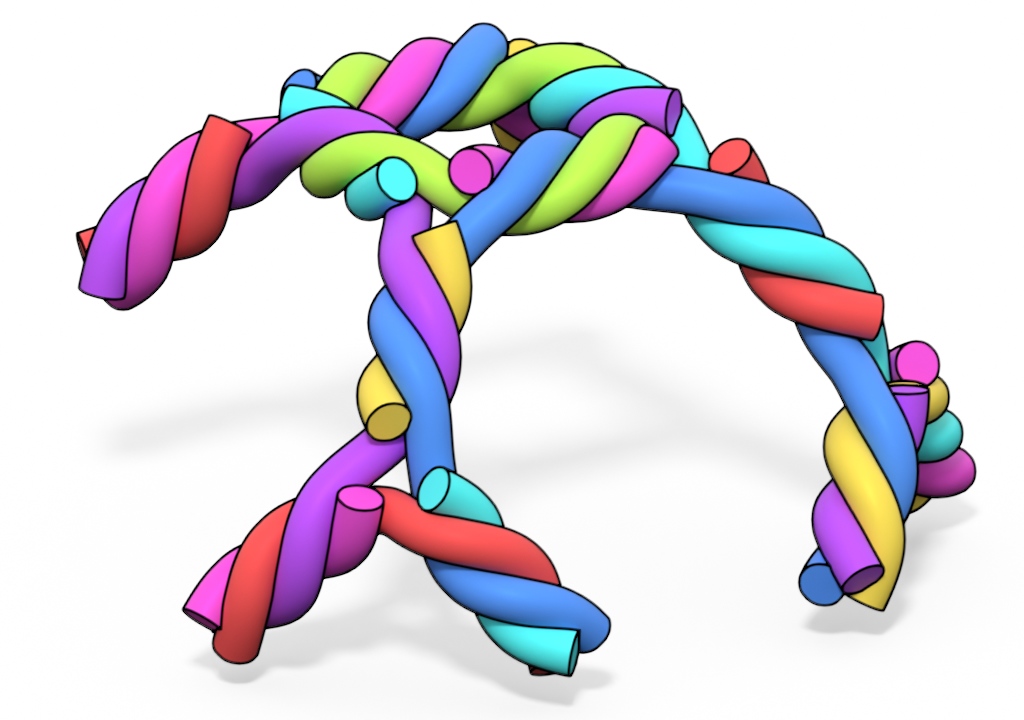}
        \put(-5,65){\centering 3}
    \end{overpic}
    \begin{overpic}[width=0.23\linewidth]{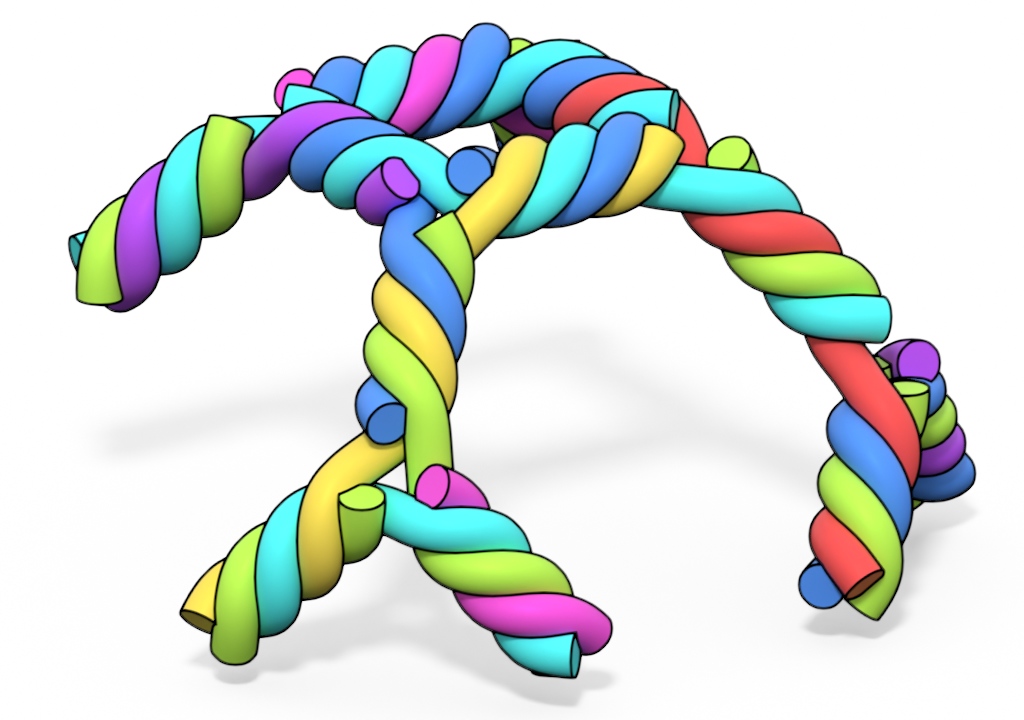}
        \put(-5,65){\centering 4}
    \end{overpic}\\
    \includegraphics[width=0.23\linewidth]{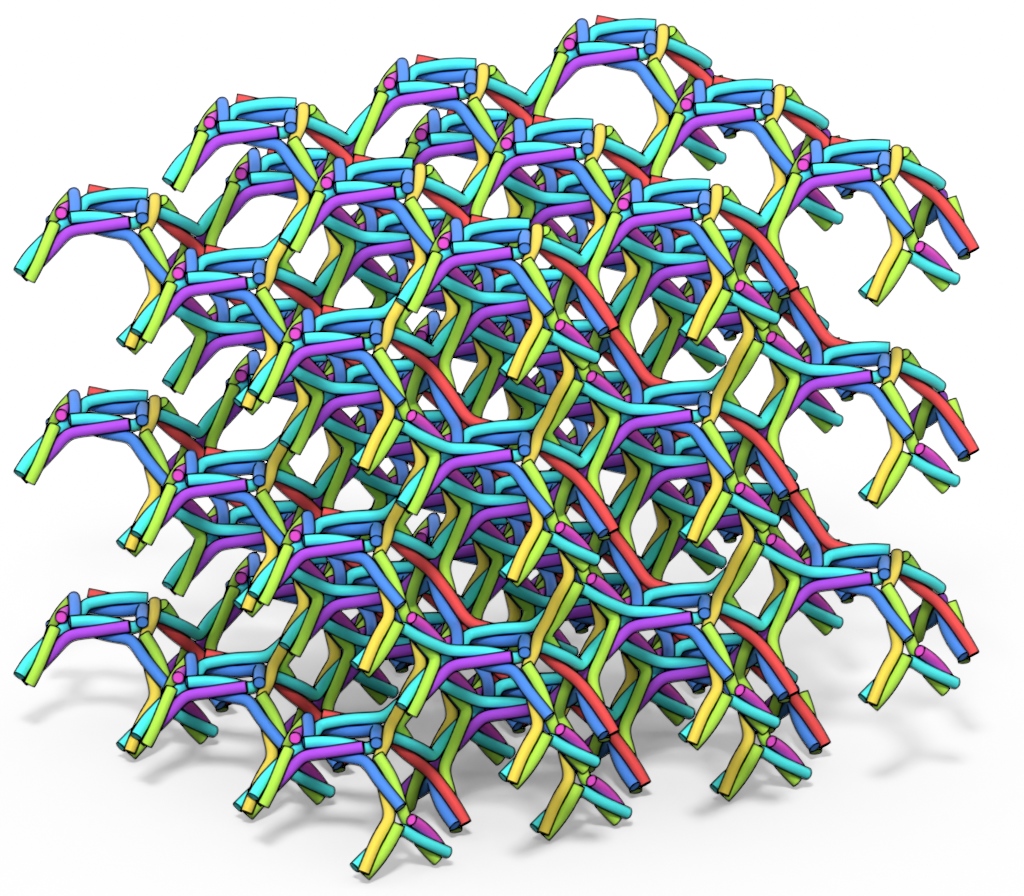}
    \includegraphics[width=0.23\linewidth]{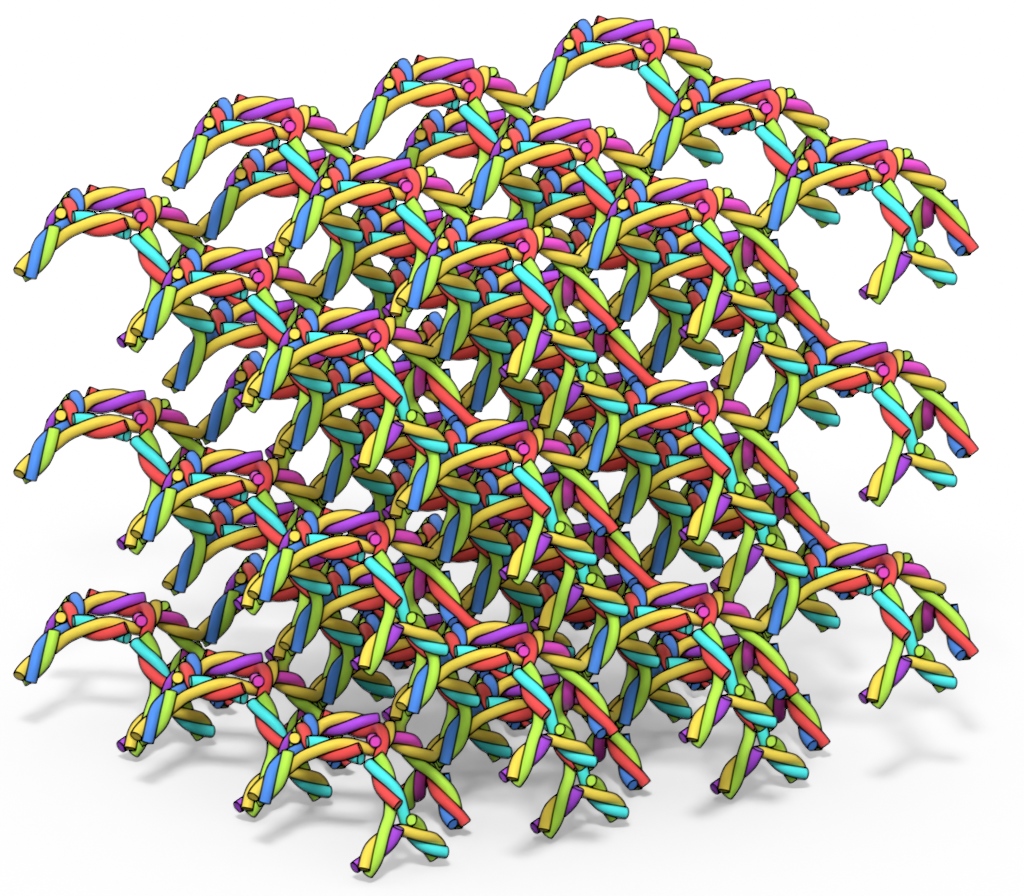}
    \includegraphics[width=0.23\linewidth]{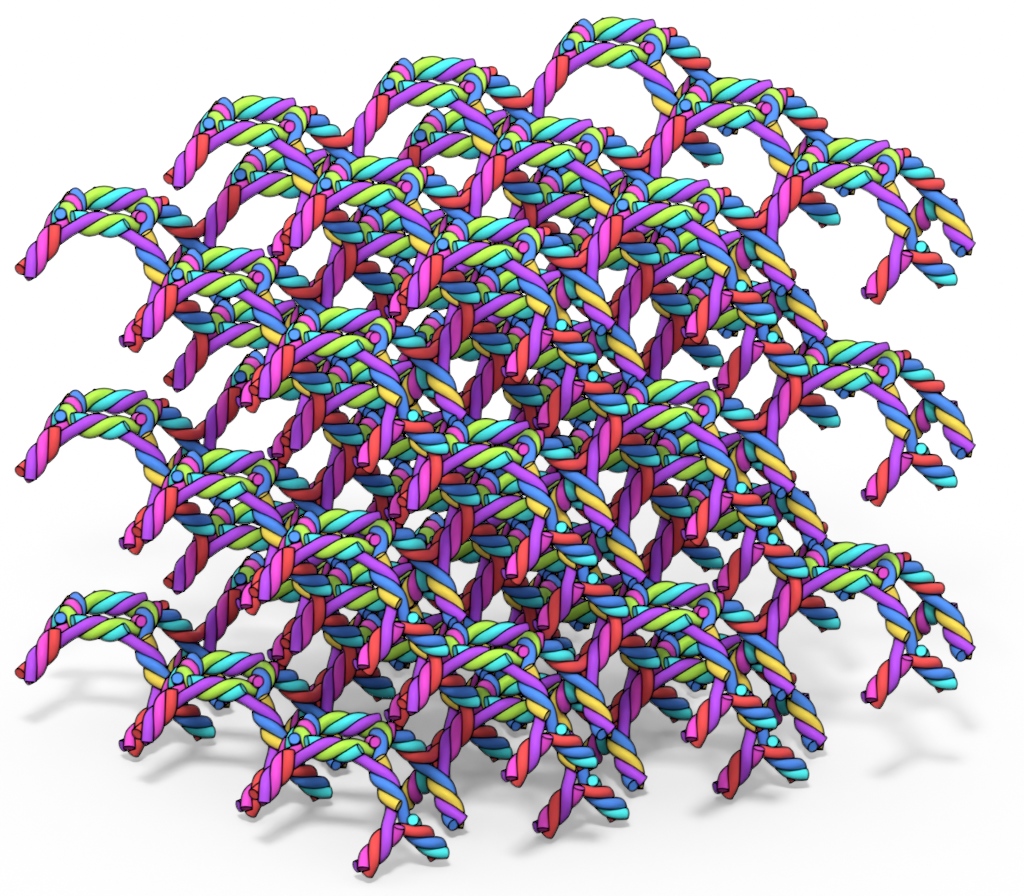}
    \includegraphics[width=0.23\linewidth]{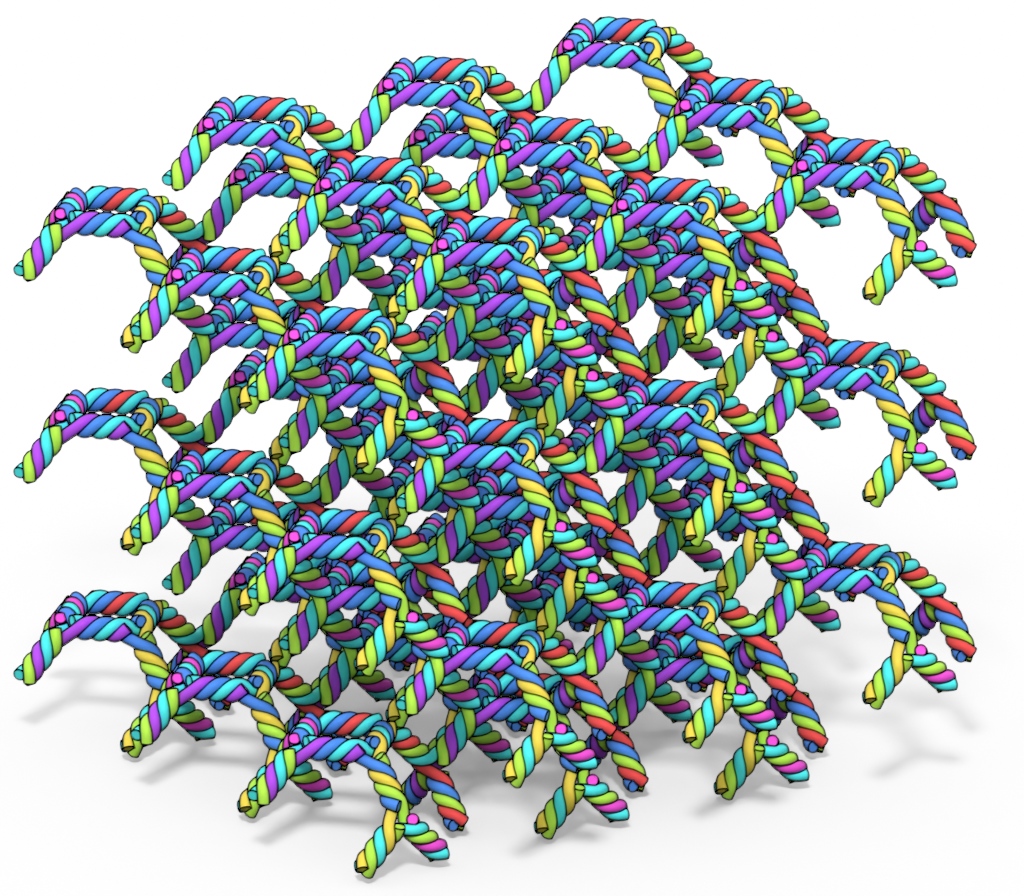}
    \caption{Here, similar to Figure~\ref{fig:bravais3D}, we are using tileable twisting modules. Instead of cubes, we are using truncated octahedra as the tiling basis, which gives us 12 unique edges, each having 3 face layout, meaning 3 threads per edges leading to 36 threads. In this example uniform twisting number gives 7 distinct connected components for different twisting numbers.}
    \Description{A figure arranged in four numbered columns labeled 1 through 4, each column containing two images stacked vertically. In the top row, each column shows a single tileable twisting module rendered as a compact cluster of thick tubular strands in multiple colors, including cyan, yellow, red, green, blue, and purple. The strands meet at several junctions and form an arched, polyhedral-like shape, with visibly different local twisting patterns across the four columns. In the bottom row, each column shows a larger repeating structure formed by tiling the corresponding module, resulting in a dense three-dimensional lattice of intertwined strands. The strand colors and overall tiling arrangement remain consistent across columns, while the degree of local twisting and the resulting global interweaving pattern vary from left to right.}
    \label{fig:bravais3D_repeated_cf}
\end{figure*}

The same uniform-twist principle extends naturally to three-dimensional periodic structures, and as in the 2D case, global connectivity is governed entirely by a single, uniformly applied edge-twist parameter.
In Figure~\ref{fig:bravais3D_repeated}, all edges of the cubic tessellation are assigned the same twist label using a tileable twisting module. Because the cube has three distinct edge classes and each edge contributes four thread segments, each periodic module contains twelve threads. As the uniform twist value varies, the number of connected components changes systematically, demonstrating that global connectivity in the cubic tessellation is controlled solely by a single, uniformly applied edge-twist parameter. Number of distinct components change in cubic tesselation, however in a truncatec octahedron tesselation uniform twisting number leads to same number of distinct components as shown in Figure~\ref{fig:bravais3D_repeated_cf}.

\begin{figure}[htb!]
    \centering
    \centering
    \begin{overpic}[width=0.24\linewidth]{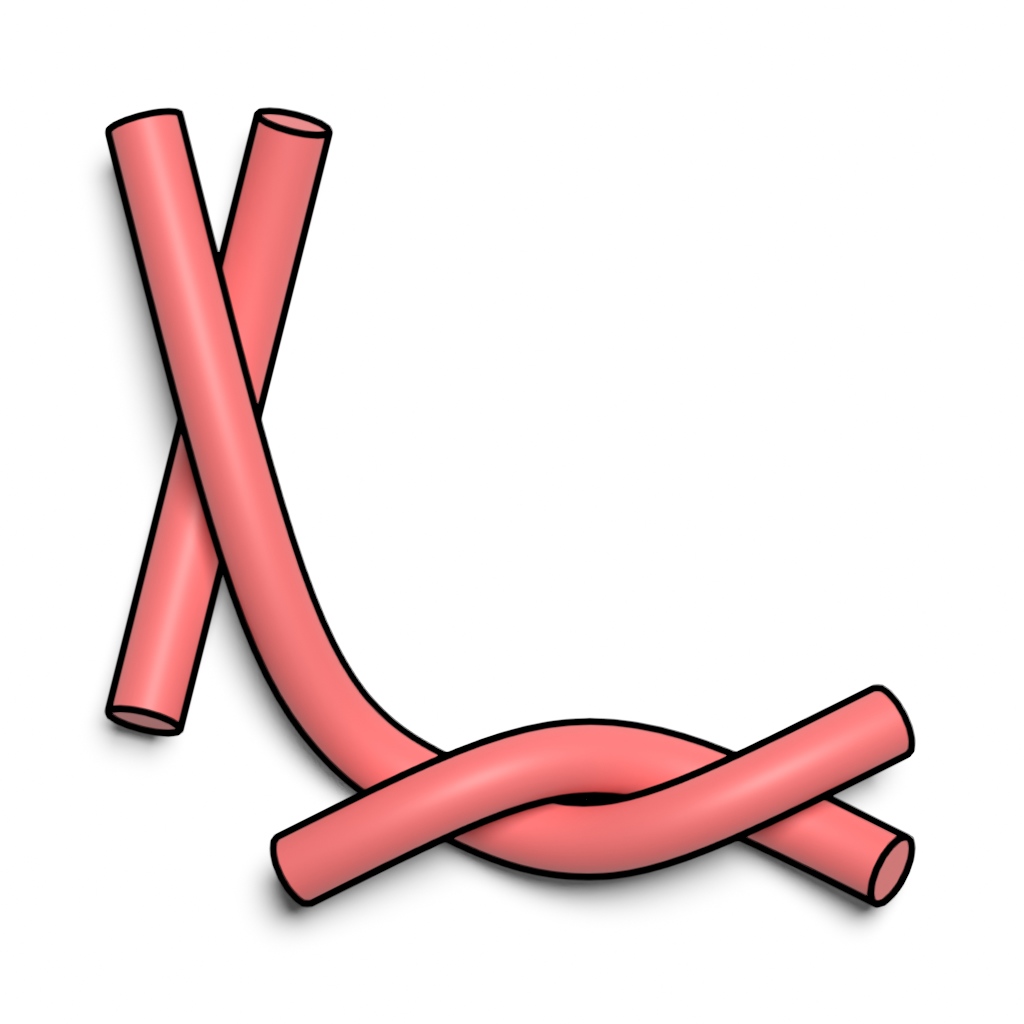}
        \put(-5,50){\rotatebox[origin=c]{90}{\centering [2,1]}}
    \end{overpic}
    \includegraphics[width=0.24\linewidth]{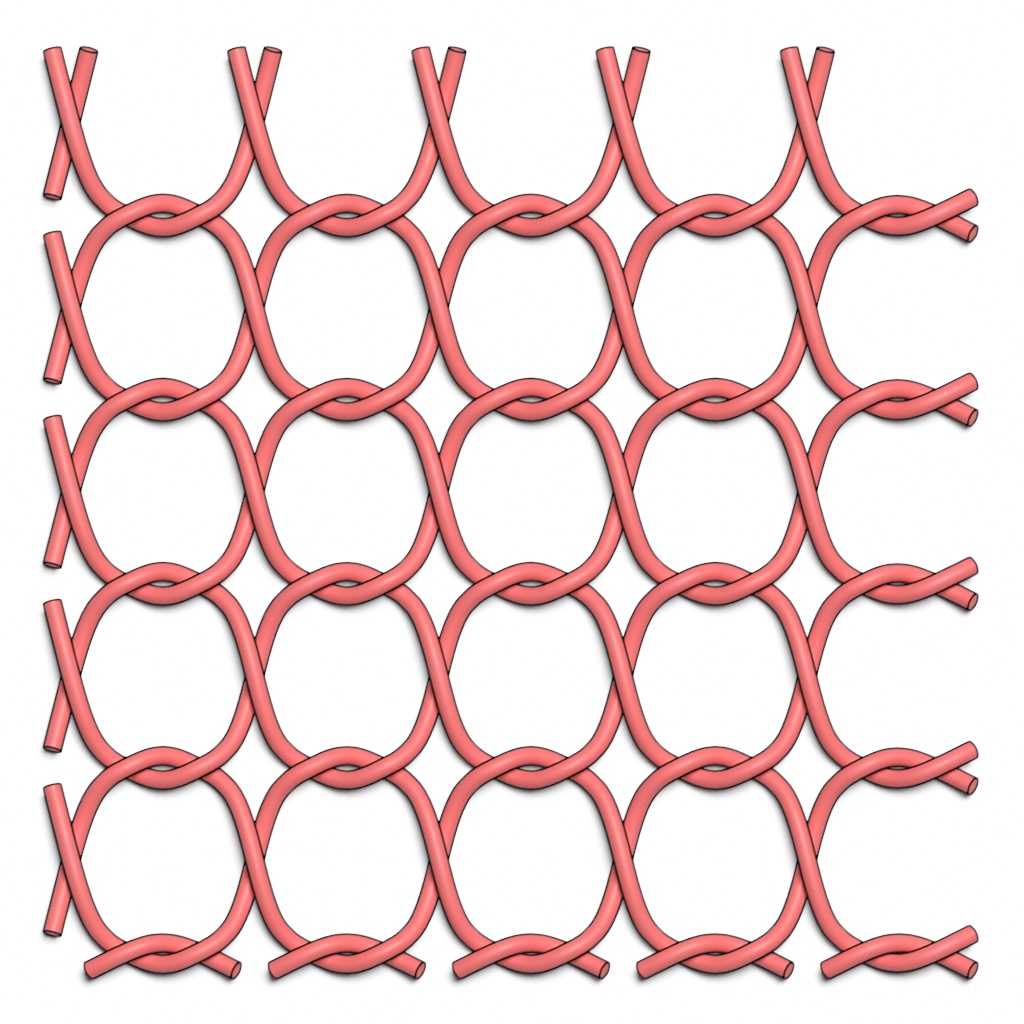}
    \begin{overpic}[width=0.24\linewidth]{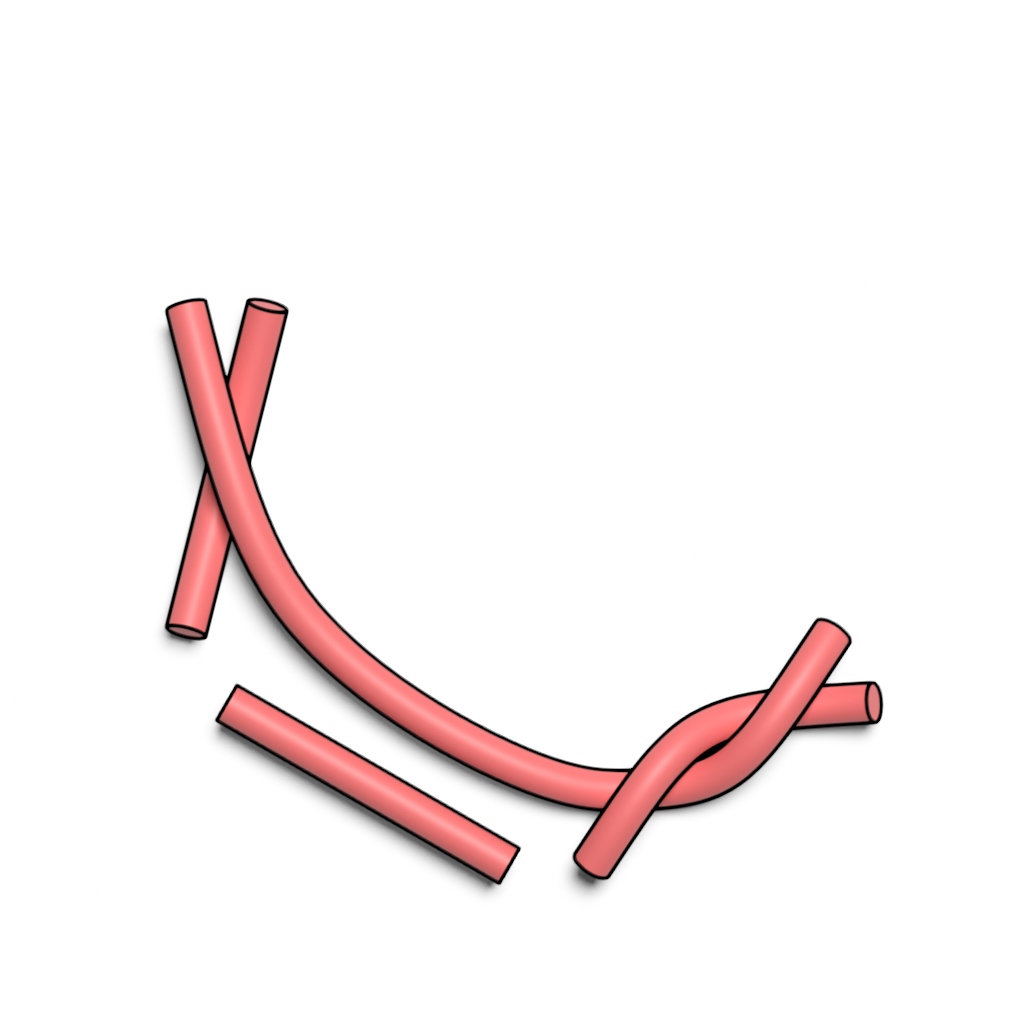}
        \put(-5,50){\rotatebox[origin=c]{90}{\centering [2,0,1]}}
    \end{overpic}
    \includegraphics[width=0.24\linewidth]{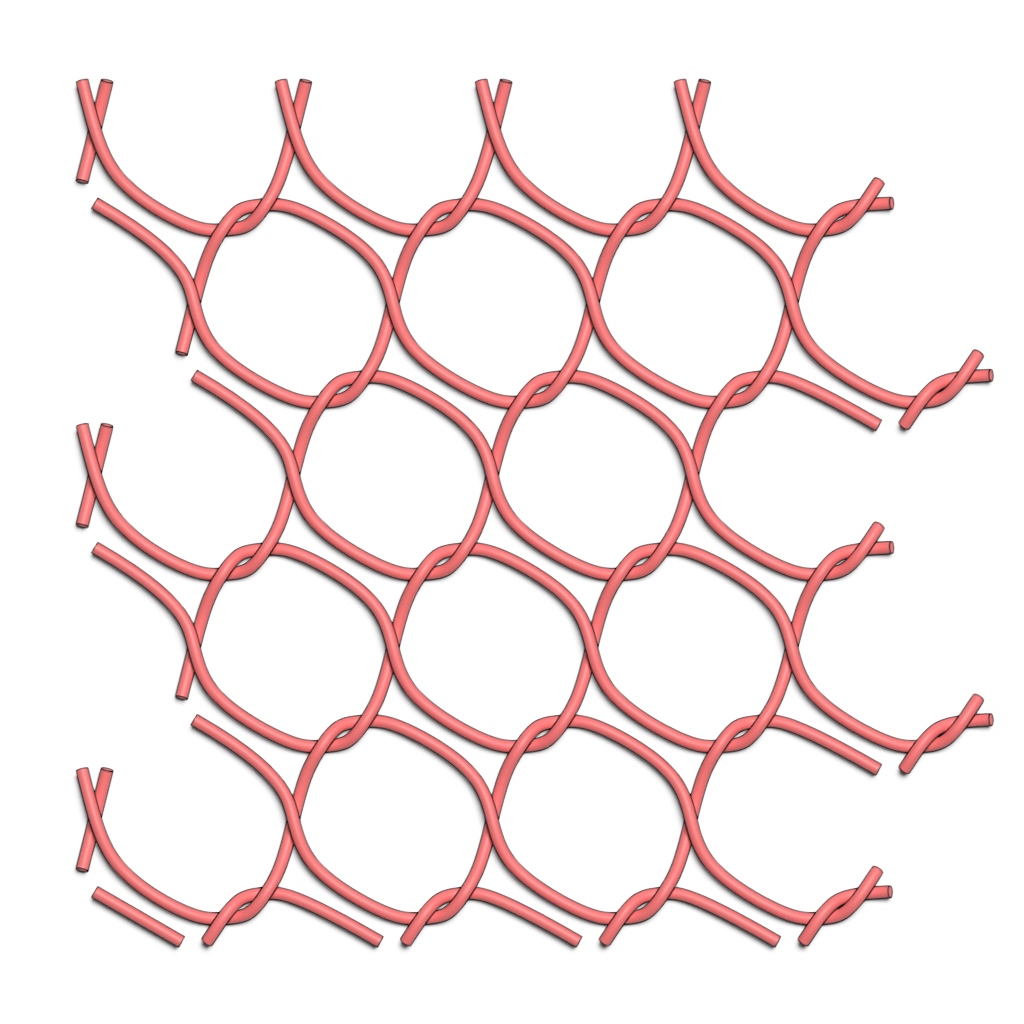}
    \begin{overpic}[width=0.24\linewidth]{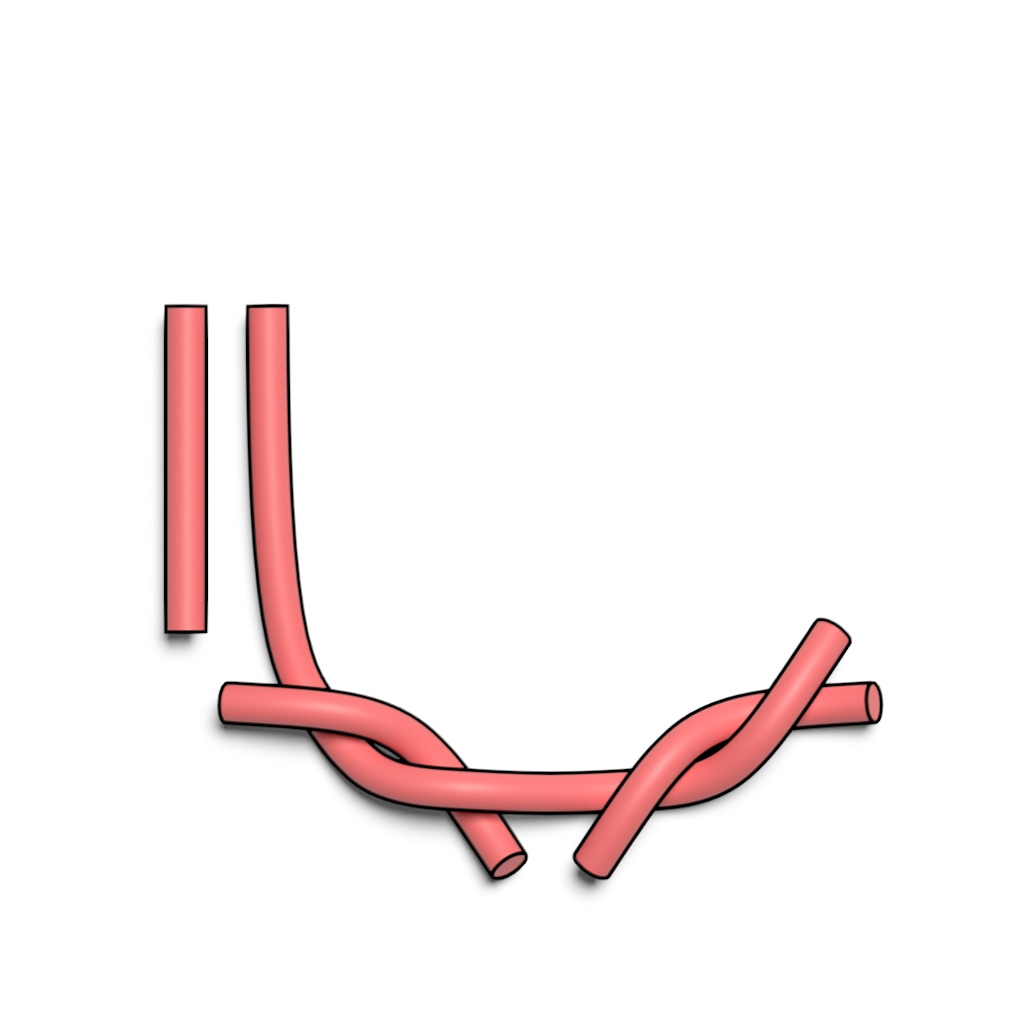}
        \put(-5,50){\rotatebox[origin=c]{90}{\centering [2,2,0]}}
    \end{overpic}    \includegraphics[width=0.24\linewidth]{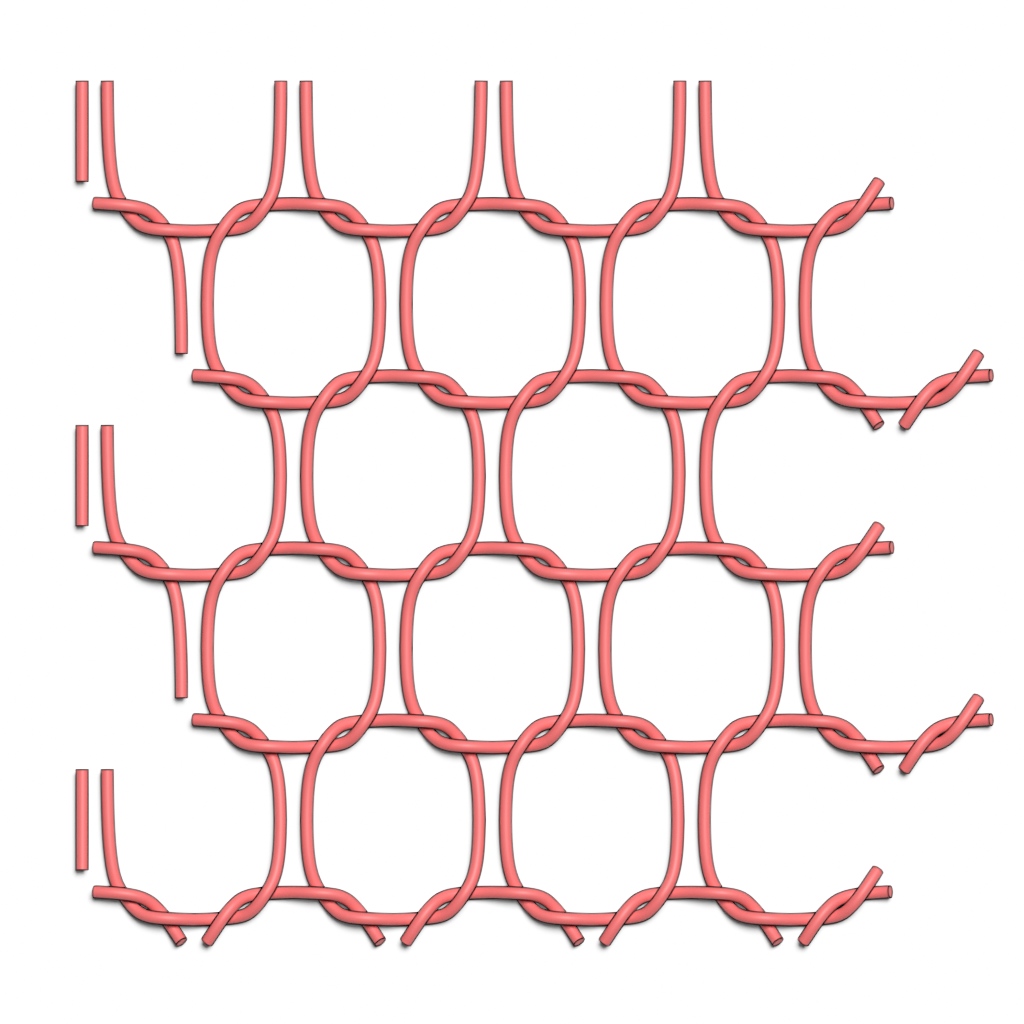}
    \begin{overpic}[width=0.24\linewidth]{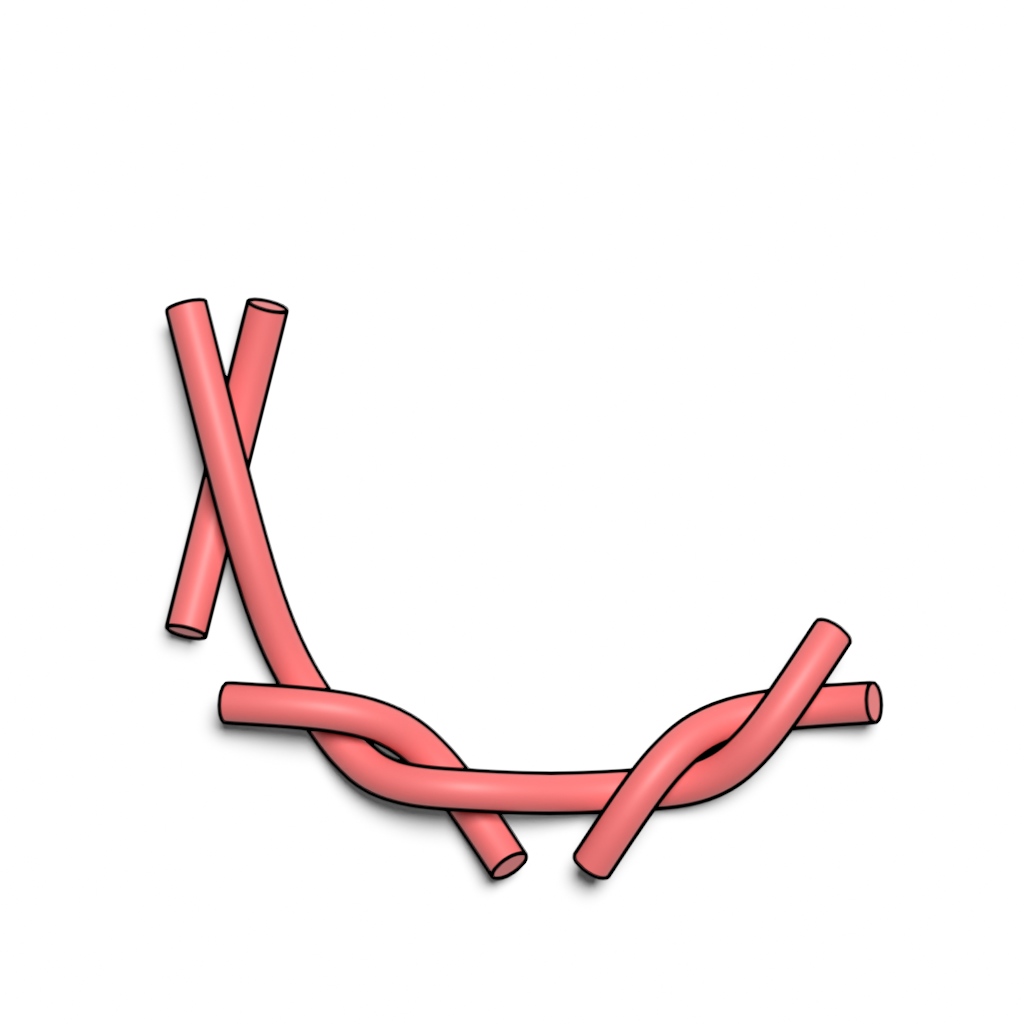}
        \put(-5,50){\rotatebox[origin=c]{90}{\centering [2,2,1]}}
    \end{overpic}    \includegraphics[width=0.24\linewidth]{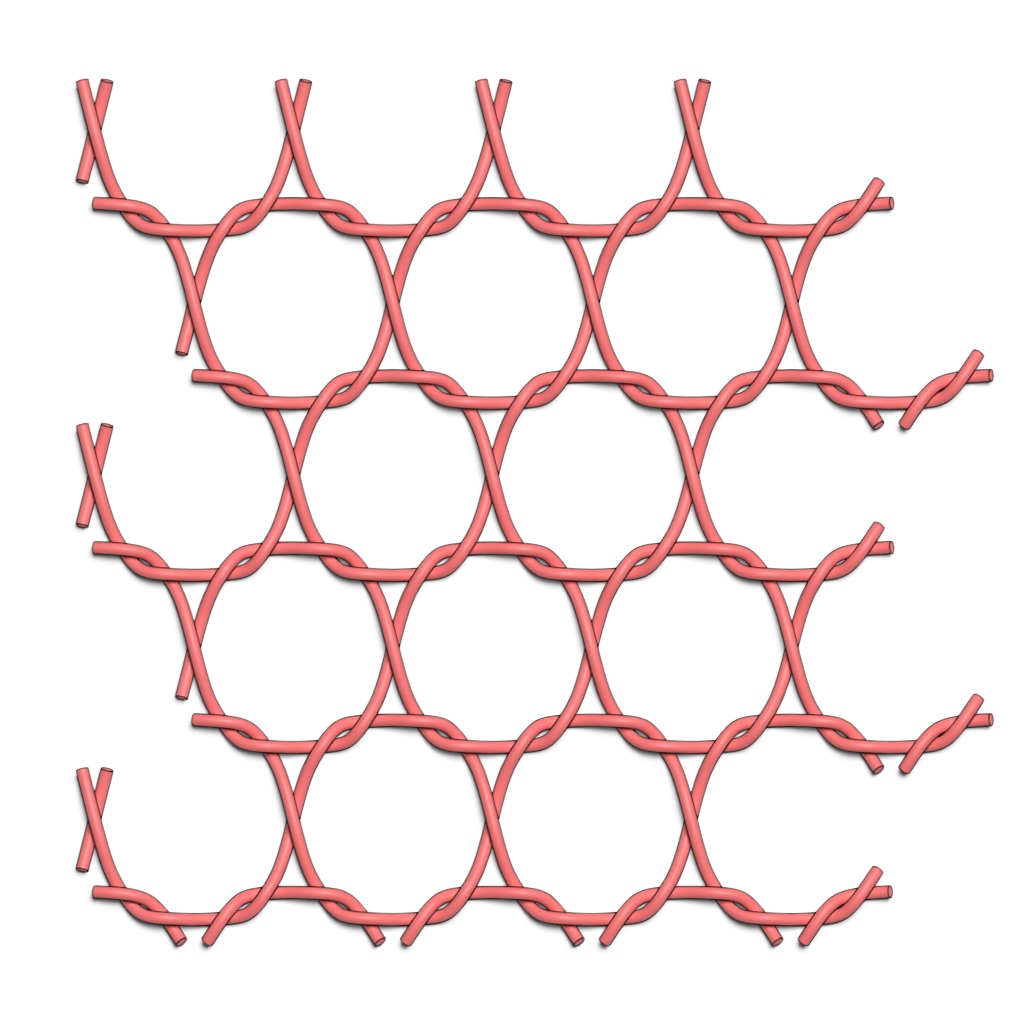}
    \caption{With non-uniform twisting for distinct edges, we are able to produce different classes of structures such as knits and chainmails.}
    \Description{A figure composed of four example cases arranged in a grid. In each case, a small local twisting configuration is shown on the left, rendered as thick red tubular segments, and a larger repeating structure generated from that configuration is shown on the right. The local configurations are labeled with bracketed numeric tuples and differ in how the tubular segments cross and twist. The corresponding larger structures show repeating planar patterns of interlinked loops, with variations in spacing, orientation, and interconnection across the four cases. All strands are rendered in a uniform red color against a white background.}  \label{fig:bravais2D_repeated_nonuniform}
\end{figure}

\begin{figure}[htb!]
    \centering
    \begin{overpic}[width=0.24\linewidth]{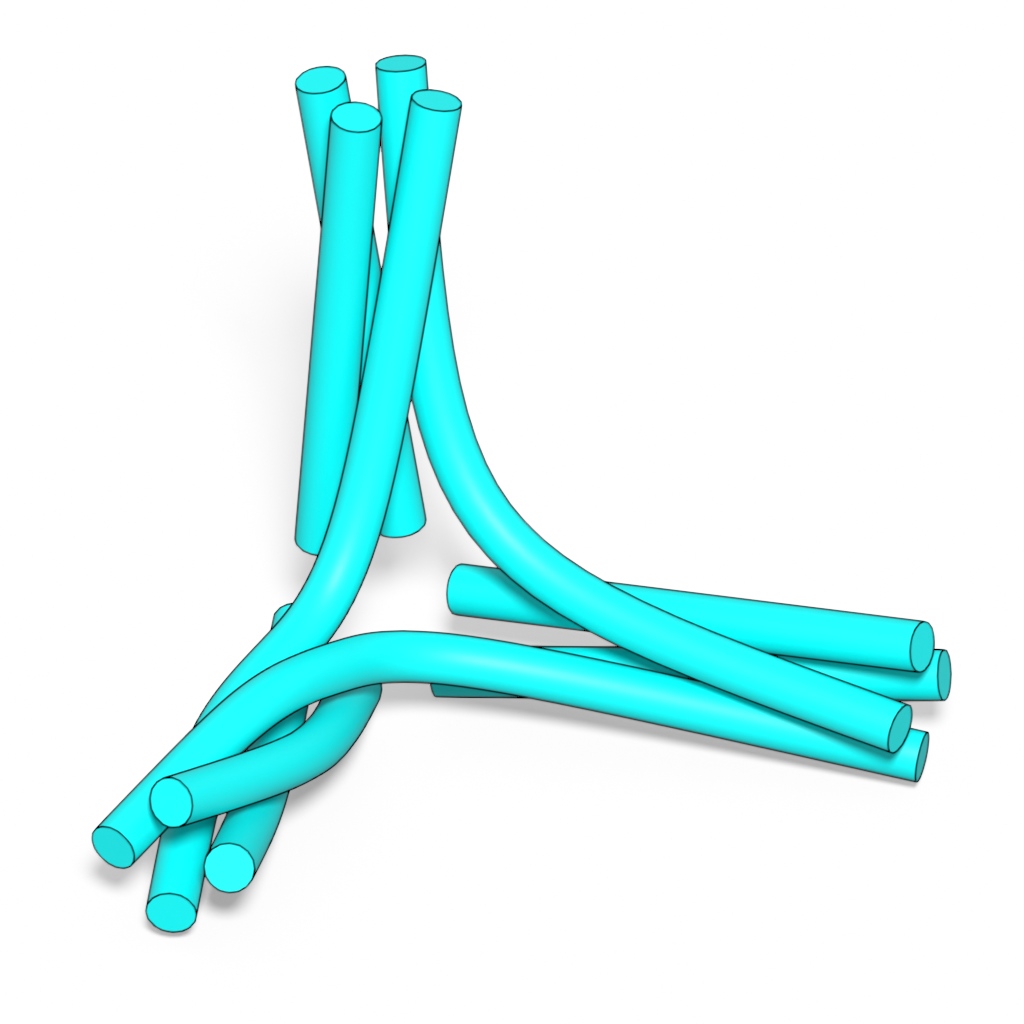}
        \put(-5,50){\rotatebox[origin=c]{90}{\centering [2,1,1]}}
    \end{overpic}
    \includegraphics[width=0.24\linewidth]{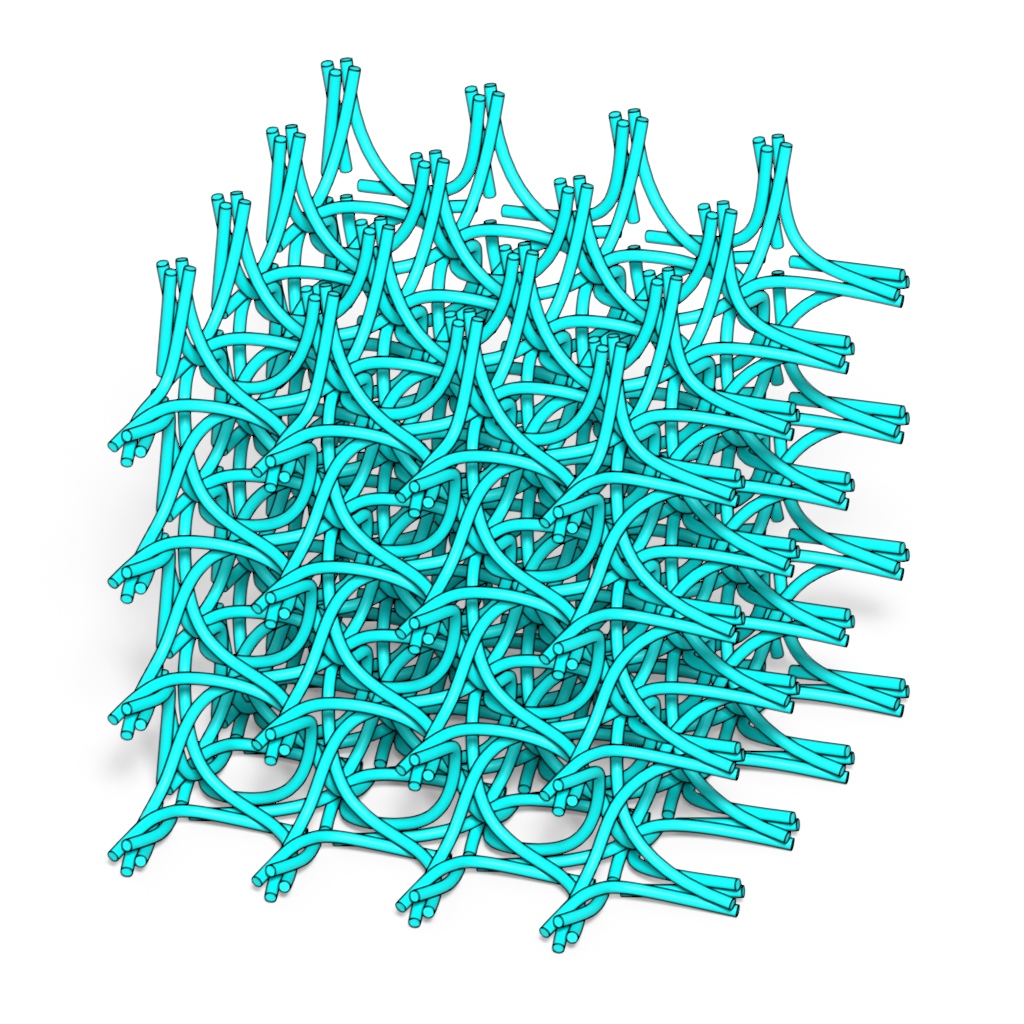}
    \includegraphics[width=0.24\linewidth]{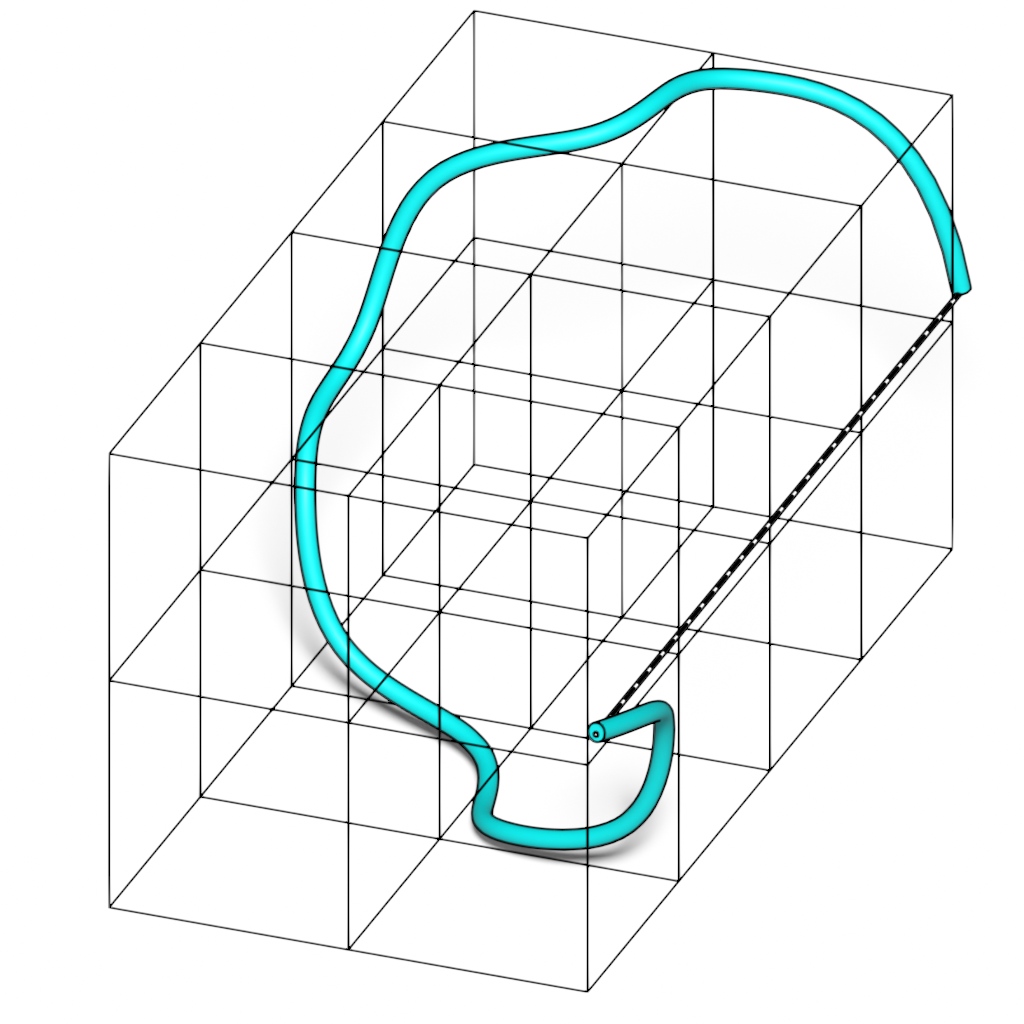}
    \includegraphics[width=0.24\linewidth]{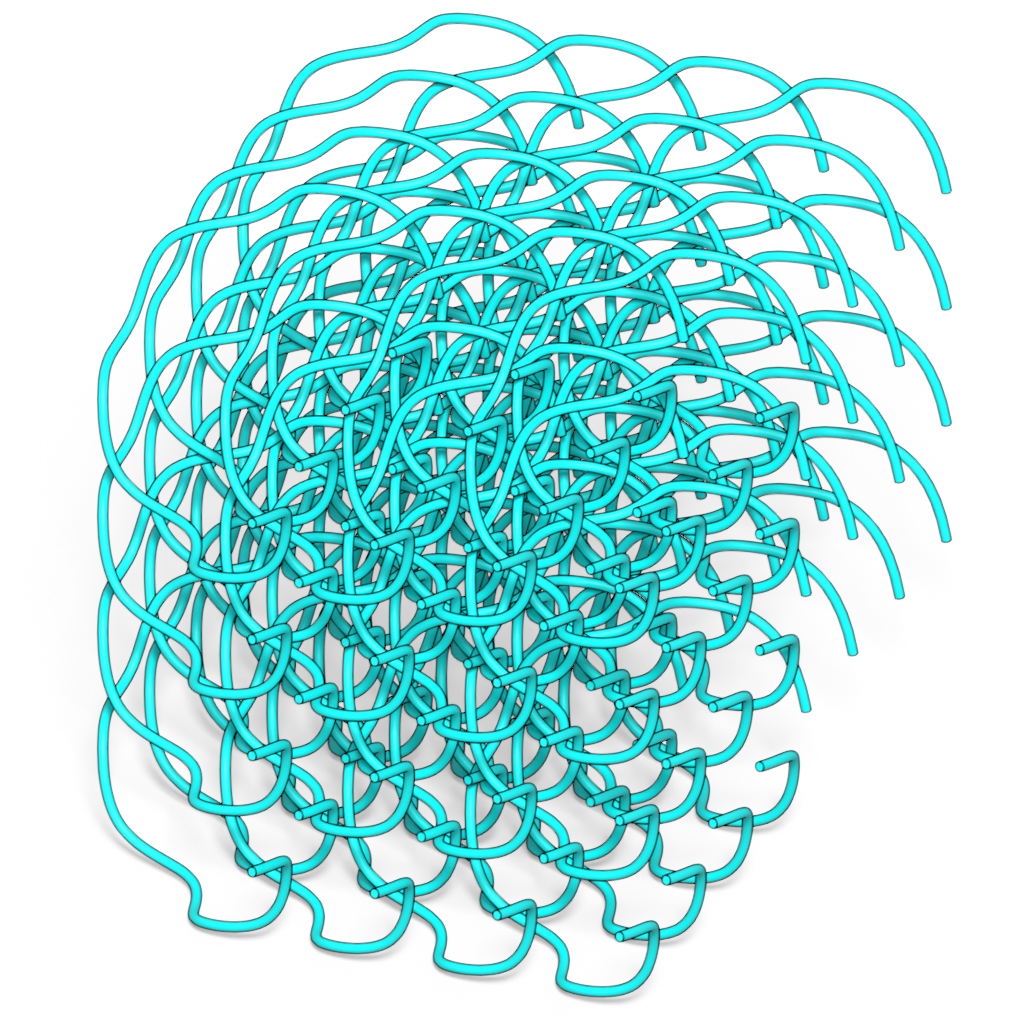}\\
    \begin{overpic}[width=0.24\linewidth]{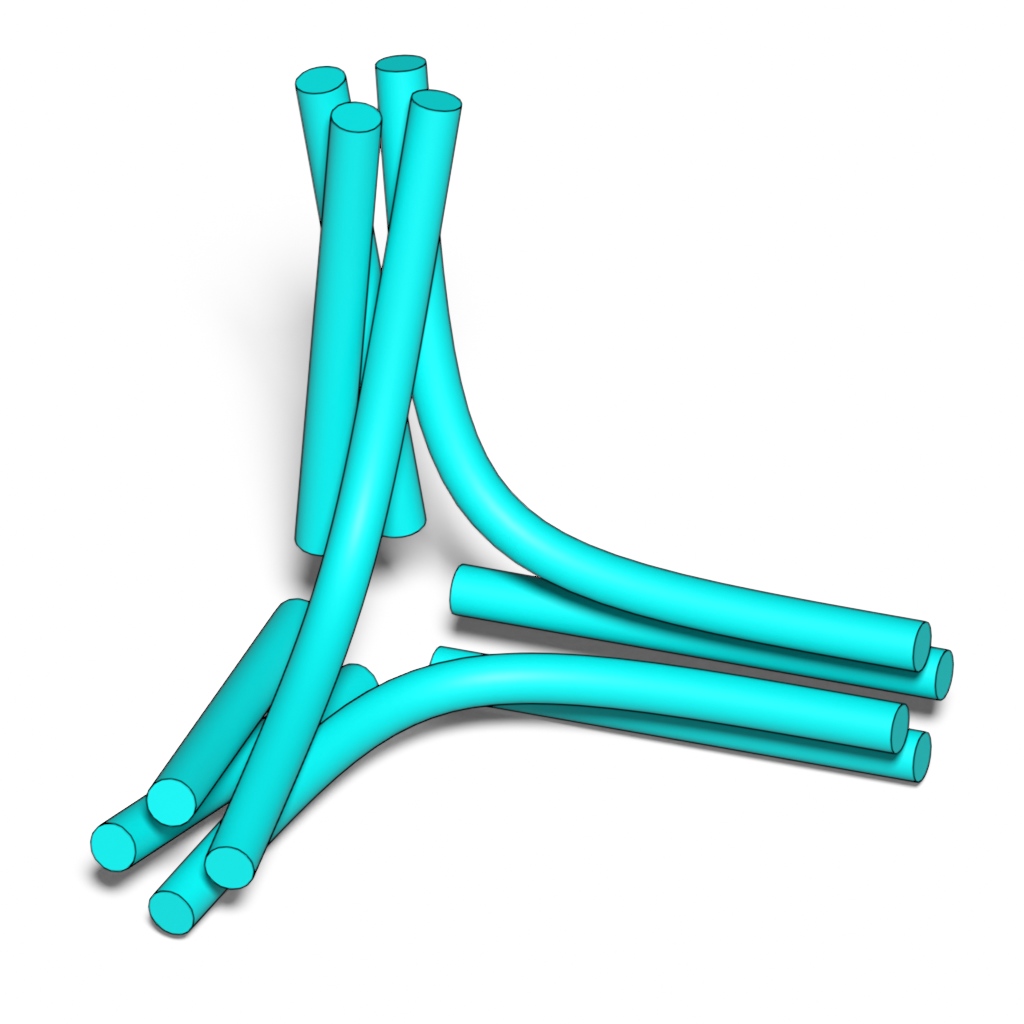}
        \put(-5,50){\rotatebox[origin=c]{90}{\centering [-1,0,1]}}
    \end{overpic}
    \includegraphics[width=0.24\linewidth]{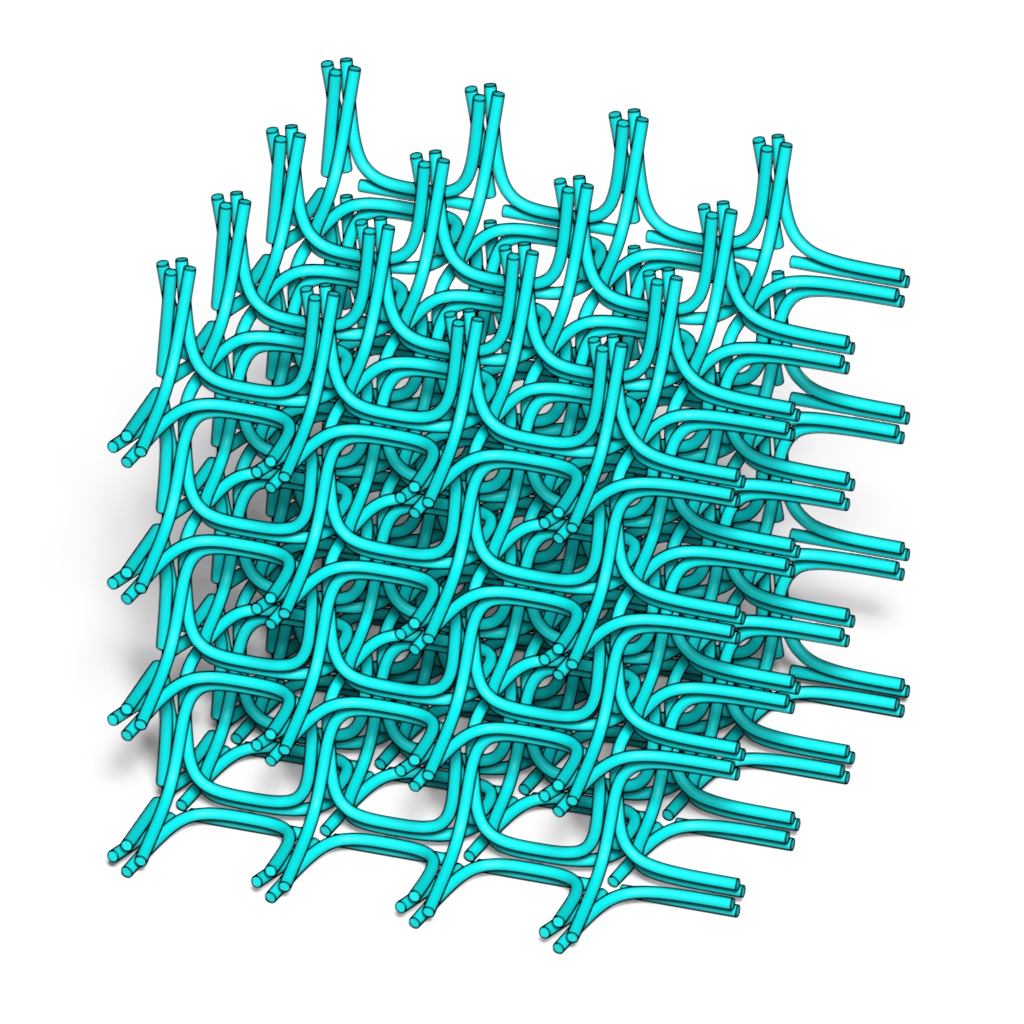}
    \includegraphics[width=0.24\linewidth]{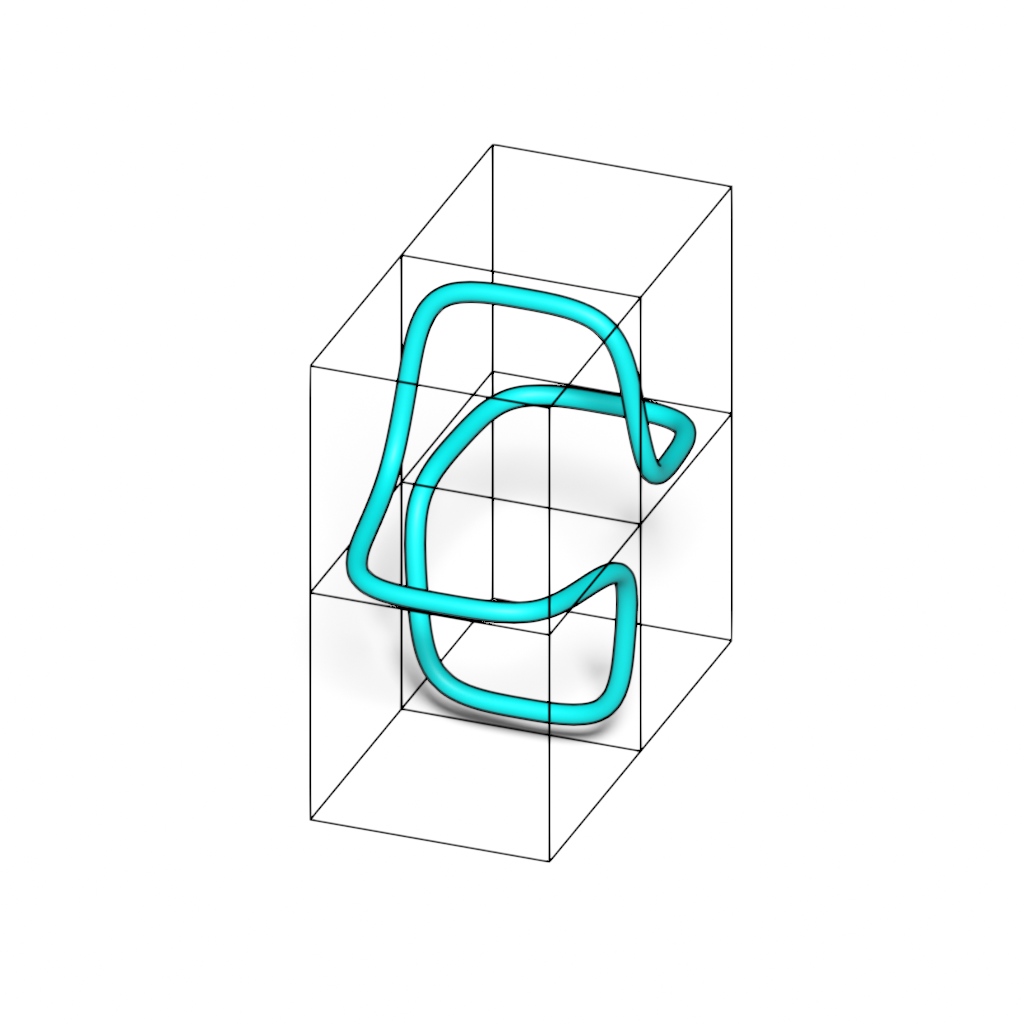}
    \includegraphics[width=0.24\linewidth]{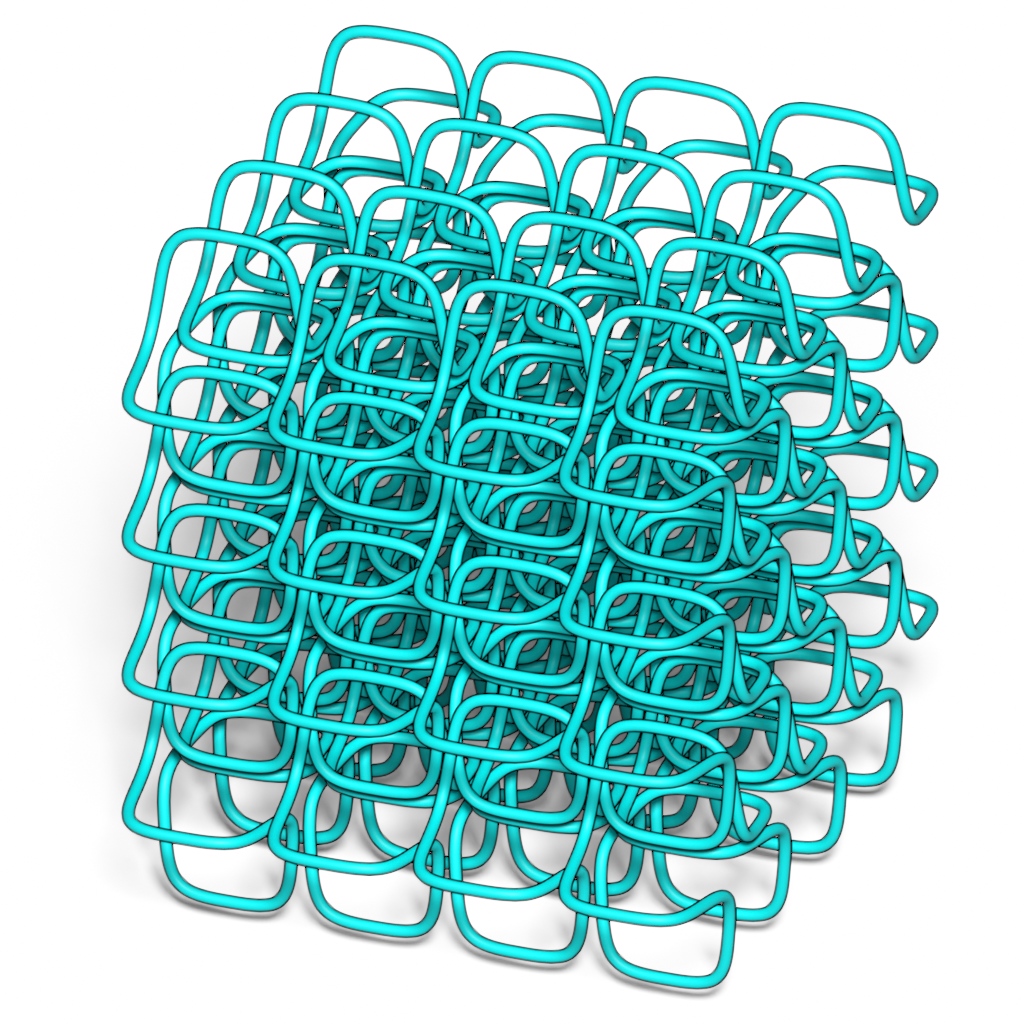}\\
    \begin{overpic}[width=0.24\linewidth]{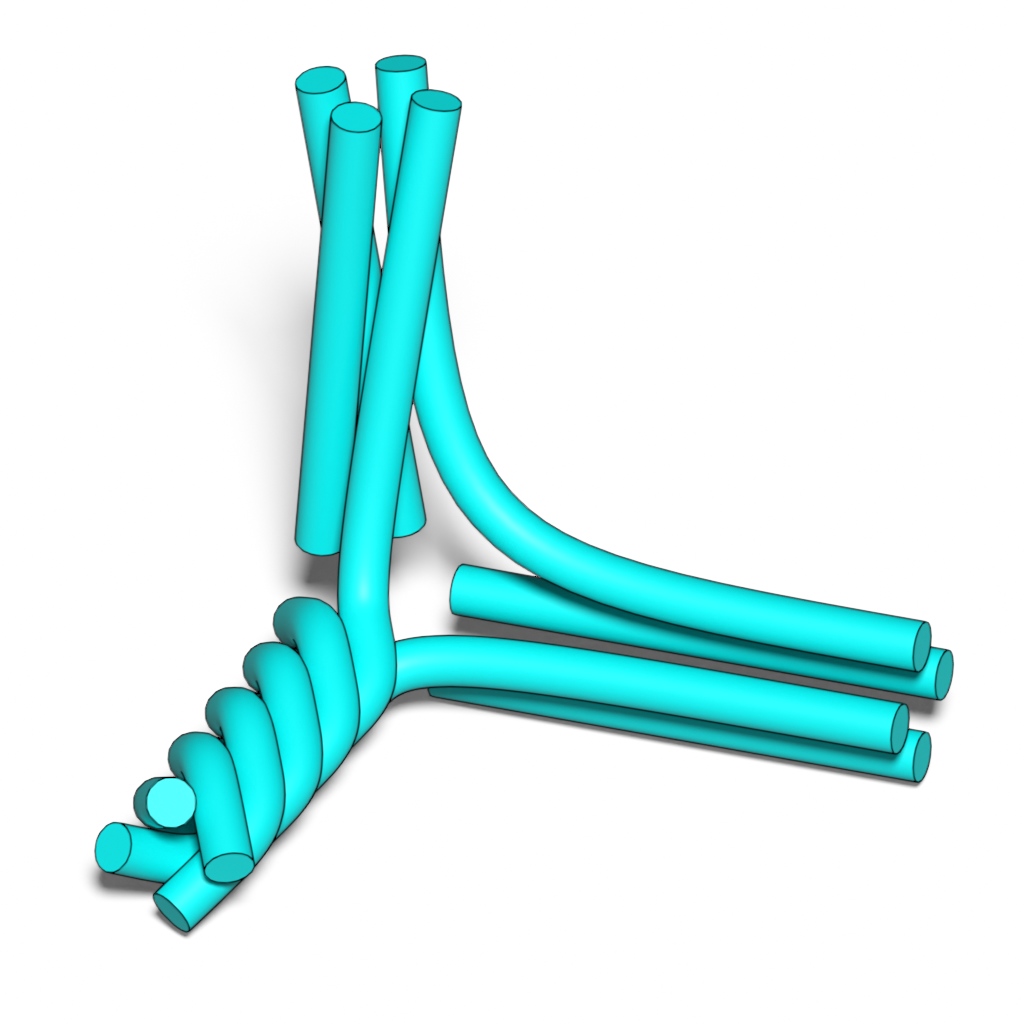}
        \put(-5,50){\rotatebox[origin=c]{90}{\centering [-5,0,1]}}
    \end{overpic}
    \includegraphics[width=0.24\linewidth]{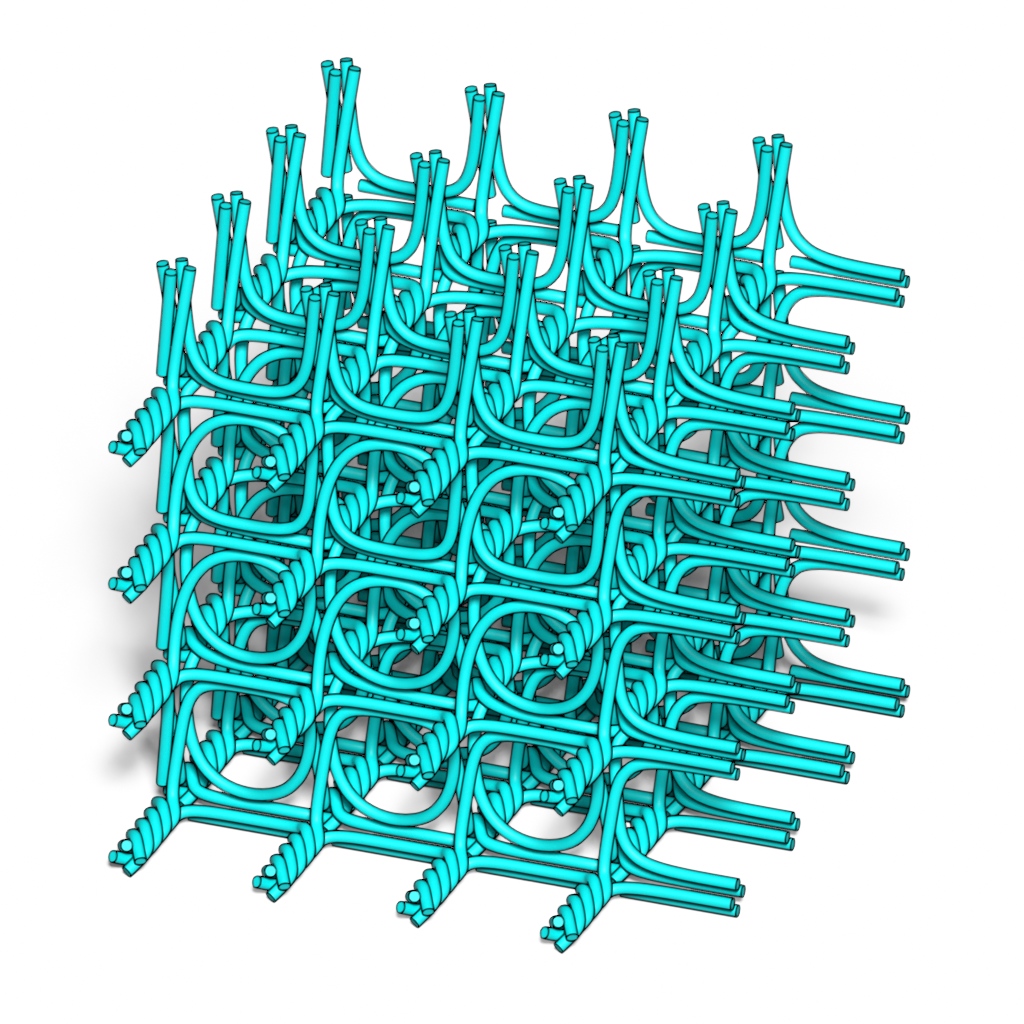}
    \includegraphics[width=0.24\linewidth]{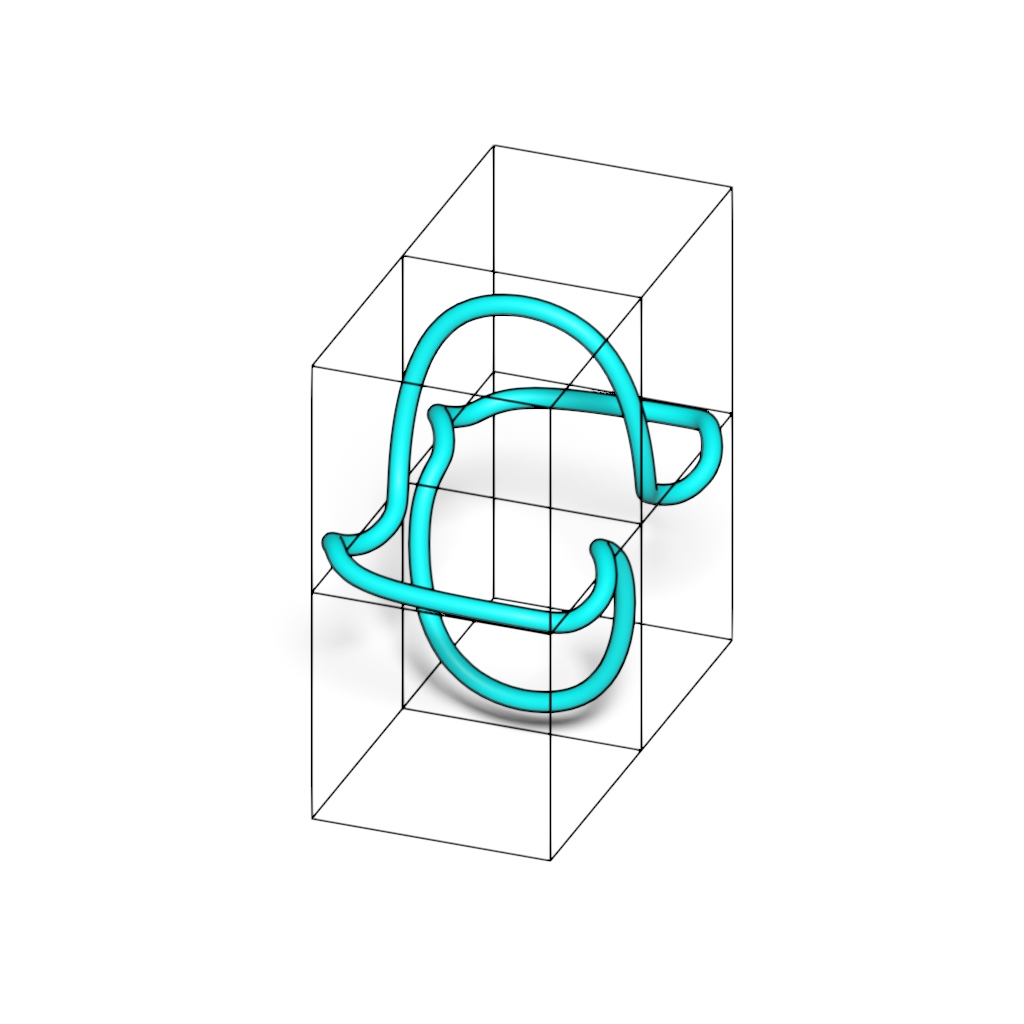}
    \includegraphics[width=0.24\linewidth]{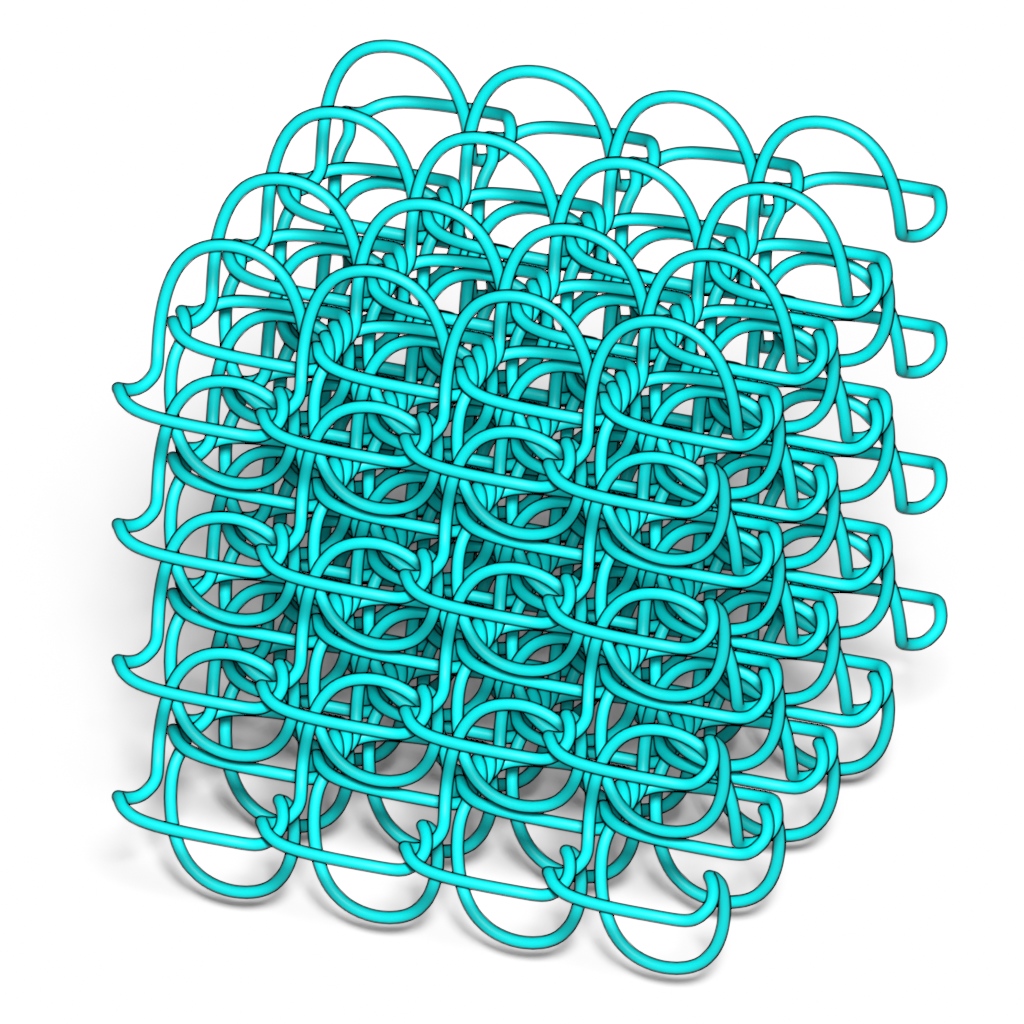}\\
    \parbox[t]{0.24\linewidth}{\centering Twisted unit}
    \parbox[t]{0.24\linewidth}{\centering Tiled twisted unit}
    \parbox[t]{0.24\linewidth}{\centering Unit thread}
    \parbox[t]{0.24\linewidth}{\centering Tiled threads}
    \caption{Using a combination of different twist numbers, it is possible to get a single unique component structures, top row shows an example where the threads are infinite and repeat in a $4\times2\times2$ domain. The middle row shows an example where the periodicity is contained within a single cycle curve, spanning $2\times1\times2$ domain. This shape creates planar chainmails that are stacked in one direction that are disconnected. Bottom row shows an example where one of the edges in the middle row is modulo twisted (-1 to -5) to form a link connection between different layers of chainmail, leading to a singly linked structure.}
    \Description{A figure arranged in three horizontal rows and four vertical columns. In each row, the leftmost image shows a local twisting configuration rendered as thick cyan tubular strands, labeled with a bracketed vector of three integers. The second column shows a periodic block formed by repeating the local configuration, resulting in a dense three-dimensional lattice of intertwined strands. The third column shows a single representative curve drawn inside a transparent rectangular bounding box, illustrating the path of one strand through the domain. The rightmost column shows a larger aggregated structure formed by repeating this curve across space. Across the three rows, the local twisting pattern changes, leading to visibly different global arrangements, including fully repeating lattices, stacked planar layers, and more densely interconnected three-dimensional configurations. All strands are rendered in the same cyan color against a white background.}
    \label{fig:cp_single}
\end{figure}

\begin{figure}[htb!]
    \centering
    \begin{overpic}[width=0.24\linewidth]{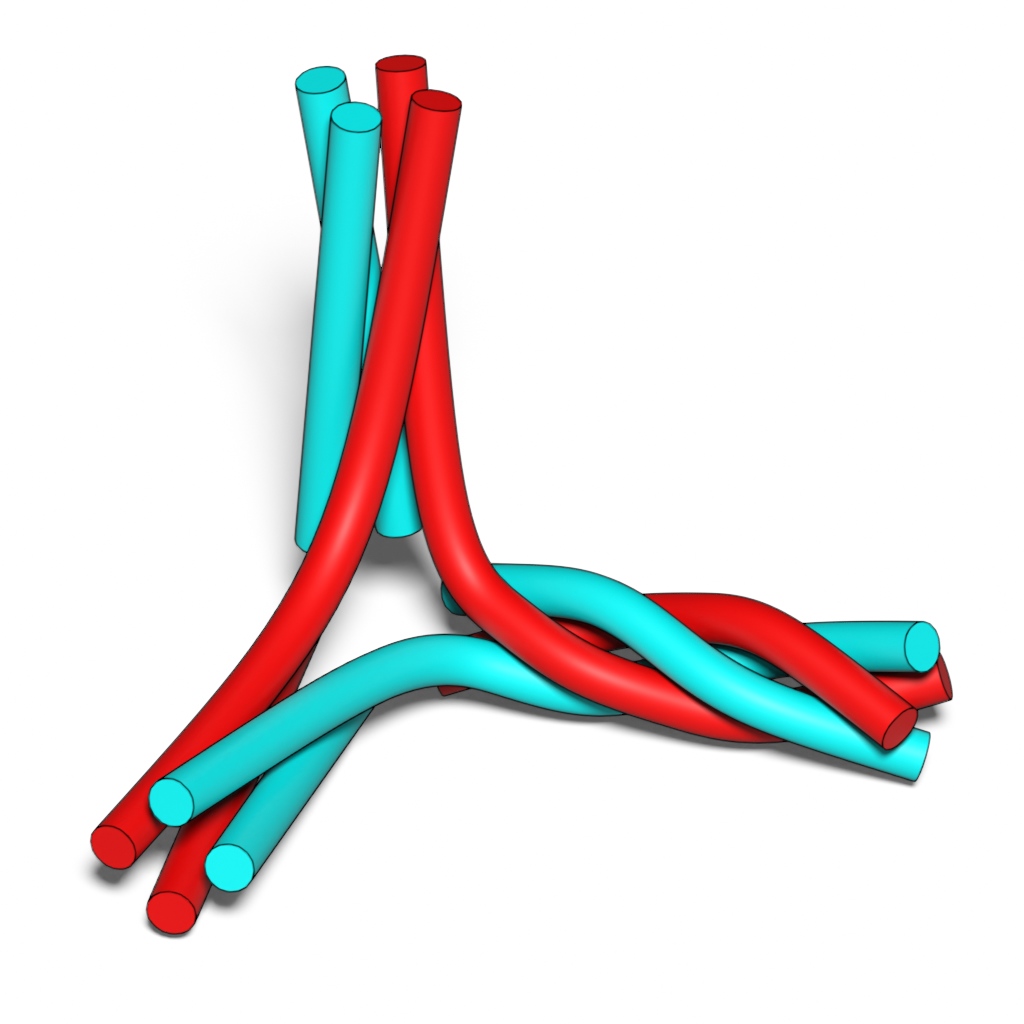}
        \put(-5,50){\rotatebox[origin=c]{90}{\centering [1,3,1]}}
    \end{overpic}
    \includegraphics[width=0.24\linewidth]{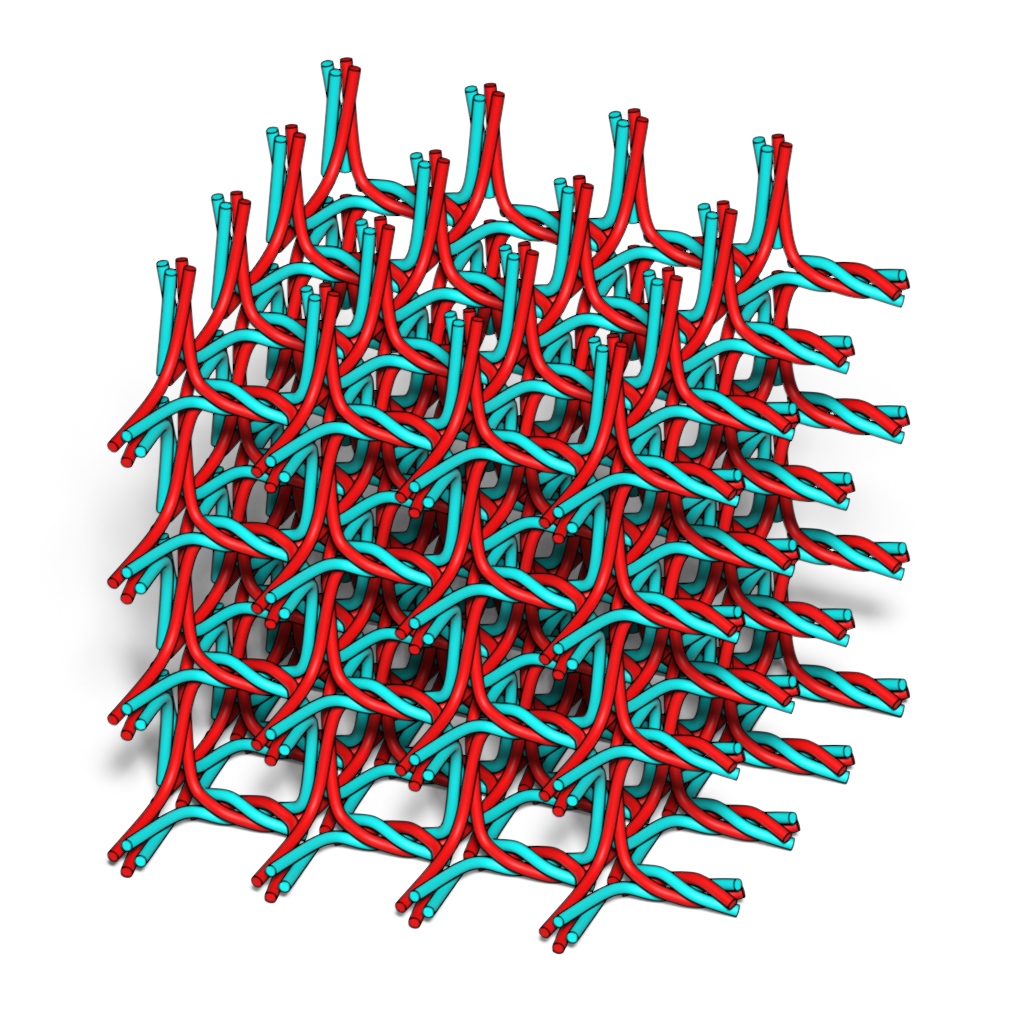}
    \includegraphics[width=0.24\linewidth]{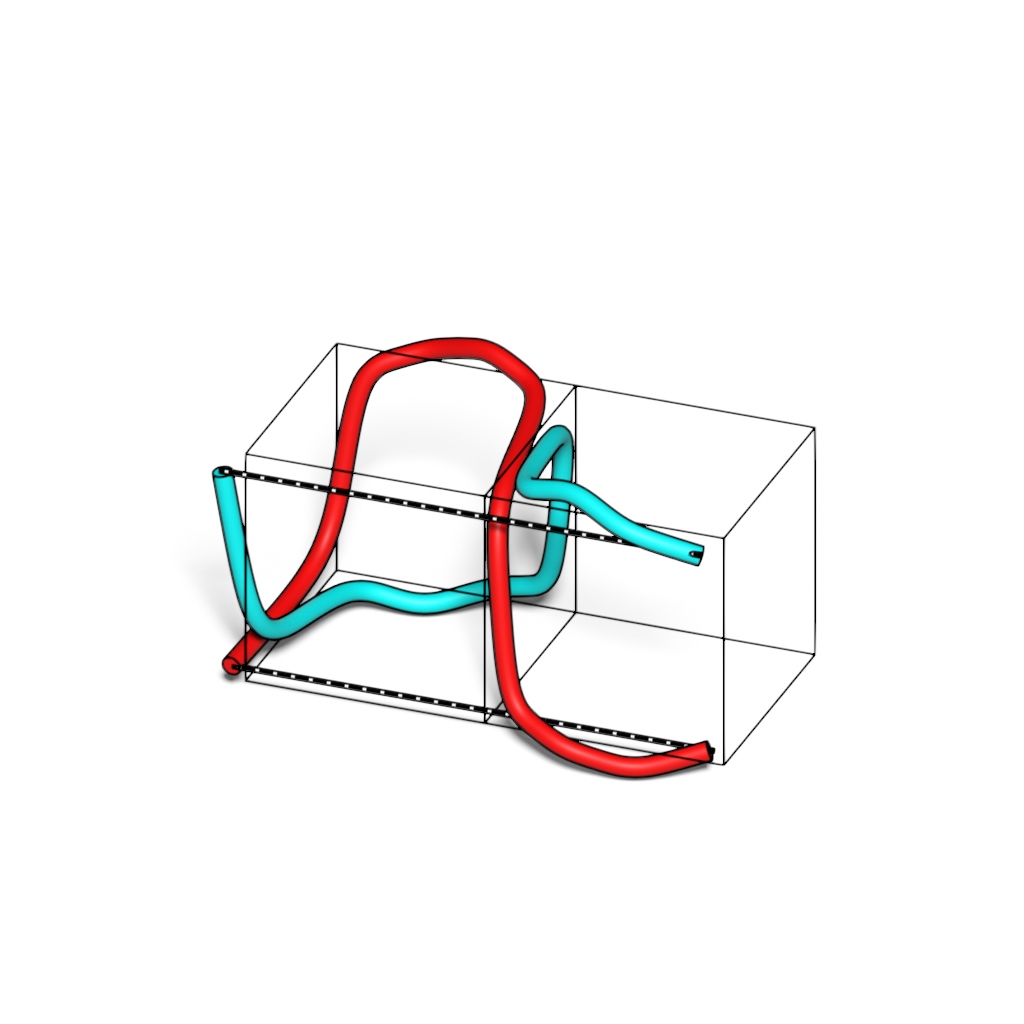}
    \includegraphics[width=0.24\linewidth]{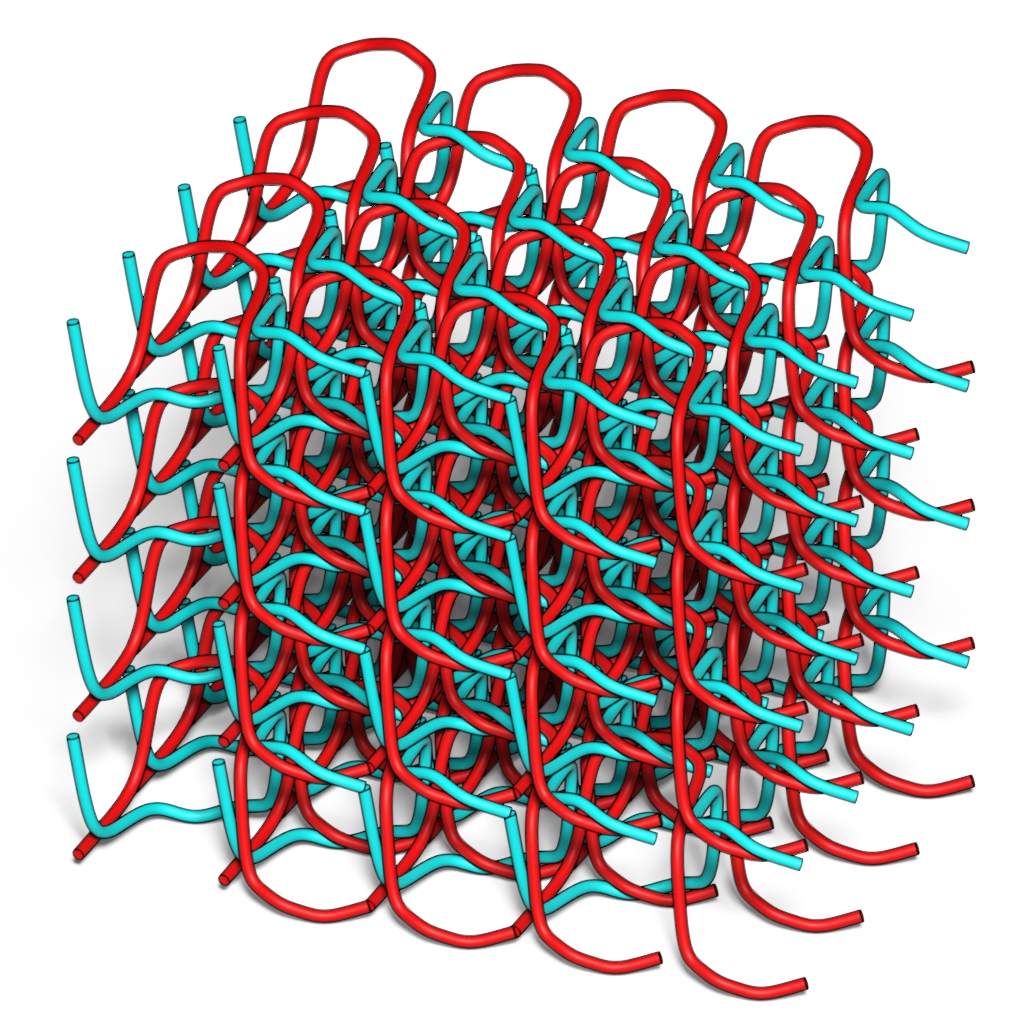}\\
    \begin{overpic}[width=0.24\linewidth]{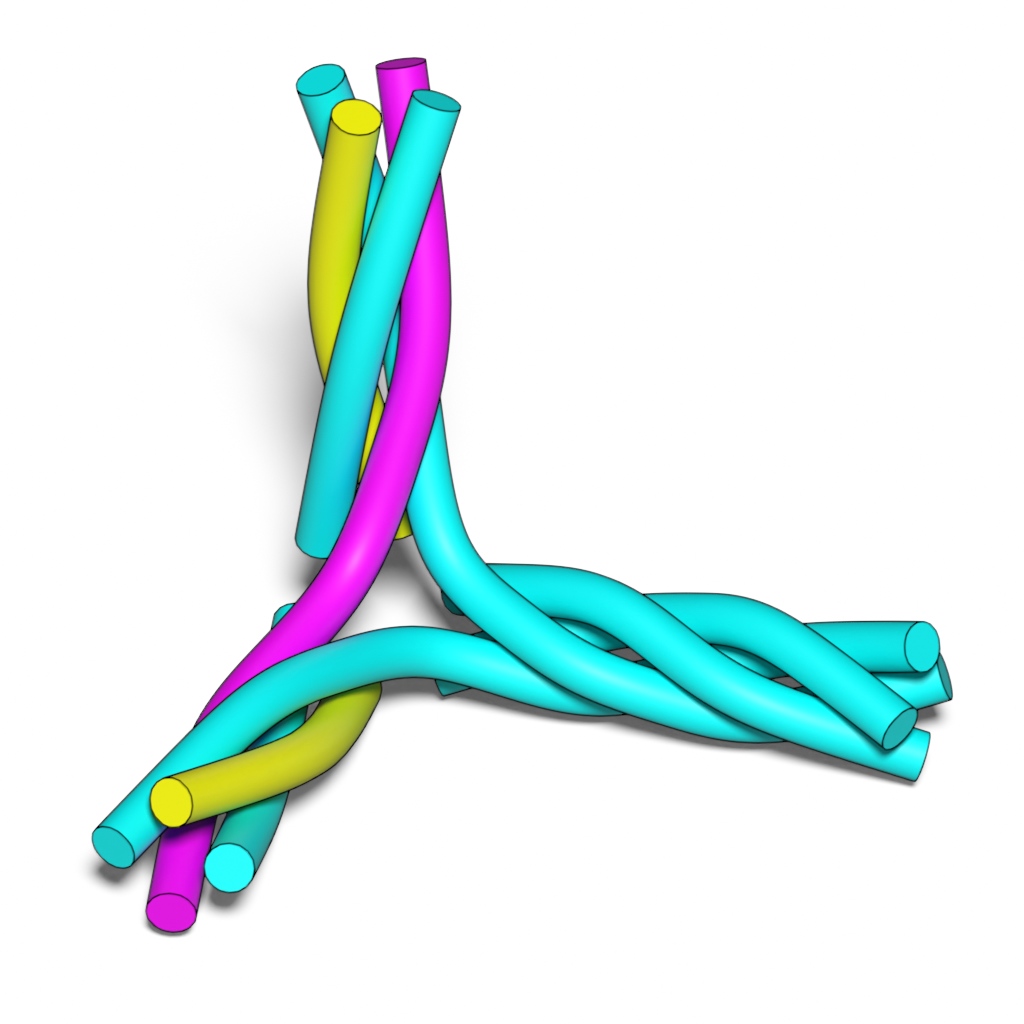}
        \put(-5,50){\rotatebox[origin=c]{90}{\centering [2,3,2]}}
    \end{overpic}
    \includegraphics[width=0.24\linewidth]{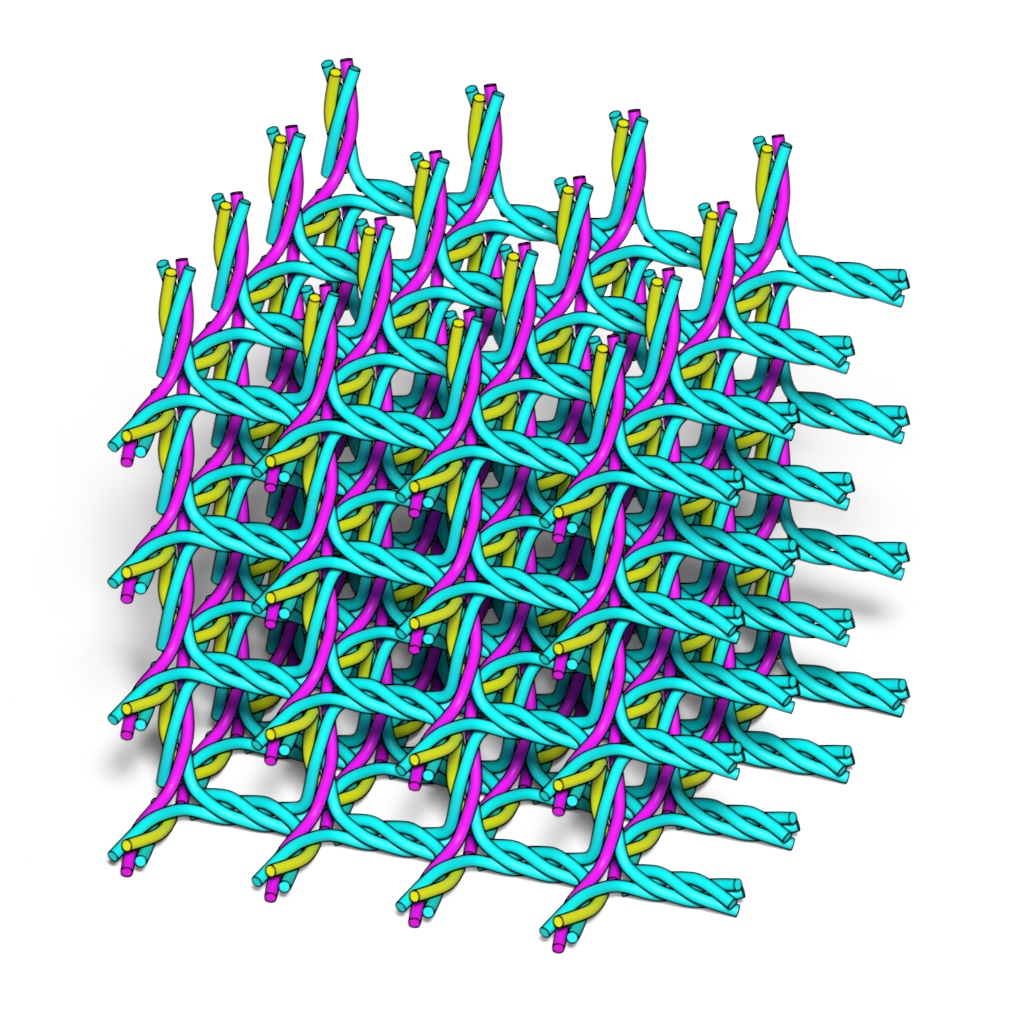}
    \includegraphics[width=0.24\linewidth]{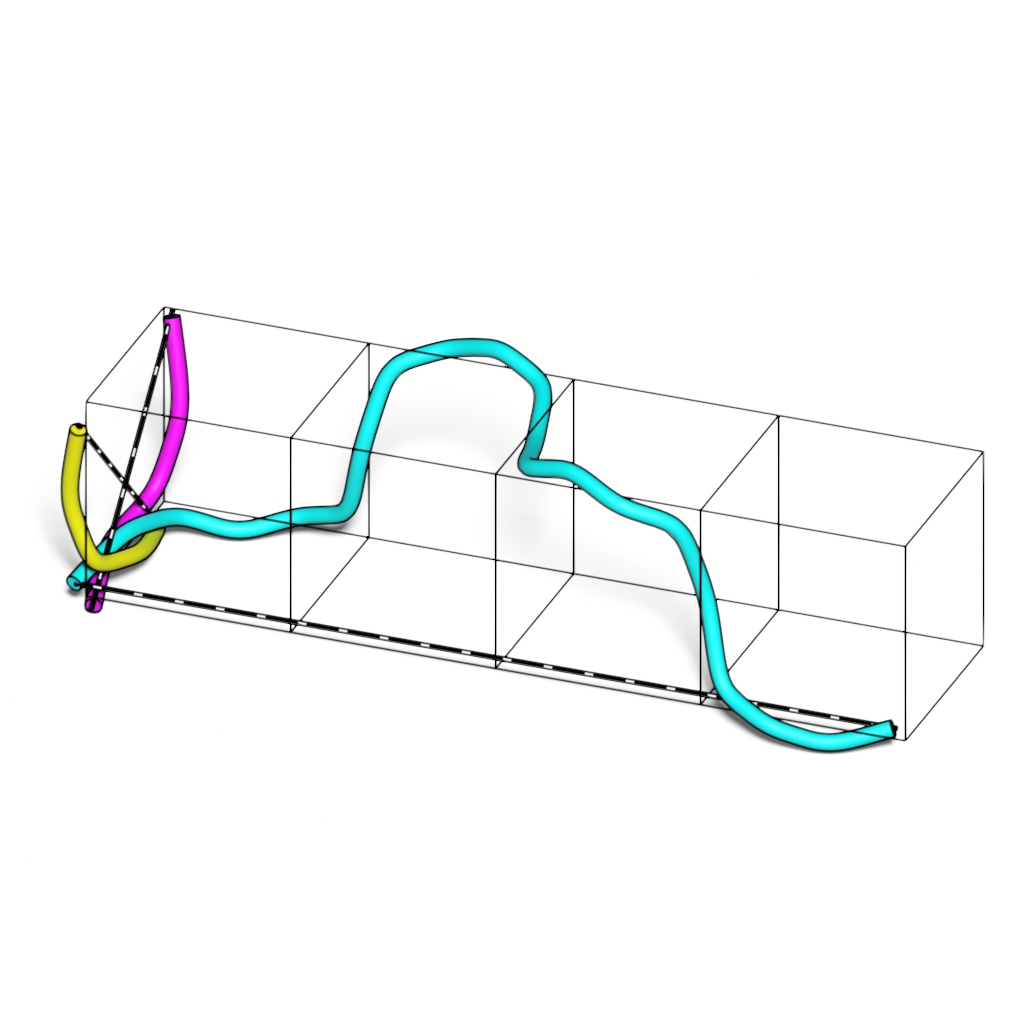}
    \includegraphics[width=0.24\linewidth]{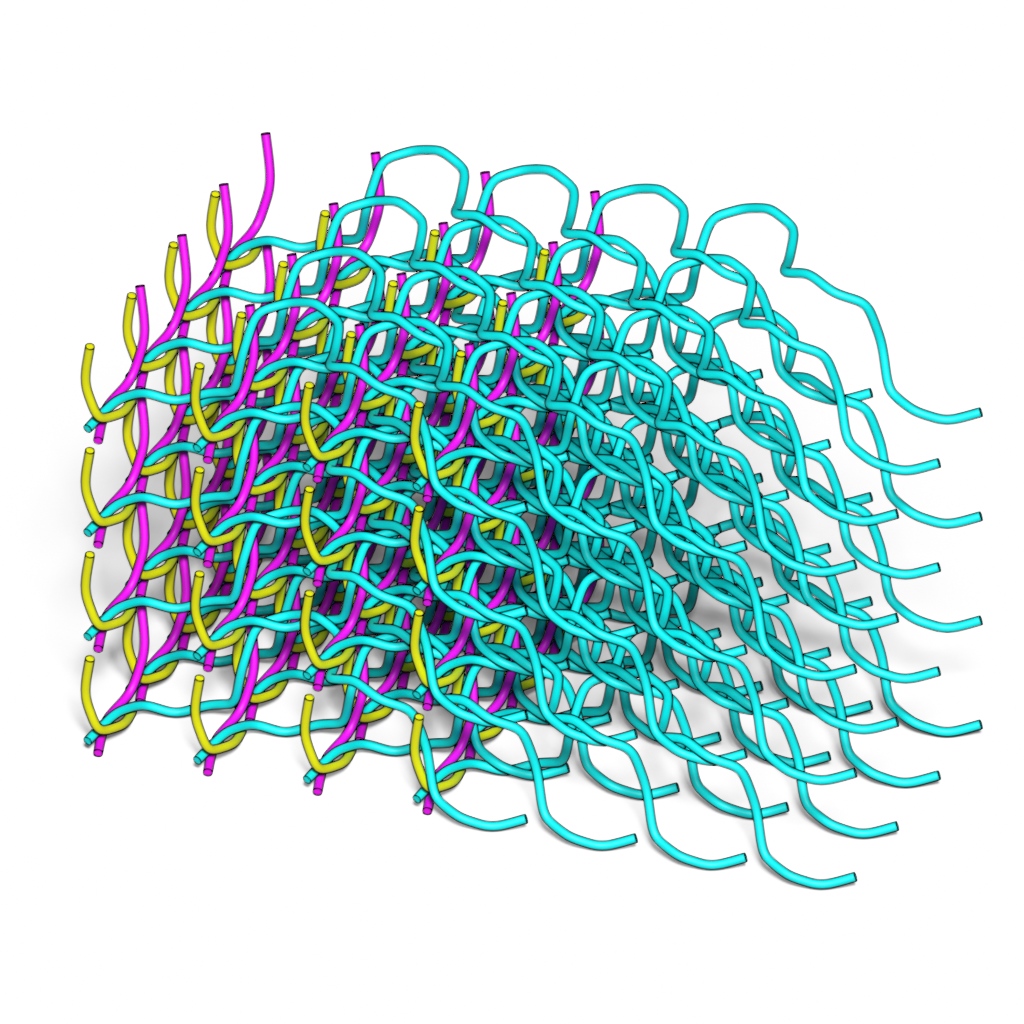}\\
    \parbox[t]{0.24\linewidth}{\centering Twisted unit}
    \parbox[t]{0.24\linewidth}{\centering Tiled twisted unit}
    \parbox[t]{0.24\linewidth}{\centering Unit thread}
    \parbox[t]{0.24\linewidth}{\centering Tiled thread}
    \caption{Different twist numbers enable control of the number of unique components in the repeating structure, the top row shows an example where there are two distinct elements with $1\times2\times1$ repeating domain. Bottom row shows an example with three distinct elements with $1\times4\times1$ repeating domain.}
    \Description{A figure with two rows and four columns, with column headings at the bottom reading “Twisted unit”, “Tiled twisted unit”, “Unit thread”, and “Tiled thread”. In each row, the first column shows a local twisting module rendered as thick tubular strands, labeled with a bracketed vector; the top module uses two strand colors (cyan and red), and the bottom module uses three strand colors (cyan, magenta, and yellow). The second column shows a periodic block formed by tiling the corresponding module, producing a dense three-dimensional lattice in the same set of colors. The third column shows one representative strand path inside a transparent rectangular bounding box, and the fourth column shows a larger structure formed by repeating that strand across space. The top row displays two interleaved strand families, while the bottom row displays three interleaved strand families, with visibly different repetition patterns and densities.}
    \label{fig:cp_multiple}
\end{figure}

Figures~\ref{fig:bravais2D_repeated_nonuniform}, ~\ref{fig:cp_single}, and ~\ref{fig:cp_multiple} illustrate periodic LK structures generated by assigning \emph{different twist labels to distinct edge classes} within a single repeating unit. In contrast to the uniform-twist cases shown earlier, not all edges are twisted in the same manner, even though the underlying periodic scaffold remains unchanged. This heterogeneous edge labeling immediately and significantly expands the design space: varying twist values locally within one Wigner-Seitz cell is sufficient to produce qualitatively new global interlacing patterns and connectivity types. As these examples demonstrate, once uniformity is relaxed, the space of achievable periodic LK structures grows rapidly, enabling the exploration of a large family of previously unexplored configurations even for a single periodic unit.

Figure~\ref{fig:cp_single} demonstrates another important design capability of the proposed framework: the ability to generate periodic LK structures from a \emph{single congruent thread}. In this example, one thread component is designed such that its rotated and translated copies under the symmetry operations of the periodic scaffold collectively tile the entire space. Depending on the chosen edge-twist assignments, the resulting threads may form either unbounded trajectories that extend infinitely through the lattice or closed cycles that remain spatially bounded. This illustrates that global coverage, connectivity, and closure properties can be controlled at the level of a single thread design, highlighting the expressive power of the framework in producing both infinite and cyclic periodic structures from minimal geometric input.

These examples are intended to illustrate the range of structures enabled by the framework, rather than to exhaustively enumerate periodic LK forms. All examples presented in this section are constructed from a single Wigner-Seitz cell serving as the periodic repeating unit. Even under this strong restriction, the framework already supports a wide variety of LK structures through uniform and non-uniform edge-twist assignments. More generally, the same construction naturally extends to two- and three-dimensional arrays of Wigner-Seitz cells with differing twist labels across cells. In two dimensions, such arrangements recover a large family of classical woven patterns and knotted configurations, while also extending beyond them. 

The design potential becomes even more pronounced in three dimensions. As shown in prior work on volumetric weaves, the number of distinct plain-weave–like configurations grows exponentially with the size of the repeating array \cite{yildiz2024volumetric}, in sharp contrast to the two-dimensional case, classical textile theory admits only a single plain weave in 2D \cite{Grunbaum80}. Although we do not enumerate such large-scale constructions here, the examples in this section make clear that the proposed framework provides the necessary combinatorial and topological control to generate them. Taken together, these observations indicate that even simple local design rules, when combined with periodic non-manifold scaffolds and edge-based twist labeling, give rise to an extraordinarily rich design space for both planar and volumetric LK structures.

Figure~\ref{fig:cp_multiple} shows that control over the number of distinct thread components can be achieved even within the \emph{smallest nontrivial repeating domain}, consisting of a $1\times2\times1$ array of Wigner-Seitz cells. Although the underlying periodic scaffold is expanded only minimally, different edge-twist assignments within this domain lead to markedly different global decompositions of the structure into thread components. In particular, threads may merge, split, or remain distinct depending on the chosen twist configuration. This demonstrates that fine-grained control over connectivity does not require large repeating arrays: even the smallest extended periodic unit provides sufficient degrees of freedom to regulate the number of thread components, further highlighting the expressive power of the framework.

The examples in the previous subsections illustrate how substantial variation in periodic LK structures can be achieved through local edge-twist assignments, even when the repeating domain is minimal. Importantly, these constructions all rely on a fixed underlying periodic scaffold. This observation suggests a complementary and equally simple way to further expand the design space: varying the periodic scaffold itself while preserving its fundamental lattice structure.

\begin{figure}[htb!]
    \centering
    \includegraphics[width=0.30\linewidth]{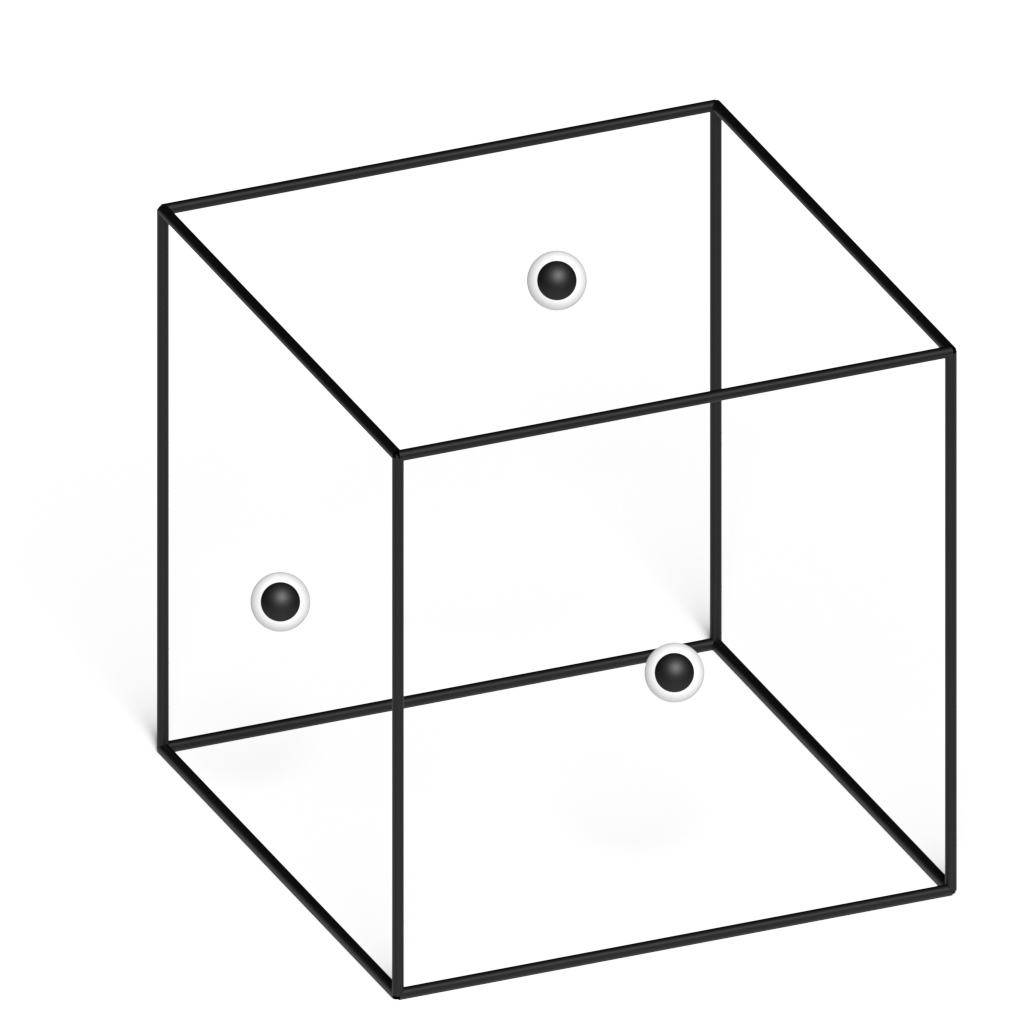}\hfill
    \includegraphics[width=0.30\linewidth]{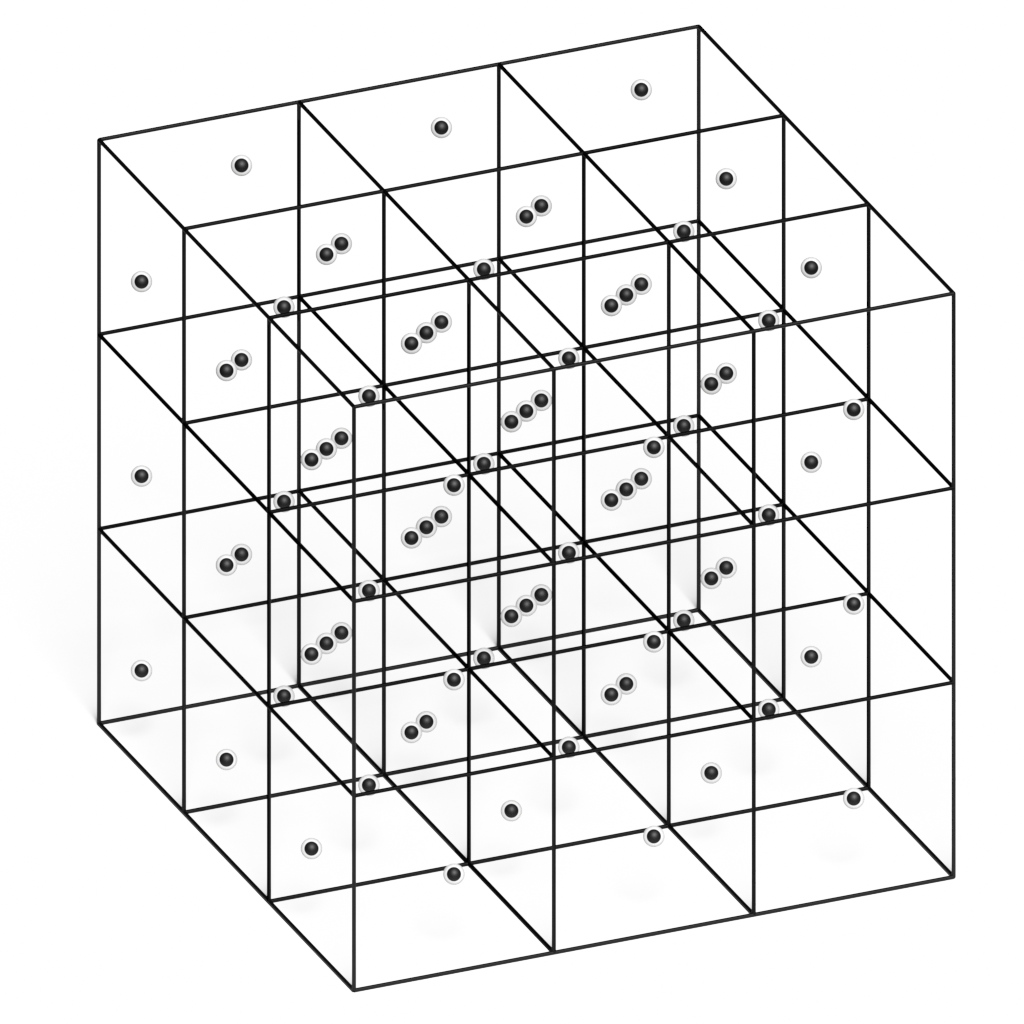}\hfill
    \includegraphics[width=0.30\linewidth]{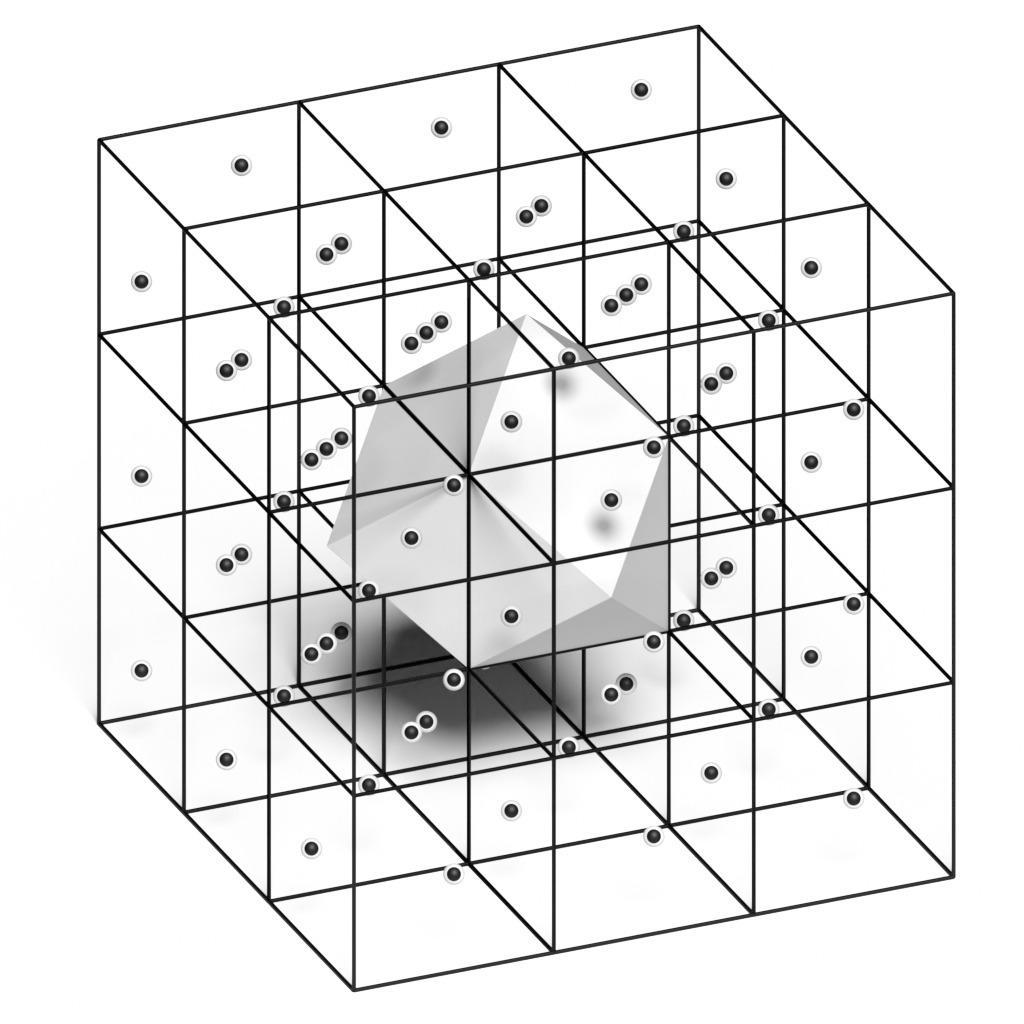}\\
    \includegraphics[width=0.30\linewidth]{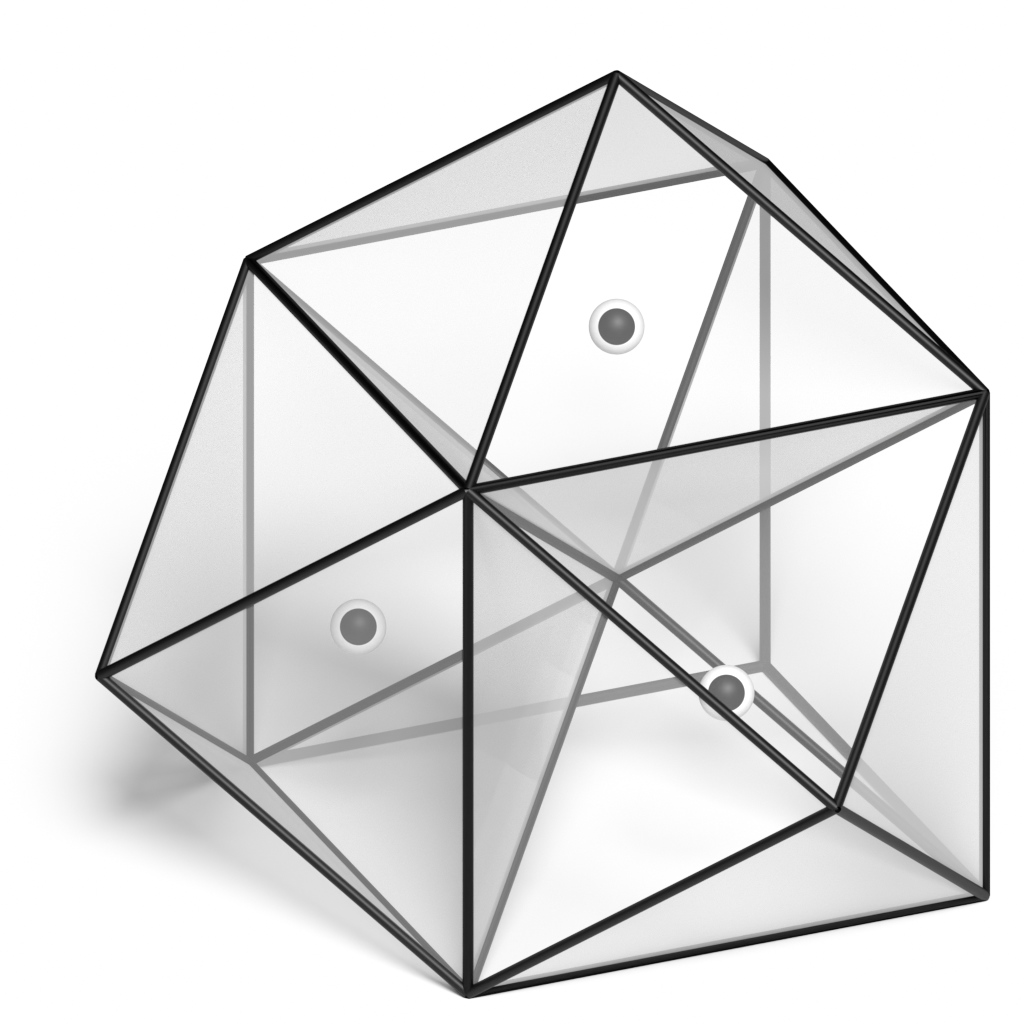}\hfill
    \includegraphics[width=0.30\linewidth]{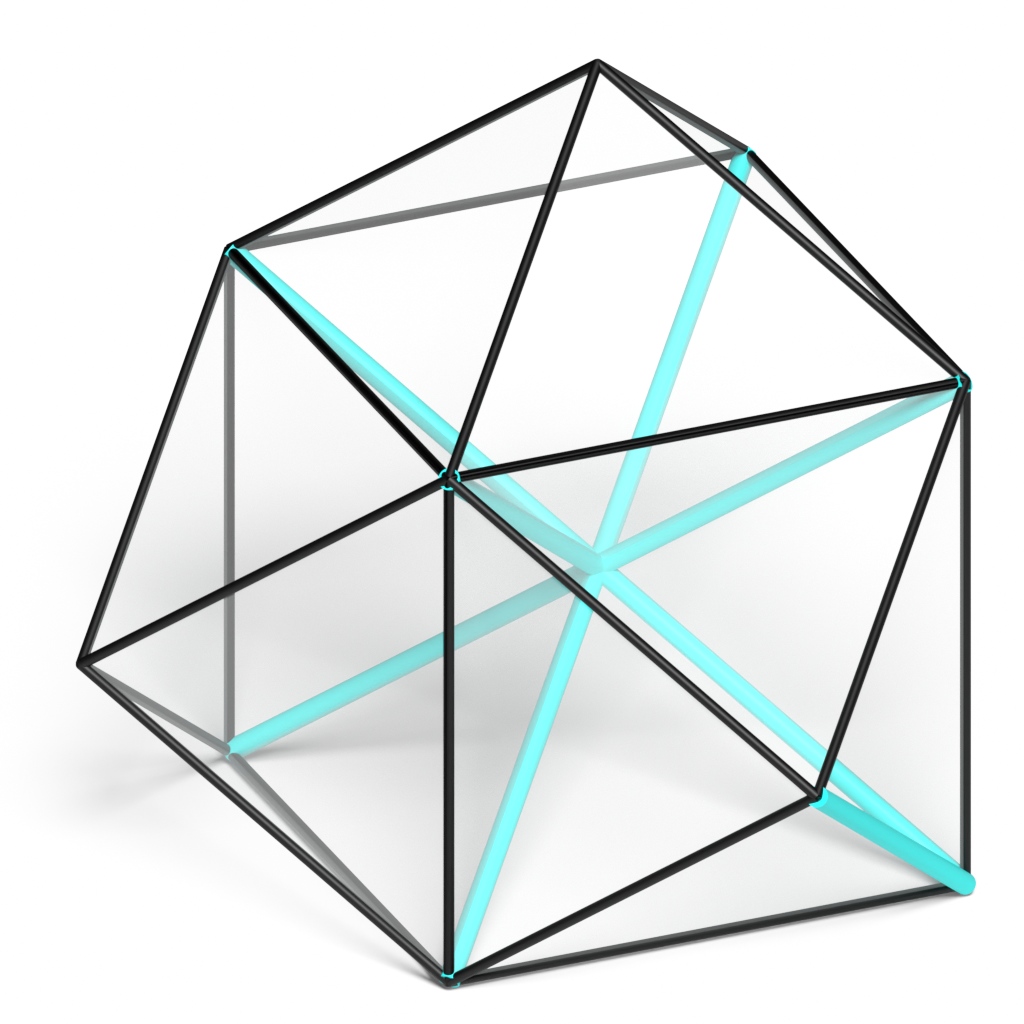}\hfill
    \includegraphics[width=0.30\linewidth]{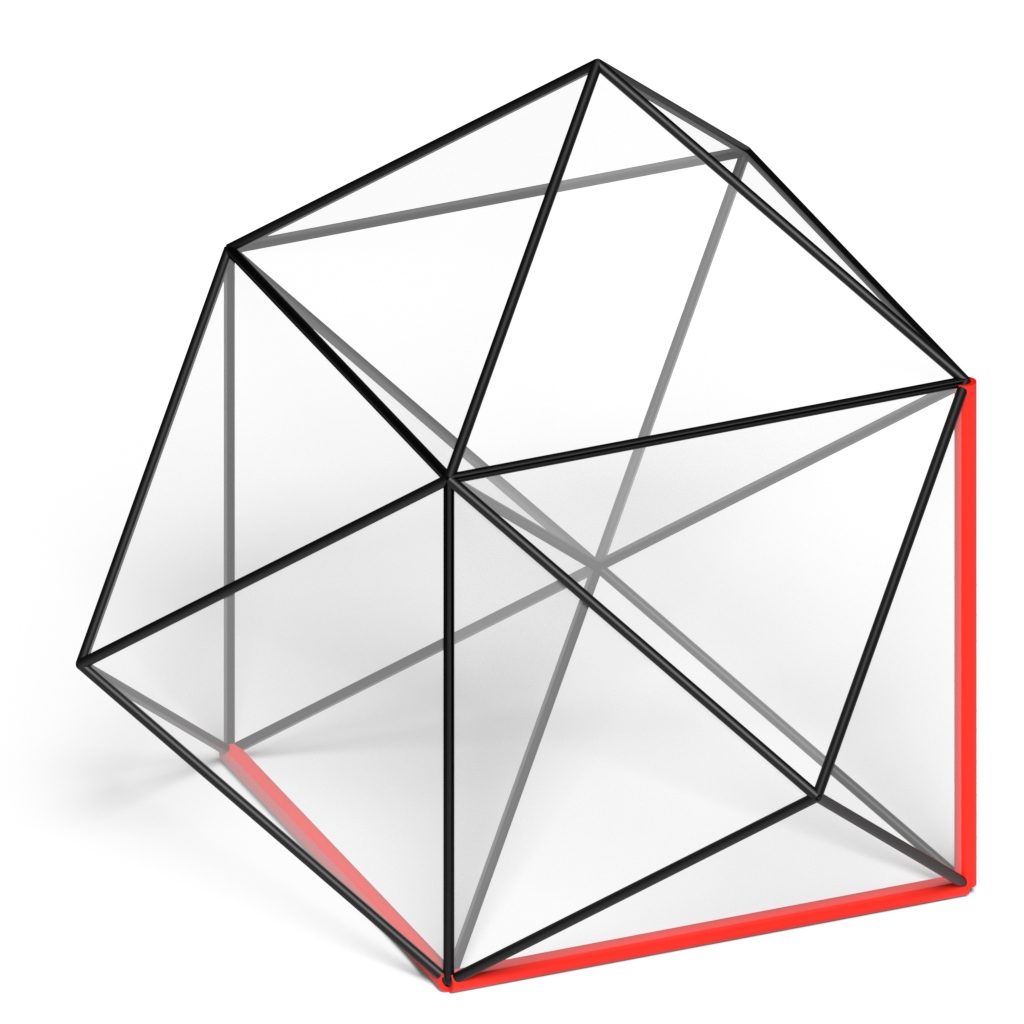}\\
    \includegraphics[width=0.30\linewidth]{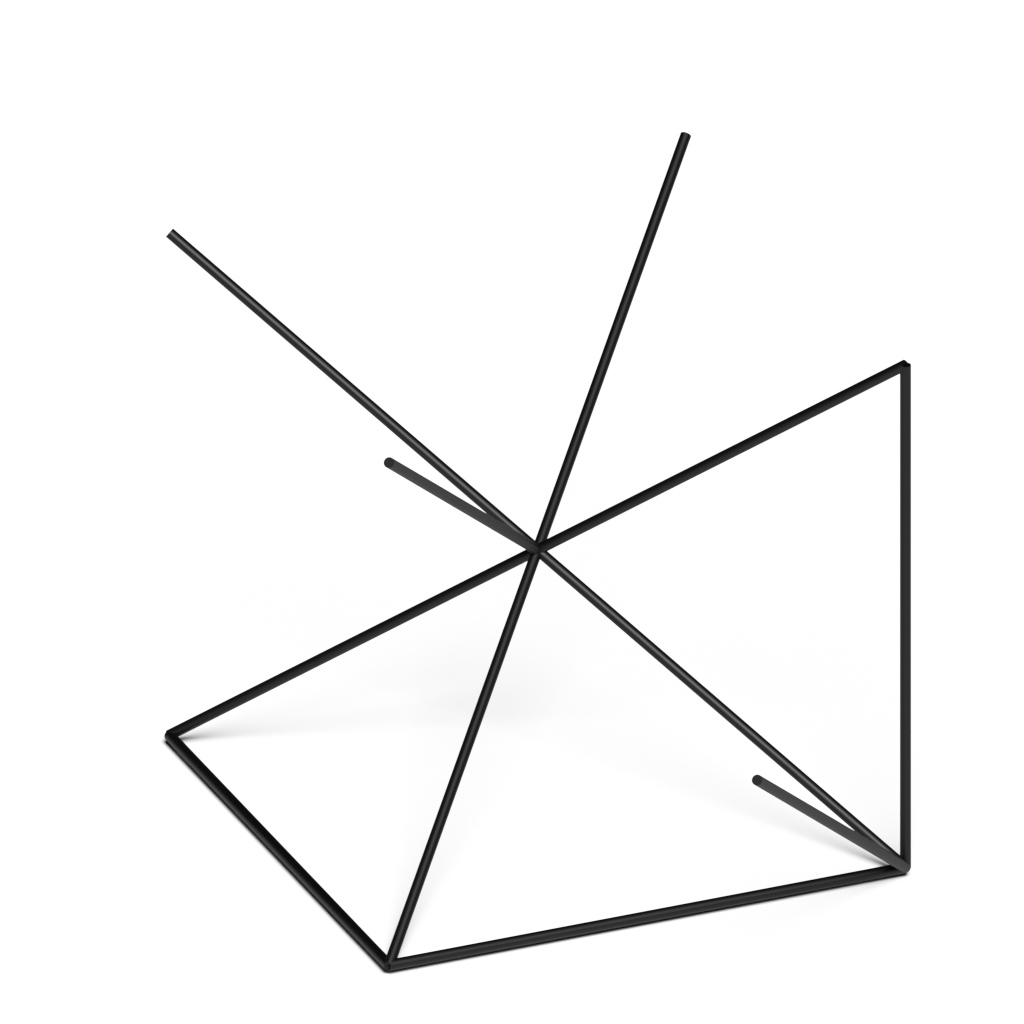}\hfill
    \includegraphics[width=0.30\linewidth]{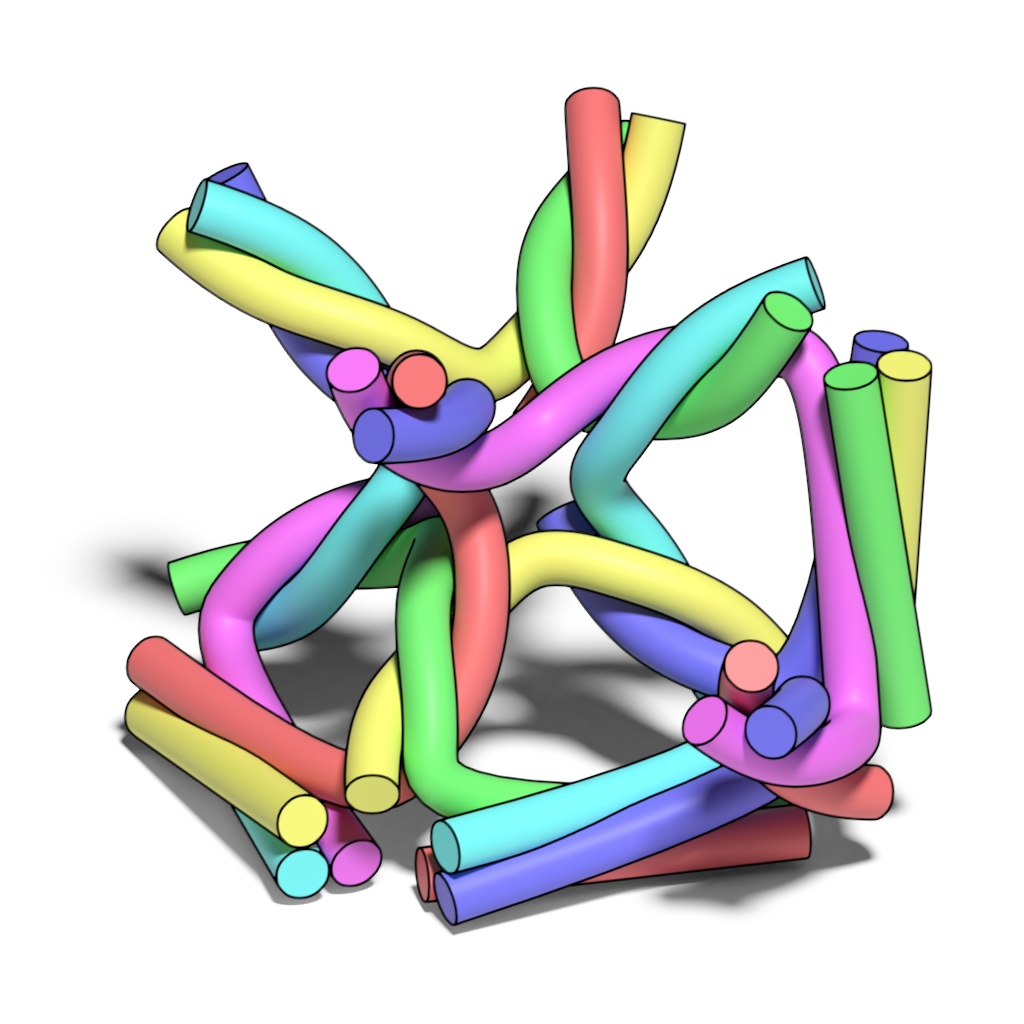}\hfill
    \includegraphics[width=0.30\linewidth]{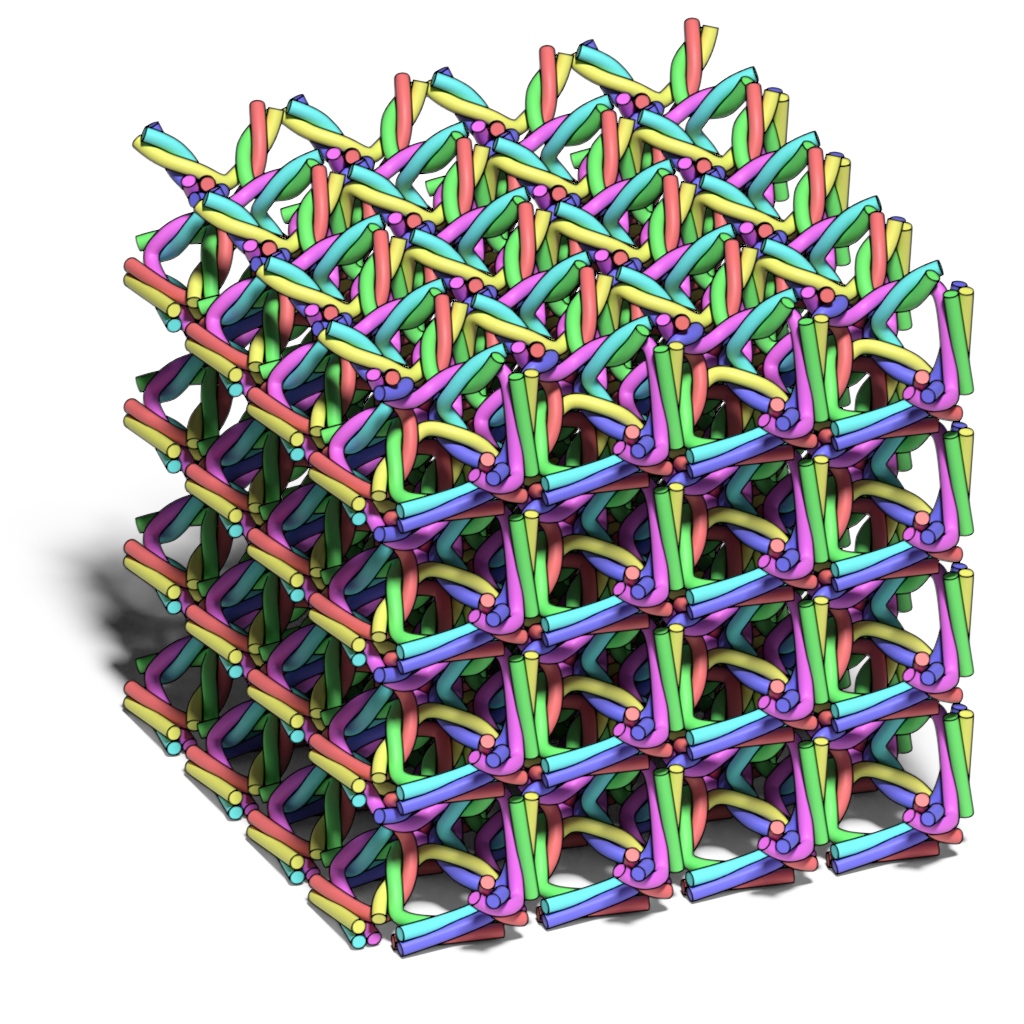}
    \caption{Arbitrary periodic cell structures generated using a repeating Voronoi domain. By placing points at the face centers of a cube, we obtain a cell structure distinct from the standard Wigner–Seitz cell. The top row illustrates the cell construction process. The middle row shows the resulting periodic structure and highlights edges with distinct valencies. These unique edges define a minimal repeating unit, which is then used to generate the twisting elements shown in the bottom row.}
    \Description{A multi-panel figure arranged in three rows. The top row shows a sequence of wireframe cubes illustrating the construction of a periodic cell: first a single cube with marked interior points, then a larger grid of repeated cubes with points at corresponding locations, and finally a shaded polyhedral cell embedded within the grid. The middle row shows a polyhedral structure with triangular facets rendered semi-transparently; in the center image, several edges are highlighted in cyan, and in the right image, a different set of edges is highlighted in red. The bottom row shows, from left to right, a planar graph-like outline, a compact cluster of thick multicolored tubular strands forming a twisting element, and a larger three-dimensional periodic assembly formed by repeating that element into a dense lattice.}
    \label{fig:voronoi_example}
\end{figure}

\subsection{Extending 3-Honeycombs with Generalization of Wigner-Seitz construction}

We now generalize the Wigner-Seitz construction to allow arbitrary periodic cell structures defined by multiple generators within a fundamental domain (see Figure~\ref{fig:voronoi_example} for an example). Rather than restricting the Voronoi diagram to a single lattice point, we consider a finite set of points embedded in a repeating lattice domain and use their induced Voronoi partition as a periodic scaffold for LK structures.

Let $\mathcal{L}(B) \subset \mathbb{R}^N$ be a full-rank Bravais lattice generated by a basis
\[
B := \{\vec{v}_0, \vec{v}_1, \ldots, \vec{v}_{N-1}\}.
\]
Let
\[
P := \{\vec{p}_0, \vec{p}_1, \ldots, \vec{p}_{M-1}\} \subset \mathbb{R}^N
\]
be a finite set of generator points contained within a single fundamental domain of $\mathcal{L}(B)$. We define the periodic point set
\[
\mathcal{P} = \bigcup_{\vec{\ell} \in \mathcal{L}(B)} \left( P + \vec{\ell} \right),
\]
and construct the Voronoi diagram induced by $\mathcal{P}$.

The resulting Voronoi cells form a periodic tiling of $\mathbb{R}^N$ whose combinatorial structure depends on both the lattice geometry and the placement of points in $P$. In contrast to Wigner-Seitz cells, which are uniquely determined by the lattice, this construction admits a continuous family of topologically distinct periodic cells, enabling the generation of non-standard cell types beyond classical crystallographic templates.

As in the Wigner-Seitz case, periodicity allows the global structure to be fully characterized by a finite subset of cells and their adjacencies. We analyze the induced cell complex within a single repeating domain and classify its edges by valency and translation equivalence. This classification yields a finite set of unique edges that define a minimal repeating scaffold under lattice symmetries.

These unique edges serve as carriers for LK twist labels. By assigning integer twists to the equivalence classes of edges, we generate twisting elements that extend consistently across the periodic domain. The resulting LK structures inherit the periodicity of the underlying Voronoi construction while exhibiting richer topological variability than lattice-derived Wigner-Seitz scaffolds. Figure~\ref{fig:voronoi_example} illustrates this process for a three-point construction on the faces of a cubic domain.

Taken together, the constructions in this section demonstrate that a remarkably large design space of periodic LK structures can be accessed through a small set of simple and local choices. Variations in edge-twist assignments, minimal extensions of the repeating domain, and straightforward modifications of the periodic scaffold each lead to substantial changes in global connectivity and structure. Importantly, these variations do not require additional geometric complexity or specialized constructions; they arise naturally from the underlying lattice-based framework. This observation motivates a broader discussion of the implications, limitations, and potential applications of such design freedom, which we address in the following section.

\section{Discussion}
\label{sec:discussion}

A key strength of the proposed representation lies in its expressive design capability rather than in classification. In classical graph-based approaches, a single abstract graph can give rise to multiple distinct knot or link embeddings, making the construction process inherently one-to-many. In contrast, our framework admits the opposite situation as well: different labeled non-manifold surface meshes may produce the same LK structure. This many-to-one behavior indicates that the representation is not suitable for direct topological classification. Achieving a one-to-one correspondence would require additional constraints or canonicalization strategies, and it remains an open question whether such restrictions can be imposed without significantly reducing the available design space.

\subsection{Limitations of Represetation}

\begin{figure}[htb!]
    \centering
    \begin{subfigure}[b]{0.46\columnwidth}  
    \includegraphics[width=0.46\textwidth]{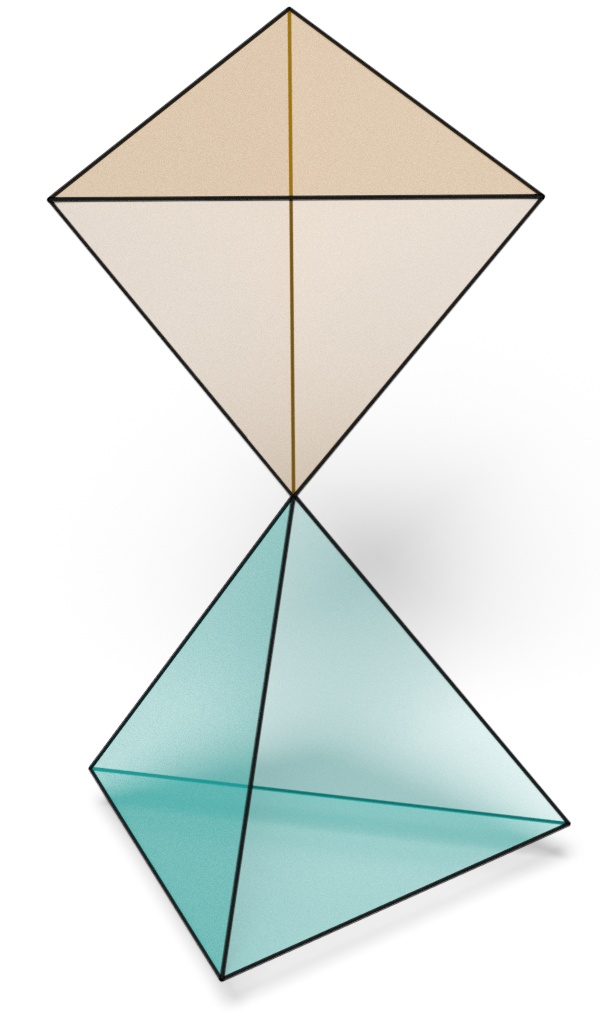}    
    \includegraphics[width=0.46\textwidth]{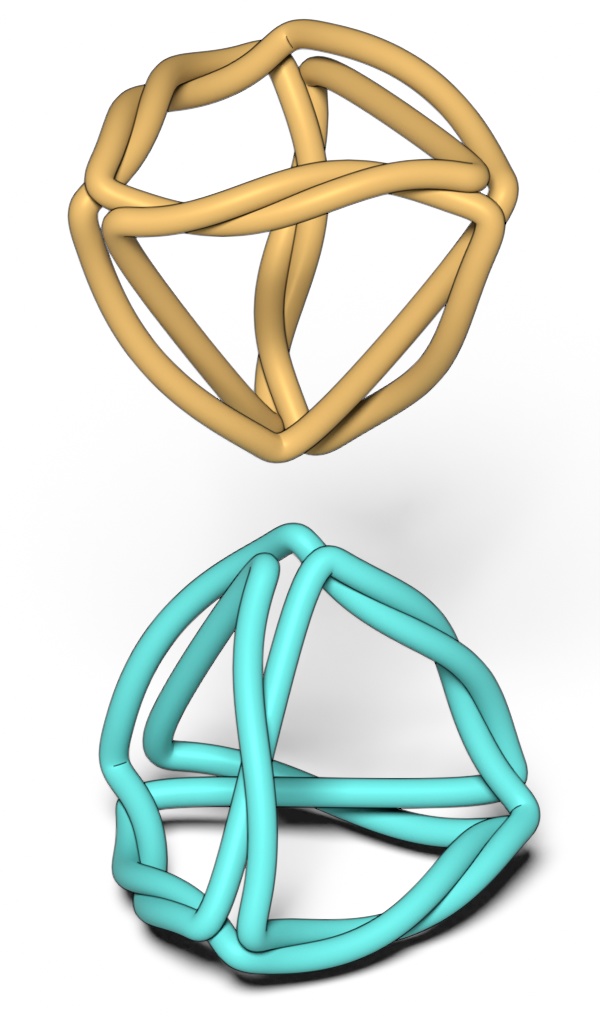}
        \caption{Vertex hinged Non-Manifold mesh and corresponding LKs.}
        \Description{Text.}
        \label{hinge0}
    \end{subfigure}
    \hfill
        \begin{subfigure}[b]{0.46\columnwidth}
        \includegraphics[width=0.46\textwidth]{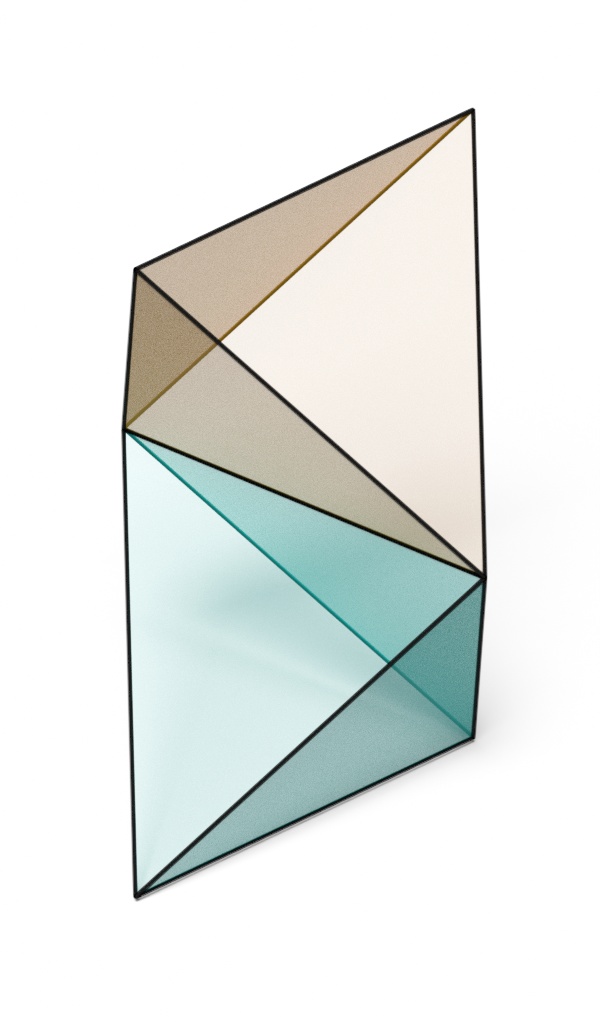}
        \includegraphics[width=0.46\textwidth]{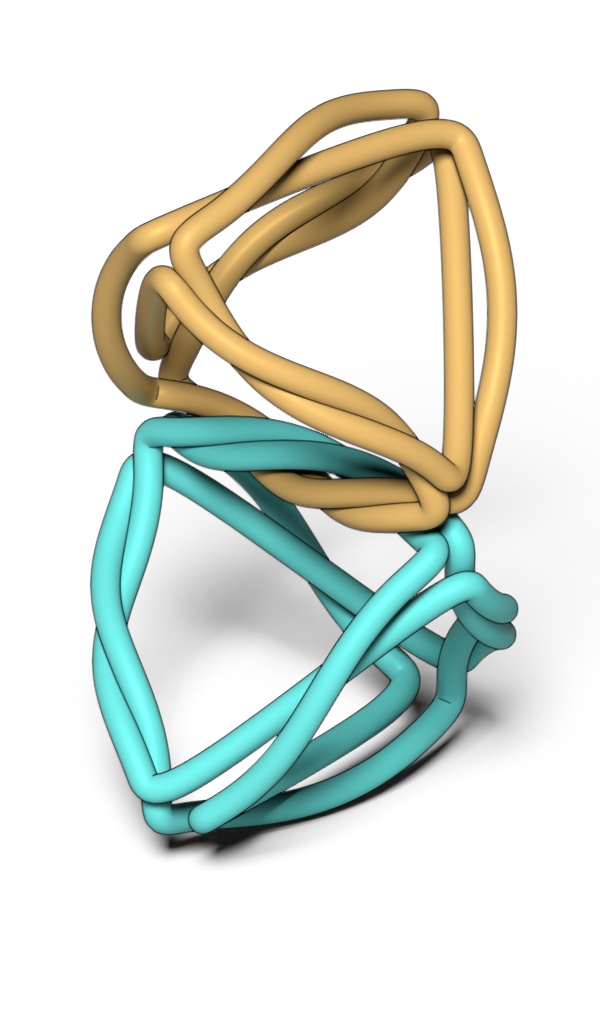}
        \caption{Edge (Piano) hinged mesh and corresponding LKs.}
        \label{hinge1}
    \end{subfigure}
    \hfill
  \caption{Our approach works on any non-manifold surface. However, if we want the final LK-structure to be connected, the parts of the initial mesh structure should not be only connected by a vertex hinge. If so, the parts will not be hung together.}
  \Description{Two side-by-side examples labeled (a) and (b), each consisting of a surface mesh shown above its corresponding linked tubular structure. In (a), the top image shows two polyhedral parts meeting at a single vertex, rendered as semi-transparent facets in beige and cyan. Below it, the corresponding LK structure appears as two separate clusters of tubular strands in matching colors that do not form a single continuous assembly. In (b), the top image shows two polyhedral parts connected along a shared edge, resembling a hinged configuration. Below it, the corresponding LK structure appears as an interlinked set of tubular strands in beige and cyan that form a single connected structure. All examples are shown against a white background at a comparable scale and viewpoint.}
  \label{fig:hinge_types}
\end{figure}

Figure~\ref{fig:hinge_types} clarifies a limitation of non-manifold representation in terms of connectivity of the final LK structure. When multiple 2-manifold mesh components are combined into a non-manifold scaffold, the resulting LK structure may become either connected or disconnected depending on \emph{how} the components are attached. In particular, attaching components only through a shared vertex (i.e., forming a vertex-connected non-manifold) is not sufficient to produce a connected LK structure: such a scaffold cannot induce a connected set of threads under edge twisting, and the output necessarily decomposes into multiple components. In contrast, when components are attached along edges (or more generally, through higher-dimensional adjacency that merges edge neighborhoods), the twisting operation can propagate connectivity between parts, yielding a connected LK structure. This figure, therefore, emphasizes that connectivity in the scaffold must be established at the level of edge neighborhoods, not merely at isolated vertices, if a connected LK structure is desired.

From a design perspective, a natural question concerns the expressive power of the representation: which classes of LK structures can, and cannot, be generated. While the framework supports a wide range of linked and knotted configurations, it does not capture all possible topological constructions. For example, general braid structures lie outside the scope of the method, as braids are fundamentally characterized by order-preserving combinatorial relations rather than by accumulated twist along edges. This limitation reflects a deliberate design choice, as the representation prioritizes twist-based control and spatial periodicity over purely combinatorial crossing descriptions.

Plaited or braided structures are primarily characterized by combinatorial crossing order~\cite{holden2016braids}, whereas twined or twisted structures require a consistent orientation and ordering along each strand. As a result, braided constructions can be fully specified by discrete over–under relationships and permutations, independent of any accumulated rotation along the strands. In contrast, the LK structures considered in this work rely on continuous twist defined along edges, where orientation and handedness are intrinsic and must be preserved globally. Consequently, while our framework naturally captures twined and twisted configurations, it does not aim to generate general braided structures, whose defining properties are combinatorial rather than twist-based.

Another limitation of the current framework concerns hierarchical constructions. In many physical systems, individual threads may themselves be composite structures, such as ropes formed by twisting smaller strands. Representing such multi-scale or hierarchical knotting would require recursive or nested twist descriptions, which are not supported by the present formulation. Extending the framework to incorporate hierarchical twist structures remains an interesting direction for future work, but would require additional abstraction layers beyond the single-level edge-based labeling considered here.

Despite these limitations, the representation exhibits substantial expressive strength. In particular, it is capable of generating a wide range of well-known knots and links that are not accessible using some existing modeling approaches, including methods that rely on rigid graph embeddings or constrained replacement rules. In Section~\ref{subsec:classical}, we demonstrate this capability by constructing several classical links and knots within the proposed framework. These examples serve not as a classification, but as evidence that the representation spans a rich and practically relevant subset of knotted and linked forms.

\subsection{Classical Knots and Links as Emergent Instances}
\label{subsec:classical}

\begin{figure}[htb!]
    \centering
    \begin{subfigure}[t]{0.23\columnwidth}  \includegraphics[width=\textwidth]{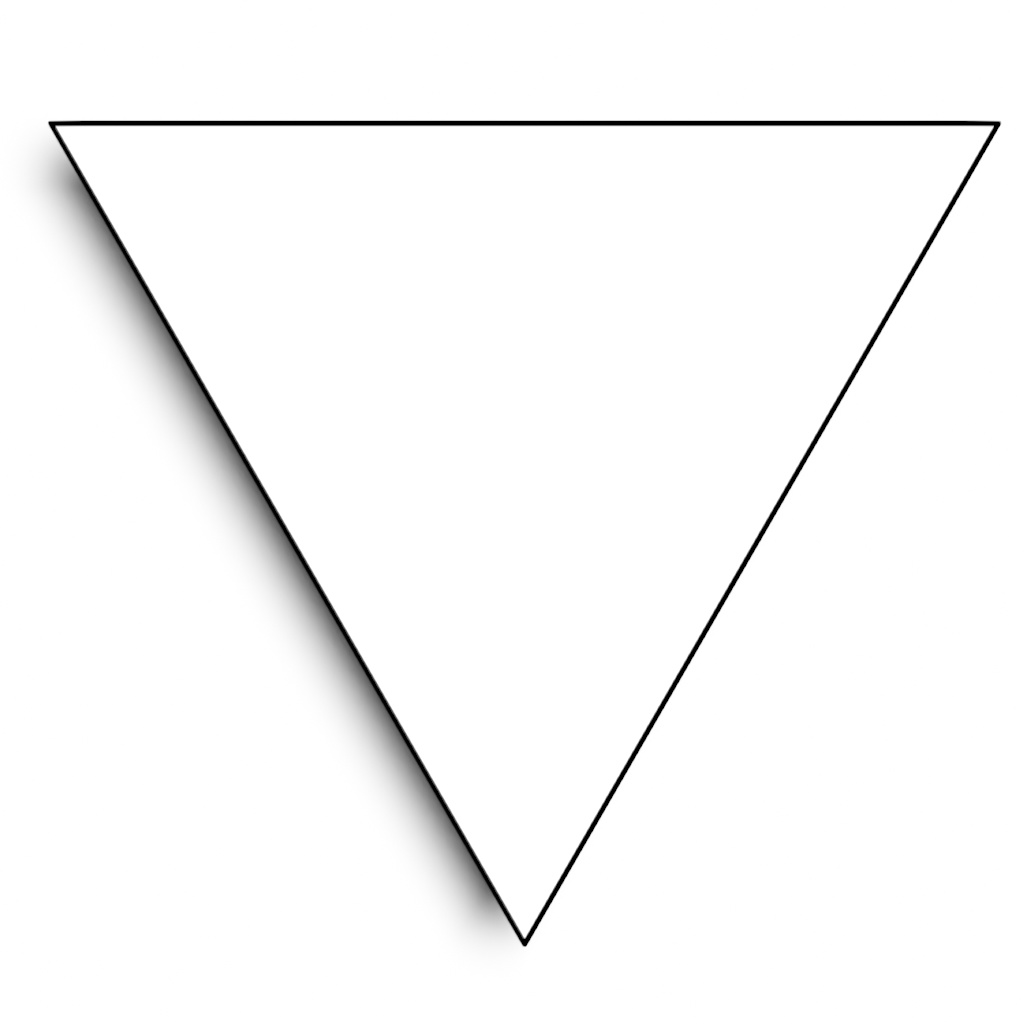}
    \includegraphics[width=\textwidth]{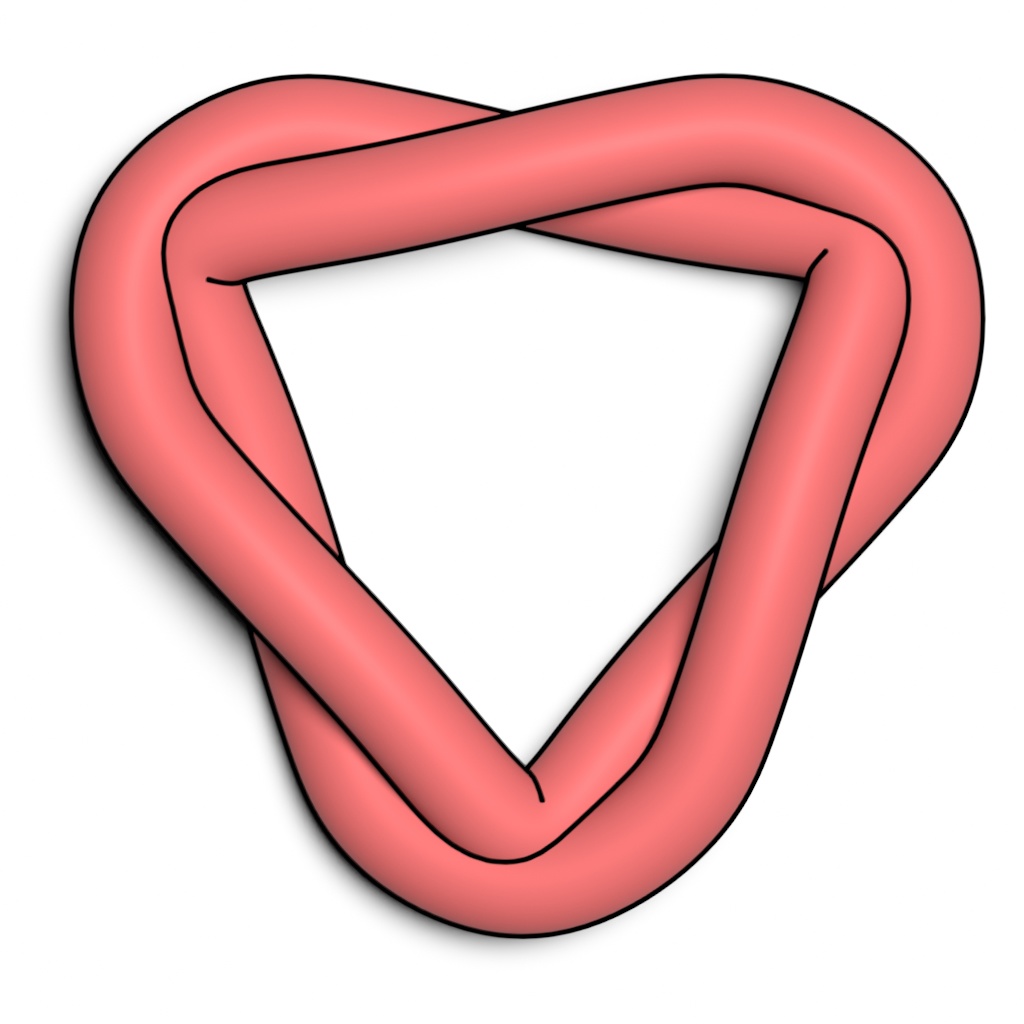}
        \caption{Trefoil knot ($3_1$).}
        \label{prime00}
    \end{subfigure}
    \hfill
    \begin{subfigure}[t]{0.23\columnwidth}       \includegraphics[width=\textwidth]{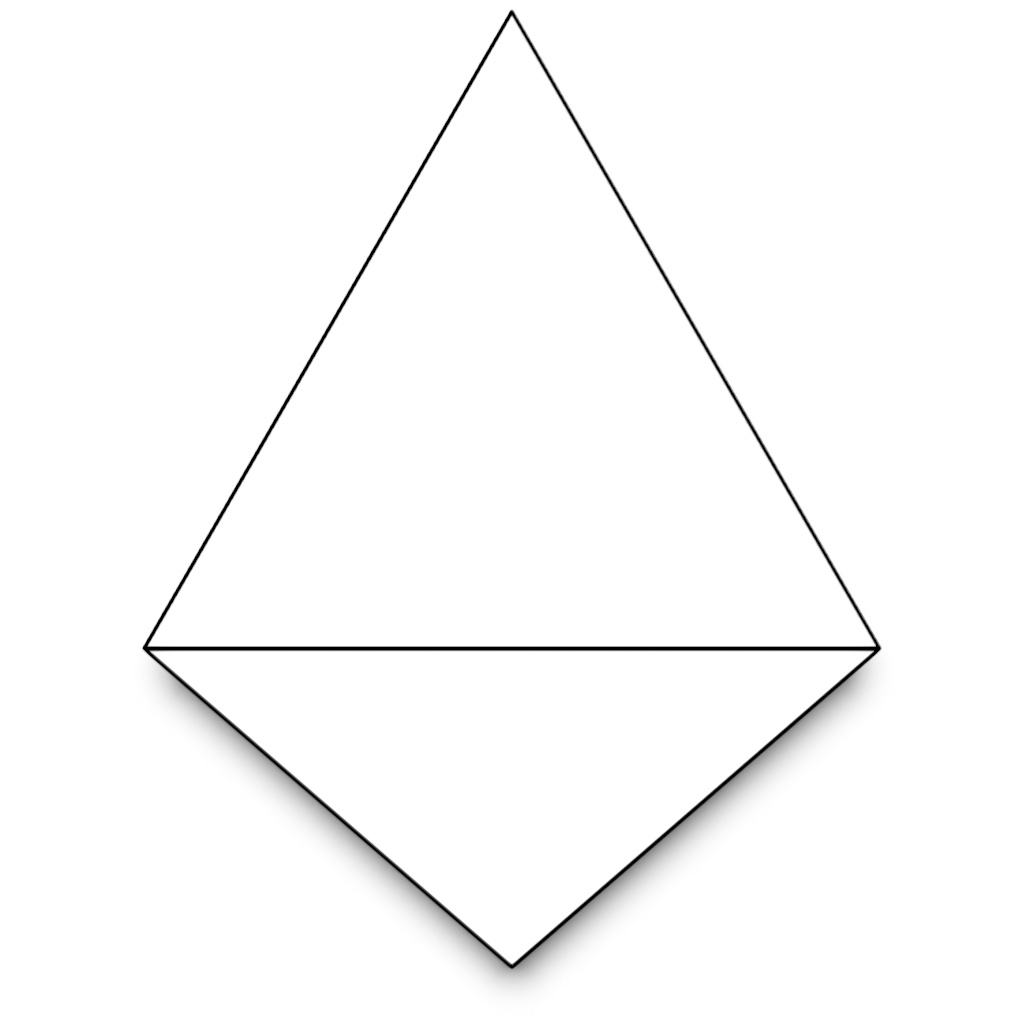}
        \includegraphics[width=\textwidth]{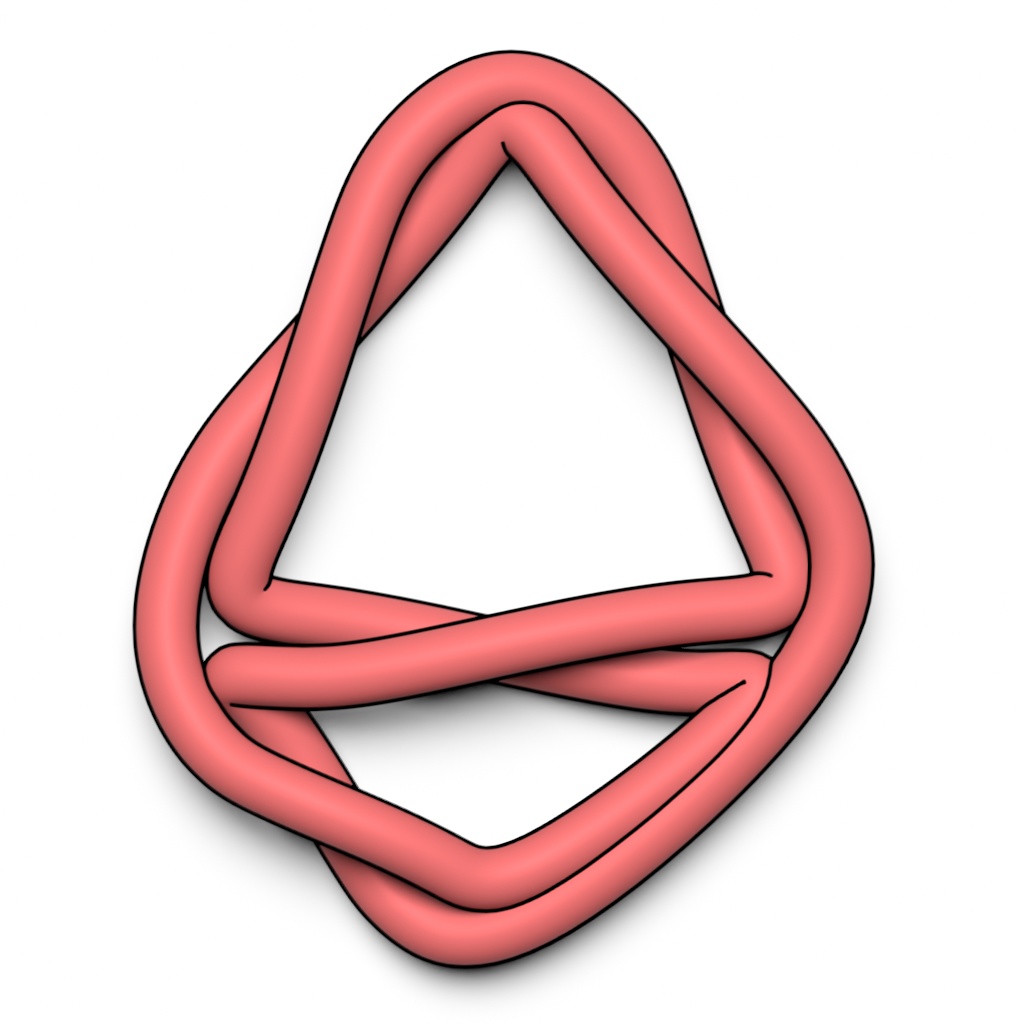}
        \caption{Figure 8 knot ($4_1$)}
        \label{prime01}
    \end{subfigure}
    \hfill
        \begin{subfigure}[t]{0.23\columnwidth}        \includegraphics[width=\textwidth]{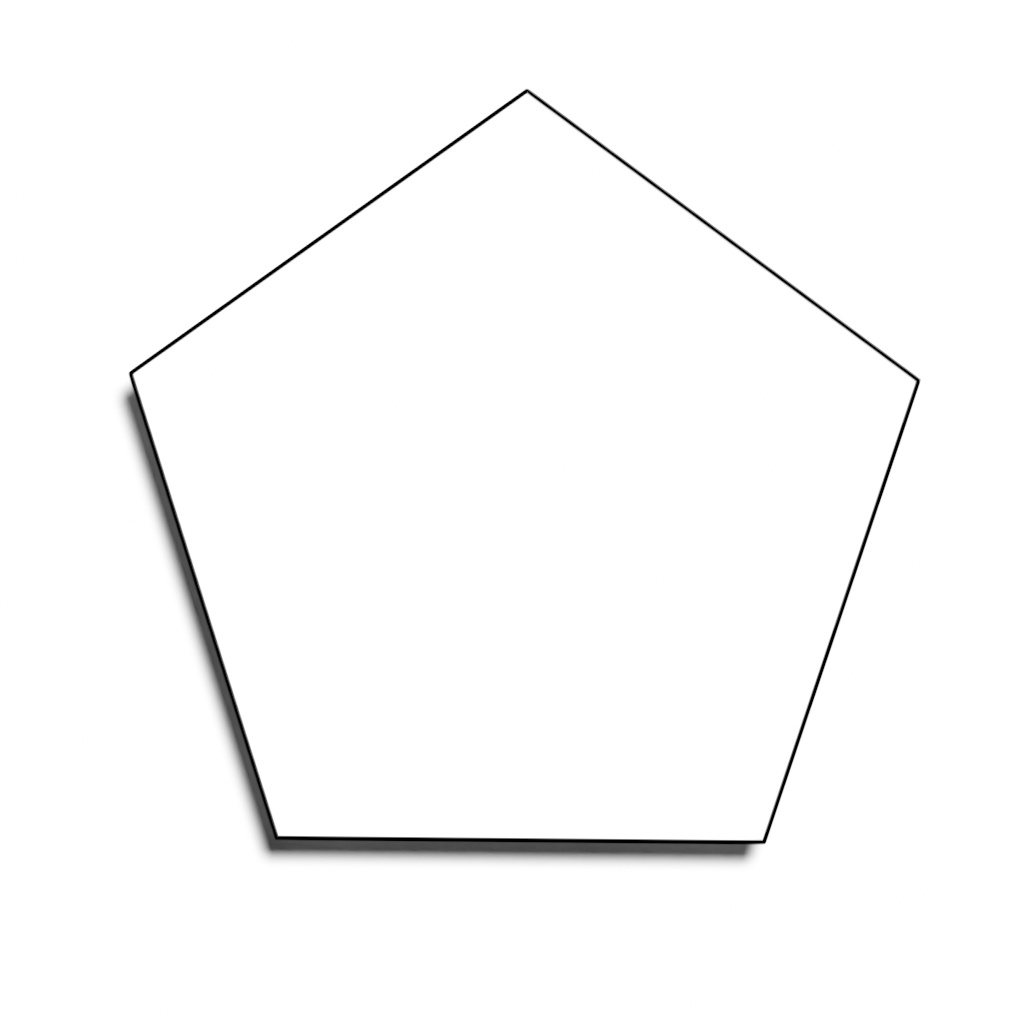}        \includegraphics[width=\textwidth]{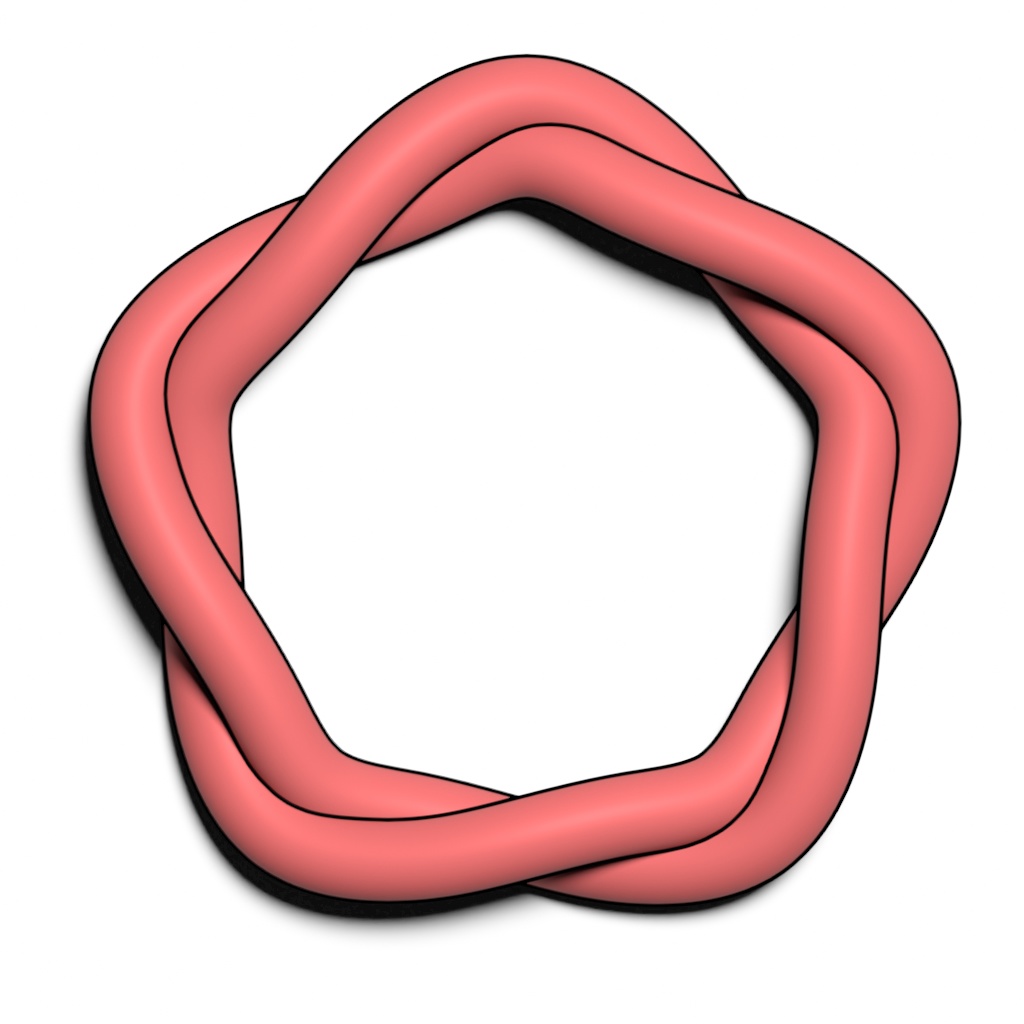}
        \caption{$5_1$ knot}
        \label{prime02}
    \end{subfigure}
    \hfill
    \begin{subfigure}[t]{0.23\columnwidth}        \includegraphics[width=\textwidth]{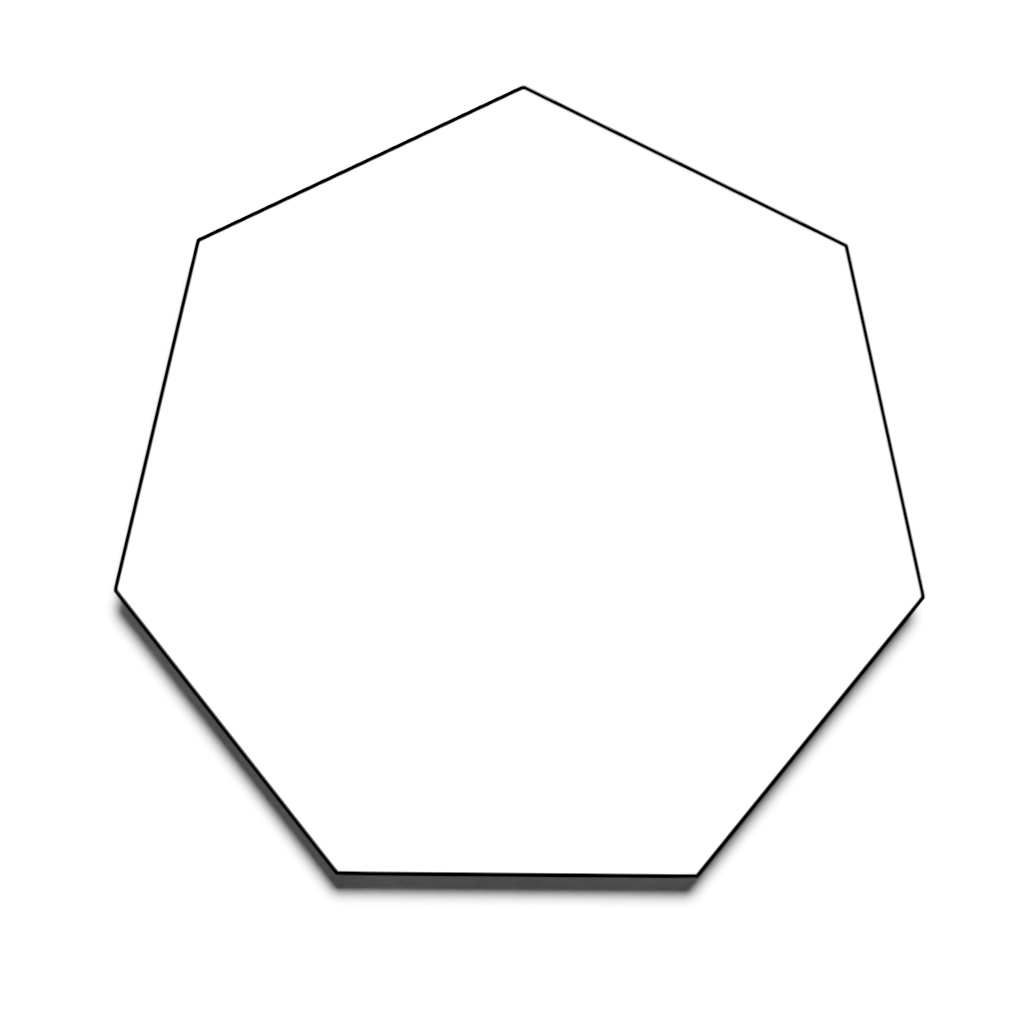}   
    \includegraphics[width=\textwidth]{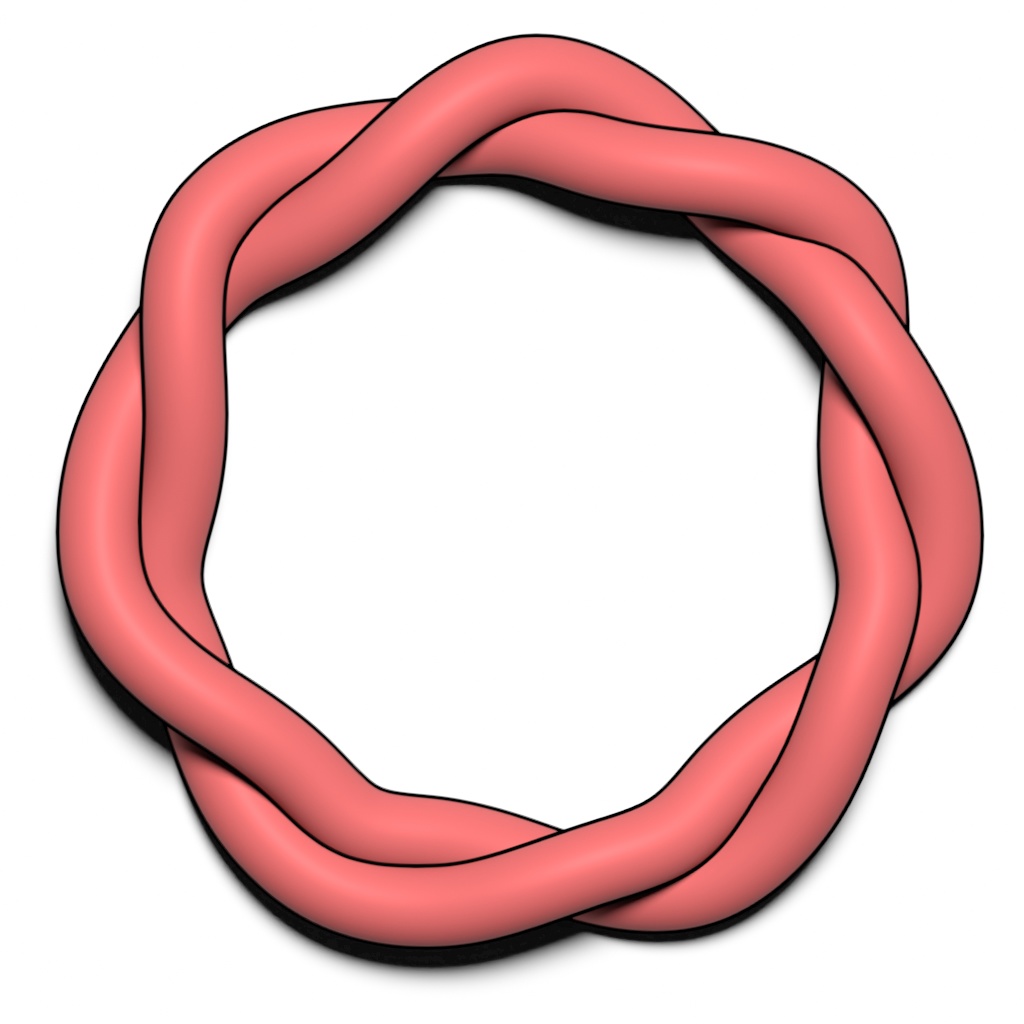}
        \caption{$7_1$ knot}
        \label{prime03}
    \end{subfigure}
    \hfill
  \caption{Examples of classical prime knots emerging from simple labeled surface meshes. }
  \Description{Four examples arranged in two rows. The top row shows simple planar polygonal shapes drawn as outlines: from left to right, a triangle, a pair of stacked triangles forming a diamond-like shape, a pentagon, and a heptagon. The bottom row shows corresponding closed tubular knot structures rendered in red. Each knot has a distinct overall shape and self-crossing pattern, with the strand looping around the interior of the polygonal form above it. The knots are labeled (a) through (d) and are shown at a similar scale and viewpoint for comparison.}
  \label{fig:prime_knots}
\end{figure}

In this part, we do not aim to catalog all knots, but to demonstrate thatsome well-known classical knot types arise naturally from simple labeled configurations. Figure~\ref{fig:prime_knots} provides examples of classical prime knots emerging from simple labeled surface meshes. These examples are not individually designed; they arise naturally from local twist-label assignments on low-complexity 2-manifold meshes, illustrating that a wide range of well-known knot types is contained within the design space of the framework. 

Three prime knots shown in Figures~\ref{prime00},~\ref{prime02}, and~\ref{prime03} are produced by twisting the edges of two-sided regular polygons with an odd number of face sides: a two-sided triangle, pentagon, and septagon. In contrast, the figure-eight knot is constructed from a 2-manifold in which one side consists of two triangles, as shown in Figure~\ref{prime01}, while the opposite side is a quadrilateral.

\begin{figure}[htb!]
    \centering
    \begin{subfigure}[t]{0.23\columnwidth}
    \includegraphics[width=\textwidth]{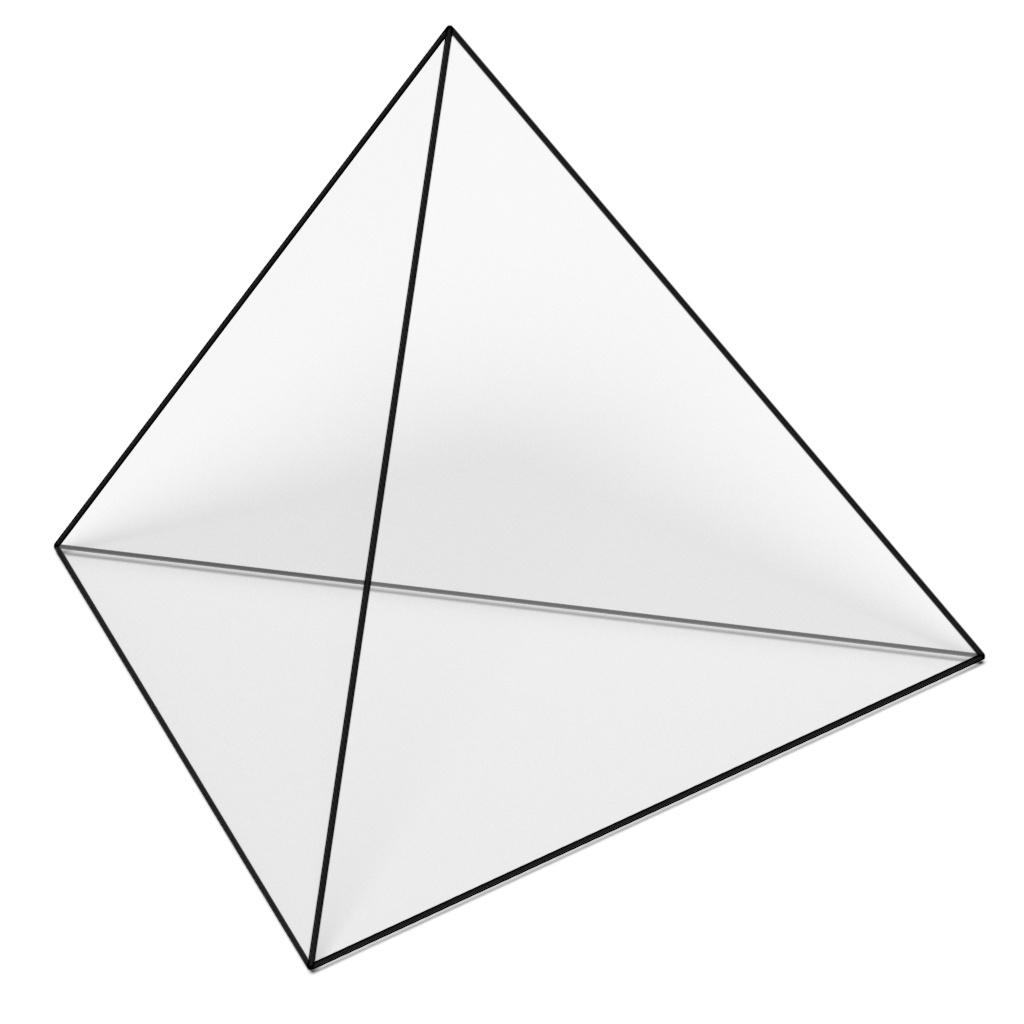}
        \includegraphics[width=\textwidth]{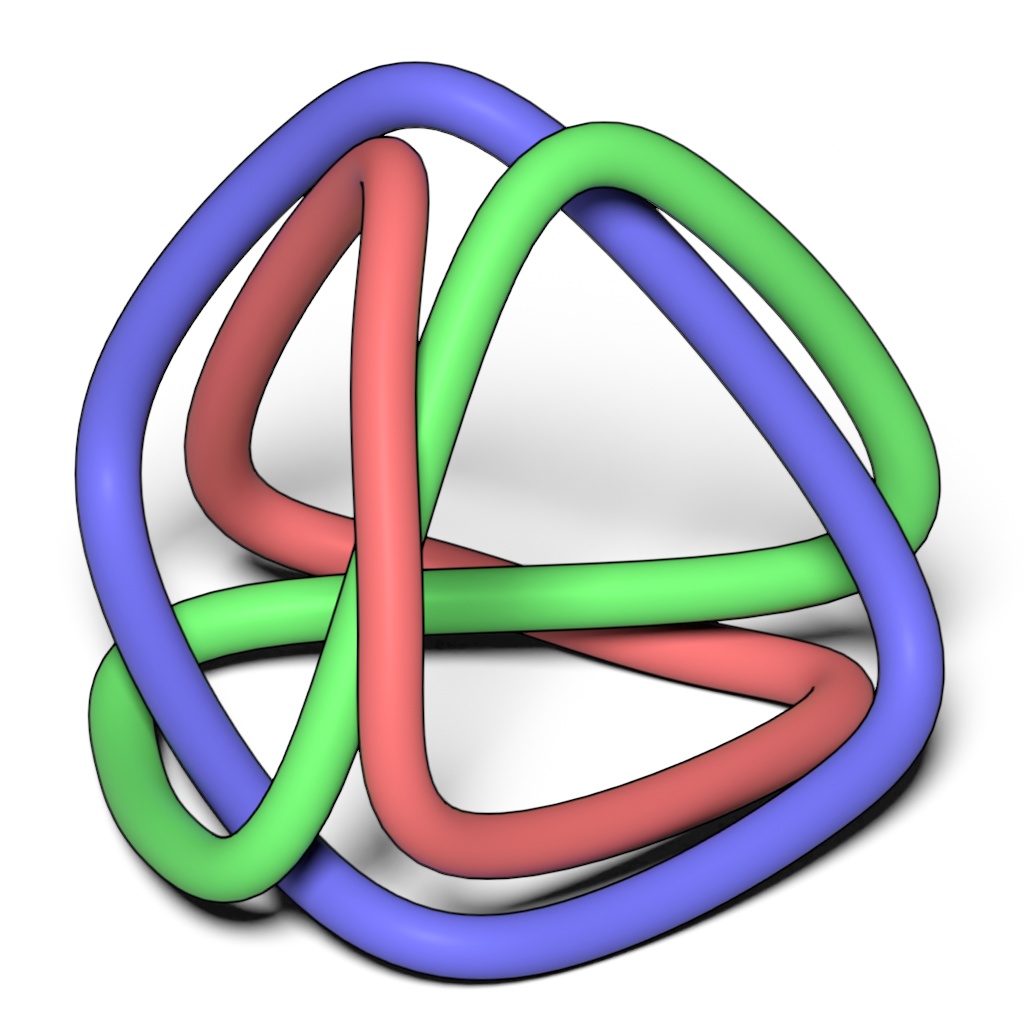}
        \caption{Borromean rings}
        \label{prime10}
    \end{subfigure}
    \hfill
    \begin{subfigure}[t]{0.23\columnwidth}
        \includegraphics[width=\textwidth]{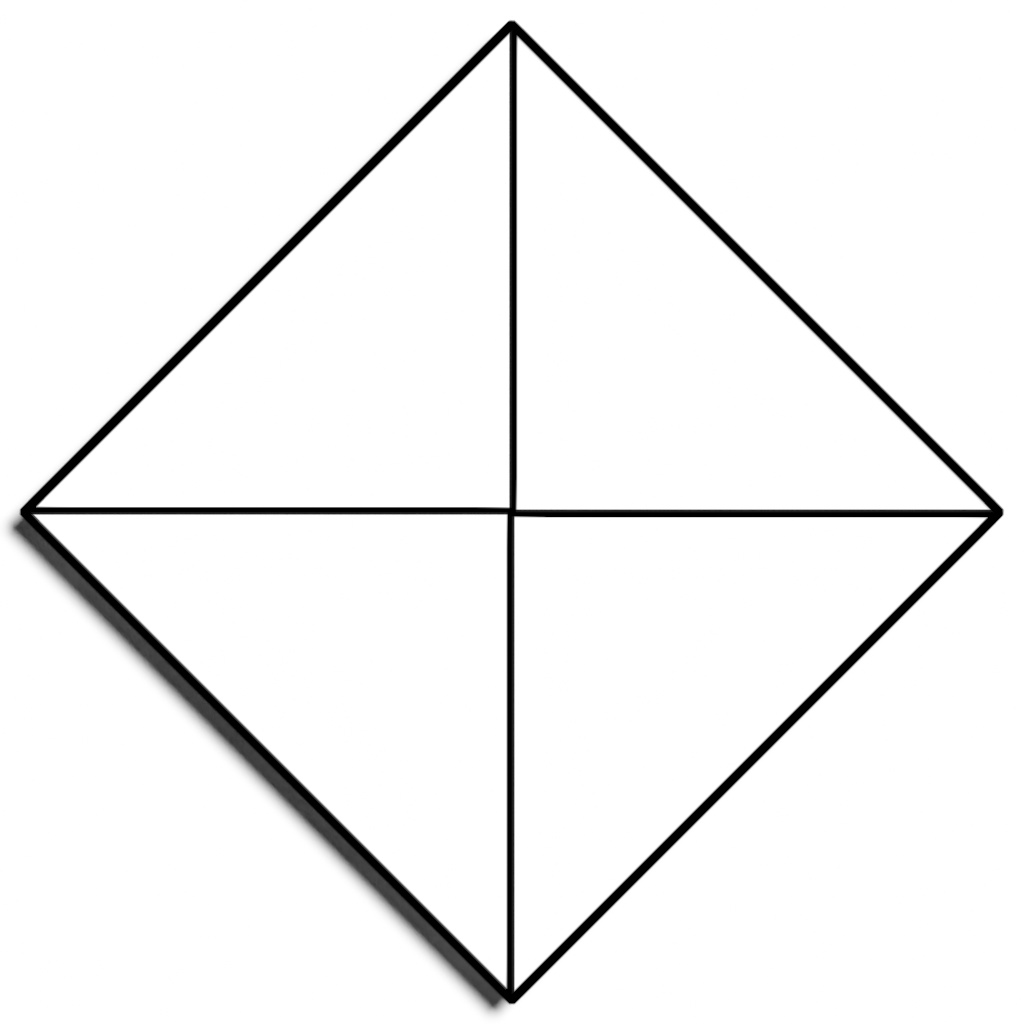}
        \includegraphics[width=\textwidth]{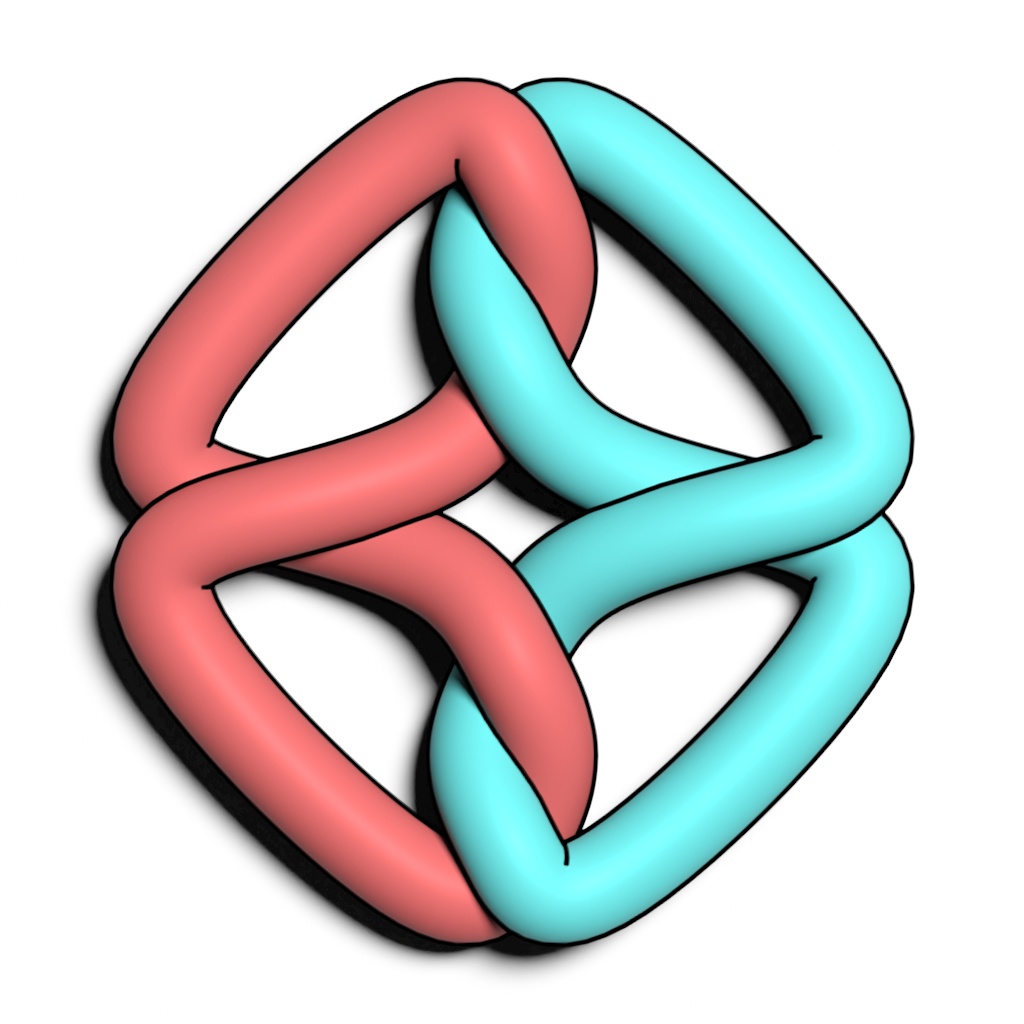}
        \caption{Whitehead link.}
        \label{prime11}
    \end{subfigure}
    \hfill
        \begin{subfigure}[t]{0.23\columnwidth}
        \includegraphics[width=\textwidth]{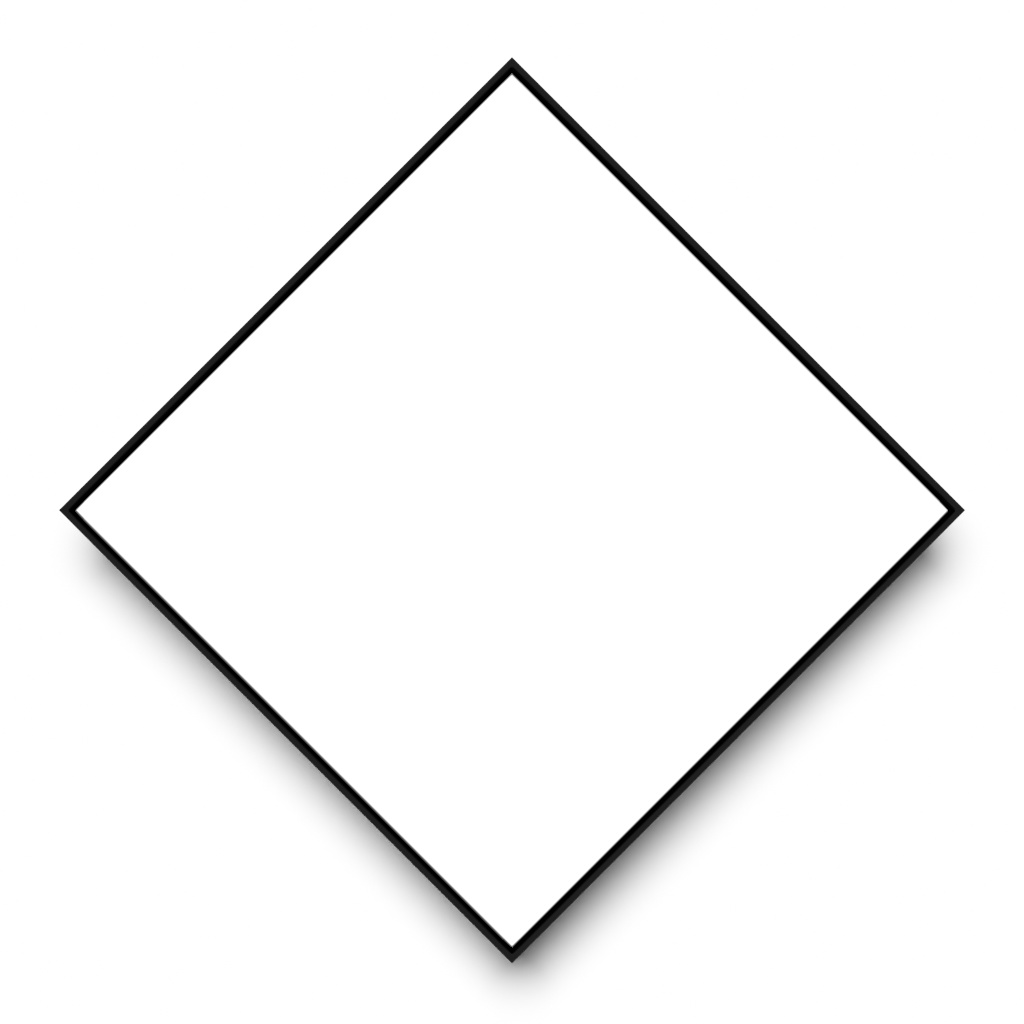}
        \includegraphics[width=\textwidth]{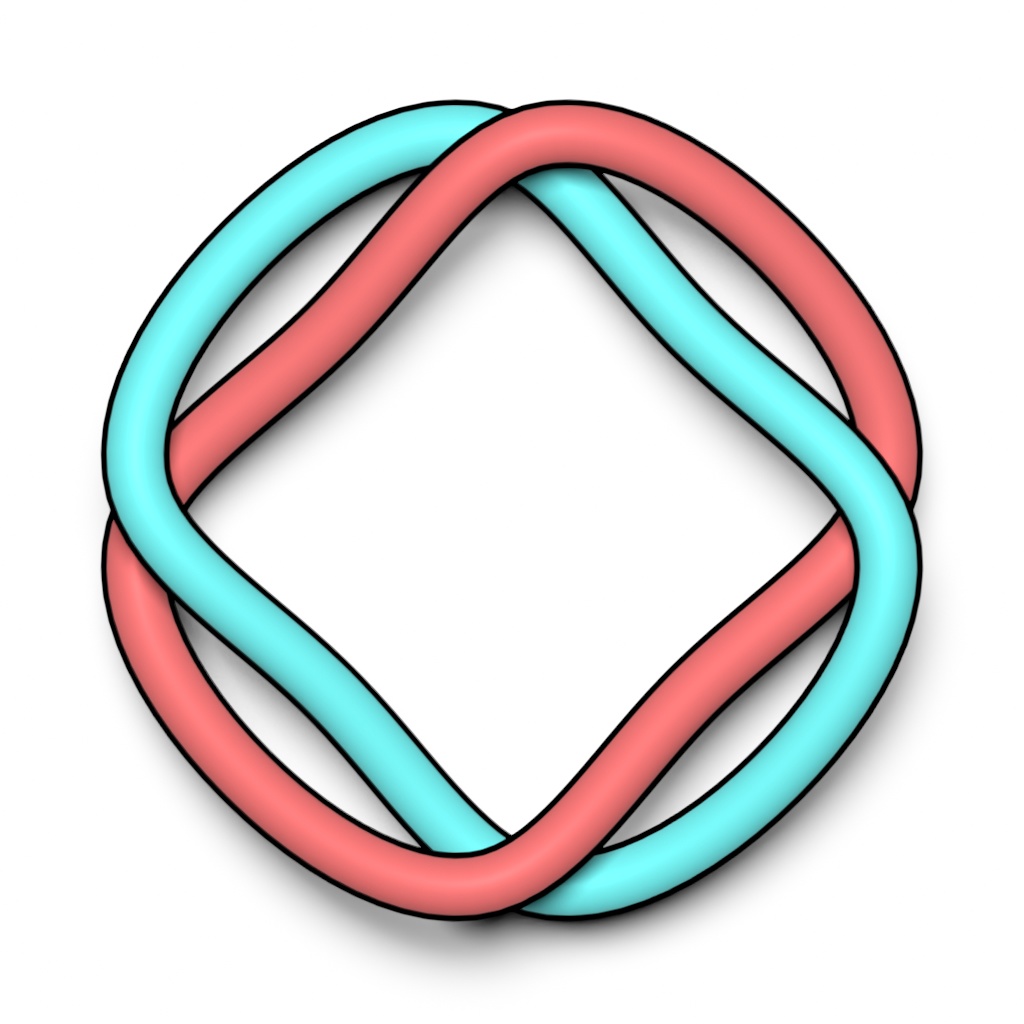}
        \caption{Solomon link.}
        \label{prime12}
    \end{subfigure}
    \hfill
    \begin{subfigure}[t]{0.23\columnwidth}
        \includegraphics[width=\textwidth]{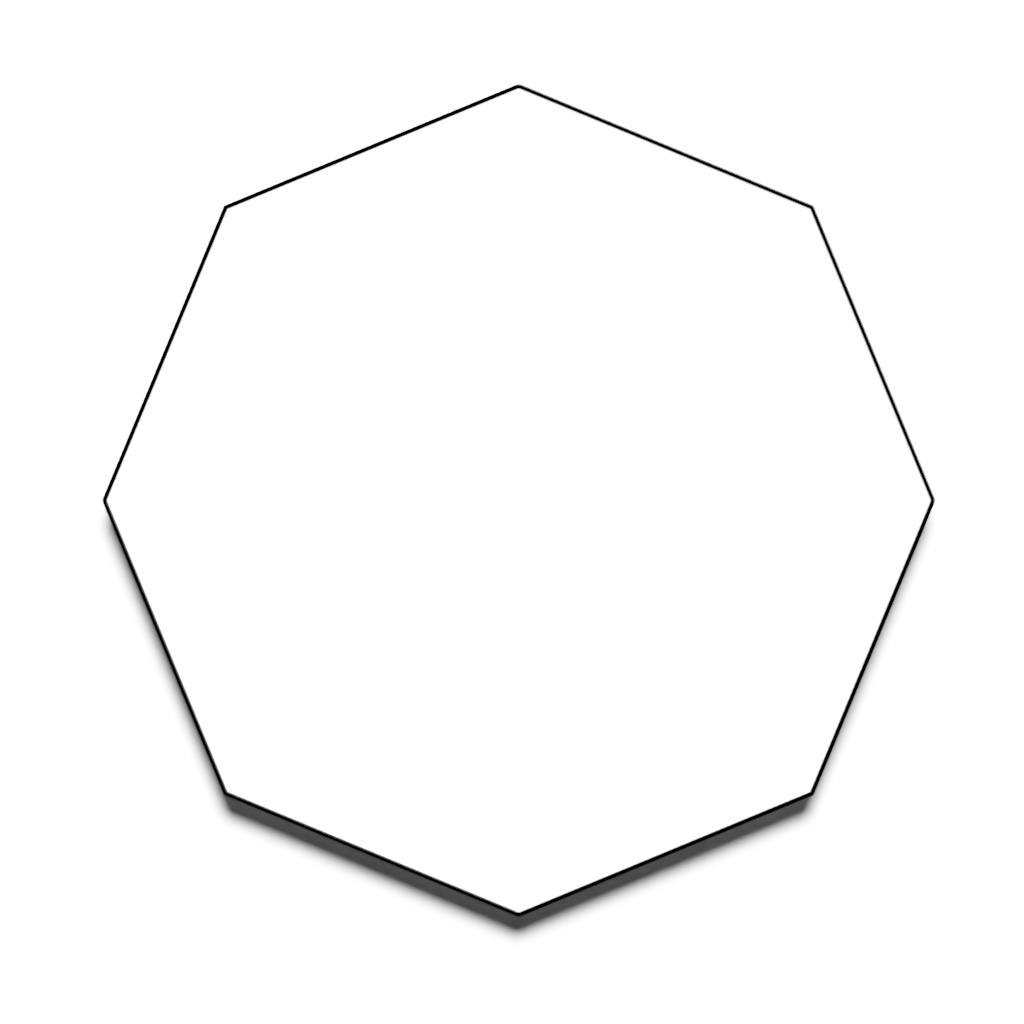}
        \includegraphics[width=\textwidth]{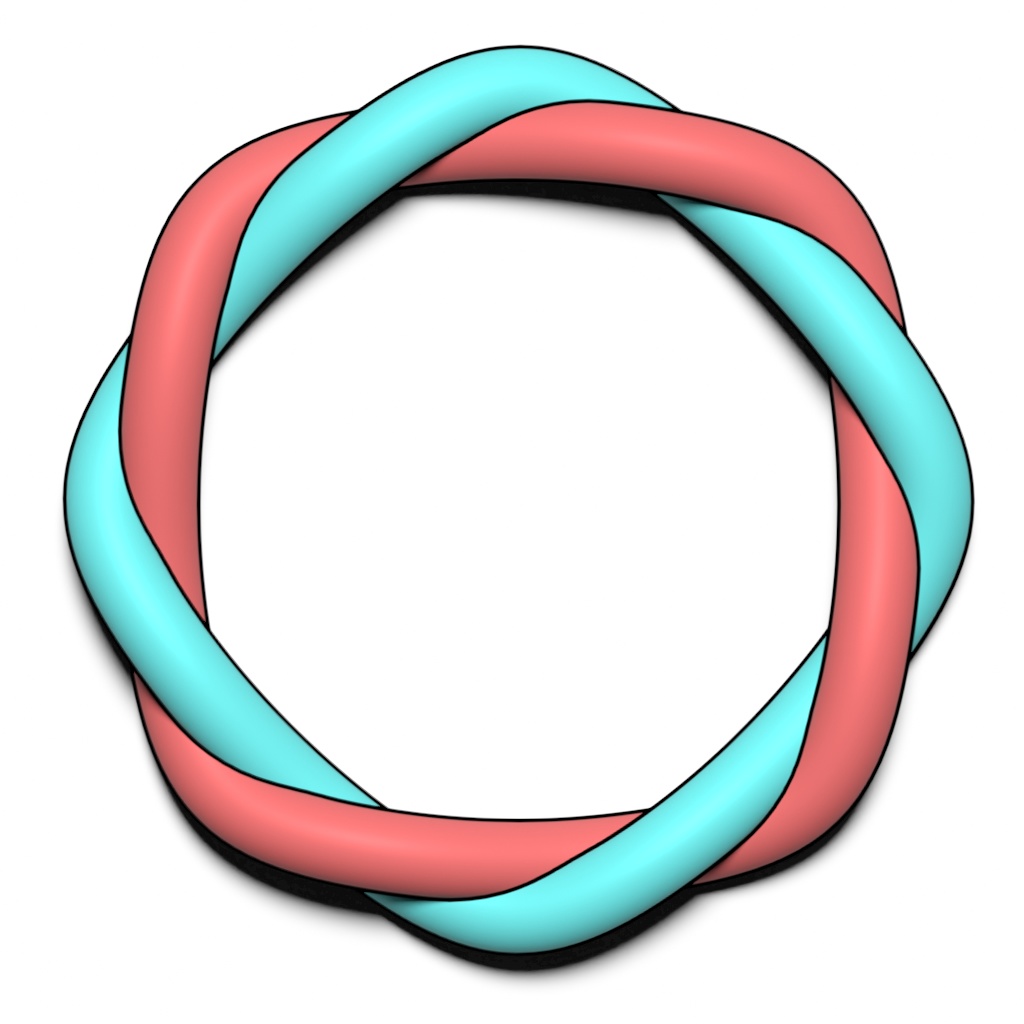}
        \caption{(2,8) torus link.}
        \label{prime13}
    \end{subfigure}
    \hfill
  \caption{Examples of well-known linked structures generated by labeled surface meshes. }
  \Description{Four examples arranged in two rows. The top row shows simple planar polygonal outlines and polyhedral projections: from left to right, a tetrahedral projection with triangular facets, a diamond-shaped quadrilateral with diagonals, a rotated square, and an octagonal outline. The bottom row shows corresponding linked tubular structures rendered in multiple colors, including red, cyan, green, and blue. Each example consists of several closed loops that are interlinked in distinct ways, with differing numbers of components and crossing patterns. The examples are labeled (a) through (d) and are shown at a similar scale and viewpoint to allow visual comparison.}
  \label{fig:prime_links}
\end{figure}

Figure~\ref{fig:prime_links} shows some well-known linked structures generated by labeled surface meshes. The Borromean rings, Whitehead link, and Solomon link appear as distinct outcomes of local combinatorial choices, demonstrating that classical linking behavior is a natural consequence of the framework rather than a special construction.

The two prime links shown in Figures~\ref{prime12} and~\ref{prime13} are produced by twisting the edges of two-sided regular polygons with an even number of sides: a two-sided square and a two-sided hexagon. In contrast, the Borromean link is constructed from a regular tetrahedron by twisting every edge once, as shown in Figure~\ref{prime10}. The Whitehead link shown in Figure~\ref{prime11} is constructed starting from a 2-manifold with boundary, with no polygon on the back side.

\subsection{Topological Representation of Physical Knot Constraints}

Physical knots, such as those occurring in ropes, cables, or polymers, differ fundamentally from mathematical knots: they are typically open-ended, material-dependent, and defined by constraints rather than closed curves. Such \textit{``physical knots''} cannot be represented directly using classical knot theory. Constructing \emph{physical knots} requires support for acyclic threads. Although the current framework does not explicitly support acyclic threads, they can be obtained by removing portions of cyclic threads, allowing some edge sides to be labeled as \emph{null}. A null edge-side remains part of the combinatorial structure of the mesh but is excluded from the geometric realization of cycles. Conceptually, null edge-sides act as \emph{cuts} that locally interrupt traversal, while preserving the surrounding connectivity needed for consistent twisting and labeling. This additional label provides practical control over whether a cycle is realized as closed or open.

This can be useful since many practical computer graphics applications, knots should not be realized as closed mathematical curves, but rather as open strands whose ends are accessible for fabrication, manipulation, or attachment. Starting from closed cycles induced by face boundaries, labeling an edge-side as null effectively removes that edge from the realized strand, thereby \emph{cutting the cycle} and producing two boundary endpoints. In this way, closed combinatorial cycles can be converted into open curves suitable for physical realization.

Importantly, null edge-sides do not alter the underlying combinatorial adjacency of the mesh. They participate in radial-edge traversal, twisting operations, and label propagation, but are ignored when embedding geometry. As a result, the local twisting behavior around adjacent edges remains well defined, even when a cycle is intentionally opened. This separation between combinatorial structure and geometric realization allows open strands to coexist naturally with closed knots and links within the same framework.

By selectively assigning null-edge-side labels, we can therefore transition seamlessly between closed knots, linked components, and open knot segments, without modifying the mesh topology or redefining twisting rules. This capability is essential for practical modeling scenarios in which knot ends must be exposed, and further demonstrates the flexibility of our labeled non-manifold mesh representation.
This mechanism modifies only the combinatorial realization of cycles and does not prescribe fabrication or material behavior.

This representation should be understood as a topological abstraction rather than a physical simulation or fabrication-ready model.

\subsection{Limitations of Design-Level Scope}

This work intentionally focuses on the conceptual and combinatorial design layer of linked and knotted structures. We do not model material behavior, physical constraints, or mechanical performance, nor do we present fabrication-oriented engineering analysis. Our goal is to establish a clean and general design framework in which topology, connectivity, and articulation are controlled through discrete labels on meshes. Such a representation is a prerequisite for any subsequent physical simulation or engineering optimization, and isolating it allows the design space itself to be explored without conflating geometric intent with material-specific assumptions.

This separation is essential: geometric embeddings alone are inherently unstable as descriptors of knot topology, since infinitesimal perturbations can change crossings. By contrast, our scaffold-based formulation guarantees that the knot and link structure is uniquely determined before any geometric realization.

A natural question concerns how the spiral geometry associated with twisted edges is computed. We intentionally do not prescribe a specific construction. Once a labeled mesh induces a set of connected cycles, generating a smooth spiral embedding along each strand is a standard geometric task and can be accomplished using a variety of existing techniques in computer graphics, differential geometry, or procedural modeling. By not committing to a particular geometric construction, we ensure that the proposed framework remains a pure topological design abstraction, compatible with multiple realization strategies and downstream objectives.

While the topological framework underlying these constructions was established in earlier work \cite{akleman2015block}, that formulation did not address how these degrees of freedom could be systematically used for design; the present work fills this gap by treating twist and reconnection as explicit design parameters.

Equally important, the framework intentionally adopts the minimal topological structure required for knot design: non-manifold surfaces provide sufficient expressive power without introducing unnecessary volumetric or orientation constraints.

Future work, therefore, may explore systematic mappings from topological designs to specific geometric or physical realizations, including optimization for fabrication, mechanics, or aesthetics; however, such mappings are orthogonal to the core design contribution presented here.
We also observe that the current design level scope can be particularly well-suited to certain classes of structures that already exist in nature such as \emph{Laves graphs}, which arise as edge graphs of space-filling polyhedra and occur ubiquitously in crystalline and intermetallic materials \cite{berry1953crystal,coxeter1955laves}. They capture the underlying connectivity of a wide range of crystalline, molecular, and architected systems, where periodicity and uniform local coordination play a central role \cite{hyde1997language,weaire1999physics}. 

\begin{figure}[tb!]
    \centering
    \begin{subfigure}[b]{0.32\columnwidth}
        \includegraphics[width=0.99\linewidth]{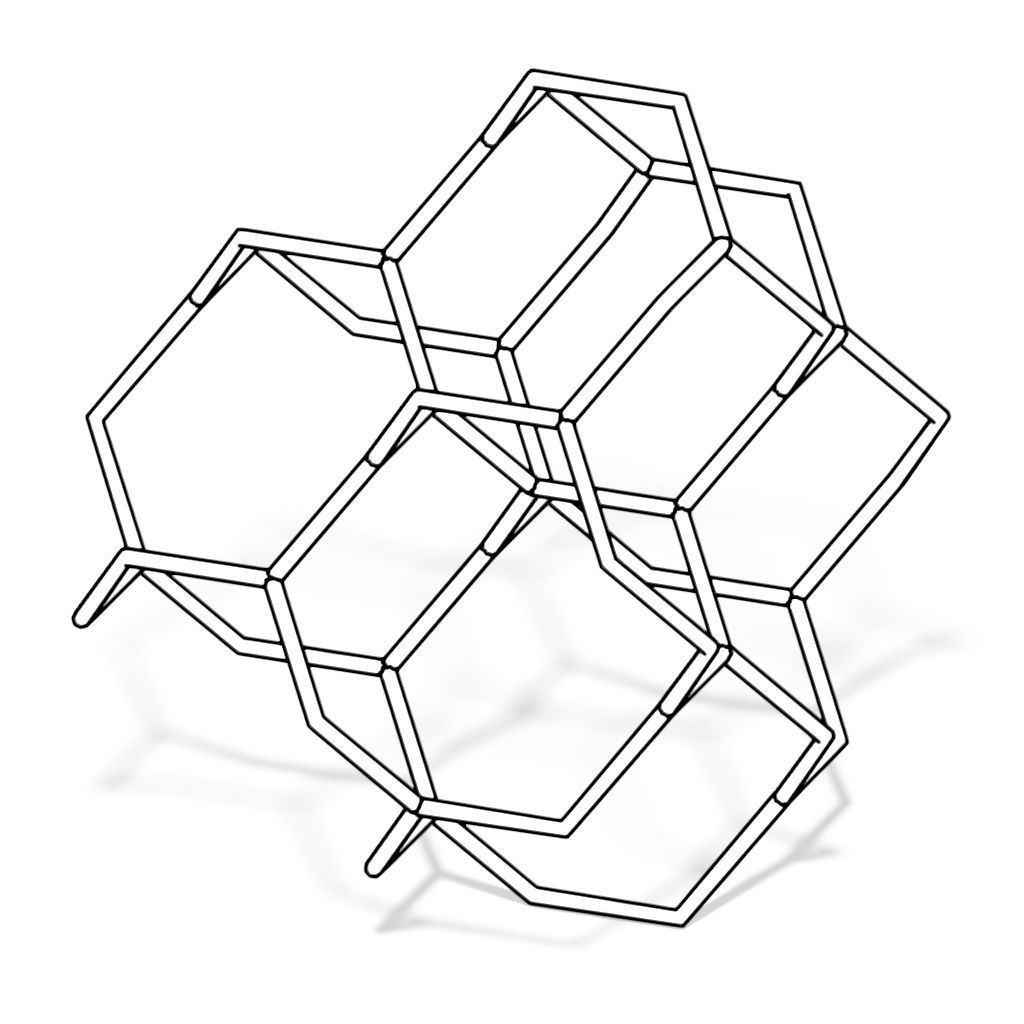}\\
        \includegraphics[width=0.99\linewidth]{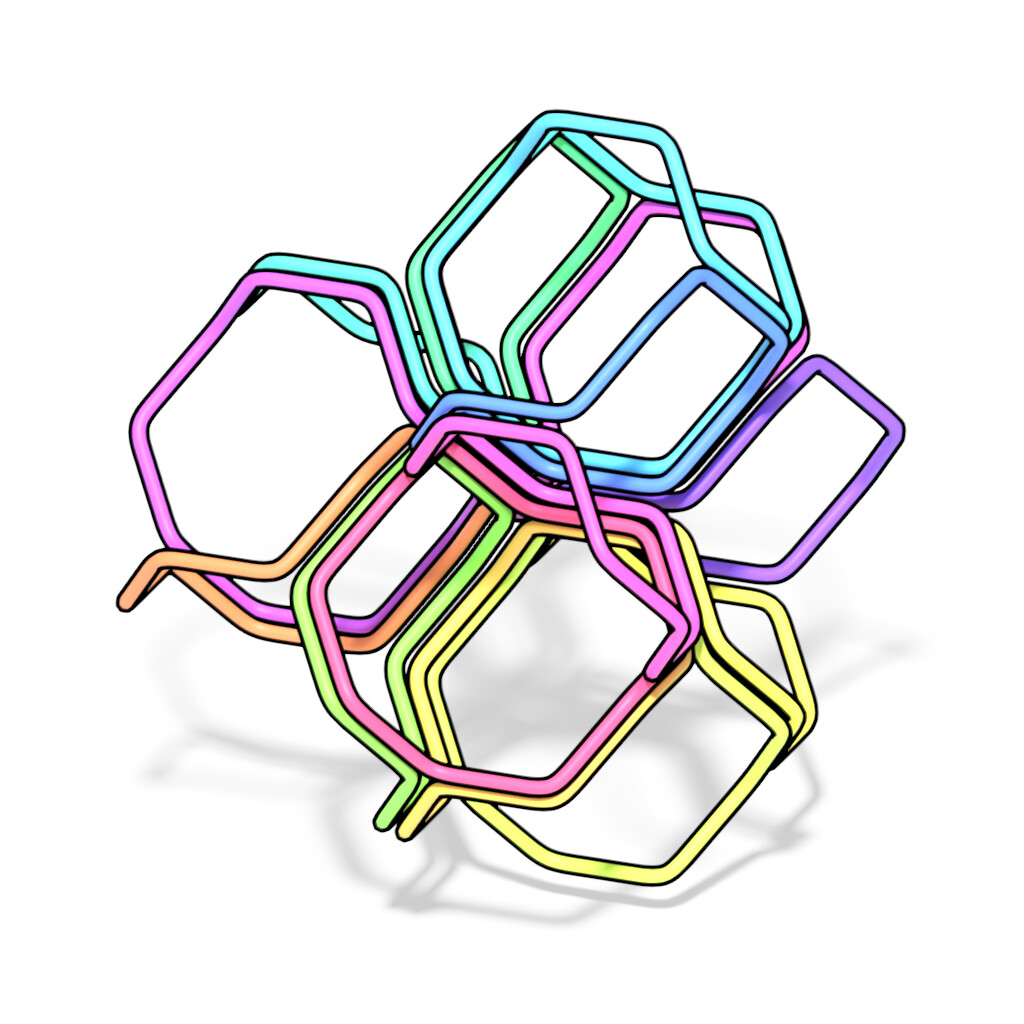}
    \end{subfigure}
    \begin{subfigure}[b]{0.64\columnwidth}
        \includegraphics[width=0.99\linewidth]{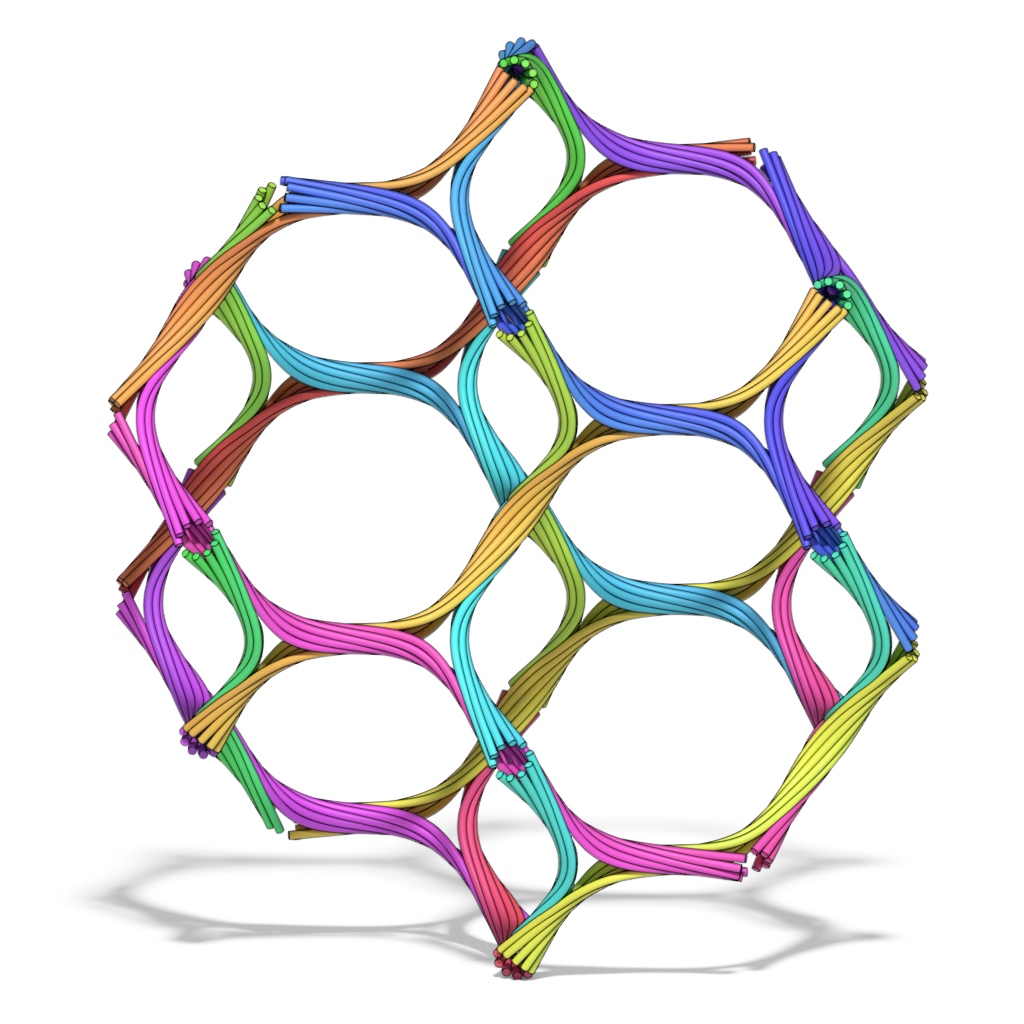}    
    \end{subfigure}
    \caption{The Laves graph and the local structure of its vertices and edges. Although the Laves graph does not admit a canonical embedding, it exhibits combinatorial features suggestive of face constructions. In this particular embedding, each edge is incident to 10 distinct length-10 loops, and each vertex is surrounded by 15 such loops. We extract these loops and use them to define face cycles. The resulting face cycles are shown in the bottom-left image, while one of the possible induced LK structures is illustrated in the right image.}
    \Description{A three-panel figure showing a Laves graph based LK structure.}
    \label{fig:laves_intro}
\end{figure}

It is important to note that graph-based scaffolds such as Laves graphs do not, by themselves, uniquely determine LK structures within our framework. A graph specifies connectivity, but the cycles required for edge twisting are defined only after the graph is embedded into a surface or, more generally, realized as a non-manifold surface mesh. A single graph can therefore admit many distinct surface embeddings, each inducing a different set of cycles and local adjacencies. When used as scaffolds for edge twisting, these different embeddings give rise to distinct LK structures, even though the underlying graph connectivity remains unchanged. This observation reveals an additional and substantial source of design freedom: the choice of surface realization for a given graph. Notably, such multiplicity is not artificial but reflects the way similar graph structures manifest in nature, where identical connectivity patterns can be realized through a wide range of geometrically and topologically distinct material organizations.

\section{Conclusion and Future Work}

In this paper, we introduced a design-oriented framework for constructing linked knot (LK) structures based on labeled non-manifold surface meshes and edge twisting. By shifting the focus from geometric embedding and classification toward topological scaffolding and controllable design parameters, the framework enables systematic exploration of large families of knotted and linked structures that are difficult to access using existing approaches. Rather than enumerating isolated examples, the method exposes a structured design space in which connectivity, periodicity, and local twist can be manipulated explicitly.

We do not claim completeness for the proposed framework. Certain classes of structures, most notably braids, are inherently defined by order-permuting relations rather than by accumulated twist and therefore fall outside a twist-based representation. Likewise, hierarchical constructions, such as ropes formed from pre-twisted sub-threads, require multi-level descriptions that are not supported by the current formulation. Rather than pursuing completeness, this work aims to provide a unified and expressive design framework for a broad and practically relevant class of LK structures. Within this scope, the framework supports the construction of many classical knots and links as well as a wide range of novel configurations. Viewed in retrospect, the proposed approach can be seen as a design-oriented realization of earlier topological modeling theory \cite{akleman2015block}, demonstrating how higher-dimensional scaffolds enable robust and expressive knot and link design beyond geometry-based approaches.

\subsection{Representation and Theoretical Extensions}

An important direction for future work is extending the framework to support braid-like structures. Braids are fundamentally characterized by order-permuting relations rather than accumulated twist, and therefore fall outside the scope of the current representation. Incorporating order permutation alongside twist-based operations would significantly broaden the range of representable structures and establish connections to braid theory.
The current framework operates at a single structural scale, where each thread is treated as a primitive entity. Many physical systems, however, exhibit hierarchical organization, such as ropes composed of already twisted sub-threads. Supporting multi-level or recursive constructions would enable modeling of such hierarchical LK structures and remains an open research direction.
While the framework intentionally allows multiple non-manifold surface realizations for a given scaffold, it may be useful to investigate restricted or canonical embedding classes. Such constraints could reduce ambiguity, support partial classification, or facilitate comparison between different constructions without sacrificing design flexibility.
Classical knot invariants are not directly suited to representations based on labeled non-manifold surfaces. Developing invariants tailored to this design-oriented framework could provide tools for reasoning about equivalence, similarity, or structural complexity within the generated design space.
The framework naturally relates to ideas from higher-dimensional topology, particularly knotted surfaces. Exploring these connections more formally may provide additional theoretical grounding and suggest extensions of the framework beyond three-dimensional LK structures.

\subsection{Design Space and Combinatorial Exploration}

Although the paper demonstrates the rapid growth of design possibilities, a systematic enumerative analysis remains future work. Quantifying how the number of distinct LK structures scales with lattice size, labeling complexity, or scaffold variation would provide deeper insight into the expressive capacity of the framework.
Beyond uniform or arbitrary non-uniform twist assignments, future work could explore structured labeling patterns, such as gradients, symmetry-preserving variations, or motif-based repetitions. Such structured variations may yield families of LK structures with controlled transitions and predictable properties.
While this work focuses on Wigner-Seitz–based honeycombs, the underlying approach is not limited to these scaffolds. Exploring alternative lattice-derived or graph-derived periodic scaffolds may further expand the accessible design space.
A single abstract graph can admit many realizations as non-manifold surface meshes, each inducing different cycles and LK structures. Formalizing and systematically exploring this graph-to-surface embedding space represents a substantial opportunity for expanding design freedom.

\subsection{Computational and Algorithmic Directions}

Developing interactive tools that allow designers to manipulate scaffolds, embeddings, and twist labels in real time would make the framework more accessible and practical for design exploration.
Algorithmic exploration of the design space, including combinatorial search or optimization-driven approaches, could also be used to discover LK structures satisfying specific geometric, topological, or functional criteria.
An inverse problem of interest is recovering scaffold configurations or twist assignments from a target LK structure. Addressing such problems would enable goal-driven design workflows within the framework.
As periodic domains grow in size and complexity, understanding the computational limits and scalability of the framework will be important for large-scale applications.

\subsection{Physical Realization and Applications}

Integrating material properties, thickness constraints, and fabrication considerations into the framework would facilitate translation from abstract designs to physical realizations. 
Importantly, the twist-based representation may be extended to model deployable or time-varying structures, including applications in 4D printing and reconfigurable materials. 
Studying the mechanical behavior of LK structures generated by the framework, including stiffness, flexibility, and failure modes, represents an important direction for applied research.
The articulated and hinge-like structures presented in this paper should not be interpreted as mechanical hinges in the traditional engineering sense. Instead, they represent \emph{topologically hinged} configurations whose articulation emerges from the connectivity and twist-label structure of the underlying mesh. Importantly, properties such as apparent stiffness, clearance, and rotational behavior are not determined by topology alone. By varying purely geometric parameters—including thread thickness, twist radius, and the relative offset between intertwined strands—a wide range of visual and kinematic behaviors can be achieved, spanning loosely coupled motions to tightly constrained hinge-like interactions.
Finally, the framework offers potential applications across architected materials, volumetric textiles, and artistic or architectural design, where periodic linked structures can be exploited for both functional and expressive purposes.

\section*{Acknowledgments}
The authors acknowledge the use of large language models, including OpenAI's ChatGPT and Google's Gemini, to assist with manuscript preparation. These tools were used for language polishing, clarity improvement, and formatting suggestions. All technical content, interpretations, and conclusions were developed and verified by the authors, who take full responsibility for the accuracy and originality of the work.    

\bibliographystyle{ACM-Reference-Format}
\bibliography{references,references1,references2}

@inproceedings{liu2025doublelayeredegridshells,
  title = {Double-Layered Elastic Gridshells with Locally Deployable Rotational-Surface Components},
  author={Liu, Yuanpeng and Suzuki, Seiichi and Isvoranu, Florin and Pauly, Mark},
  booktitle={Advances in Architectural Geometry (AAG 2025)},
pages={35--46},
  year={2025},
publisher={AAG},
address={Vienna, Austria},
  note={Presented at Advances in Architectural Geometry (AAG 2025), Cambridge, MA, USA. Paper available on EPFL Infoscience repository.}
}

@inproceedings{yang2025topologicalwovennodes,
  title = {Topological generation of woven nodes for double-layered elastic gridshells built from deployable cylindrical components},
  author={Yang, Hong-Bin and Suzuki, Seiichi and Pauly, Mark},
  booktitle={Advances in Architectural Geometry (AAG 2025)},
  pages={47--58},
  year={2025},
publisher={AAG},
address={Vienna, Austria},
  note={Presented at Advances in Architectural Geometry (AAG 2025), Cambridge, MA, USA. Paper available on EPFL Infoscience repository.}
}

@article{ren2021curvedribbons,
author = {Ren, Yingying and Panetta, Julian and Chen, Tian and Isvoranu, Florin and Poincloux, Samuel and Brandt, Christopher and Martin, Alison and Pauly, Mark},
title = {3D weaving with curved ribbons},
year = {2021},
issue_date = {August 2021},
publisher = {Association for Computing Machinery},
address = {New York, NY, USA},
volume = {40},
number = {4},
journal = {ACM Trans. Graph.},
month = jul,
articleno = {127},
numpages = {15},
}

@article{glaser2021dnananostructures,
AUTHOR = {Glaser, Martin and Deb, Sourav and Seier, Florian and Agrawal, Amay and Liedl, Tim and Douglas, Shawn and Gupta, Manish K. and Smith, David M.},
TITLE = {The Art of Designing DNA Nanostructures with CAD Software},
JOURNAL = {Molecules},
VOLUME = {26},
YEAR = {2021},
NUMBER = {8},
ARTICLE-NUMBER = {2287},
URL = {https://www.mdpi.com/1420-3049/26/8/2287},
PubMedID = {33920889},
pages = {2287}
}

@phdthesis{weiler1986radialedge,
author={Weiler,Kevin J.},
year={1986},
title = {Topological structures for geometric modeling (boundary representation, manifold, radial edge structure)},
school = {Rensselaer Polytechnic Institute},
journal = {ProQuest Dissertations and Theses},
pages={340},
note={Copyright - Database copyright ProQuest LLC; ProQuest does not claim copyright in the individual underlying works; Last updated - 2023-02-19},
isbn={979-8-206-01827-1},
language={English},
url={http://proxy.library.tamu.edu/login?url=https://www.proquest.com/dissertations-theses/topological-structures-geometric-modeling/docview/303506577/se-2},
}

@article{yildiz2025linked,
title = {Triply periodic mesh-based linked-structures: Design and construction of a family of periodic links based on Bravais lattices and corresponding Wigner-Seitz cells},
journal = {Computers \& Graphics},
volume = {132},
pages = {104369},
year = {2025},
url = {https://www.sciencedirect.com/science/article/pii/S0097849325002109},
author = {Tolga Yildiz and Ergun Akleman},
}

@article{yildiz2025woven,
    author = {Yildiz, Tolga T and Niemeyer, Alice C and Krishnamurthy, Vinayak R and Akleman, Ergun},
    title = {A constructive framework for discovery, design, and classification of volumetric Bravais weaves},
    journal = {PNAS Nexus},
    volume = {4},
    number = {8},
    pages = {pgaf219},
    year = {2025},
    month = {07},   
    url = {https://doi.org/10.1093/pnasnexus/pgaf219},
    eprint = {https://academic.oup.com/pnasnexus/article-pdf/4/8/pgaf219/63794488/pgaf219.pdf},
}

@article{hyde2022tangledpolyhedra,
author = {Stephen T. Hyde  and Myfanwy E. Evans },
title = {Symmetric tangled Platonic polyhedra},
journal = {Proceedings of the National Academy of Sciences},
volume = {119},
number = {1},
pages = {e2110345118},
year = {2022},
URL = {https://www.pnas.org/doi/abs/10.1073/pnas.2110345118},
eprint = {https://www.pnas.org/doi/pdf/10.1073/pnas.2110345118},
}

@article{evans2013periodiceentanglement1,
author = "Evans, Myfanwy E. and Robins, Vanessa and Hyde, Stephen T.",
title = "Periodic entanglement i: networks from hyperbolic reticulations",
journal = "Acta crystallographica section a",
year = "2013",
volume = "69",
number = "3",
pages = "241--261",
month = "May",
url = {https://doi.org/10.1107/S0108767313001670},

}

@article{evans2013periodiceentanglement2,
author = "Evans, Myfanwy E. and Robins, Vanessa and Hyde, Stephen T.",
title = "Periodic entanglement ii: weavings from hyperbolic line patterns",
journal = "Acta crystallographica section a",
year = "2013",
volume = "69",
number = "3",
pages = "262--275",
month = "May",
url = {https://doi.org/10.1107/S0108767313001682},

}

@article{jessop2025butterfly,
author = {Anna-Lee Jessop  and Peta L. Clode  and Martin Saunders  and Myfanwy E. Evans  and Stephen T. Hyde  and James N. McPherson  and Kasper S. Pedersen  and Jacob J. K. Kirkensgaard  and Nipam H. Patel  and Kyle A. DeMarr  and W. Owen McMillan  and Bodo D. Wilts  and Gerd E. Schr{\"o}der-Turk },
title = {Hierarchical woven fibrillar structures in developing single gyroids in butterflies},
journal = {Proceedings of the National Academy of Sciences},
volume = {122},
number = {40},
pages = {e2507297122},
year = {2025},
URL = {https://www.pnas.org/doi/abs/10.1073/pnas.2507297122},
eprint = {https://www.pnas.org/doi/pdf/10.1073/pnas.2507297122},
}

@inproceedings{sequin2021polyhedral,
  author      = {S\'{e}quin, Carlo H.},
  title       = {Polyhedral-Edge Knots},
  pages       = {63--70},
  booktitle   = {Proceedings of Bridges 2021: Mathematics, Art, Music, Architecture, Culture},
  year        = {2021},
  editor      = {Swart, David and Farris, Frank and Torrence, Eve},
  isbn        = {978-1-938664-39-7},
  issn        = {1099-6702},
  publisher   = {Tessellations Publishing},
  address     = {Phoenix, Arizona},

  url         = {http://archive.bridgesmathart.org/2021/bridges2021-63.html}
}

@article{zhang1994dnatroctahedron,
author = {Zhang, Yuwen and Seeman, Nadrian C.},
title = {Construction of a DNA-Truncated Octahedron},
journal = {Journal of the American Chemical Society},
volume = {116},
number = {5},
pages = {1661-1669},
year = {1994},
doi = {10.1021/ja00084a006},
URL = { 
        https://doi.org/10.1021/ja00084a006
},
eprint = { 
        https://doi.org/10.1021/ja00084a006
}
}

@inproceedings{verhoeff2011chainlinkfence,
  author      = {Verhoeff, Tom and Verhoeff, Koos},
  title       = {From Chain-link Fence to Space-Spanning Mathematical Structures},
  pages       = {73--80},
  booktitle   = {Proceedings of Bridges 2011: Mathematics, Music, Art, Architecture, Culture},
  year        = {2011},
  editor      = {Sarhangi, Reza and S\'{e}quin, Carlo H.},
  isbn        = {978-0-9846042-6-5},
  issn        = {1099-6702},
  publisher   = {Tessellations Publishing},
  address     = {Phoenix, Arizona},

  url         = {http://archive.bridgesmathart.org/2011/bridges2011-73.html}
}

@inproceedings{holden2016braids,
  author      = {Holden, Joshua and Holden, Lana},
  title       = {Modeling Braids, Cables, and Weaves with Stranded Cellular Automata},
  pages       = {127--134},
  booktitle   = {Proceedings of Bridges 2016: Mathematics, Music, Art, Architecture, Education, Culture},
  year        = {2016},
  editor      = {Torrence, Eve and Torrence, Bruce and S\'{e}quin, Carlo and McKenna, Douglas and Fenyvesi, Krist\'{o}f and Sarhangi, Reza},
  isbn        = {978-1-938664-19-9},
  issn        = {1099-6702},
  publisher   = {Tessellations Publishing},
  address     = {Phoenix, Arizona},

  url         = {http://archive.bridgesmathart.org/2016/bridges2016-127.html}
}

@inproceedings{lipschutz2022linkedknots,
  author      = {Lipsch\"{u}tz, Henriette and Skrodzki, Martin and Reitebuch, Ulrich and Polthier, Konrad},
  title       = {Linked Knots from the \textit{gyro} Operation on the Dodecahedron},
  pages       = {175--182},
  booktitle   = {Proceedings of Bridges 2022: Mathematics, Art, Music, Architecture, Culture},
  year        = {2022},
  editor      = {Reimann, David and Norton, Douglas and Torrence, Eve},
  isbn        = {978-1-938664-42-7},
  issn        = {1099-6702},
  publisher   = {Tessellations Publishing},
  address     = {Phoenix, Arizona},

  url         = {http://archive.bridgesmathart.org/2022/bridges2022-175.html}
}

@inproceedings{gailiunas2022tplinks,
  author      = {Gailiunas, Paul},
  title       = {Triply Periodic Links},
  pages       = {191--198},
  booktitle   = {Proceedings of Bridges 2022: Mathematics, Art, Music, Architecture, Culture},
  year        = {2022},
  editor      = {Reimann, David and Norton, Douglas and Torrence, Eve},
  isbn        = {978-1-938664-42-7},
  issn        = {1099-6702},
  publisher   = {Tessellations Publishing},
  address     = {Phoenix, Arizona},

  url         = {http://archive.bridgesmathart.org/2022/bridges2022-191.html}
}

@inproceedings{roelofs2007rings,
  author      = {Roelofs, Rinus},
  title       = {Entwined Circular Rings},
  pages       = {81--90},
  booktitle   = {Bridges Donostia: Mathematics, Music, Art, Architecture, Culture},
  year        = {2007},
  editor      = {Sarhangi, Reza and Barrallo, Javier},
  isbn        = {0-9665201-8-1},
  issn        = {1099-6702},
  publisher   = {Tarquin Publications},
  address     = {London},

  url         = {http://archive.bridgesmathart.org/2007/bridges2007-81.html}
}

@article{tait1877knots,
  title={On Knots I},
  author={Tait, Peter G.},
  journal={Transactions of the Royal Society of Edinburgh},
  volume={28},
  number={3},
  pages={145-190},
  year={1877},
  doi={10.1017/S0080456800090633},
  url={https://api.semanticscholar.org/CorpusID:171186257},
  publisher={Royan Society, Edinburg}
}

@article{tait1877links,
  title={On Links},
  author={Tait, Peter G.},
  journal={Transactions of the Royal Society of Edinburgh},
  volume={28},
  number={9},
  pages={ 321–332},
  year={1877},
  doi={10.1017/S0080456800090633},
  publisher={Royan Society, Edinburg}
}

@article{akleman2009cyclic,
  title={Cyclic plain-weaving on polygonal mesh surfaces with graph rotation systems},
  author={Akleman, Ergun and Chen, Jianer and Xing, Qing and Gross, Jonathan L},
  journal={ACM Transactions on Graphics (TOG)},
  volume={28},
  number={3},
  pages={1--8},
  year={2009},
  publisher={ACM New York, NY, USA}
}

@inproceedings{xing2010single,
  author    = {Xing, Qing and Akleman, Ergun and Chen, Jianer and Gross, Jonathan L.},
  title     = {Single-cycle Plain-Woven Objects},
  booktitle = {2010 Shape Modeling International Conference},
  pages     = {90--99},
  year      = {2010},
  publisher = {IEEE},
  address   = {Piscataway, New Jersey, USA}
}

@article{akleman2011cyclic,
  title={Cyclic twill-woven objects},
  author={Akleman, Ergun and Chen, Jianer and Chen, YenLin and Xing, Qing and Gross, Jonathan L},
  journal={Computers \& Graphics},
  volume={35},
  number={3},
  pages={623--631},
  year={2011},
  publisher={Elsevier}
}

@article{akleman2015extended,
  title={Extended graph rotation systems as a model for cyclic weaving on orientable surfaces},
  author={Akleman, Ergun and Chen, Jianer and Gross, Jonathan L},
  journal={Discrete Applied Mathematics},
  volume={193},
  pages={61--79},
  year={2015},
  publisher={Elsevier}
}

@article{akleman2015block,
  title={Block meshes: Topologically robust shape modeling with graphs embedded on 3-manifolds},
  author={Akleman, Ergun and Chen, Jianer and Gross, Jonathan L},
  journal={Computers \& Graphics},
  volume={46},
  pages={306--326},
  year={2015},
  publisher={Elsevier}
}

@article{akleman2020topologically,
  title={A topologically complete theory of weaving},
  author={Akleman, Ergun and Chen, Jianer and Gross, Jonathan L and Hu, Shiyu},
  journal={SIAM Journal on Discrete Mathematics},
  volume={34},
  number={4},
  pages={2457--2480},
  year={2020},
  publisher={SIAM}
}

@article{yildiz2024volumetric,
  title={Volumetric nonwoven structures: An algebraic framework for systematic design of infinite polyhedral frames using nonwoven fabric patterns},
  author={Yildiz, Tolga and Akleman, Ergun},
  journal={Computers \& Graphics},
volume={4},
number={8},
  pages={103979},
  year={2024},
  publisher={Elsevier}
}

@article{moestopo2023knots,
  title={Knots are not for naught: Design, properties, and topology of hierarchical intertwined microarchitected materials},
  author={Moestopo, Widianto P and Shaker, Sammy and Deng, Weiting and Greer, Julia R},
  journal={Science Advances},
  volume={9},
  number={10},
  pages={eade6725},
  year={2023},
  publisher={American Association for the Advancement of Science}
}

@article{fielden2017molecular,
  title={Molecular knots},
  author={Fielden, Stephen DP and Leigh, David A and Woltering, Steffen L},
  journal={Angewandte Chemie International Edition},
  volume={56},
  number={37},
  pages={11166--11194},
  year={2017},
  publisher={Wiley Online Library}
}

@article{craik2001cystine,
  title={The cystine knot motif in toxins and implications for drug design},
  author={Craik, David J and Daly, Norelle L and Waine, Clement},
  journal={Toxicon},
  volume={39},
  number={1},
  pages={43--60},
  year={2001},
  publisher={Elsevier}
}

@article{neuwirth1979theory,
  title={The theory of knots},
  author={Neuwirth, Lee},
  journal={Scientific American},
  volume={240},
  number={6},
  pages={110--125},
  year={1979},
  publisher={JSTOR}
}

@article{forgan2011chemical,
  title={Chemical topology: complex molecular knots, links, and entanglements},
  author={Forgan, Ross S and Sauvage, Jean-Pierre and Stoddart, J Fraser},
  journal={Chemical Reviews},
  volume={111},
  number={9},
  pages={5434--5464},
  year={2011},
  publisher={ACS Publications}
}

@inproceedings{kozlov2018knots,
  author    = {Kozlov, Dmitri},
  title     = {Knots as a Principle of Form in Modern Art and Architecture},
  booktitle = {2nd International Conference on Art Studies: Science, Experience, Education (ICASSEE 2018)},
  pages     = {367--372},
  year      = {2018},
  publisher = {Atlantis Press},
  address   = {Paris, France}
}

@article{horner2016knot,
  title={Knot theory in modern chemistry},
  author={Horner, Kate E and Miller, Mark A and Steed, Jonathan W and Sutcliffe, Paul M},
  journal={Chemical Society Reviews},
  volume={45},
  number={23},
  pages={6432--6448},
  year={2016},
  publisher={Royal Society of Chemistry}
}

@article{kozlov2013structures,
  title={Structures of periodical knots and links as geometric models of complex surfaces for designing},
  author={Kozlov, Dmitri},
  journal={Nexus Network Journal},
  volume={15},
  number={2},
  pages={241--255},
  year={2013},
  publisher={Springer}
}

@inproceedings{borhani2016emergence,
  title     = {The emergence of a new material culture: forging unprecedented alliances between design and engineering},
  author    = {Borhani, Alireza and Kalantar, Negar and others},
  booktitle = {DS 83: Proceedings of the 18th International Conference on Engineering and Product Design Education (E\&PDE16), Design Education: Collaboration and Cross-Disciplinarity, Aalborg, Denmark, 8th--9th September 2016},
  pages     = {417--422},
  year      = {2016},
  publisher = {The Design Society},
  address   = {Glasgow, United Kingdom}
}

@inproceedings{borhani2016material,
  title     = {Material active geometry---constituting programmable materials for responsive building skins},
  author    = {Borhani, A. and Kalantar, N.},
  booktitle = {Complexity \& Simplicity: Proceedings of the 34th eCAADe Conference},
  volume    = {1},
  pages     = {639--648},
  year      = {2016},
  publisher = {eCAADe (Education and Research in Computer Aided Architectural Design in Europe) and Oulu School of Architecture, University of Oulu},
  address   = {Oulu, Finland}
}

@book{gross2001topological,
  author    = {Gross, Jonathan L. and Tucker, Thomas W.},
  title     = {Topological Graph Theory},
  publisher = {Courier Corporation},
  address   = {Mineola, New York, USA},
  year      = {2001}
}

@inproceedings{conway1970enumeration,
  author    = {Conway, John H.},
  title     = {An enumeration of knots and links, and some of their algebraic properties},
  booktitle = {Computational Problems in Abstract Algebra},
  editor    = {Leech, John},
  pages     = {329--358},
  year      = {1970},
  publisher = {Elsevier},
  address   = {New York, USA}
}

@book{rolfsen1976knots,
  author    = {Dale Rolfsen},
  title     = {Knots and Links},
  publisher = {Publish or Perish},
  address   = {Berkeley, CA},
  year      = {1976}
}

@book{adams1994knot,
  author    = {Colin C. Adams},
  title     = {The Knot Book},
  publisher = {W. H. Freeman and Company},
  address   = {New York},
  year      = {1994}
}

@phdthesis{edmonds1960combinatorial,
  author       = {Edmonds, John Robert, Jr.},
  title        = {A Combinatorial Representation for Oriented Polyhedral Surfaces},
  school       = {University of Maryland, College Park},
  year         = {1960},
  type         = {Ph.D. thesis},
  note         = {Department of Mathematics},
  url          = {https://api.drum.lib.umd.edu/server/api/core/bitstreams/68532bbd-8dee-41a9-97cc-d1d72e133b85/content}
}

@book{kauffman1987knots,
  title={On Knots},
  author={Kauffman, LH},
  publisher={Princeton University Press},
  address={Princeton Nj},
  year={1987}
}

@article{akleman1999,
  author = {E. Akleman and J. Chen},
  title = {Guaranteeing the 2-manifold property for meshes with
     doubly linked face list},
  journal = {International Journal of Shape Modeling},
  volume = 5,
  number = 2,
  pages = {149-177},
  year = 1999,
}

@mastersthesis{baumgart1972,
  author = {B. J. Baumgart},
  title = {Winged-edge polyhedron representation},
  school = {Technical Report CS-320, Stanford University},
  year = 1972,
}

@article{guibas1985,
  title={Primitives for the manipulation of general subdivisions and the computation of Voronoi},
  author={Guibas, Leonidas and Stolfi, Jorge},
  journal={ACM transactions on graphics (TOG)},
  volume={4},
  number={2},
  pages={74--123},
  year={1985},
  publisher={ACM}
}

@book{hoffmann1989,
  author    = {Hoffmann, Claude M.},
  title     = {Geometric \& Solid Modeling: An Introduction},
  publisher = {Morgan Kaufmann Publishers},
  address   = {San Mateo, California, USA},
  year      = {1989}
}

@book{mantyla1988,
  author    = {M{\"a}ntyl{\"a}, Martti},
  title     = {An Introduction to Solid Modeling},
  publisher = {Computer Science Press},
  address   = {Rockville, Maryland, USA},
  year      = {1988}
}

@article{yoshimoto1971cube,
  author  = {Yoshimoto, Akira},
  title   = {An Amusing Cube},
  journal = {Mathematics Magazine},
  volume  = {44},
  number  = {2},
  pages   = {87--89},
  year    = {1971}
}

@book{frederickson2002hinged,
  author    = {Frederickson, Greg N.},
  title     = {Hinged Dissections: Swinging and Twisting},
  publisher = {Cambridge University Press},
  address   = {Cambridge, United Kingdom},
  year      = {2002}
}

@article{stone1942flexagons,
  author  = {Stone, Bryant Tuckerman},
  title   = {Flexagons},
  journal = {Scientific American},
  volume  = {167},
  number  = {6},
  pages   = {246--250},
  year    = {1942}
}

@book{sherman1994flexagons,
  author    = {Sherman, Joseph and Stewart, Paul},
  title     = {The Flexagon Book},
  publisher = {Cambridge University Press},
  address   = {Cambridge, United Kingdom},
  year      = {1994}
}

@book{schattschneider1987escher,
  author    = {Schattschneider, Doris and Walker, Wallace},
  title     = {M. C. Escher Kaleidocycles},
  publisher = {Ballantine Books},
  address   = {New York, NY, USA},
  year      = {1987},
}

@misc{wikipediaMedialGraph,
  author       = "{Wikipedia contributors}",
  title        = "{Medial graph --- Wikipedia{,} The Free Encyclopedia}",
  year         = "2025",
  howpublished = "\url{https://en.wikipedia.org/wiki/Medial_graph}",
  note         = "[Online; accessed 17-January-2026]"
}

@article{Grunbaum80,
 author = {B. Grunbaum and G. Shephard},
 title = {Satins and twills: an introduction to the geometry of fabrics},
 journal = {Mathematics Magazine},
 volume = {53},
 number = {},
 pages = {139-161},
 year = {1980},
}

@article{Grunbaum85,
 author = {B. Grunbaum and G. Shephard},
 title = {A catalogue of isonemal fabrics},
 journal = {Annals of the New York Academy of Sciences},
 volume = {440},
 number = {},
 pages = {279-298},
 year = {1985},
}

@article{Grunbaum86,
 author = {B. Grunbaum and G. Shephard},
 title = {An extension to the catalogue of isonemal fabrics},
 journal = {Discrete Mathematics},
 volume = {60},
 number = {},
 pages = {155-192},
 year = {1986},
 url ={http://portal.acm.org/citation.cfm?id=10139.10148},
}

@article{grunbaum1988isonemal,
  title={Isonemal fabrics},
  author={Gr{\"u}nbaum, Branko and Shephard, Geoffrey C},
  journal={The American Mathematical Monthly},
  volume={95},
  number={1},
  pages={5--30},
  year={1988},
  publisher={Taylor \& Francis}
}

@software{Roosendaal1995blender,
  title        = {Blender},
  author       = {Roosendaal, Ton and Blender Foundation},
  year         = {1995},
  url          = {https://www.blender.org},
  note         = {Open-source 3D creation suite},
}

@book{freedman1990topology,
  author    = {Freedman, Michael H. and Quinn, Frank},
  title     = {Topology of 4-Manifolds},
  publisher = {Princeton University Press},
  address   = {Princeton, New Jersey, USA},
  year      = {1990}
}

@book{kamada2002surfaces,
  author    = {Kamada, Seiichi},
  title     = {Surfaces in 4-Space},
  publisher = {Springer},
  address   = {Berlin, Germany},
  year      = {2002}
}

@book{guillemin1974differential,
  author    = {Guillemin, Victor and Pollack, Alan},
  title     = {Differential Topology},
  publisher = {Prentice-Hall},
  address   = {Englewood Cliffs, New Jersey, USA},
  year      = {1974}
}

@book{milnor1965topology,
  author    = {Milnor, John},
  title     = {Topology from the Differentiable Viewpoint},
  publisher = {University of Virginia Press},
  address   = {Charlottesville, Virginia, USA},
  year      = {1965}
}

@book{hyde1997language,
  author    = {Hyde, Stephen T. and O'Keeffe, Michael and Proserpio, Davide M. and Seshadri, Ram},
  title     = {The Language of Shape},
  publisher = {Elsevier},
  address   = {Amsterdam, Netherlands},
  year      = {1997},
}

@book{weaire1999physics,
  author    = {Weaire, Denis and Phelan, Robert},
  title     = {The Physics of Foams},
  publisher = {Oxford University Press},
  address   = {Oxford, UK},
  year      = {1999},
}

@article{coxeter1955laves,
  title={On Laves' graph of girth ten},
  author={Coxeter, HSM},
  journal={Canadian Journal of Mathematics},
  volume={7},
  pages={18--23},
  year={1955},
  publisher={Cambridge University Press}
}

@article{berry1953crystal,
  title={The crystal chemistry of the Laves phases},
  author={Berry, RL and Raynor, GV},
  journal={Acta Crystallographica},
  volume={6},
  number={2},
  pages={178--186},
  year={1953},
  publisher={International Union of Crystallography}
}

@article{singal2024programming,
  title={Programming mechanics in knitted materials, stitch by stitch},
  author={Singal, Krishma and Dimitriyev, Michael S and Gonzalez, Sarah E and Cachine, A Patrick and Quinn, Sam and Matsumoto, Elisabetta A},
  journal={Nature Communications},
  volume={15},
  number={1},
  pages={2622},
  year={2024},
  publisher={Nature Publishing Group UK London}
}

@article{klotz2024chirality,
  title={Chirality effects in molecular chainmail},
  author={Klotz, Alexander R and Anderson, Caleb J and Dimitriyev, Michael S},
  journal={Soft Matter},
  volume={20},
  number={35},
  pages={7044--7058},
  year={2024},
  publisher={Royal Society of Chemistry}
}

@article{niu2025geometric,
  title={Geometric modeling of knitted fabrics},
  author={Niu, Lauren and Dion, Genevi{\`e}ve and Kamien, Randall D},
  journal={Proceedings of the National Academy of Sciences},
  volume={122},
  number={7},
  pages={e2416536122},
  year={2025},
  publisher={National Academy of Sciences}
}

@article{knittel2020modelling,
  title={Modelling textile structures using bicontinuous surfaces},
  author={Knittel, Chelsea E and Tanis, Michael and Stoltzfus, Amy L and Castle, Toen and Kamien, Randall D and Dion, Genevieve},
  journal={Journal of Mathematics and the Arts},
  volume={14},
  number={4},
  pages={331--344},
  year={2020},
  publisher={Taylor \& Francis}
}
\end{document}